\documentclass[onecolumn,amsmath,amssymb,nofootinbib,12pt]{article}
\usepackage{jheppub}
\usepackage{ifpdf}

\usepackage[bb=boondox]{mathalfa}

\usepackage{graphicx,subcaption}
\graphicspath{ {figures/} }
\usepackage{amsfonts}
\usepackage{amsmath}
\usepackage{amssymb}
\usepackage{epsfig}
\usepackage{pdfpages}
\usepackage{bbm}
\usepackage{graphicx,epstopdf}
\usepackage[numbers]{natbib} 
\usepackage[makeroom]{cancel}
\usepackage{hyperref}
\hypersetup{pdftex,colorlinks=true,allcolors=blue}
\usepackage{array}
\usepackage[export]{adjustbox}

\usepackage[normalem]{ulem}

\newcommand{\RN}[1]{%
  \textup{\uppercase\expandafter{\romannumeral#1}}%
}

\newcommand{\del}{\partial}

\newcommand{\phys}{\rm phys}

\definecolor{limegreen}{rgb}{0.2, 0.8, 0.2}
\definecolor{deepcarrotorange}{rgb}{0.91, 0.41, 0.17}
\definecolor{darkviolet}{rgb}{0.58, 0.0, 0.83}	
\definecolor{cyan}{rgb}{0.0, 0.72, 0.92}

\newcommand{\eg}{{\it e.g.,}\ }
\newcommand{\ie}{{\it i.e.,}\ }
\newcommand{\reef}[1]{(\ref{#1})}

\newcommand{\mt}[1]{\textrm{\tiny #1}}

\newcommand{\Gn}{G_\mt{N}}

\newcommand{\mA}{\mathcal{A}}
\newcommand{\mB}{\mathcal{B}}

\newcommand{\beq}{\begin{equation}}
\newcommand{\eeq}{\end{equation}}
\newcommand{\beqa}{\begin{eqnarray}}
\newcommand{\eeqa}{\end{eqnarray}}

\newcommand{\bea}{\begin{eqnarray}}
\newcommand{\eea}{\end{eqnarray}}

\newcommand{\vk}{{\vec k}}

\newcommand{\cev}[1]{\reflectbox{\ensuremath{\vec{ \reflectbox{\ensuremath{#1}}}}}}
\newcommand{\tr}{{\rm tr}}

\newcommand{\bra}[1]{\left< #1 \right|}
\newcommand{\ket}[1]{\left| #1 \right>}

\newcommand{\braket}[2]{\left<#1|#2\right>}

\renewcommand{\(}{\left(}
\renewcommand{\)}{\right)}
\renewcommand{\[}{\left[}
\renewcommand{\]}{\right]}

\def\tr{{\text{Tr}}}

\newcommand{\psT}{\psi_\mt{T}}
\newcommand{\psR}{\psi_\mt{R}}
\newcommand{\s}{\sigma}

\newcommand{\mC}{\mathcal{C}}
\newcommand{\mO}{\mathcal{O}}

\newcommand{\mL}{\mathcal{L}}
\newcommand{\vq}{{\vec q}}
\newcommand{\vqp}{{\vec q}^{\,\, \prime}}

\newcommand{\RNum}[1]{\uppercase\expandafter{\romannumeral #1\relax}}

\usepackage[thinlines]{easytable}
\usepackage{textcomp}

\title{Complexity of Mixed States in QFT and Holography}

\author[a]{Elena Caceres,}
\author[b]{Shira Chapman,}
\author[a]{Josiah D. Couch,}
\author[c,d]{Juan P. Hernandez,}
\author[c]{Robert C. Myers}
\author[c,d]{and Shan-Ming Ruan}

\affiliation[a]{Theory Group, Department of Physics, University of Texas, Austin, TX 78712, USA}
\affiliation[b]{Institute for Theoretical Physics Amsterdam and Delta Institute for Theoretical Physics,
University of Amsterdam, Science Park 904, 1098 XH Amsterdam, The Netherlands}
\affiliation[c]{Perimeter Institute for Theoretical Physics, Waterloo, ON N2L 2Y5, Canada}
\affiliation[d]{Department of Physics $\&$ Astronomy, University of Waterloo, Waterloo, ON N2L 3G1, Canada}

\emailAdd{elenac@utexas.edu}
\emailAdd{s.chapman@uva.nl}
\emailAdd{josiah.couch@utexas.edu}
\emailAdd{jhernandez@pitp.ca}
\emailAdd{rmyers@pitp.ca}
\emailAdd{sruan@pitp.ca}

\date{\today}

\abstract{We study the complexity of Gaussian mixed states in a free scalar field theory using the `purification complexity'. The latter is defined as the lowest value of the circuit complexity, optimized over all possible purifications of a given mixed state. We argue that the optimal purifications only contain the essential number of ancillary degrees of freedom necessary in order to purify the mixed state. We also introduce the concept of `mode-by-mode purifications' where each mode in the mixed state is purified separately and examine the extent to which such purifications are optimal. We explore the  purification complexity for thermal states of a free scalar QFT in any number of dimensions, and for subregions of the vacuum state in two dimensions. We compare our results to those found using the various holographic proposals for the complexity of subregions. We find a number of qualitative similarities between the two in terms of the structure of divergences and the presence of a volume law. We also examine the `mutual complexity' in the various cases studied in this paper.}

\begin{document}

\maketitle

\section{Introduction}
Quantum information concepts and their embedding in gravitational holography \cite{Maldacena} have proved very useful for developing our understanding of the bulk-boundary map, \eg see \cite{RTreview,QECC,Swingle,Harlow}. One particular notion, which has captured increasing attention, is computational complexity. The complexity of a quantum state is defined as the minimal number of simple operations required in order to construct the state starting from a simple unentangled product state \cite{Encyclopedia,Aaronson:2016vto}. There exist several proposals for the holographic dual of computational complexity \cite{Volume1,Volume2,Volume3,Action1,Action2,Couch:2016exn}, however, at the moment, we can only test them at a phenomenological level due to the absence of a well-posed definition for the complexity for quantum field theory states. One front, in which progress has been made is that of Gaussian and nearly Gaussian states, \eg \cite{qft1,qft2,Fermions1,Fermions2,ComplexityRG}. Most of those studies, however, focused on pure states, and very little is known about the complexity of mixed states. Several proposals were made to define mixed-state complexity in \cite{BrianMixedComplexity} and our goal here is to examine one of these, the purification complexity, in detail for mixed  Gaussian states. Let us also mention that in holography, several proposals have been made for the gravitational dual of the complexity of mixed states associated with reduced density matrices on subregions of the boundary of asymptotically AdS spaces \cite{Alishahiha:2015rta,Carmi:2016wjl} and we will also compare our QFT results with those coming from holography, at least at the qualitative level.

\begin{figure}[htbp]
    \centering\includegraphics[width=6.1in]{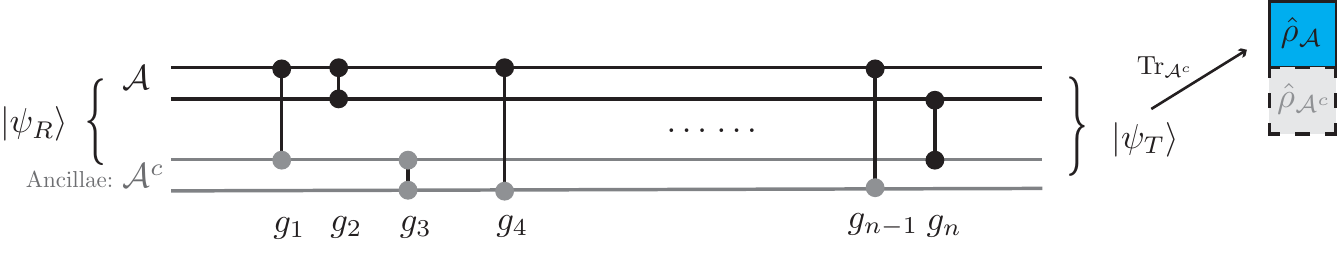}
    \caption{Circuit with the ancillary degrees of freedom.
The mixed state $\hat{\rho}_\mA$ that we want to prepare is obtained at the final step after tracing out the ancillae.}\label{circuit_purification}
\end{figure}

{\bf Circuits with Ancillae and Purification Complexity:}
Preparing a mixed state $\hat \rho_\mathcal{A}$ on some Hilbert space $\mathcal{A}$, starting from a pure reference state, cannot be achieved using only unitary gates. Instead, we should think of preparing the state using a set of allowed universal (non-unitary) gates, which consist of completely positive trace-preserving maps acting on the reference state. However, this approach is equivalent to extending the Hilbert space to include {\bf ancillary} degrees of freedom and working with unitary gates acting on this extended Hilbert space, \eg see \cite{dilaton,watrous2009quantum,Aharonov:1998zf} and chapter 8 in \cite{NielsenChuang}. One can think that the set of unitary gates is extended to include {\bf ancillary gates}, which introduce a new ancillary degree of freedom (in some simple product state) as needed, and {\bf erasure gates}, which erase or trace out a single degree of freedom whenever is convenient. Alternatively, as illustrated in figure \ref{circuit_purification}, we can think that the reference state is an unentangled product state on all of the needed or available
auxiliary degrees of freedom, as well as the physical degrees of freedom, \ie the reference state (and all of the intermediate pure states) live on an extended Hilbert space $\mA \otimes \mA^c$.  Then after applying a unitary circuit to this extended state, the ancillae are all traced out of the final pure state to produce the desired mixed state on the physical Hilbert space $\mathcal{A}$ alone.

Following this discussion, we can define the complexity of mixed states by considering the complexity of pure states which purify them. Obviously, the purifications of a given mixed state are not unique. However, a natural definition of mixed state complexity -- the so-called {\bf purification complexity}  \cite{BrianMixedComplexity}  is defined as the minimal pure state complexity among all possible purifications of our mixed state, \ie as usual, we are optimizing over the circuits which take the reference state to a target state $\ket{\Psi_{\mA\mA^c}}$, which is a purification of the desired mixed state $\hat \rho_\mA$, but we must also optimized over the possible purifications of $\hat \rho_\mA$, \ie
\begin{equation}\label{def_pureC}
\mC \( \hat \rho_\mathcal{A} \)  \equiv \text{min}_{\mA^c} \,  \mC\( \ket{\Psi_{\mA\mA^c}} \), \quad \text{such that} \quad \hat \rho_\mathcal{A} = \text{Tr}_{\mA^c}\ket{\Psi_{\mA\mA^c}} \bra{\Psi_{\mA\mA^c}} \ \,.
\end{equation}
Recall that we are applying this analysis to study mixed Gaussian states. A simplifying assumption in our analysis will be that the purified states are also Gaussian. This allows us to use the prescription of \cite{qft1} for evaluating the complexity of the possible purifications, and we then minimize over the parameters of the purifications, as in eq.~\reef{def_pureC} above.\footnote{We might mention that this assumption also appeared in a recent
discussion \cite{pure2} of the entanglement of purification \cite{pure0,pure1,pure2} for Gaussian states.}

Completing our complexity model requires specifying the cost function.\footnote{The cost functions assign a cost to different trajectories in the space of unitary transformations between the different states --- see section \ref{subsec:compure} for further details.} A variety of cost-functions have been considered in the literature for the complexity of pure Gaussian states (\eg see \cite{qft1,qft2,Fermions1,coherent}). As was pointed out in \cite{qft1,qft2}, the $F_1$ cost function (see eq.~\reef{func2})
seems most closely related to complexity in holography because the structures of the UV divergences match. Hence we will focus our analysis on this choice in the following. However, the precise results are also found to depend on the basis chosen for the fundamental gates. For example, a recent study of the complexity of the thermofield double (TFD) state \cite{Chapman:2018hou} has shown the importance of choosing a basis which is not entirely diagonal when two systems are involved.\footnote{The TFD state is a purification of the thermal density matrix on a given QFT$_L$ (the ``Left'' copy) in terms of another identical QFT$_R$ (the ``Right'' copy). When studying the complexity of this state, it is important to work with a basis which distinguishes the ``Left'' and ``Right'' degrees of freedom to reproduce qualitative features of the holographic complexity of the double-sided AdS black hole.} Hence, we also explore the possibility of working in a basis which distinguishes the ancillary degrees of freedom from the physical degrees of freedom of the original reduced density matrix. We refer to such basis as the {\bf physical basis}, as opposed to the {\bf diagonal basis} which mixes the two sets of degrees of freedom.

At this point, let us add that it is natural to think of the auxiliary degrees of freedom as a resource in the preparation of the desired mixed states and hence in differentiating possible purifications, one would assign an additional cost for including more ancillae, \ie
we can assign an extra cost for the ancillary and erasure gates commented on above. However, we will not consider the effect of such an additional cost for the bulk of our analysis, but we return to this issue briefly in the discussion section \ref{sec:disc}.\\

{\bf Outline and summary of main results:} We start in section \ref{warmup}, by exploring the purification complexity for mixed states with a single harmonic oscillator, purified by the addition of a single extra ancillary degree of freedom. In the diagonal basis, we can obtain an analytic result which is given by  eq.~\eqref{complexity_one_mode}, while the physical basis complexity requires some numerical treatment. We prove that the diagonal basis complexity is smaller than the physical basis complexity for these small systems.

We proceed in section \ref{sec:manyho}, by exploring the optimal purifications of multi-mode Gaussian states. We generalize the various notions of diagonal and physical basis complexities to the case of mixed states of more than one mode. In this section, we also explore the optimality of essential and mode-by-mode purifications. {\bf Essential purifications} are purifications with the minimal number of new degrees of freedom needed to purify the state. For the case of a single oscillator, we compare purifications with a single additional degree of freedom to purifications with two additional degrees of freedom and show that the optimal purifications are obtained without the use of the extra ancilla. This motivates us to make a conjecture that optimal purifications will be essential purifications, even for a larger number of oscillators. This conclusion holds both in the diagonal and in the physical basis. We explain how to bring a general mixed state to the form of a tensor product of one mode mixed states and define the concept of {\bf mode-by-mode} purifications where each mode is purified separately. We demonstrate that this subset of purifications is optimal when the original state is simply a product of one mode mixed states. For a general mixed state, we show that these purifications are not optimal, but that they give a good approximation for the complexity. Some technical details and extensions related to the topics of this section have been left for appendices \ref{app:purification} and \ref{app:numerics}.

In sections \ref{apply01} and \ref{apply02}, we examine the purification complexity for mixed states for two examples in a free scalar field theory: a thermal density matrix and the reduced density matrix for a subregion of the vacuum. In both cases, we examine a quantity denoted by the {\bf mutual complexity}. The latter is defined by beginning with a pure state $|\Psi_{\mA \mB}\rangle$ on an extended Hilbert space.\footnote{Mutual complexity could just as easily be defined with an initially mixed state $\rho_{\mA\mB}$ (see discussion in section \ref{sec:disc}). However, the initial state is a pure state in the two examples that we consider here.} Now tracing over the $\mB$ degrees of freedom yields the mixed state $\rho_{\mA}$, whereas integrating out the $\mA$ degrees of freedom yields $\rho_{\mB}$. Then the mutual complexity is given by comparing the complexities of these three states with
\begin{equation}\label{MutualComplexityDefSub}
\Delta\mC = \mC(\rho_{\mA}) + \mC(\rho_{\mB}) - \mC(|\Psi_{\mA \mB}\rangle)\,.
\end{equation}
The complexity is said to be subadditive when $\Delta\mC>0$ and superadditive when $\Delta\mC<0$. That is, subadditivity indicates that the complexity of the state on the full system is less than the sum of the complexities for the states on the two subsystems (and vice versa for superadditivity).

We start in section \ref{apply01}, by applying our previous results to the case of a thermal state in the free scalar field theory. We show that the result for the purification complexity is simply the sum of the results for the complexities of the various momentum space modes. We ask the question of whether the thermofield double state (TFD) provides an optimal purification of the thermal state. For the individual modes, we find that this is the case only for a very small range of frequencies, and so the TFD state does not correspond to the optimal purification for the thermal state in the field theory.  Further, the UV divergences associated with the optimal purification are less (by a factor of 2) than those of the TFD. This turns out to be essential in order to recover the same structure of divergences as in holography. In this case, we can evaluate the mutual complexity for the TFD state as
\begin{equation}\label{sushi}
\Delta\mC =  2\, \mC\!\(\hat\rho_{th}(\beta)\) - \mC\!\( \ket{\rm TFD}\)   \,,
\end{equation}
where $\hat\rho_{th}(\beta)$ is the thermal density matrix. Given the previous comment, we find that $\Delta\mC$ is UV finite and further that it is always positive in the diagonal basis, \ie the complexity is subadditive with this choice. In the physical basis, we find that the mutual complexity can be either positive and negative, depending on the reference state frequency and the temperature of thermal state.

In section \ref{apply02}, we study the complexity of subregions of the vacuum in a two-dimensional free scalar QFT. We study the dependence of the purification complexity on the UV cutoff in the diagonal and physical bases. We find that in either case, the leading divergence scales with the volume of the subsystem and depends on the cutoff and reference state frequency in a manner that is similar to that observed in holography, once the relevant parameters are identified. We also find that the subleading divergence of the complexity of subregions of the vacuum in the diagonal basis is logarithmic, similarly to what is found for holographic complexity using the subregion-CA and subregion-CV2.0 proposals.
We also study the mutual complexity in both the diagonal basis and the physical basis. The mutual complexity in the physical basis has two possible definitions, which we state in eq.~\eqref{twoDeltaCs}, depending on what the interpretation of the physical basis complexity of the vacuum should be. Evaluating the mutual complexity \reef{MutualComplexityDefSub}
for the vacuum state, we observe that it is positive and logarithmically divergent in the diagonal basis, as well as for one of the physical basis definitions. For the other physical basis definition, it is negative and linearly divergent. Further, it reaches a broad maximum (minimum) when the subsystem size equals to half of the system for all cases, and is symmetric about that point. The distinction between the various bases is highlighted in appendix \ref{app:4HO} with a simple example of mixed states in subregions of a free field theory on a lattice with four sites, \ie of a system of four coupled harmonic oscillators.

Section \ref{sec:holo} reviews and extends various results from the holographic literature regarding the complexity of mixed states. We review the two holographic proposals for subregion complexity using volume and action \cite{Alishahiha:2015rta,Carmi:2016wjl} of bulk regions naturally associated with the Ryu-Takayanagi surface \cite{Ryu:2006bv,Ryu:2006ef} and the entanglement wedge \cite{EW1,EW2,EW3} of the subregion. In addition, we propose a natural extension of the CV2.0 of \cite{Couch:2016exn} for subregions.  We then present the results for the holographic complexity of a thermal state living on a single boundary of a two-sided black hole, as well as results for subregions of the vacuum in various dimensions and for various boundary geometries. The relevant calculations for this section are found in appendix \ref{app:appsubCA4}.

The results obtained for the mutual complexity, which determine the additivity properties of the complexity in the various cases, which we studied both in the free scalar QFT and in holography,  are summarized here in table \ref{tab.comp}. Generally, we found that the complexity is superadditive in the holographic setting, while in contrast, we found that it is subadditive in the setting of the free scalar theory using the diagonal basis.  However, for the QFT complexity in the physical basis, we found that the complexity is superadditive for the thermal state for reference frequencies that are very far from the other physical scales of the problem, \ie when $\mu$ is a deep IR or a deep UV scale. Further, for subregions of the vacuum, the physical basis complexity is superadditive if the complexity of the vacuum is considered in a basis where the degrees of freedom on either side of the partition are distinguished. However, if we remove that constraint, the resulting mutual complexity is positive, and the complexity is instead subadditive.

\begin{table}
\begin{center}
{
\renewcommand{\arraystretch}{1.2}
\begin{tabular}{ | c || c| c | c | c |}
\hline
{}  & Thermal state & Subregions of the vacuum  \\
\hline
\hline
QFT (diagonal basis) & $\Delta\mC>0$ &  $\Delta\mC>0$ \\
\hline
QFT (physical basis) & $\Delta\mC<0$ for $\beta\mu\gg 1$ or $\beta\mu\ll 1$ &  $\Delta\mC<0$, $\Delta \tilde{\mC} >0\,{}^{\textcolor{blue}{\diamond}}$\\
\hline
Holography (CV) & $\Delta \mathcal{C}_V\le 0\,{}^{\textcolor{blue}{\S}}$  & $\Delta \mathcal{C}_
V\le 0\,{}^{\textcolor{blue}{\S}}$ \\
\hline
Holography (CA) & $\Delta \mathcal{C}_
A< 0\,{}^{\textcolor{blue}{\dagger}}$ &
$\Delta \mathcal{C}_A  < 0\,{}^{\textcolor{blue}{\ddagger}}$   \\
\hline
Holography (CV$2.0$) & $\Delta \mathcal{C}_{V2.0}<0\,{}^{\textcolor{blue}{\dagger}}$ & $\Delta \mathcal{C}_{V2.0}<0\,{}^{\textcolor{blue}{\ddagger}}$  \\
\hline
\end{tabular}
}
\caption{Comparison of the mutual complexity in field theory and in holography for the various cases studied in this paper.  Above, $\mu$ is the characteristic frequency of the reference state while $\beta$ is the inverse temperature. \\
${}^{\textcolor{blue}{\diamond}}$ there are two possible definitions for mutual complexity in the physical basis for subregions of the vacuum, see discussion around eq.~\eqref{twoDeltaCs} for more details; ${}^{\textcolor{blue}{\S}}$ the inequality is saturated (\ie $\Delta \mathcal{C}_{V}=0$) when evaluated for $t_L=0=t_R$ for the TFD state and for $t=0$ for the vacuum state, as was done in the preceding QFT calculations;
${}^{\textcolor{blue}{\dagger}}$ in both cases, $\Delta\mC$ was proportional to the entropy of the thermal state;
${}^{\textcolor{blue}{\ddagger}}$ in both cases, the leading contribution to $\Delta \mathcal{C}$ had the same form as the leading divergence in the entanglement entropy of the subregions.}\label{tab.comp}
\end{center}
\end{table}

We conclude with a brief review and discussion of our results in section \ref{sec:disc}. This discussion includes a review of other possible definitions for the complexity of mixed states, and exploring the possibility of using different cost functions in evaluating the purification complexity. Further, we examine the relation between mixed state complexity and entanglement entropy, and consider what additional information is contained in the purification complexity beyond that contained in the entanglement entropy. 
We also make a detailed comparison between our results for mixed state complexity in the free scalar QFT and those determined with the various holographic proposals. In  the context of AdS$_3$/CFT$_2$, we numerically identify an exact formula for the subleading divergences in  holographic complexity of a subregion of the vacuum on a finite circle. The latter is inspired by the close relation between the mutual complexity and the entanglement entropy, and by the formula for entanglement entropy of an interval in CFT$_2$ on a finite circle \cite{Calabrese:2004eu,Calabrese:2005zw}. This result also motivates using the same formula to numerically fit our results for the mutual complexity (as a function of the subregion size) in the diagonal basis for the free scalar QFT.

Before proceeding further, we must acknowledge that the purification complexity of a single oscillator is briefly considered using the $F_2$ cost function (see eq.~\reef{func2}) in  \cite{Camargo:2018eof}. This overlaps somewhat with the discussion in section \ref{warmup}, where we consider the purification complexity for the same system but focus on the $F_1$ cost function.

\section{Purification Complexity of a Single Harmonic Oscillator}\label{warmup}
Our aim in this paper is to explore the complexity of mixed states. In particular, we will examine the so-called ``purification complexity" \cite{BrianMixedComplexity}, defined as the minimal complexity of a pure state which purifies our mixed state, see eq.~\eqref{def_pureC}. Our analysis will focus on Gaussian mixed states and so before we plunge into the details of the purification complexity, we begin with a brief review of the construction in \cite{qft1} to evaluate the complexity of pure
Gaussian states --- see also discussions in \cite{Fermions1,coherent}.

\subsection{Complexity of Pure Gaussian States}\label{subsec:compure}
The authors of \cite{qft1} proposed a framework for evaluating the complexity of Gaussian states of bosonic field theories. The idea was to discretize the field theory on a spatial lattice such that one obtains a chain of coupled harmonic oscillators with position operators $\hat x_a$ and momentum operators $\hat p_b$ satisfying usual commutation relations $[\hat x_a,\hat p_b]=i \delta_{ab}$, where $a,b=1,\ldots,N$ indicate the positions on the lattice. The wavefunction of a pure Gaussian state with vanishing first moments (\ie $\langle \hat x_a\rangle = 0=\langle \hat p_a\rangle$) which will serve as our target state takes the following form in the position-space representation
\begin{equation}\label{wavematrix}
\langle x_a | \psT \rangle \equiv \psT(x_a) = {\cal N}_{\mt{T}} \exp \left[-\frac{1}{2}\sum_{a,b=1}^{N} M^{ab}_\mt{T}\, x_a\, x_b \right].
\end{equation}
The normalization constant is given by ${{\cal N}_{\mt{T}}}^4 = {\rm det}\left(\frac{M_T}{\pi}\right)$. For simplicity, we will focus on cases where the matrix $M^{ab}$ is real (and of course, symmetric).\footnote{We will describe below how to evaluate the complexity of such states according to \cite{qft1}. For cases where $M^{ab}$ is complex, a more general treatment is needed where the $GL(N,\mathbb{R})$ group of gates, appearing below in eq.~\reef{gates}, must be extended to $Sp(2N,\mathbb{R})$, \eg see \cite{Chapman:2018hou}.}
The matrix $M^{ab}$ can be diagonalized by an orthogonal transformation in terms of a set of ``normal mode'' coordinates $\tilde x_k$ and characteristic frequencies $\omega_k$\footnote{In the following, we are taking the normal modes $\tilde{x}_k$ to be real linear combinations of the position basis modes $x_a$. Later we will find that for applications in QFT it is easier to consider complex normal modes $x_k$ (see, \eg  eqs.~\eqref{Thefreq} and \eqref{eq:Fourier}). In this case we should replace $\tilde{x}_k^2 \to |x_k|^2$ in eq.~\eqref{TargetGaussianPure}.}
\begin{equation}\label{TargetGaussianPure}
\langle \tilde  x_k | \psT \rangle =\psT(\tilde x_k) =  {\cal N}_{\mt{T}} \exp \left[-\frac{1}{2}\sum_{k=1}^{N} \omega_k \,\tilde x_k^2 \right].
\end{equation}
The latter can be viewed as the Gaussian wavefunction
\begin{equation}\label{Odiag}
\psT(\tilde x_k) =  {\cal N}_{\mt{T}}\exp \left[-\frac{1}{2}\sum_{k,k'=1}^{N} \tilde M^{k k'}_\mt{T}\, \tilde x_k\, \tilde x_{k'} \right]\quad{\rm with}\quad
\tilde M_\mt{T}  = O^T M_\mt{T}\, O  = \text{diag}(\omega_1, \cdots, \omega_N)\,,
\end{equation}
and where the orthogonal matrix $O$ produces the change of basis $x_a = O_a{}^k \tilde x_k$ which diagonalizes the matrix $M_\mt{T}$. As an example, one might think of the ground state of a chain of coupled harmonic oscillators with normal mode frequencies $\omega_k$, where the mass of the harmonic oscillators has been set to one. In fact, to be consistent with dimensional analysis, we have assumed that all the equations above also contain a characteristic mass which we will set to one from now on.

A natural reference state is the factorized Gaussian state\footnote{The normalization constant of the reference state is given by ${\cal N}_{\mt{R}}^4 = {\rm det} \left( \frac{M_R}{\pi}\right)= \left(\frac{\mu}{\pi}\right)^N $.}
\begin{equation}\label{ref_stateO}
\langle x_a | \psR \rangle \equiv \psR(x_a)  =  {\cal N}_{\mt{R}} \exp \left[-\frac{1}{2}\sum_{a=1}^{N} \mu \, x_a^2 \right]\,,
\end{equation}
where the degrees of freedom are completely disentangled in the position basis. Note that we are choosing the same reference frequency $\mu$ for each $x_a$ so that the degrees of freedom are all on the same footing, \ie
\begin{equation}
M_\mt{R}=\mu\ \text{diag}(1,1, \cdots,1).
\end{equation}
Hence for the example of a chain of oscillators, the reference state is translation invariant.\footnote{Similarly, the ground state of any translation invariant Hamiltonian will be translation invariant. This would be reflected in the entries of the parameter matrix $M_{ab}$ in eq.~\eqref{wavematrix} which will be a function of $a-b$.} With this simple reference state, the change of basis introduced in eq.~\reef{Odiag} yields
\begin{equation}\label{ref_state}
\langle \tilde x_k | \psR \rangle= \psR(\tilde x_k)  = {\cal N}_{\mt{R}} \exp \left[-\frac{1}{2}\sum_{k=1}^{N} \mu \,\tilde x_k^2 \right]\,.
\end{equation}
That is, in the diagonal basis, the reference state remains a factorized Gaussian  with $\tilde M_\mt{R}=M_\mt{R}$.

Now, the target state \reef{TargetGaussianPure} can be produced by acting with a unitary transformation on this reference state \reef{ref_state}, \ie $\ket{\psT}= U_\mt{TR}\, \ket{\psR}$ where $U_\mt{TR}$  is constructed as a string of fundamental gates,
\begin{equation}
g_{ab}= e^{i\frac{\varepsilon}{2} (\hat x_a \hat p_b + \hat p_b \hat x_a)}\,.\label{gates}
\end{equation}
These gates produce a $GL(N,\mathbb{R})$ group of transformations. Those with $a\ne b$ introduce entanglement between the different oscillators, while with $a=b$, the gates scale the coefficients of the corresponding coordinate -- see \cite{qft1} for further details.
Generally, there will be an infinite number of such ``circuits," \ie sequences of fundamental gates, which will accomplish the desired transformation. The complexity is defined as the minimum number of gates needed to construct the desired target state \reef{TargetGaussianPure} from the reference state \reef{ref_state}.

To identify the optimal circuit, Nielsen and his collaborators \cite{nielsen2006quantum,nielsen2008,Nielsen:2006} developed a geometric method, which was adapted to evaluate the complexity of QFT states in \cite{qft1}.  This construction is based on a continuum representation of  the unitary transformations
    \begin{equation}\label{unitaries}
    U(\s) = \cev{\mathcal{P}}\, \exp \!\[ -i \int^\s_0\!\!\! d s\, \mathcal{H}( s)\], \quad \text{where} \quad \mathcal{H}(s)= \sum_I Y^I(s)\,\mathcal{O}_I
    \end{equation}
where $s$ parametrizes the circuit and $\cev{\mathcal{P}} $ signifies a path ordering along $s$ from right to left. The ``Hamiltonian" $\mathcal{H}(s)$ is constructed from the (Hermitian) generators $\mathcal{O}_I$ of the fundamental gates, \eg $\mO_{ab}=-\frac12(\hat x_a \hat p_b + \hat p_b \hat x_a)$ in eq.~\reef{gates}. The coefficients $Y^I(s)$ are control functions  specifying which gates (and how many times they) are applied at any particular point $s$ in the circuit. In eq.~\reef{unitaries}, we have actually specified a path $U(\s)$ through the space of unitaries, or through the space of states with $\ket{\psi(\s)}=U(\s)\ket{\psi_\mt{R}}$. We then fix the boundary conditions for the circuits of interest, with $0\le\s\le1$, as
    \begin{equation}
    U(\s=0)= \mathbbm{1}\,, \qquad  U(\s=1)= U_\mt{TR}\,,
    \end{equation}
where $U_\mt{TR}$ is the desired unitary producing $\ket{\psT}= U_\mt{TR}\, \ket{\psR}$. From this perspective, the $Y^I(s)$ can also be interpreted as the components of the tangent vector to this trajectory.

Nielsen's approach identifies the optimal circuit by minimizing the cost defined as
    \begin{equation}\label{gen-cost}
    \mathcal{D}(U(\s))\equiv \int^1_0 ds ~ F \( U(s), Y^I(s)  \),
    \end{equation}
where $F$ is a local functional of the position $U(s)$ and the tangent vector $Y^I(s)$ along the trajectory.\footnote{When this functional only depends on $Y^I(s)$ as in eq.~\reef{func2}, the cost (and the underlying geometry) is right invariant, \eg \cite{Brown:2016wib,Brown:2017jil}.} Two simple examples of such cost functions are
    \begin{equation}\label{func2}
    F_1(U,Y)=\sum_I \left|Y^I\right|~,\qquad\qquad
    F_2(U,Y)=\sqrt{\sum_I  \(Y^I\)^2}~.
    \end{equation}
With the $F_2$ measure, the cost \reef{gen-cost} is simply the proper distance in a Riemannian geometry, and hence identifying the optimal circuit is equivalent to finding the shortest geodesic connecting the reference and target states in this geometry. With the $F_1$ measure, the cost essentially counts the number of gates, and so this choice comes closest to the original concept of complexity. However, in contrast with the $F_2$ measure, a disadvantage of the $F_1$ cost function is that it is not ``covariant", \ie the corresponding complexity $\mC_1$ depends on the choice of the basis for the generators $\mO_I$.\footnote{In~\cite{Fermions1,coherent}, a basis-independent alternative was proposed using the Schatten norm. For Gaussian states with vanishing first moments, \ie $\langle x_a\rangle = 0=\langle p_a\rangle$, the complexity found using the ($p=1$) Schatten cost function is identical with ${\cal C}_1$, as shown in eq.~\reef{complexity_pure}. However, we note that this Schatten complexity does not yield the desired complexity of formation for the TFD states studied in \cite{Chapman:2018hou}.} However, the structure of the UV divergences for the $\mC_1$ complexity was found to be similar to that for holographic complexity \cite{qft1,qft2}. Further, the basis dependence played an important role in \cite{Chapman:2018hou}, which studied the complexity of thermofield double (TFD) states for a free scalar. In particular, the complexity of formation was found to match that for holographic systems \cite{Chapman:2016hwi}, \ie $\widetilde{\Delta\mC}_{\text{formation}}  \propto S_\mt{th}$ in the massless limit,\footnote{Here, $S_\mt{th}$ is the thermal entropy of the thermal mixed state living on either side of the TFD, or equivalently the entanglement entropy between the two copies of the field theory.} when the gates were chosen to act on the physical degrees of freedom corresponding to the two separate copies of the field theory, \ie the Left-Right basis \cite{Chapman:2018hou}. In contrast, if the basis of gates were chosen to act on the diagonal modes (with which the TFD state could be expressed as a simple product state), the $\mC_1$ complexity produced $\widetilde{\Delta\mC}_{\text{formation}} \simeq 0$ to leading order.

We reviewed the results above to motivate that in this paper, we will focus entirely on studying the purification complexity of mixed states using the $F_1$ measure. Further, we will test the sensitivity of our $\mC_1$ complexity to the choice of basis. In particular, in each case, we will examine the results for the physical basis and the diagonal basis.  As a further review of key results, let us add the following:

For a broad variety of cost functions including those in eq.~\reef{func2}, the optimal circuit taking eq.~\reef{ref_state} to eq.~\reef{TargetGaussianPure} is simply a straight-line path which only applies the scaling gates \reef{gates} (with $a=b$) to each of the corresponding normal modes $\tilde x_k$ \cite{qft1}.
In fact, \cite{qft1} recasts the discussion of circuits in terms of a matrix representation. In particular, the trajectory through the space of states is described by
\begin{equation}
\tilde{M}(\sigma)=U(\sigma)\, \tilde M_\mt{R}\, U^T(\sigma)\,,
\end{equation}
where the $\tilde M^{ab}$ define Gaussian wavefunctions in terms of the normal modes, as in eq.~\reef{Odiag}. For the case in hand, the optimal trajectory is simply
\begin{equation}\label{Hgen}
U(\sigma) = e^{\tilde H \sigma}, \qquad {\rm with}\quad
\tilde H =\frac{1}{2}\, \text{diag}\!\left(\ln(\omega_1/\mu), \cdots, \ln(\omega_N/\mu)
\right)\,.
\end{equation}
For this linear trajectory, the complexity is given in terms of the  elements of  $\tilde H$, and in particular, the $\mC_1$ complexity becomes
\begin{equation}\label{complexity_pure}
\begin{split}
\mC_{1}^{\mt{diag}} & =
\frac 12\, \sum_{k=1}^N \left|\ln \frac{\omega_k}{\mu}\right|  \,.
\end{split}
\end{equation}
We make repeated use of this result in the following and so the interested reader is invited to see \cite{qft1} for a detailed derivation. As we noted above, the $\mC_1$ complexity is sensitive to the choice of basis for the gates (or generators), and the superscript `diag' above is added to indicate that the complexity was evaluated using gates acting on the normal-mode coordinates
$\tilde x_k$.

However, as noted in the previous discussion, it is interesting to  consider different choices of basis in certain cases. This is simply done by rotating the generator $\tilde H$ to the relevant basis and summing over (the absolute values of) its elements
\begin{equation}\label{rotate2pos}
 H = O\, \tilde H\, O^T\qquad{\rm and}\qquad \mathcal{C}_1 = \sum_{a,b=1}^N |H^{ab}|\,.
\end{equation}
Implicitly, we have assumed here that the straight-line circuit \eqref{Hgen} remains optimal in the new basis. However, in general (and for our examples below), it is difficult to prove that this simple trajectory is still optimal. Nevertheless, evaluating the cost of the trajectory \eqref{Hgen} provides a bound on the $\mC_1$ complexity for the new basis.
In examining mixed state complexity below, we will consider the {\bf physical basis} which distinguishes between the two classes of oscillators in purifications of a mixed state, \ie the original physical oscillators and the auxiliary degrees of freedom. We will indicate when our calculations refer to this basis by using the superscript `phys'. More details on different interesting bases and the distinction between them can be found in section~\ref{subsec3phys} and appendix~\ref{app:4HO}.

In closing here, let us add that an alternative approach to the complexity of QFT states based on the Fubini-Study metric was developed in \cite{qft2}. For Gaussian states with vanishing first moments and an appropriate definition of the measure, this alternative approach produces precisely the same complexity as in eq.~\reef{complexity_pure}. Hence we expect that many of our results for the purification complexity of mixed states in the following can be easily extended to the Fubini-Study approach.

\subsection{Gaussian Purifications of One-Mode Mixed States}\label{sec:onemodepuri}

Turning to the purification complexity of mixed states, we begin by considering Gaussian density matrices for a single oscillator and explore their purifications. Consider a single harmonic oscillator in a mixed state $\hat \rho$, such that
\beq\label{dense1}
\rho(x,x')\equiv \bra{x} \hat \rho \ket{x'} = \left(\frac{a-b}\pi\right)^{1/2}\, e^{-\frac{1}{2} \left( a x^2 + a x'^2 - 2 b x x' \right) }
\eeq
where we will assume that $a$ and $b$ are real. Note that this is compatible with $\rho$ being a Hermitian operator, \ie $\rho^\dagger=\rho$ or $\rho^*(x',x)=\rho(x,x')$. The overall normalization constant was chosen to ensure ${\rm Tr}[\rho]=\int dx\, \rho(x,x)=1$. In order for the Gaussian integral in this norm to be well defined, we need $a>b$. Further, in order that the density matrix be positive semi-definite (\ie $\bra{\psi} \hat{\rho} \ket{\psi}\geqslant 0$ for arbitrary wavefunctions $\psi(x)$) we should require that $b\geqslant 0$.\footnote{Since probabilities are all either zero or positive, the density matrix is positive semidefinite, \eg see section III of \cite{mann1993gaussian}. We will see below that $b\geqslant 0$ ensures that the purifying wavefunction also has real parameters.}

Next, we consider purifications of the density matrix \eqref{dense1} by pure Gaussian states with two degrees of freedom
\beq\label{wfunction1}
\psi_{12}(x,y)\equiv \braket{x,y}{\psi}  = \left(\frac{\omega_1\omega_2-k^2}{\pi^2}\right)^{1/4} e^{-\frac{1}{2}\left(\omega_1 x^2 + \omega_2 y^2 + 2 k\, xy \right)}
\eeq
where again we will assume for simplicity that $\omega_{1,2}$ and $k$ are all real. For this wavefunction to be normalizable, \ie $1=\int dx\,dy\ |\psi(x,y)|^2$, we need $\omega_2>0$ and $\omega_1\, \omega_2 -k^2 >0$.
The density matrix corresponding to $\ket{\psi}$ is simply given by
\begin{equation}
\rho_{12}(x,y,x',y') = \left(\frac{\omega_1\omega_2-k^2}{\pi^2}\right)^{1/2} e^{-\frac{1}{2}\left(\omega_1 x'^2 + \omega_2 y'^2 + 2 k x'y' \right)} e^{-\frac{1}{2}\left(\omega_1 x^2 + \omega_2 y^2 + 2 k xy \right)}\,.
\end{equation}
Tracing out the auxiliary oscillator, we find
\begin{align}
\begin{split}
\rho_{1}(x',x)
=\int dy\, \rho_{12}(x,y,x',y)
= \frac{\sqrt{\omega_1\omega_2-k^2}}{\sqrt{\pi\,\omega_2}}\, e^{-\frac{1}{2} \left[ \left(\omega_1-\frac{k^2}{2\omega_2}\right) (x^2 +x'^2) - \frac{k^2}{\omega_2 }\,xx'\right]}.
\end{split}
\end{align}
Therefore comparing the above density matrix to eq.~\reef{dense1}, we find
\beq\label{ABA}
a = \omega_1 - \frac{k^2}{2\omega_2}\,,\qquad
b = \frac{ k^2}{2\omega_2 }\,.
\eeq
From the second equation, we see that $b\geq0$ ensures a real purification. Note that for $b=0$, we simply get
\beq
a=\omega_1\,, \qquad k=0
\eeq
and $\omega_2$ is unconstrained. That is, for the density matrix \reef{dense1} of an already pure state (\ie  $\rho(x,x')=\psi_1(x)\psi^\dagger_1(x')$), the purification in eq.~\reef{wfunction1} is itself simply the product of two decoupled wavefunctions (\ie $\psi_{12}(x,y) = \psi_1(x)\psi_2(y)$). For non-zero $b$, we may solve for $\omega_1$ and $\omega_2$ in terms of $a$, $b$ and $k$ to find
\beq
\omega_1=a + b\,,\qquad
\omega_2 = \frac{ k^2}{2b}\,.
\eeq
Hence we arrive at the one-parameter family of wavefunctions
\begin{equation}\label{pure1}
\psi_{12}(x,y)  = \left(\frac{(a-b)}{2b}\frac{k^2}{\pi^2}\right)^{1/4}\, e^{-\frac{1}{2}\left[\left(a+b\right) x^2 + \frac{ k^2}{2b} y^2 + 2 k xy \right]}\,,
\end{equation}
all of which produce the same density matrix \reef{dense1} upon tracing out the auxiliary position $y$. The purification complexity is then found by optimizing the usual pure state complexity over the free parameter $k$ distinguishing these different purifications.


\subsection{Alternative Description of the Purifications} \label{subsec:altdesc}

Before we evaluate the purification complexity of the density matrix in eq.~\reef{dense1}, it will be convenient to introduce a second representation of the Gaussian states in order to simplify the optimization and to make clear the role of the ancillae for our Gaussian examples. Hence let us work in terms of the energy eigenstates of a given Hamiltonian
\begin{equation}\label{Hamiltonian}
H = \frac{1}{2} \hat p^2+ \frac{1}{2} \omega^2 \hat x^2 = \omega \left(a^\dagger a +\frac{1}{2}\right)\,,
\end{equation}
where we have set the mass to one.\footnote{The frequency $\omega$ of the oscillator is an arbitrary choice here, but of course, the result of our analysis will only depend on this choice through the parameters of the density matrix \reef{dense1}.}
The annihilation and creation operators are defined as usual with
\begin{equation}
a\equiv\sqrt{\frac{ \omega}{2}}\left(\hat x+i \frac{\hat p}{ \omega}\right), \qquad
a^{\dagger} \equiv \sqrt{\frac{ \omega}{2}}\left(\hat x-i \frac{\hat p}{ \omega}\right) \label{ac}
\end{equation}
and satisfy the commutation relations $[a,a^{\dagger}]=1$. The corresponding energy eigenstates can be written as
\begin{equation}
|n\rangle = \frac{(a^{\dagger})^n}{\sqrt{n!}} |0\rangle
\end{equation}
where $|0\rangle$ is the vacuum state of the Hamiltonian \eqref{Hamiltonian}.

It is well known in the literature of quantum information, \eg see  \cite{RevModPhys.84.621,ferraro2005gaussian,serafini2017quantum}, that Gaussian states can be decomposed in terms of standard operators defined using these creation and annihilation operators.
In particular, the most general real
density matrix of a one-mode Gaussian state can be decomposed according to\footnote{In this paper, we only consider Gaussian states with $\langle x\rangle=0=\langle p\rangle$, which implies that the exponent of the Gaussian wavefunction does not contain a term linear in $x$. If such terms were present, we would have to extend eq.~\reef{Gaussian_decom} by conjugating with the displacement operator, \eg see the discussion of complexity of coherent states in \cite{coherent}.}
\begin{equation}\label{Gaussian_decom}
\begin{split}
\hat{\rho}_1 &= \hat S_1(r) \, \hat{\upsilon}_{th} (\beta,\omega)\,\hat S_1^\dagger(r) \,.
\end{split}
\end{equation}
The operator $\hat S_1(r)$ is the one-mode squeezing operator, acting on our oscillator which we denote by the subscript 1 (in anticipation for introducing a second oscillator for the purification, which we will denote by a subscript 2), which for real values of $r$ reads\footnote{Note that the frequency $\omega$ from the definition of $a,\,a^\dagger$ in eq.~\reef{ac} does not appear here. The infinitesimal version of this squeezing operator is simply the scaling gate (with $a=b$) in eq.~\reef{gates}. }
 \begin{equation}\label{onemode_squeezed}
 \begin{split}
\hat S_1(r) \equiv e^{-\frac{r}{2}\left({a_1^\dagger}{}^2-a_1^2\right)}
= e^{i \frac{r}2 \left(\hat{x}_1\hat{p}_1 + \hat{p}_1 \hat{x}_1 \right)}\,.
\end{split}
\end{equation}
This squeezing operator acts on the wavefunction $\psi(x) \equiv \langle x|\psi \rangle$ by rescaling the coordinate $x$ according to
$\langle x|\hat S_1(r)|\psi \rangle = e^{r/2}\, \psi(e^r x)$.
The remaining operator $ \hat{\upsilon}_{th} (\beta,\omega )$ is a thermal density matrix for the canonical ensemble with temperature $1/\beta$, \ie
\begin{equation}\label{density_thermal}
\hat{\upsilon}_{th}(\beta, \omega)
 \equiv \frac{e^{-\beta \omega\, a^\dagger a}}{\tr (e^{-\beta \omega\, a^\dagger a})}=  \(1- e^{-\beta\omega}\) \sum_{n=0}^\infty e^{-\beta \omega\,n} \ket{n}\!\bra{n}\,.
\end{equation}

We can evaluate the position space representation of the density matrix $\hat{\rho}$ in eq.~\eqref{Gaussian_decom}, \ie $\langle x|\hat{\rho}| x' \rangle$, using Mehler's formula~\cite{erdelyi1953higher}, \eg
\begin{equation}
 \sum_{n=0}^\infty \frac{u^n}{2^n n!} H_n(x)H_n(y)=\frac{1}{\sqrt{1-u^2}} \exp \( -\frac{u^2(x^2+y^2)-2u xy }{1-u^2} \).
\end{equation}
Of course, this yields a Gaussian density matrix of the form in eq.~\eqref{dense1} with the following parameters
\begin{equation}\label{parameters}
a = \frac{e^{2r}\, \omega\, \cosh \beta \omega}{\sinh
\beta\omega}>0\,, \qquad  b =\frac{e^{2r}\, \omega}{\sinh \beta \omega} >0 \,, \qquad  \frac{a}{b}= \cosh \beta \omega \ge 1 \,.
\end{equation}
Demanding that the temperature and frequency are positive is then equivalent to the previous restrictions, $a>b\geqslant0$, discussed around eq.~\reef{dense1}. We note that while the parameter $\omega$ was introduced as a dimensional scale here, our result for the complexity will only depend on the dimensionless combinations $\beta \omega$ and $\mu/\omega$, as well as the (dimensionless) squeezing parameter $r$.\footnote{Below, we will see that the complexity only depends on two parameters, namely $\beta\omega$ and a particular combination of $\mu/\omega$ and $r$. The latter reduction can be traced back to a symmetry of complexity, \ie the `distance' between the reference state and target state is left unchanged if we rescale $\mu$ and shift $r$ simultaneously.} However, the parameter $\omega$ will still play an important role later on when considering different modes of a free QFT on the lattice in sections \ref{apply01} and \ref{apply02}.\footnote{As we noted above, $\omega$ does not appear in the squeezing operator and further, $\omega$ only appears in the dimensionless combination $\beta\omega$ in the thermal density matrix \reef{density_thermal} (and implicitly in the definition of $|n\rangle$ in that same equation). However, from eq.~\reef{parameters}, we see that it sets the scale of the dimensionful parameters, $a$ and $b$, in eq.~\reef{dense1}. Further, it will set the scale of the dimensionful parameters in the purified state \reef{wfunction1} --- see eq.~\reef{transform_paras} below.} When the temperature is set to zero, \ie $\beta\omega\to\infty$, eq.~\reef{Gaussian_decom} reduces to a pure state. From eq.~\reef{parameters}, we see that this corresponds to the limit $b/a \rightarrow 0$.

The decomposition \eqref{Gaussian_decom} suggests that in order to purify this mixed state, one must purify the thermal part $\hat{\upsilon}_{th}$ of the density matrix.\footnote{In section \ref{sec_EE}, we explicitly demonstrate that the thermal part of eq.~\reef{Gaussian_decom} is also the component which determines the (entanglement) entropy of the mixed Gaussian state.} This can be done in terms of the thermofield double state, \eg see \cite{Chapman:2018hou}
\begin{equation}\label{two_mode}
\begin{split}
\ket{\text{TFD}}_{12}
\equiv
S_{12}(\alpha)\, \ket{0}_1\ket{0}_2
= \left(1-e^{-\beta \omega} \right)^{1/2}\, \sum_{n=0}^\infty e^{-\frac{1}{2}\,\beta \omega \,n } \ket{n}_1\ket{n}_2
\end{split}
\end{equation}
where we have introduced the two-mode squeezing operator which entangles the two degrees of freedom,
\begin{equation}\label{2squeezed_operator}
\begin{split}
S_{12}(\alpha) &\equiv e^{\alpha\, \left(a_1^\dagger a_2^\dagger - a_1 a_2\right)}
= e^{-i \alpha \(\hat{x}_1\hat{p}_2 + \hat{p}_1 \hat{x}_2\) }\,.
\end{split}
\end{equation}
The (real) squeezing parameter $\alpha$ for eq.~\reef{two_mode} is given by
\begin{equation}\label{hope}
\tanh \alpha = e^{-\beta \omega/2} \,,\qquad \alpha= \frac{1}{2}\ln \frac{1 +e^{-\beta \omega/2}}{1-e^{-\beta \omega/2}}\,.
\end{equation}
The thermal density matrix $\hat \upsilon_{th}$ in eq.~\eqref{density_thermal} is then produced by tracing out the auxiliary degree of freedom
\begin{equation}\label{TFDpureonemode}
\tr_2( \ket{\text{TFD}}_{12} \bra{\text{TFD}}_{12}) =  \(1- e^{-\beta\omega}\) \sum_n^\infty e^{-\beta \omega\,n} \ket{n}_1\!\bra{n}_1= \hat{\upsilon}_{th}(\beta,\omega) \,.
\end{equation}
However, we may also act with any unitary operator on the second oscillator in eq.~\reef{two_mode} and then this trace would yield an identical thermal density matrix. Hence we can write the most general two-mode purification of eq.~\eqref{Gaussian_decom} as
\begin{equation}\label{Fock_psi12}
\ket{\psi}_{12} = S_1(r)\,S_2(s) \, S_{12}(\alpha) \,\ket{0}_1\ket{0}_2\,,
\end{equation}
where we have introduced a second one-mode squeezing operator $S_2(s)$ to account for the freedom noted above in defining the purification of $\hat{\upsilon}_{th}(\beta,\omega)$. Eq.~\reef{Fock_psi12} is the most general two-mode purification using Gaussian states with real parameters. This can be seen by writing the position-space wavefunction
\begin{equation}\label{Pure12}
\begin{split}
&\qquad \qquad\qquad\quad \psi_{12}(x,y) \equiv \langle{x,y}\ket{\psi}_{12}=
\\
= &\sqrt{\frac{ \omega}{\pi}}\, e^{\frac{r+s}{2}} \,
\exp\[-\frac{\omega }{2}\(\cosh 2\alpha\,  (e^{2r}x^2+  e^{2 s}y^2) - 2\,e^{r+ s} x\,y \sinh 2\alpha\)\]\,.
\end{split}
\end{equation}
This wavefunction has precisely the same form as given in eq.~\eqref{wfunction1}, and we identify the parameters as
\beq\label{transform_paras}
\omega_1 =\omega\, e^{2r} \cosh 2\alpha\,,  \qquad \omega_2= \omega\, e^{2 s} \cosh 2\alpha\,, \qquad  k= -\omega \,e^{r+s} \sinh 2\alpha\,.
\eeq
Of course, substituting these relations into eq.~\eqref{ABA} yields the same values for $a,b$ as shown in eq.~\eqref{parameters}, where we have used the following identities following from eq.~\eqref{hope}
\beq
\cosh2\alpha=\frac{1}{\tanh(\beta\omega/2)}\,,\qquad
\sinh2\alpha=\frac1{\sinh(\beta\omega/2)}\ ,\qquad
\tanh^2\!\alpha=e^{-\beta\omega}\,.
\label{hope2}
\eeq
In the representation \eqref{Pure12}, the squeezing parameter $s$ encodes the freedom in defining the purification, which was previously captured by $k$ in eq.~\reef{pure1}. Hence with this description,
the purification complexity will be found by optimizing the usual pure state complexity over $s$.

To close here, we note that the expressions in eqs.~\reef{Pure12} and \reef{transform_paras}, as well as throughout the next section, can easily be written in terms of the parameter $\beta\omega$, which appears in the thermal density matrix \reef{density_thermal} using the relations \eqref{hope2}.
However, we continue to write our results in terms of the squeezing parameter $\alpha$ appearing in the purification \reef{Fock_psi12}.  One reason for this is that it simplifies the expressions for the limits of validity of the different regimes in our final result for the purification complexity --- see eq.~\reef{complexity_one_mode}. Further, $\alpha$ will also be a convenient parameter in our discussion of the purification complexity of a thermal density matrix (and in comparing it to the complexity of the thermofield double state \cite{Chapman:2018hou}) in section \ref{apply01}.

\subsection{Purification Complexity in the Diagonal Basis}
\label{sec:onemode}

According to the definition of purification complexity \cite{BrianMixedComplexity}, see also eq.~\eqref{def_pureC}, we evaluate the complexity of the mixed state by optimizing the purification to have the minimal circuit complexity as a pure state. We emphasize that we are simplifying this problem here by focusing on Gaussian mixed states and constraining ourselves to only considering Gaussian purifications. As mentioned in section \ref{subsec:compure}, throughout the following, we focus on the complexity defined with the $F_1$ cost function \reef{func2}. Recall that the $\mC_1$ complexity for Gaussian states was found to replicate the behaviours of holographic complexity most closely \cite{qft1,qft2,Chapman:2018hou}. However, as was also mentioned above, the $F_1$ cost function is basis dependent, and so we must specify that in this subsection, we evaluate the $\mC_1$ complexity in the diagonal basis. We will explore the results using the physical basis, which does not mix the original degree of freedom with the ancilla, in the next subsection.

The coefficient matrix $M_\mt{T}^{ab}$ in eq.~\eqref{wavematrix} for the purifying wavefunction $\ket{\psi_{12}}$ in eq.~\eqref{Pure12} is given by
\beq
M_\mt{T}^{ab}=\omega  \left(
\begin{array}{cc}
\ e^{2 r} \cosh 2 \alpha & -e^{r+ s} \sinh 2 \alpha \\
-e^{r+s} \sinh 2 \alpha & \ e^{2 s} \cosh 2 \alpha \\
\end{array}
\right).
\label{house}
\eeq
Again, the free parameter $s$ specifies a family of purifications of the same mixed state $\hat{\rho}_1$ in eq.~\eqref{Gaussian_decom}.
The prescription for evaluating the complexity of pure states was briefly reviewed in section \ref{subsec:compure}, and the $\mC_1$ complexity was given in eq.~\reef{complexity_pure}. Hence, the complexity of the Gaussian state \eqref{Fock_psi12}  becomes\footnote{We note again that the superscript `diag' indicates that we are working with the diagonal basis, \ie with gates acting on the eigenmodes  which mix the physical and auxiliary degrees of freedom.}
\begin{equation}
\mathcal{C}_{1}^{\text{diag}}\left(\ket{\psi}_{12}\right) = \frac 12 \left|\ln \frac{\omega_+}{\mu}\right|+\frac 12\left|\ln \frac{\omega_-}{\mu}\right|\,,
\label{walk1}
\end{equation}
where $\omega_\pm$ are the eigenvalues of the matrix $M^{ab}$,
\ie
\begin{equation}\label{eigenvalues_2modes}
\omega_{\pm}= \omega\, e^{r+s}\( \cosh\! 2\alpha\, \cosh (r-s) \pm \sqrt{\cosh^2 2\alpha\,\cosh^2 (r-s) -1} \)\,.
\end{equation}
Now according to the definition of purification complexity \eqref{def_pureC}, the complexity of the corresponding mixed state \reef{Gaussian_decom} is given by\footnote{Note that we only optimize over the purification of the target state. We assume that the reference state is fixed as a factorized Gaussian, where both the physical and auxiliary degrees of freedom appear with the same reference frequency. \label{foot7}}
\begin{equation}
\mC_{1}^{\text{diag}}\left(\hat{\rho}_1\right) = \text{min}_{s}\,  \mC_{1}^{\text{diag}} \left(\ket{\psi}_{12}\right) \,,
\label{walk2}
\end{equation}
where the dependence on the squeezing parameter $s$ is hidden in the eigenfrequencies $\omega_\pm$ in eq.~\reef{eigenvalues_2modes}.

Before proceeding, we must consider that there are three possibilities in eq.~\reef{walk1} depending on the relative magnitudes of the frequencies,
\begin{equation}\label{eq:cases1}
\begin{split}
\text{case 1: }~ \mathcal{C}_{1}^{\text{diag}} &= \frac 12 \ln \frac{\mu^2}{\omega_+\omega_-} = -( \bar r + \bar s)\,, \qquad \qquad\qquad \qquad\quad\, \mu \ge \omega_\pm\,,\\
\text{case 2: }~  \mathcal{C}_{1}^{\text{diag}} &=\frac 12 \ln \frac{\omega_+}{\omega_-}
= \cosh^{-1}\! \big[\cosh 2\alpha \cosh (\bar r - \bar s)  \big]\,,
\qquad  \omega_- \le\mu \le \omega_+,\\
\text{case 3: }~  \mathcal{C}_{1}^{\text{diag}} &=\frac 12  \ln \frac{\omega_+\omega_-}{\mu^2} = \bar r  + \bar s\, , \qquad \qquad\qquad \qquad\qquad\quad \mu \le \omega_\pm\,.
\end{split}
\end{equation}
These results have been simplified by the introduction of the shifted
squeezing parameters,
\begin{equation}\label{rbar}
 \bar r  \equiv r+\frac 12  \ln \frac{\omega }{\mu}\qquad
 {\rm and}\qquad
 \bar s \equiv s+\frac 12  \ln \frac{\omega }{\mu}\, .
\end{equation}
Now in order to perform the minimization in eq.~\reef{walk2}, we must identify the different regimes in eq.~\eqref{eq:cases1} in terms of the parameters of the purifying wavefunction,\footnote{This is done by analyzing the functional dependence of $\frac{\omega_\pm}{\mu}$ on $\cosh(2\alpha)\cosh(\bar r -\bar s)$ separately for each sign of $\bar r  + \bar s$.}
\begin{equation}\label{eq:limits23}
\begin{split}
&\text{case 1: }~ \tanh^2\! \alpha\le \tanh\bar r\, \tanh\bar  s \quad \text{and} \quad \bar r + \bar s \le 0\,, \\
&\text{case 2: }~\tanh^2\! \alpha \ge  \tanh\bar r\, \tanh\bar  s\,, \\
&\text{case 3: }~ \tanh^2\! \alpha \le  \tanh\bar r\, \tanh\bar  s \quad \text{and} \quad \bar r + \bar s \ge 0
\,.
\end{split}
\end{equation}
We see immediately that for case 1, both $\bar r$ and $\bar s$ will be negative, while for case 3, both will be positive. Let us next identify the value of $\bar s$ which yields a minimal complexity within each regime. For case 1, the complexity in eq.~\eqref{eq:cases1} is monotonically decreasing as a function of $\bar s$, and hence the minimal complexity is obtain by the maximal allowed value of $\bar s$, which can be found from eq.~\eqref{eq:limits23}. Similarly for case 3, the complexity is monotonically increasing with $\bar s$, and so the minimal complexity is associated with the minimal value of $\bar s$ allowed according to the inequalities in eq.~\eqref{eq:limits23}.\footnote{Recall that the boundary of the allowed values for $\bar s$ in each of these cases are precisely those for which $\omega_+=\mu$ or $\omega_-=\mu$ for case 1 and 3 respectively. Thus, the optimal purification in case 1 will have $\omega_+=\mu$, and similarly the optimal purification in case 3 will have $\omega_-=\mu$.} Incidentally, these two critical values of $\bar s$ coincide and are given by\footnote{Let us note that when $\bar{r}<0$ and $e^{2 \bar r} \cosh(2 \alpha)>1$, $\bar{s}_{\rm crit}$ is pushed to minus infinity. Therefore case 1 is not valid for any value of $\bar s$ and we are left with case 2 only. Similarly, for $\bar{r}>0$ and $e^{-2 \bar r} \cosh(2 \alpha)>1$, $\bar{s}_{\rm crit}$ is pushed to infinity, case 3 is not valid for any value of $\bar s$ and we are once again left with case 2 only.}
\begin{equation}\label{case13r2}
\text{case 1,3: }~\bar s_{\text{crit}} = \tanh^{-1}\!\left(\frac{\tanh^2 \alpha}{\tanh \bar r }\right)=
\frac 12  \ln\! \left(  {\frac{{e^{2 \bar{r} } \cosh 2 \alpha-1}}{{e^{2 \bar{r} }-\cosh 2 \alpha}}}\right)\,.
\end{equation}
Hence the minimal complexity in these two regimes is given by
\begin{equation}
\text{case 1,3: }~\mathcal{C}_{1}^{\text{diag}} =\pm \frac 12  \ln \!\left(    \frac{{1-e^{-2 \bar{r} }\cosh 2 \alpha}}{{e^{2 \bar{r} } \cosh 2 \alpha-1}}\right).
\end{equation}
For case 2, the minimal complexity is obtained by minimizing the function in eq.~\eqref{eq:cases1}, which leads to
\begin{equation}\label{case2compl}
\text{case 2: }~\bar s_{\text{min}} = \bar r  \qquad \longrightarrow \qquad\mathcal{C}_{1}^{\text{diag}}=2 \alpha\,.
\end{equation}

Now the final step is to clarify which one of these minimal complexities is the relevant one for given values of $\bar r $ and $\alpha$. If $\bar r <0$ for instance, both cases 1 and 2 could be in principle relevant, as long as $e^{2\bar r}\cosh(2\alpha)<1$. However for $0 > \bar{r} > -\alpha$, the lowest complexity is that in case 2 and hence the final answer for the purification complexity is given by eq.~\eqref{case2compl}. A similar argument can be given in the overlapping regime of cases 2 and 3. We finally arrive at the purification complexity \reef{walk2} for the one-mode Gaussian mixed states \reef{Gaussian_decom},
\begin{equation}\label{complexity_one_mode}
\mC_{1}^{\text{diag}}\(\hat{\rho}_1\)  =
\left\{
\begin{array}{lr}
\frac 12 \, \ln \!\left(    \frac{{e^{-2 \bar{r} }\cosh 2 \alpha }-1}{1-{e^{2 \bar{r} } \cosh 2 \alpha }}\right), &~~~ 0\le  \alpha  \le -\bar{r} \,,\\
\\
2\alpha , &~~~ \alpha \ge |\bar{r} |\,,\\
\\
\frac 12\, \ln\! \left(    \frac{{e^{2 \bar{r} } \cosh 2\alpha-1}}{{1-e^{-2 \bar{r} }\cosh 2\alpha}} \right) , &~~~ 0\le \alpha \le \bar{r} \,.
\end{array}
\right.
\end{equation}
One interesting point about this result is that the complexity of the mixed state $\hat{\rho}_1$ generally depends on both the thermal parameter $\beta\omega$ (or alternatively, $\alpha$), and the shifted squeezing parameter $\bar{r}$ (which has absorbed the ratio $\mu/\omega$), whereas  the  (entanglement)  entropy of this state only depends on the combination $\beta\omega$. We return to this point in section \ref{sec_EE}.

At this point, we can also point out the various benefits of the parametrization introduced in subsection \ref{subsec:altdesc}. First, $\alpha$ and  $r$ are natural dimensionless parameters associated with the thermal state and its squeezing. The state described by those parameters is always physical, which means we do not need to impose extra constraints on those parameters. In particular, the density matrix is automatically positive semi-definitive and hermitian for any positive temperature and frequency. For $r=0$, the density matrix corresponds to a thermal state at temperature $1/\beta$ for a single harmonic oscillator of frequency $\omega$. More generally, for non-zero $r$, one can think of it as the thermal density matrix with an inverse temperature $\beta'=e^{-2r}\beta$ for a harmonic oscillator of frequency $\omega' = e^{2r}\omega$. That is, using eq.~\eqref{parameters}, one can easily show that
\begin{equation}\label{scaling_rels}
\hat{\rho}_1 = \hat S_1(r) \, \hat{\upsilon}_{th} (\beta,\omega)\,\hat S_1^\dagger(r) = \hat{\upsilon}_{th} (e^{-2r}\beta, e^{2r}\omega)\,.
\end{equation}
In addition, these parameters simplify the analytical analysis of the minimization, and bring the final result for the complexity and, in particular, the limits of validity of each regime into a (much more) compact form. Further, the physical meaning of the purification becomes clear --- in order to purify the Gaussian state, we only need to purify its thermal component, and the extra freedom in the optimization comes from the squeezing operator $S_2(s)$ on the ancilla. Finally, the parametrization is closely related to the thermofield double state at temperature $1/\beta$ which is defined by $r =s=0$; and for $r=s\neq0$, it is the thermofield double at temperature $1/\beta'$ of a harmonic oscillator of frequency $\omega'$ (where $\beta'$ and $\omega'$ are the same as defined above).

\subsection{Purification Complexity in the Physical Basis} \label{fizz}

Next, we explore the sensitivity of our previous results to the choice of the basis. In particular, we re-examine the purification complexity of the one-mode mixed Gaussian state, defined in eq.~\reef{dense1} or \eqref{Gaussian_decom},  with the $F_1$ cost function but using the physical basis. That is, here the gates implicitly act directly on the original and auxiliary degrees of freedom, rather than on the linear combinations comprising the eigenmodes of $M_\mt{T}$ describing the purification. This change of basis is accomplished with the orthogonal transformation described in eq.~\reef{rotate2pos}.

To begin, we re-express the wavefunction matrix \eqref{house} for the purification  $\ket{\psi_{12}}$ in terms of the shifted squeezing parameters in eq.~\eqref{rbar} as follows
\begin{equation}\label{MabPos}
M_\mt{T}^{ab}=\mu \left(
\begin{array}{cc}
\ e^{2 \bar{r}} \cosh 2 \alpha & -e^{\bar{r}+ \bar{s}} \sinh 2 \alpha \\
-e^{\bar{r}+ \bar{s}} \sinh 2 \alpha & \ e^{2 \bar{s}} \cosh 2 \alpha \\
\end{array}
\right)\,.
\end{equation}
Similarly, the eigenvalues \eqref{eigenvalues_2modes} become
\begin{equation}\label{round2}
\omega_{\pm}= \mu\, e^{\bar{r}+ \bar{s}}\( \cosh\! 2\alpha\, \cosh (\bar{r}- \bar{s}) \pm \sqrt{\cosh^2 2\alpha\,\cosh^2 (\bar{r}- \bar{s}) -1} \)\,.
\end{equation}
Now, in order to evaluate the $\mathcal{C}_1^{\mt{phys}}$ complexity as in eq.~\eqref{rotate2pos}, we need to determine the orthogonal transformation which brings the matrix \eqref{MabPos} to its diagonal form, see eq.~\eqref{Odiag}. That is,
\begin{equation}\label{thetadef}
 \tilde M_\mt{T}=\left(
\begin{array}{cc}
\omega_- & 0  \\
0 & \omega_+  \\
\end{array}
\right)=
O^T \, M_\mt{T} \, O\qquad{\rm with}\quad
 O \equiv\left(
\begin{array}{cc}
\cos \theta & -\sin \theta  \\
\sin \theta & \cos \theta  \\
\end{array}
\right)\,,
\end{equation}
where $\theta \in [0,\frac{\pi}{2}]$ and
\begin{equation}\label{rounder}
\begin{split}
& \sin \theta =\frac{1}{\sqrt{X^2+1}}\,,\qquad
\cos \theta =\frac{X}{\sqrt{X^2+1}}\,, \\
&X\equiv \frac{1}{\sinh 2\alpha} \left(\sqrt{\cosh ^22 \alpha  \cosh ^2(\bar r-\bar s)-1}-\cosh 2 \alpha \sinh (\bar r- \bar s)\right)  \ge 0\,.
\end{split}
\end{equation}
The next step is to rotate the generator $\tilde H$ in eq.~\eqref{Hgen}, {\it i.e.,}
\begin{equation}
\tilde{H}= \frac{1}{2} \left(
\begin{array}{cc}
\ln \frac{\omega_-}{\mu} & 0\\
0 & \ln \frac{\omega_+}{\mu} \\
\end{array}
\right)\, ,
\end{equation}
as in eq.~\eqref{rotate2pos}, which defines the circuit generator in the physical basis\footnote{As an aside, we note that the circuit generator $H$ is easily expressed in terms of the ``relative wavefunction" matrix $M_{\mt T}  M_\mt{R}^{-1}$  directly in the physical basis as
$H =  \frac{1}{2}\ln \left(M_{\mt T}  M_\mt{R}^{-1} \right)$.}
\begin{equation}\label{position_H}
H= O \, \tilde{H} \, O^T = \frac{1}{2} \left(
\begin{array}{cc}
\cos^2\theta \ln \frac{\omega_-}{\mu} + \sin^2 \theta \ln \frac{\omega_+}{\mu}   &- \sin \theta\cos \theta\,\ln \frac{\omega _+}{\omega _-}\\
-\sin \theta\cos \theta\,\ln \frac{\omega _+}{\omega _-}&  \cos^2\theta \ln \frac{\omega_+}{\mu} + \sin^2 \theta \ln \frac{\omega_-}{\mu}
\end{array}
\right)\,.
\end{equation}
Again using eq.~\eqref{rotate2pos},  $\mC_1$ for the purified state corresponding to the wavefunction matrix \eqref{MabPos} in the physical basis becomes
\begin{equation}
\begin{split}\label{complexityPosition123}
\mC_1^{\mt{phys}} (\ket{\psi_{12}})
=\frac 14\, \bigg( 2\, \sin 2\theta &  \ln  \frac{\omega_+}{\omega_-}  + \left|  \ln \frac{\omega_+\omega_-}{\mu^2}-\cos 2\theta \ln  \frac{\omega_+}{\omega_-}  \right|
\\ &\left.
+\left|  \ln \frac{\omega_+\omega_-}{\mu^2}+\cos 2\theta \ln  \frac{\omega_+}{\omega_-}  \right| \, \right) . \\
\end{split}
\end{equation}

It will be convenient to optimize the purification by varying the angle $\theta$ rather than working with the squeezing parameter $s$.  Hence we use eq.~\reef{rounder} to replace
\begin{equation}
\label{eq:stotheta}
\sinh(\bar r-\bar s)=-
\tanh 2 \alpha \, \cot 2 \theta\,.
\end{equation}
Note that the sign of $\sinh(\bar r-\bar s)$ will be positive for $\theta>\pi/4$ and negative for $\theta<\pi/4$. Combining this expression with eqs.~\reef{round2} and \reef{rounder}, we can also express the other factors in eq.~\eqref{complexityPosition123} in terms of $\theta$ as follows
\begin{equation}
\begin{split}
\label{eq:omegatotheta}
&\frac 12\, \ln \frac{\omega_+}{\omega_-}
= \cosh^{-1}\! \(\cosh 2\alpha \cosh (\bar r-\bar s) \)
=\sinh ^{-1}(\sinh 2 \alpha \, \csc 2 \theta ) \,,
\\
&\frac 12\,  \ln \frac{\omega_+\omega_-}{\mu^2} =\bar{r}+\bar{s}= 2\bar{r} +\sinh ^{-1}(\tanh 2 \alpha \, \cot 2 \theta )\,.\\
\end{split}
\end{equation}
Using these expressions and examining eq.~\eqref{complexityPosition123} according to the different possible signs in the absolute values, we obtain
\beqa
(a)~--\,:&& \mC_1^{\mt{phys}}=- 2\bar{r}+\sin 2 \theta \,\sinh ^{-1}\!\( \frac{\sinh 2 \alpha }{\sin 2\theta} \)  -\sinh ^{-1}(\tanh 2 \alpha \, \cot 2 \theta )
\nonumber\\
(b) ~+-\,:&& \mC_1^{\mt{phys}}= \sqrt{2} \sin\( 2\theta -\frac{\pi}{4}\)\, \sinh ^{-1}\!\( \frac{\sinh 2 \alpha }{\sin 2\theta} \)   \label{complexitypositionabcd}\\
(c) ~-+\,:&& \mC_1^{\mt{phys}}= \sqrt{2} \sin\( 2\theta +\frac{\pi}{4}\)\, \sinh ^{-1}\!\( \frac{\sinh 2 \alpha }{\sin 2\theta} \)  \nonumber\\
(d) ~++\,:&& \mC_1^{\mt{phys}}=  2\bar{r}+\sin 2 \theta \,\sinh ^{-1}\!\( \frac{\sinh 2 \alpha }{\sin 2\theta} \) +\sinh ^{-1}(\tanh 2 \alpha \, \cot 2 \theta ) \nonumber
\eeqa
where for instance $+-$ indicates that the sign of the expression inside the first absolute value in eq.~\eqref{complexityPosition123} is positive and the sign of the expression inside the second absolute value is negative.
Finally, the purification complexity in the physical basis for the one-mode Gaussian mixed state is given by minimizing this expression with respect to the free parameter $\theta$
\begin{equation}\label{purification_C_pos}
\mC_1^{\mt{phys}} \( \hat{\rho}_1\)
=\text{min}_{\theta}\, \mC_{1}^{\mt{phys}} \left(\ket{\psi}_{12}\right)\,.
\end{equation}

Unfortunately, the exact analytical minimization of eq.~\eqref{purification_C_pos} is not possible since it would require solving a transcendental equation. Hence, in order to develop some intuition, let us consider the simple case $\mu=\omega e^{2r}$, \ie $\bar{r}=0$  where the purification complexity reduces to
\begin{equation}\label{SimpleCasePos}
\mC_1^{\mt{phys}}  =
\left\{
\begin{array}{lr}
 \sqrt{2} \sin\( 2\theta -\frac{\pi}{4}\) \,\sinh ^{-1}\!\( \frac{\sinh 2 \alpha }{\sin 2\theta} \)  \qquad :&\! +-\,,\\
\\
 \sqrt{2} \sin\( 2\theta +\frac{\pi}{4}\) \,\sinh ^{-1}\!\( \frac{\sinh 2 \alpha }{\sin 2\theta} \) \qquad :&\  -+\,.\\
\end{array}
\right.
\end{equation}
That is, $\mC_1^{\mt{phys}}$ is given be either cases (b) or (c) in eq.~\reef{complexitypositionabcd}. We are able to rule out cases (a) and (d) (\ie $++$ and $--$) by verifying that the product of the terms in the absolute values in eq.~\eqref{complexityPosition123} is negative using the identity\footnote{This identity can be verified separately in each region $0<\theta<\pi/4$ and $\pi/4<\theta<\pi/2$ by using the fact that for $\alpha=0$ we obtain an equality together with the fact that the derivative of the left hand side with respect to $\alpha$ has a definite sign in each region, namely, it is negative  for $0<\theta<\pi/4$ and positive for $\pi/4<\theta<\pi/2$.}
\begin{equation}
\sinh^{-1}\left(\tanh 2\alpha\,|\cot 2 \theta\,|\,\right)-\,|\cos 2\theta\,| \,\sinh^{-1}\left(\sinh 2\alpha\,|\csc 2\theta\,|\,\right)
<0\,.
\end{equation}

To proceed further, let us point out an interesting way to identify which set of signs of the terms in the absolute values is relevant for the evaluation of complexity. We can regard the expressions for each of the cases in eq.~\reef{complexitypositionabcd} as evaluating the expression in eq.~\reef{complexityPosition123}, but without the absolute values, rather we are inserting the specified signs in front of the last two terms. Hence for a given value of $\theta$, we can evaluate all four of these expressions. However, the correct result will correspond to the largest value because in this case with the specified signs, both of the second and third terms must be making a positive contribution to the complexity, as required by the absolute values in eq.~\reef{complexityPosition123}. Using this reasoning in eq.~\eqref{SimpleCasePos} with $\bar r=0$, we can see that when $\theta<\pi/4$, case (c) is the correct choice, while for $\theta>\pi/4$, the relevant case is (b). This fact will also be useful when performing the numerical analysis of more general cases later on.
We may also use the identity $ a \sinh^{-1}(x)>\sinh^{-1} (a x)$ for $a>1$, $x>0$,\footnote{This is due to the fact that $\sinh^{-1}(x)$ is concave down.} with $a=\sin(2\theta)\pm \cos(2\theta)>1$ for $0<\theta<\pi/4$ and $\pi/4<\theta<\pi/2$ respectively, as well as the monotonicity of $\sinh^{-1}(x)$, in order to demonstrate that the minimal value for the complexity is obtained for $\theta=\frac{\pi}{4}$ (which corresponds to $\bar{s}=\bar{r}=0$), see eq.~\eqref{eq:stotheta}. This yields the following purification complexity
\begin{equation}
\mC_1^{\mt{phys}} \( \hat{\rho}_1(\bar{r}=0)\) = \text{min}_{\theta}\,  \mC_{1}^{\mt{phys}} \left(\ket{\psi}_{12}\right)= 2\alpha \,.
\end{equation}
We may also point out, that for $r=0$, this is simply the TFD purification of a state with temperature $\beta$ and frequency $\omega=\mu$. The addition of the squeezing parameter $r$ leads to the TFD purification of a state with temperature $\beta'=e^{-2r }\beta$ and frequency $\omega'=\omega e^{2r}$ which is equal to the reference frequency $\mu$, according to the logic described around eq.~\eqref{scaling_rels}.

\begin{figure}[htbp]
    \includegraphics[width=2.8in]{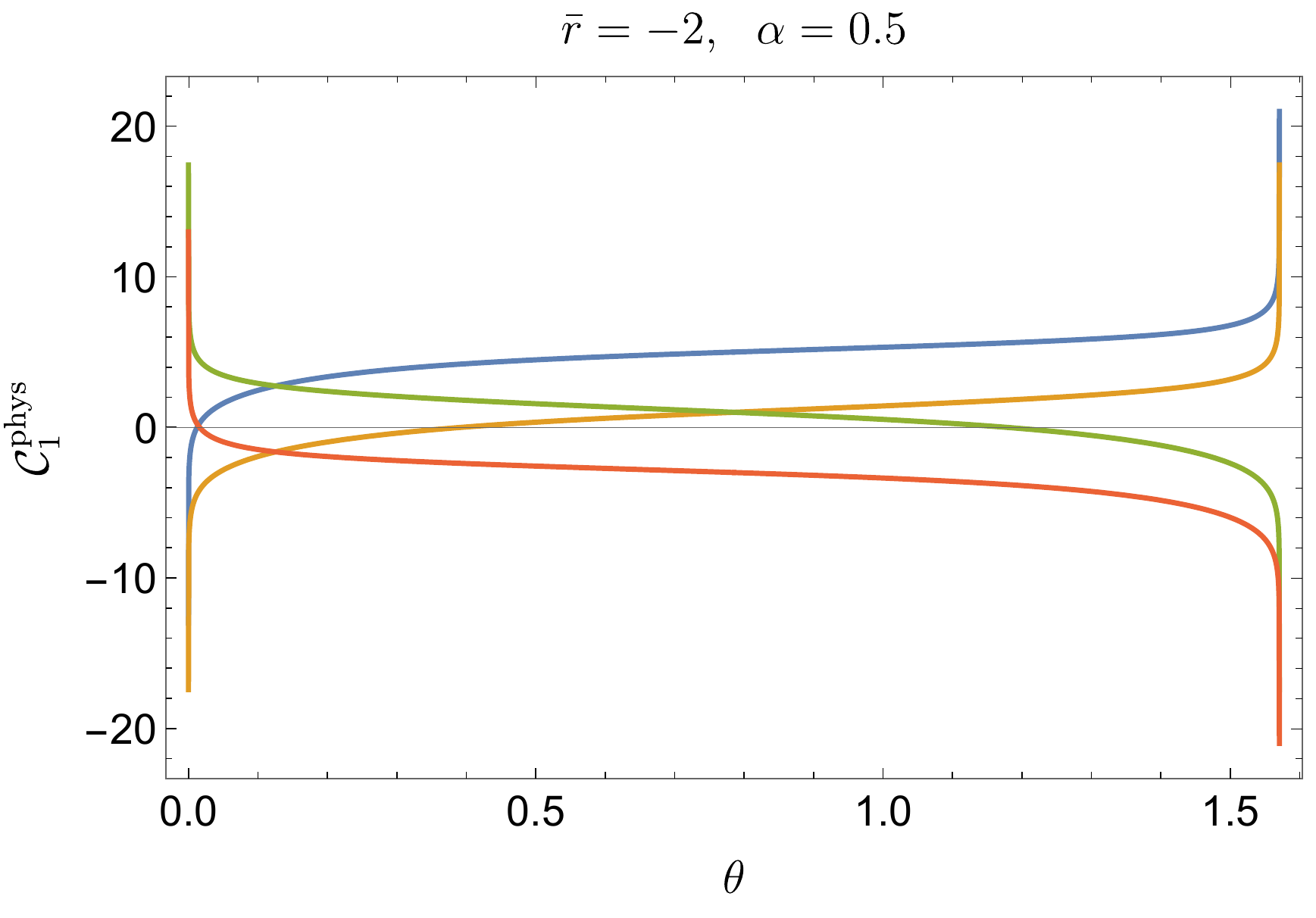}
    \includegraphics[width=2.8in,left]{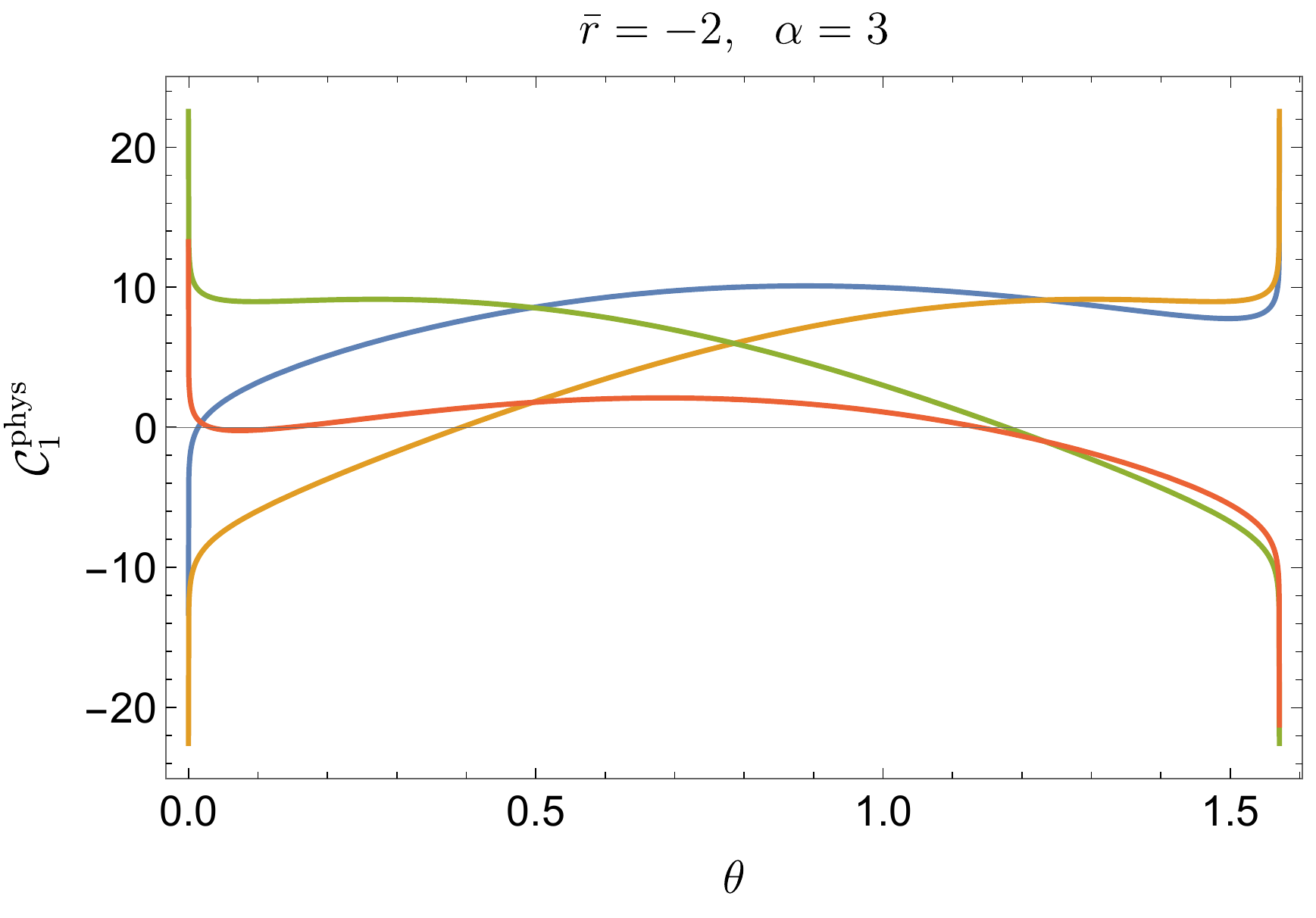}
   \includegraphics[width=2.8in]{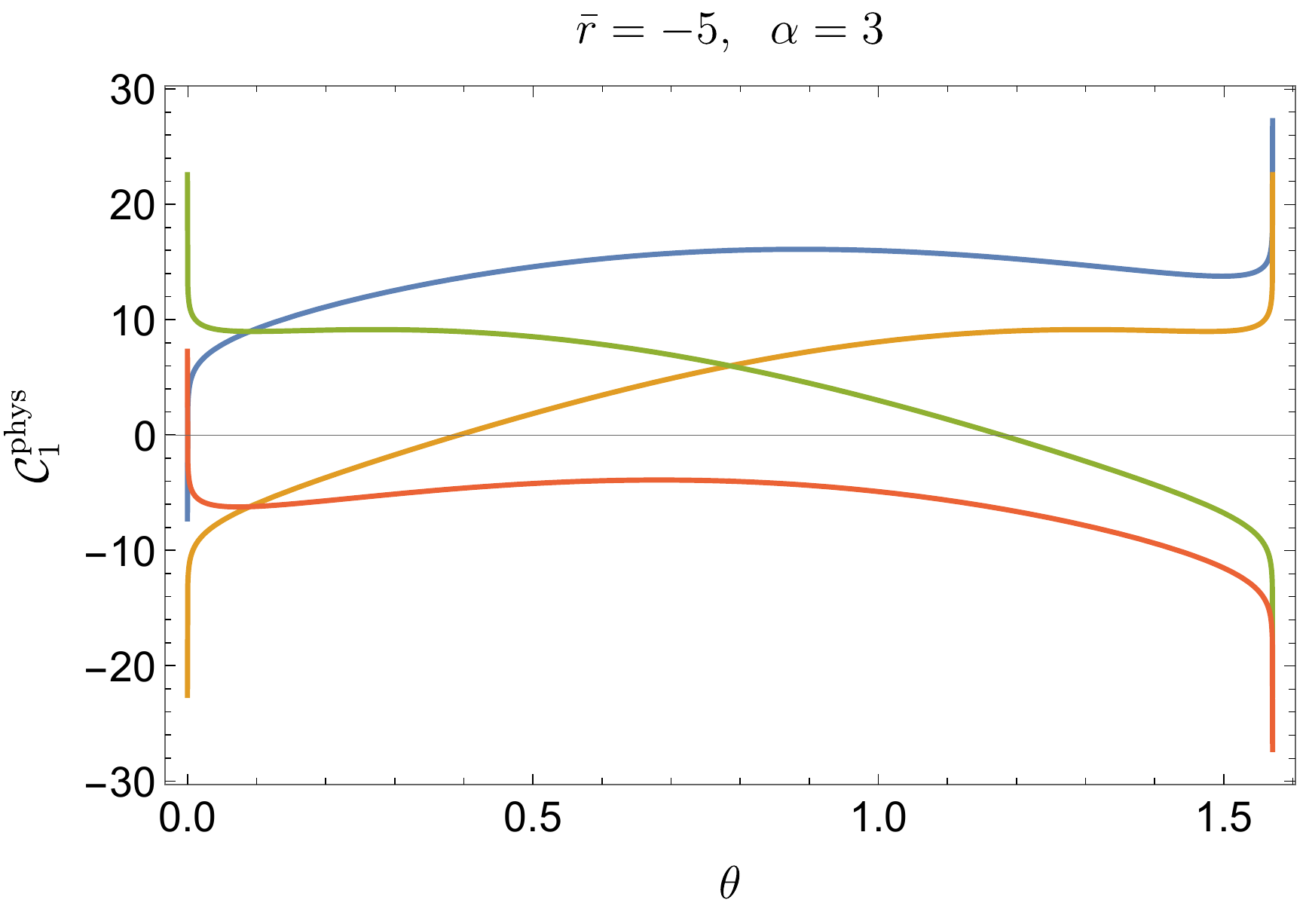}
    \includegraphics[width=3.2in]{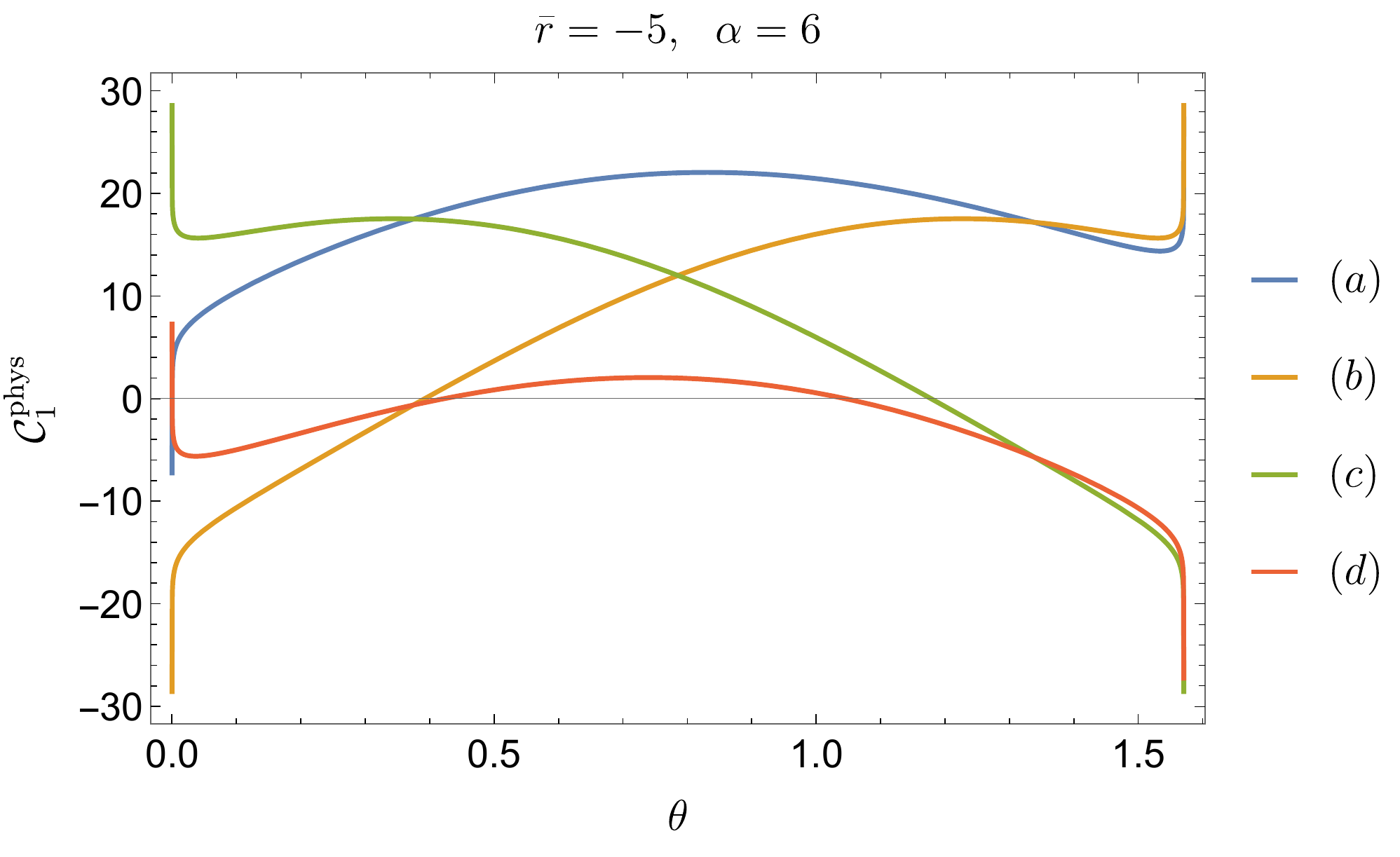}
    \caption{Possible values for the pure state complexity $\mC_1^{\mt{phys}}(|\psi_{12}\rangle)$ in the physical basis as a function of $\theta$, for all possible sign combinations according to eq.~\eqref{complexitypositionabcd} for fixed values of $\bar{r}$ and $\alpha$. The complexity of the mixed state purified by $|\psi_{12}\rangle$ is obtained by minimizing over the uppermost envelope of each of these plots.}\label{fig:complexity_pos}
\end{figure}

Next, we return to the general case for which we examine the optimization \reef{purification_C_pos} numerically. Without loss of generality, let us assume that $\mu \ge \omega e^{2r}$, or equivalently $\bar{r}<0$.\footnote{Note that the system is symmetric under the exchange $\bar r \rightarrow -\bar r$, $\theta \rightarrow \frac{\pi}{2}-\theta$, and (a)$\leftrightarrow$(d), (b)$\leftrightarrow$(c). As a consequence, though the details of the analysis will slightly vary, the value of the complexity obtained by minimizing \eqref{complexityPosition123} will only depend on the absolute value of $\bar r$.}  We will try to use the same logic as above in order to identify the ranges of $\theta$ in which the different sets of signs in eq.~\eqref{complexitypositionabcd} are valid. It is useful to start by looking at a plot of all possible sign combinations given by the four cases (a)--(d), for all values of $0<\theta<\pi/2$ --- see figure \ref{fig:complexity_pos}. As noted before, the relevant sign combination for the complexity will always be the highest of the four lines, since that possibility takes into account the correct (positive) signs for all the absolute values. Therefore, we must minimize the complexity over the uppermost envelope of the plots in figure \ref{fig:complexity_pos}. Let us proceed with this graphical understanding in mind. For a non-zero value of $\bar r<0$, the different cases in eq.~\eqref{complexitypositionabcd} are shifted up (case (a)), down (case (d)) or not modified (cases (b) and (c)). Using the same inequalities mentioned above, it is straightforward to see that case (d) becomes irrelevant and is smaller than at least one of the other cases for all values of $\theta$.
Therefore, in each region of $\theta$, we should consider two competing sign combinations:
\begin{equation}
\mC_1^{\mt{phys}}\left(\ket{\psi}_{12}\right) =\left\{
\begin{array}{lr}
{\rm case}\ (a)\ {\rm or}\ (b)\quad{\rm for}\ \frac{\pi}{4}\le \theta \le \frac{\pi}{2} \,,\\
\\
{\rm case}\ (a)\ {\rm or}\ (c)\quad{\rm for}\ 0\le \theta \le \frac{\pi}{4} \,.\\
\end{array}
\right.
\end{equation}

We have examined these cases numerically, see figure \ref{fig:complexity_pos}. The minimal purification complexity is obtained for a value of $\theta$ that either lies at minimal points of the curves (a), (b) or (c) or at the intersections of the curves (a) and (c) or of the curves (a) and (b) depending on the values of $\bar r$ and $\alpha$ considered. These values can be identified by solving transcendental equations. For example, in the regime where $\alpha $ is small or $-\bar{r}$ is large,  the minimal complexity is obtained at the point where the curves for cases (a) and (c) intersect, which corresponds to solving the equation
\begin{equation}\label{gumby7}
\begin{split}
-2\bar{r}= \sinh ^{-1}(\tanh 2 \alpha \, \cot 2 \theta_c )+ \cos 2\theta_c\,\sinh ^{-1}\!\(\frac{\sinh 2 \alpha }{\sin 2\theta_c} \)\,,
\end{split}
\end{equation}
and the purification complexity reads
\begin{equation}\label{pos_ac}
\mC_{1,c}^{\mt{phys}} \( \hat{\rho}_1\)=(\sin 2\theta_c + \cos 2\theta_c)\,\sinh ^{-1}\!\(\frac{\sinh 2 \alpha}{\sin 2\theta_c} \) \,.
\end{equation}

\begin{figure}[htbp]
    \centering\includegraphics[width=4.5in]{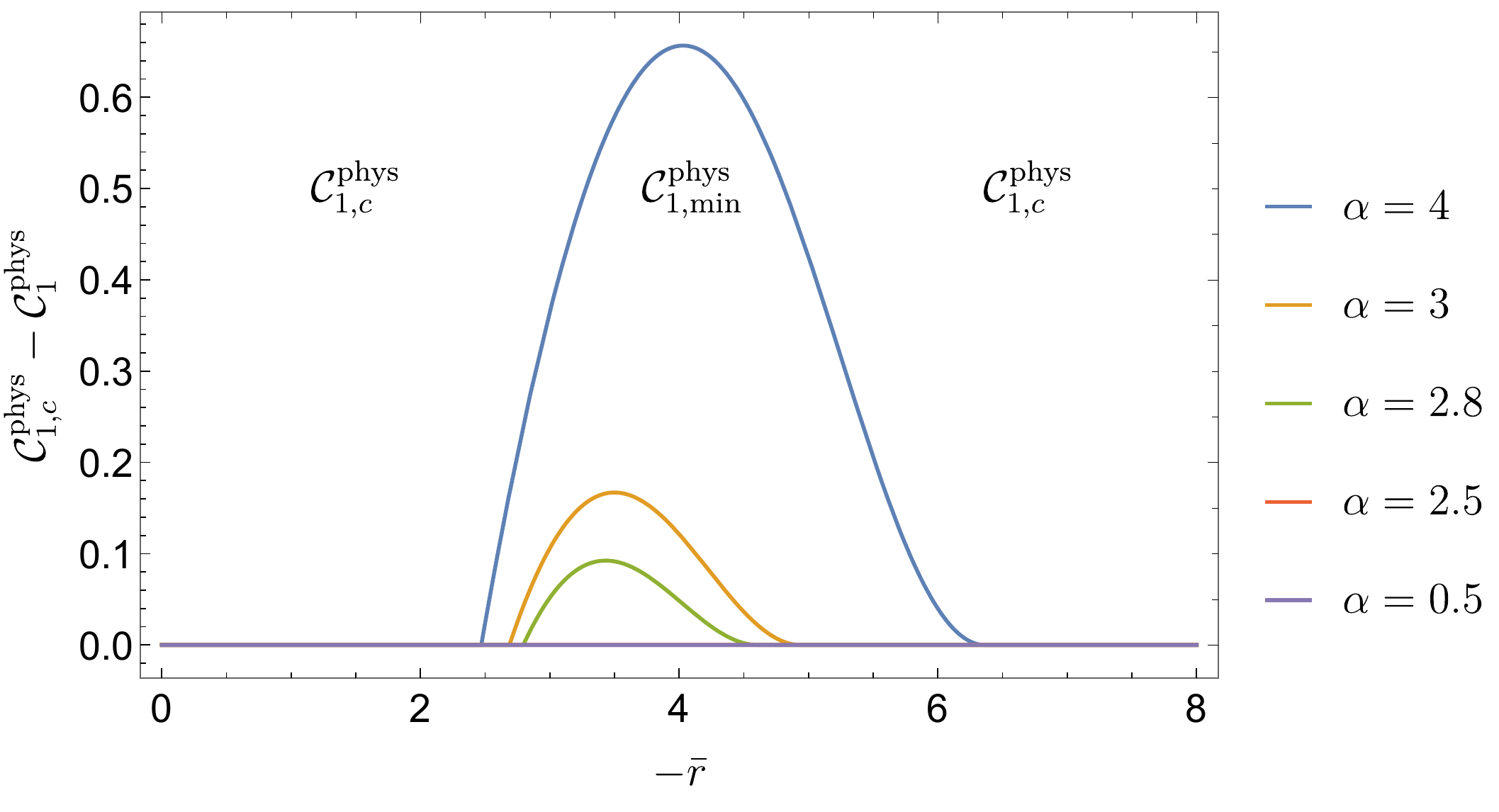}
    \caption{The difference between the complexity obtained for $\theta_c$ at the intersection of cases (a) and (c) and the exact purification complexity of one-mode Gaussian states in the physical basis $\mC_{1,c}^{\mt{phys}} \( \hat{\rho}_1\)-\mC_1^{\mt{phys}} \( \hat{\rho}_1\)$ as a function of $\bar{r}$ for some fixed values of $\alpha$. We see that the complexity obtained at the intersection between cases (a) and (c) with $\mC_{1,c}^{\mt{phys}} \( \hat{\rho}_1\)$ in eq.~\eqref{pos_ac} ceases to be optimal for some region of the parameter $\bar{r}$ for large enough values of $\alpha$.}\label{Cc-C1}
\end{figure}

When the parameter $\alpha$ is large enough, we can find that the minimal complexity corresponds to the minimal point along the curve (c) rather than to the intersection of curves (c) and (a). This is illustrated in figure \ref{Cc-C1} which plots the difference $\mC_{1,c}^{\mt{phys}}-\mC_{1}^{\mt{phys}}$. The non-zero values in the middle of this plot mean that the minimization is obtained at the local minimal point of curve (c) where the complexity is given by
\begin{equation}
\mC_{1}^{\mt{phys}} \( \hat{\rho}_1\) = \sqrt{2} \sin\( 2\theta_{\mt{min}} +\frac{\pi}{4}\)\, \sinh ^{-1}\!\( \frac{\sinh 2 \alpha }{\sin 2\theta_\mt{min}} \)
\end{equation}
where
\begin{equation}
 \partial_{\theta} \( \sin\( 2\theta +\frac{\pi}{4}\)\, \sinh^{-1}\!\( \frac{\sinh 2 \alpha }{\sin 2\theta} \) \) \bigg|_{\theta_{\mt{min}}} =0, \qquad 0 \le  \theta_{\text{min}} \le \theta_c  \, .
\end{equation}

\begin{figure}[htbp]
    \centering\includegraphics[width=4.5in]{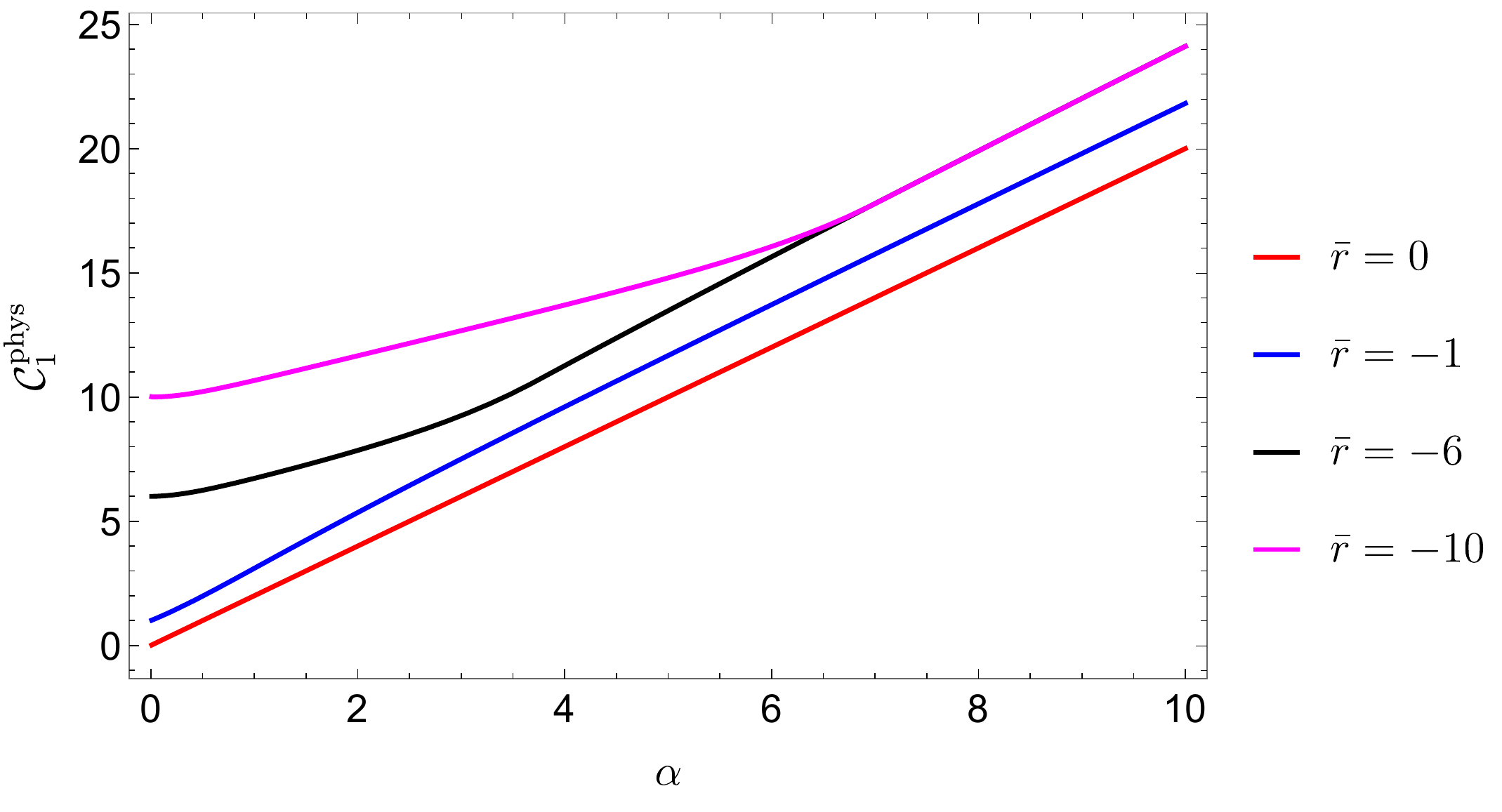}
    \caption{Purification complexity  of one-mode Gaussian states in the physical basis $\mC_{1}^{\mt{phys}} \( \hat{\rho}_1\)$ as a function of $\alpha$ for some fixed values of $\bar r$. The fact that the curves with $\bar r=-6$ and $\bar r=-10$ coincide after a certain value of $\alpha$ is due to the fact that this minimization is obtained at the minimum of case (c) which is $\bar r$ independent.}\label{complexity_LR}
\end{figure}

Although we cannot solve for $\theta_c$ or $\theta_\mt{min}$ analytically, we may evaluate them numerically. Similar equations can be written for other possible positions of the minimum.
Figure \ref{complexity_LR} contains results for $\mC_1^{\mt{phys}} \( \hat{\rho}_1 \)$ from numerical minimization with fixed value of $\bar{r}$.

\subsubsection{Differences between the two bases}

We must stress again that with the physical basis, the gates act directly on the original and auxiliary degrees of freedom. This contrasts with the diagonal basis where the gates act on the linear combinations comprising the eigenmodes of $M_\mt{T}$ describing the purification. In particular, then, one of the diagonal generators is precisely aligned with the generator $\tilde H$ of the optimal circuit in eq.~\reef{Hgen}. As a consequence, one expects that with other choices of basis, the purification complexity of mixed states (as well as the complexity of pure states) will not be smaller than in the diagonal basis.

Comparing our results for of the one-mode Gaussian mixed states   in the physical basis \eqref{complexityPosition123}  to those in the diagonal basis \eqref{eq:cases1}, we can show
\begin{equation}
\begin{split}
\mC_1^{\mt{phys}}
&=\frac 12 \(  \sin 2\theta \ln  \frac{\omega_+}{\omega_-}  +\left| \cos 2\theta \frac 12\ln  \frac{\omega_+}{\omega_-} -\frac 12 \ln \frac{\omega_+\omega_-}{\mu^2} \right| +\left| \frac 12 \ln \frac{\omega_+\omega_-}{\mu^2}+\cos 2\theta \frac 12\ln  \frac{\omega_+}{\omega_-}  \right|  \) \,, \\
&\ge  (\sin 2\theta +|\cos 2\theta |)\,\frac 12  \ln  \frac{\omega_+}{\omega_-}   \ge \mC^\mt{diag}_1 (\text{case 2})\,,
\end{split}
\end{equation}
and
\begin{equation}
\begin{split}
\mC_1^{\mt{phys}}
&=\frac 12 \(  \sin 2\theta \ln  \frac{\omega_+}{\omega_-}  +\left| \frac 12 \ln \frac{\omega_+\omega_-}{\mu^2}-\cos 2\theta \frac 12\ln  \frac{\omega_+}{\omega_-}  \right| +\left| \frac 12 \ln \frac{\omega_+\omega_-}{\mu^2}+\cos 2\theta \frac 12\ln  \frac{\omega_+}{\omega_-}  \right|  \) \,, \\
&\ge  \sin 2\theta \,\frac 12  \ln  \frac{\omega_+}{\omega_-}  +\frac{1}{2} \left|  \ln  \frac{\omega_+\omega_-}{\mu^2}  \right|  \ge \mC^\mt{diag}_1 (\text{case 1,3})\,,
\end{split}
\end{equation}
where we used the inequality $|a-c|+|c-b| \ge |a-b|$. Hence, we conclude
\begin{equation}
\mC_1^{\mt{phys}} (\ket{\psi_{12}}) \ge  \mC^\mt{diag}_1 (\ket{\psi_{12}}) \,, \qquad  \mC_1^{\mt{phys}} \( \hat{\rho}_1\) \ge \mC^\mt{diag}_1 \( \hat{\rho}_1\) \,,
\end{equation}
as expected. It is also easy to demonstrate that the latter inequality holds in various examples by numerical minimization. 

\section{Optimal Purification of Mixed Gaussian States}\label{sec:manyho}

In this section, we generalize the discussion of Gaussian mixed states to systems with an arbitrary number of (bosonic) modes. We also examine some fundamental issues related to the purification of such mixed states. The definition of purification complexity \reef{def_pureC} suggests that we should optimize the cost over all possible purifications, however, the procedure that we adopted in the previous section is only to optimize over Gaussian purifications.\footnote{Further as in footnote \ref{foot7}, we will assume that the reference state is a fixed factorized Gaussian, where both the physical degrees of freedom and the ancillae appear with the same reference frequency.} We maintain this same approach here and throughout the following, and leave it to a future project to test whether more general purifications yield a lower complexity.

In the previous section, our mixed state \reef{dense1} described a single physical degree of freedom, and it was purified by introducing a single ancilla. When trying to evaluate the purification complexity for a mixed state with many modes, one must ask the question of how many ancillae are needed to produce the minimal complexity. In subsection \ref{sec:purificationX}, we begin by identifying the minimum number of extra degrees of freedom that are needed to purify a given mixed state. We will refer to such purifications with only the essential number of ancillae as {\bf essential purifications}. Note that as we introduce additional ancillae, the number of free parameters over which one would optimize increases, and so one might expect that this will also reduce the corresponding purification complexity. However, we will argue that this intuition is incorrect for Gaussian mixed states and that the optimal purification should be an essential purification.
Further, in identifying the minimum number of ancillae, our approach is to construct a `diagonal' basis in which the density matrix takes a canonical form where each eigenmode is separately either in a mixed or pure state. Each of the mixed state modes can then be purified by a single ancilla in a purification, which we refer to as a {\bf mode-by-mode purification}. We identify the circumstances in which these mode-by-mode purifications are optimal and also find that they still give a good approximation to the true optimal purification in certain situations.

Our discussion is divided into two parts: In subsection \ref{optimal}, we analyze the complexity in the diagonal basis, and in subsection \ref{subsec3phys}, we repeat the same analysis in the physical basis. This last discussion also includes a precise definition of what we mean by physical basis complexity in systems with an arbitrary number of modes. The conclusions about optimality are similar in both bases. Some technical details and extensions related to the topics of this section appear in two appendices: appendix \ref{app:purification} explains how to extend the alternative parametrization of section \ref{subsec:altdesc} to the case of multi-mode Gaussian states and appendix \ref{app:numerics} contains a numerical check of the analytic results of section \ref{threeA}.

This section then provides useful background for section \ref{apply01} and \ref{apply02} where we consider the purification complexity in some examples of Gaussian mixed states in quantum field theory. In particular, the purifications over which we optimize there will be both essential and mode-by-mode purifications. However, let us add that the reader might simply read section \ref{sec:purificationX} as well as the introduction for section \ref{subsec3phys}. These sections will introduce the indispensable elements of our notation, which will be needed to read sections \ref{apply01} and \ref{apply02}, where we discuss applications to a free scalar field theory.

\subsection{Purifying General Gaussian States}\label{sec:purificationX}
In this subsection, we study Gaussian purifications of Gaussian density matrices with an arbitrary number of modes. The discussion will follow closely the one in \cite{pure2}, and as before, we will focus on density matrices and wavefunctions with real parameters for simplicity. We start with the wavefunction of a pure Gaussian state
\begin{equation}\label{Gaussian_AB}
\Psi_{\mA\mA^c} = \mathcal{N}_{\mA\mA^c}\ \exp\! \[  -\frac{1}{2} (\vec q_{\mA},\vec q_{\mA^c}) \left(
\begin{array}{cc}
\Gamma & K \\
K^T & \Omega \\
\end{array}
\right) \left(
\begin{array}{c}
\vec q_{\mA} \\
\vec q_{\mA^c} \\
\end{array}
\right)        \]\,,
\end{equation}
where the degrees of freedom were divided into the ``inside'' region $\mA$ containing the $N_{\mA}$ coordinates $\vec q_{\mA}$, and the ``outside'' region $\mA^c$ containing the $N_{\mA^c}$ coordinates $\vec q_{\mA^c}$. The wavefunction matrix in eq.~\eqref{Gaussian_AB} has to be positive definite in order for the wavefunction to be normalizable. The square matrices $\Gamma$ and $\Omega$ are real, symmetric and positive definite.\footnote{Note that sub-matrices of positive definite matrices are also positive definite. It will also be important that positive definite matrices are invertible.} Further, the rectangular $N_{\mA}\times N_{\mA^c}$ matrix $K$ is also real\footnote{The restriction to real matrices here and above are a choice that we impose to simplify our analysis. In contrast, the positivity of $\Gamma$ and $\Omega$ is required to ensure that the wavefunction is normalizable.} and $\mathcal{N}_{\mA\mA^c}$ is the normalization factor (ensuring that the wavefunction has unit norm). The reduced density matrix describing the mixed state on the subsystem ${\cal A}$ is obtained by tracing out the degrees of freedom in the outside region ${\mA^c}$, as follows
\begin{equation}
\hat{\rho}_{\mA} = \tr_{\mA^c} \big(  \ket{\Psi_{\mA\mA^c}}\bra{\Psi_{\mA\mA^c}} \big)\,.
\end{equation}
This amounts to the Gaussian integral
\begin{equation}\label{densityFun_A}
\begin{split}
\hspace{-2pt} \rho_{\mA}\(\vec q_{\mA}, {\vec q}_{\mA}^{\,\, \prime}\) &= \int d\, q_{\mA^c} \ \Psi_{\mA\mA^c}(\vq_{\mA}, \vq_{\mA^c})\, \Psi^\dagger_{\mA\mA^c}(\vqp_{\mA},\vq_{\mA^c})  \\
&= \mathcal{N}_{\mA}\ \exp\! \[  -\frac{1}{2} (\vq_{\mA},\vqp_{\mA}) \left(
\begin{array}{cc}
\Gamma- \frac{1}{2}K\Omega^{-1}K^T& - \frac{1}{2}K\Omega^{-1}K^T \\
-\frac{1}{2}K\Omega^{-1}K^T& \Gamma- \frac{1}{2}K\Omega^{-1}K^T \\
\end{array}
\right) \left(
\begin{array}{c}
\vq_{\mA} \\
\vqp_{\mA} \\
\end{array}
\right)        \]\,.
\end{split}
\end{equation}
Following the reverse logic, let us start with a general mixed Gaussian state of $N_\mA$ modes with the (real) density matrix
\begin{equation}\label{density_fun_A}
\rho_{\mA}\(\vq_{\mA}, \vqp_{\mA}\) = \mathcal{N}_{\mA}\
\exp\! \[  -\frac{1}{2} (\vq_{\mA},\vqp_{\mA}) \left(
\begin{array}{cc}
A & - B \\
-B& A \\
\end{array}
\right) \left(
\begin{array}{c}
\vq_{\mA} \\
\vqp_{\mA} \\
\end{array}
\right)        \]\,,
\end{equation}
where the ${N_{\mA} \times N_{\mA} }$ matrices
$A$ and $B$ are both real and symmetric. Further, we must require $B$ to be positive semi-definite to ensure that the density matrix is non-negative, and $A-B$ to be a strictly positive matrix to ensure that the density matrix can be normalized.\footnote{This also implies that $A$ is a strictly positive  matrix, since the sum of two positive definite matrices is also positive definite.} In this case, a wavefunction of the form \eqref{Gaussian_AB} will purify $\rho_{\mA}$ if the two following constraints are satisfied
\begin{equation}\label{pure_constrians}
\Gamma= A+B\,, \qquad   \frac{1}{2}K\Omega^{-1}K^T=B \,.
\end{equation}
In this situation, the $\vq_\mA$ are the physical degrees of freedom while the $\vq_{\mA^c}$ are now auxiliary degrees of freedom.
While $\Gamma$ is completely fixed by the first constraint above, it should be clear that the second constraint leaves a great deal of freedom in the choice of $\Omega$ and $K$. Assuming $K$ has a left inverse  (and $B$ is invertible),\footnote{We stress that these conditions are not achieved for generic purifications. For example,
a linear transformation $K: {\cal A}^c \to {\cal A}$ has left inverse if and only if it is injective (\ie one-to-one). This immediately implies that $N_{{\cal A}^c} = {\rm dim}({\cal A}^c) \leq {\rm dim}({\cal A}) = N_{\cal A}$. This constraint does not hold in general since we can introduce as many ancillae as we wish in purifying a given mixed state. However, it does hold for essential mode-by-mode purifications, which will be the focus of our analysis in the following. Similar comments apply for the conditions under which $B$ is invertible.}
we can rewrite the constraints \reef{pure_constrians} as
\begin{equation}\label{constrain2}
\Gamma= A+B\,, \qquad   \Omega= \frac12\, K^{T} B^{-1} K \,,
\end{equation}
where $\Omega$ is completely determined by $B$ and $K$. Hence we can think of the freedom in choosing the purification as being parameterized by the choice of the $N_{\mA}N_{\mA^c}$ components of $K$.
Of course, this is the multi-mode generalization of the freedom found in eq.~\reef{pure1}, where the single parameter $k$ parameterized the purifications of the density matrix \reef{dense1} for a single degree of freedom. Hence with many modes (and ancillae), the purification complexity will be found by optimizing the usual complexity of the purification \reef{Gaussian_AB} over the freedom in choosing the matrix $K$.

However, it is natural to first ask what is the minimum number of ancillae $N_{\mA^c}$ required  to purify the mixed state ${\rho}_{\mA}$. A priori, we cannot be sure that such purifications, with  only the essential number of additional modes, will lead to the minimal value of the purification complexity, however, we will provide evidence for this later in this section. In order to count the degrees of freedom needed for the purification, we start by bringing the matrices $A$ and $B$ in eq.~\eqref{density_fun_A} to a canonical form by performing a sequence of coordinate transformations:
First, we find an orthogonal matrix $O_A$ that diagonalizes $A$, \ie $D_A = O^T_A\cdot A\cdot O_A$. We then rescale the coordinates $\vq_A$ such that $A$ becomes the unit matrix. Finally, we diagonalize the transformed $B$ matrix with a second orthogonal transformation $O_B$. The complete coordinate transformation reads
\begin{equation}\label{nonorthogonal_trans}
\vq_{\mA}= O_A \cdot D_A^{-1/2}\cdot O_B \cdot \tilde \vq_{\mA}\,,
\end{equation}
and of course, the same equation holds for $\vqp_\mA$. In this basis,\footnote{As an aside, we note that eq.~\eqref{nonorthogonal_trans} is not an orthogonal transformation and as a consequence, the reference state \eqref{ref_state}, which we are implicitly choosing for the purified $\mA\mA^c$ system,
\begin{equation}
\Psi_{R}\(\vq_{\mA},\vq_{\mA^c}\)= \mathcal{N}_{R}\ \exp\! \[ -\frac{\mu}{2} (\vq_{\mA},\vq_{\mA^c}) \left(
\begin{array}{cc}
\mathbb{I}_{N_{\mA}} & 0 \\
0 &   \mathbb{I}_{N_{\mA^c}} \\
\end{array}
\right) \left(
\begin{array}{c}
\vq_{\mA} \\
\vq_{\mA^c} \\
\end{array}
\right)        \]\,,
 \label{ref_stat}
\end{equation}
transforms nontrivially. The transformed reference state becomes
\begin{equation}\label{new_referenece}
\Psi_{R}\(\tilde{\vq}_{\mA},\vq_{\mA^c}\)= \mathcal{N}_{R}\,{\rm det}(D_A)\ \exp\! \[  -\frac{\mu}{2} (\tilde{\vq}_{\mA},\vq_{\mA^c}) \left(
\begin{array}{cc}
O_B^T \cdot D_A^{-1}\cdot O_B & 0 \\
0 &   \mathbb{I}_{N_{\mA^c}} \\
\end{array}
\right) \left(
\begin{array}{c}
\tilde{\vq}_{\mA} \\
\vq_{\mA^c} \\
\end{array}
\right)        \]\,,
\end{equation}
which is no longer an unentangled product state. However, this point is irrelevant for our argument determining the minimal value of degrees of freedom  $N_{\mA^c}$ required for the purification.} the quadratic form describing the density matrix \eqref{density_fun_A} is given in terms of matrices $\tilde A$ and $\tilde B$ which read
\begin{equation}\label{new_AB}
\tilde{A}=\mathbb{I}_{N_{\mA}}, \quad \tilde{B}= O_B^T\cdot {D_A^{-1/2}}\cdot O_A^T \cdot B \cdot O_A \cdot {D_A^{-1/2}}\cdot O_B= D_B\,.
\end{equation}
In this canonical form, the matrix $B$ has become
\begin{equation}\label{matrix_B}
\tilde{B}=D_B=\left(
\begin{array}{ccccccc}
b_1 &  &  &  &  &  &  \\
 & b_2 &  &  &  &  &  \\
 &  & \ddots &  &  &  &  \\
 &  &  & b_{n_B} &  &  &  \\
 &  &  &  &  0 &  &  \\
 &  &  &  &  & \ddots &  \\
 &  &  &  &  &  & 0 \\
\end{array}
\right),
\end{equation}
with $n_B=\text{rank}(\tilde B)=\text{rank}(B)$\ non-zero components. Therefore written in terms of the transformed coordinates $\tilde\vq_\mA$, the density matrix ${\rho}_{\mA}$ has been decomposed into $n_B$ two-by-two blocks describing modes in a mixed state, \ie
\beq
\left(
\begin{array}{cc}
1& b_i \\
b_i& 1 \\
\end{array}
\right) \,,
\eeq
and $N_{\mA}-n_B$ two-by-two unit matrices describing modes in a pure state. Now it is possible to follow the procedure in section \ref{sec:onemodepuri} to purify each of the mixed-state modes with a single ancilla, and finally transform back with eq.~\eqref{nonorthogonal_trans} to obtain a purification of the density matrix ${\rho}_{\mA}$ in the original $\vq_\mA$ basis. We refer to such purifications as { \bf mode-by-mode purifications}.

It is also straightforward to show that we cannot purify ${\rho}_{\mA}$ with less than $n_B$ additional degrees of freedom, namely $N_{\mA^c} \ge n_B$. Towards this goal, we consider the following theorem regarding the rank of the product of two matrices
\begin{equation}
\text{rank}(M \cdot N) \le \text{min}(\text{rank}(M),\text{rank}(N))\,.
\end{equation}
Hence applying this theorem to the second constraint in eq.~\eqref{pure_constrians}, \ie $\frac{1}{2}K\,\Omega^{-1}K^T=B$, we see that if a solution exists then we must have $\text{rank}(B) \le \text{min}(\text{rank}(\Omega^{-1}),\text{rank}(K))$. Next we observe that since the $N_{\mA^c}\times N_{\mA^c}$ matrix $\Omega$ is invertible, $\text{rank}(\Omega^{-1})=\text{rank}(\Omega)=N_{\mA^c}$. Furthermore,  $\text{rank}(K)\le \text{min}(N_\mA,N_{\mA^c})\le N_{\mA^c}$ since $K$ is an $N_\mA\times N_{\mA^c}$ matrix. Hence we arrive at
\begin{equation}\label{rank}
N_{\mA^c} \ge n_B\,,
\end{equation}
where  $n_B\equiv \text{rank}(B)$. That is, we will need at least
$n_B$ ancillae in the $\mA^c$ system in order to purify the mixed Gaussian state ${\rho}_{\mA}$. However, having explicitly constructed a purification with $N_{\mA^c}=n_B$ above, we know that it is possible to saturate this bound and we may conclude that this is the minimum number of extra degrees of freedom needed for the purification.

We refer to these purifications containing only the essential number of ancillae as {\bf essential purifications}.\footnote{We chose this name to distinguish this class of purifications from the {\it optimal} purifications, which are defined to be the purifications yielding the minimal complexity.}

\subsection{Optimal Purification in the Diagonal Basis} \label{optimal}
In the previous subsection, we found the minimum number of ancillae required to purify a mixed Gaussian state (with a Gaussian pure state). However, we still need to find the {\bf optimal purification} for the mixed state ${\rho}_{\mA}$ according to the definition of purification complexity \reef{def_pureC}, a question which we examine in the diagonal basis here. While we do not have a general solution for this question, we will argue that the optimal purification has a relatively simple form at least in certain interesting cases. First, we demonstrate that the optimal purification is, in fact, the essential purification for the case of a single physical degree of freedom. It would become very cumbersome to extend our proof to higher numbers of physical modes, but we believe that our result suggests that the same should hold more generally.

On the other hand, as we demonstrated above, even if we fix the number of ancillae, there are many ways to purify ${\rho}_{\mA}$ when the system $\mA$ contains many modes. Finally, we argue that at least for some simple but interesting Gaussian states in physical problems, the optimal purification can be found by optimizing the purifications of the individual diagonals. However, before proceeding with these questions, we begin by showing that there is a symmetry amongst the Gaussian purifications, which leads to same purification complexity from a family of distinct purifications all of which produce the same mixed state.

\subsubsection{Degenerate purifications} \label{genre}
Here, we will demonstrate that there is a degeneracy amongst the purifications \reef{Gaussian_AB} defined by eq.~\reef{pure_constrians}. That is, we will show that for a fixed mixed state, there are many distinct purifications, all of which have the same diagonal spectra and hence, they have the same complexity using eq.~\reef{complexity_pure}. This introduces a symmetry which will be useful to simplify our analysis in the following.

Beginning with a purification described by eq.~\eqref{Gaussian_AB}, we can perform a coordinate transformation on the ancillae
\begin{equation}
\left(\begin{array}{c}
\vq_{\mA} \\
\hat{\vq}_{\mA^c} \\
\end{array}
\right)
=  \left(
\begin{array}{cc}
\mathbb{I}_{N_{\mA}}& 0 \\
0 & R_{\mA^c}^{-1} \\
\end{array}
\right)
\left(
\begin{array}{c}
\vq_{\mA} \\
\vq_{\mA^c} \\
\end{array}
\right) \,, \label{transf99}
\end{equation}
where in general, $R_{\mA^c}\in GL(N_{\mA^c},\mathbb{R})$. Of course, the transformed wavefunction is characterized by the matrix
\begin{equation}
\left(
\begin{array}{cc}
\hat{\Gamma} & \hat{K} \\
\hat{K}^T & \hat{\Omega} \\
\end{array}
\right) =
\left(
\begin{array}{cc}
\Gamma &  K\,R_{\mA^c} \\
R_{\mA^c}^T\,K^T & R_{\mA^c}^T\,\Omega\, R_{\mA^c}\\
\end{array}
\right) \,. \label{newmat}
\end{equation}
Integrating out the $\hat{\vq}_{\mA^{c}}$ still yield precisely the same density matrix. Now in considering complexity, we may require that the reference state \reef{ref_stat} remains unchanged by the transformation \reef{transf99}, which imposes the constraint
\beq
 R_{\mA^c}^T\, R_{\mA^c}=\mathbb{I}_{N_{\mA^c}}\,,
\eeq
\ie $R_{\mA^c}  \in  SO(N_{\mA^c})$. That is, restricting eq.~\reef{transf99} to be an orthogonal transformation leaves the reference state unchanged, but further, such a transformation will also leave the diagonal spectrum, \ie the eigenvalues of eq.~\reef{newmat}, unchanged. Hence, evaluating the complexity with the expressions in eq.~\reef{complexity_pure}, we would find that all of these distinct purifications yield precisely the same complexity, and hence the same purification complexity for the corresponding mixed state.
This degeneracy will allow us to reduce the number of parameters in searching for the optimal purification below.

Perhaps we should add that since the complexity is a scalar function on the $N_\mA N_{\mA^c}$-dimensional space of purifications, we expect that for a generic value of the complexity, a $(N_\mA N_{\mA^c}-1)$-dimensional subspace will be degenerate, \ie have the same complexity. Of course, this is a much larger subspace than that defined by the $SO(N_{\mA^c})$ transformations above, \ie the latter defines a subspace of dimension $\frac12 N_{\mA^c}(N_{\mA^c}-1)$. The key feature distinguishing these purifications is that the diagonal spectrum is left invariant by the $SO(N_{\mA^c})$ transformations. In contrast, for a typical purification on the degenerate subspace, the spectrum will be different even though the complexity $\mC_1$ is unchanged.

\subsubsection{Essential Purifications} \label{threeA}
In general, one would expect that increasing the number of ancillae might help in reducing the complexity of the corresponding purifications for a fixed density matrix.  In this section, we will demonstrate that this is not the case for the Gaussian states in which we are interested. More precisely, we will consider the mixed state \reef{dense1} for a single harmonic oscillator and show that purifying this Gaussian state with two ancillae does not improve the purification complexity over the previous complexity \reef{complexity_one_mode} found with a single ancilla. Further, we will take this result for a single oscillator as an indication that adding extra ancillae does not improve the purification complexity for Gaussian mixed states in general.

We begin with the following Gaussian state for three modes,
\beq\label{pure33}
\psi_{123}(x,y,z) = \langle{x,y,z}\ket{\psi_{123}}
=   \left({\rm det}\frac{M_3}{\pi}\right)^{1/4} {\rm exp}\left[-\frac12\, \vec x^{\, T}{\cdot} M_3{\cdot}\vec x\right]\,,
\eeq
where as before, $M_3$ is chosen to be real, and $\vec x^{\, T}=(x,y,z)$, where $x$ corresponds to the physical degree of freedom while $y$ and $z$ are the ancillae. In order for this state to be a purification of the single-mode density matrix in eq.~\reef{dense1}, \ie
\beq
\hat{\rho}_1(x)=\tr_{y,z}\big( \ket{\psi_{123}}\bra{\psi_{123}}\big) \,,
\label{trace44}
\eeq
we must constrain the parameters in $M_3$ appropriately.
To understand these constraints, we write
\beq
\label{eq:polar}
M_3 = \big(R^\phi_{13}\big)^T \big(R^\theta_{12}\big)^T
\begin{pmatrix}
    \lambda_1 & 0 & 0 \\
    0 & \lambda_2 & 0 \\
    0 & 0 & \lambda_3
\end{pmatrix}
R^\theta_{12}\, R^\phi_{13}\,,
\eeq
where
\beq
R^\theta_{12} =
\begin{pmatrix}
    \ {\rm cos}\theta & {\rm sin}\theta & 0\\
    -{\rm sin}\theta & {\rm cos}\theta & 0\\
    0 & 0 & 1
\end{pmatrix} \qquad{\rm and}\qquad
R^\phi_{13} = \begin{pmatrix}
    \ {\rm cos}\phi & 0 & {\rm sin}\phi\\
    0 & 1 & 0 \\
    -{\rm sin}\phi & 0 & {\rm cos}\phi
\end{pmatrix}\,.
\eeq
That is, we have parameterized the matrix $M_3$ in terms of the three eigenvalues, $\lambda_i$ with $i={1,2,3}$, and two angles, $\theta$ and $\phi$. In principle, $M_3$ should be described by six independent parameters, but we have discarded the last rotation angle because of the degeneracy described in the previous subsection. Now eq.~\reef{trace44} imposes two constraints (cf. eq.~\eqref{pure_constrians} with $A=a$ and $B=b$) with which we can solve for the angles as
\beqa
\label{eq:phi}
\sin^2\phi&=&\frac{\lambda_3}{(\lambda_3-\lambda_1)(\lambda_3-\lambda_2)}\left(a+b-(\lambda_1+\lambda_2)+
\frac{\lambda_1\lambda_2}{a-b}\right)\,,\\
{\rm sin}^2\theta &=& \lambda_2 \frac{\lambda_1-\lambda_3}{\lambda_1-\lambda_2} \frac{ a+b-(\lambda_1+\lambda_3)+\frac{\lambda_1\lambda_3}{a-b}}{(a+b-\lambda_3)\lambda_3-\lambda_1\lambda_2
+\frac{\lambda_1\lambda_2\lambda_3}{a-b}}
\label{eq:theta}\\
&= &\frac{\lambda_2}{\lambda_3} \frac{\lambda_1-\lambda_3}{\lambda_1-\lambda_2} \left(1+ \frac{(a-b-\lambda_3)\lambda_1(\lambda_2-\lambda_3)}{(a-b)((a+b-\lambda_3)\lambda_3-\lambda_1\lambda_2)
+\lambda_1\lambda_2\lambda_3} \right)\,.
\nonumber
\eeqa
Note that these expressions restrict the three eigenvalues to lie within an allowed space where eqs.~\reef{eq:phi} and \reef{eq:theta} yield $0\le \sin^2\phi\le1$ and  $0\le \sin^2\theta\le1$.

Now using the $F_1$ cost function, the complexity of the Gaussian state \eqref{pure33} becomes
\begin{equation}
\mathcal{C}^{\text{diag}}_{1}\left(\ket{\psi}_{123}\right) = \frac{1}{2} \( \left|{\ln}\frac{\lambda_1}{\mu}\right|+\left|{\ln}\frac{\lambda_2}{\mu}\right|+\left|{\ln}\frac{\lambda_3}{\mu}\right| \)\,,
\label{330x}
\end{equation}
where $\mu$ is the frequency characterizing the corresponding reference state \reef{ref_state}. Of course, because of the absolute values, the form of $\mC_1$ will depend on whether the ratios $\lambda_i/\mu$ are bigger or smaller than one (similar to what was found in eq.~\reef{eq:cases1} with a single ancilla). Hence, there are eight distinct branches and we note that they intersect at the three planes defined by $\lambda_i=\mu$. That is, $\mC_1=\ln f$ where $f$ is any of eight combinations of products of the ratios $\lambda_i/\mu$ or their inverses (whichever is greater than one), \eg $f=\frac{\lambda_1\lambda_2}{\mu\,\lambda_3}$ in the octant where $\lambda_1,\lambda_2>\mu>\lambda_3$.

Given eq.~\reef{330x}, the purification complexity is given by optimizing the eigenvalues to minimize the result. One can argue that this minimum will not appear at some point inside one of the octants as follows: Firstly, we recall that the eigenvalues $\lambda_i$, as well as the reference frequency $\mu$, are all positive quantities. Now within any of the branches (or octants), $f$ has a simple functional dependence on the eigenvalues. In particular, when $\lambda_i<\mu$, $f$ contains a factor of $1/\lambda_i$ and so $\partial_{\lambda_i}\mC_1 =-1/\lambda_i <0$. Now, naively, the minimum along this direction would appear at $\lambda_i\to \infty$, but this is inconsistent with the constraint that  $\lambda_i<\mu$. Therefore there are no local extrema within the corresponding octants. Similarly, with $\lambda_i>\mu$, $f$ contains a factor of $\lambda_i$ and $\partial_{\lambda_i}\mC_1 =1/\lambda_i >0$. In this case, the derivative again vanishes with $\lambda_3\to\infty$, but the corresponding extrema would be a maximum of the complexity. Again, we conclude that no local extrema appear within these octants. Therefore, we are led to conclude that the minima for the complexity \reef{330x} must appear either (1) on the planes where the branches intersect or (2) at the boundaries of the allowed parameter space for the $\lambda_i$.

Next, we consider the boundaries of the allowed parameter space. The latter arise where either of the expressions in eqs.~\reef{eq:phi} and \reef{eq:theta} reaches zero or one, \ie ${\rm sin}^2\theta = 0$ or 1, or ${\rm sin}^2\phi = 0$ or 1. At these boundaries, we find that only two of the degrees of freedom are entangled:
\beqa
M_3|_{\sin\theta=0} &=& \begin{pmatrix}
    \lambda_1 \cos^2\phi + \lambda_3\sin^2\phi & 0 & (\lambda_1-\lambda_3)\cos\phi\sin\phi\\
    0 & \lambda_2 & 0 \\
    (\lambda_1-\lambda_3)\cos \phi\sin \phi & 0 & \lambda_3 \cos^2\phi + \lambda_1\sin^2\phi
\end{pmatrix} \,,
\nonumber\\
M_3|_{\sin^2\theta=1} &=& \begin{pmatrix}
    \lambda_2 \cos^2\phi + \lambda_3\sin^2\phi & 0 & (\lambda_2-\lambda_3)\cos\phi\sin\phi\\
    0 & \lambda_1 & 0 \\
    (\lambda_2-\lambda_3)\cos \phi\sin \phi & 0 & \lambda_3 \cos^2\phi + \lambda_2\sin^2\phi
\end{pmatrix} \,,
\label{hospital}\\
M_3|_{\sin\phi=0} &=& \begin{pmatrix}
    \lambda_1 \cos^2\theta + \lambda_2\sin^2\theta & \ (\lambda_1-\lambda_2)\cos\theta\sin\theta & 0\\
    (\lambda_1-\lambda_2)\cos \theta\sin \theta & \ \lambda_2 \cos^2\theta + \lambda_1\sin^2\theta & 0\\
    0 & 0 & \lambda_3
\end{pmatrix} \,,
\nonumber\\
    M_3|_{\sin^2\phi=1} &=& \begin{pmatrix}
        \lambda_3 & 0 & 0\\
        0 & \lambda_2 \cos^2\theta + \lambda_1 \sin^2\theta & \ (\lambda_1-\lambda_2)\cos\theta\sin\theta\\
        0 & (\lambda_1-\lambda_2)\cos \theta\sin \theta &\ \lambda_1 \cos^2\theta + \lambda_3\sin^2\theta \\
    \end{pmatrix} \,.
    \nonumber
    \eeqa
We may discard the last case (\ie $\sin^2\phi=1$) because the corresponding purification \reef{pure33} only involves entanglement between the two ancillae. Hence this will always leave the physical oscillator in a pure state when tracing out the two ancillae.  In the other three cases, the physical oscillator couples to one of the two ancillae. In any of these situations, the complexity will be minimized by setting the eigenvalue for the unentangled degree of freedom to the reference frequency, \ie $\lambda_i=\mu$, in which case its contribution to the complexity vanishes. Hence with this choice, we can discard the second ancilla, and the problem reduces to determining the purification complexity with a single ancilla. That is, the minimum complexity on any of these three edges will be precisely the same as that found with a single ancilla in section \ref{warmup}.\footnote{Following the results of section \ref{warmup}, to find the optimal two harmonic oscillator purification of our density matrix, we would find three cases again depending on the relation of the parameters of the density matrix. For cases 1 and 3 of section \ref{sec:onemode}, one of the eigenmodes of the optimal two harmonic oscillator purification is equal to the reference frequency. The fact that the unentangled eigenmode of the optimal three harmonic oscillator purification is also equal to the reference frequency implies that the optimal three harmonic oscillator purification has a degenerate eigenvalue. This makes one of the angles in the polar decomposition~\eqref{eq:polar} degenerate in the same way that the angles of radial coordinates are degenerate at the origin of the coordinate system.}

This leaves us to consider the intersection planes between the various branches of eq.~\reef{330x}. Of course, on any of these intersections, one of the eigenvalues is again set to the reference frequency, \eg $\lambda_3=\mu$. Hence we note that
the minima identified above arise at the intersection of one of the intersection planes with one of the boundaries of the allowed parameter space. However, on the `interior' of the intersection plane, we still have the freedom to optimize two independent eigenvalues (rather than just one on the boundary), and so one might wonder if the complexity finds a lower minimum in the interior. However, one may use analogous arguments to those above examining $\partial_{\lambda_i}\mC_1$ to argue that again on any of the intersection planes the minimum must be where this plane meets the boundary or one of the other planes where another eigenvalue reaches $\mu$. Hence the first possibility is already covered by the previous analysis of the complexity on the boundary of the allowed parameter space. Repeating the derivative argument for the intersection of two planes, one is lead to the possibility that the minimum may lie at the intersection of all three planes, \ie at $\lambda_1=\lambda_2=\lambda_3=\mu$. However, this possibility is ruled out since eq.~\reef{eq:phi} makes clear that this point is not within the allowed parameter space.\footnote{Actually, if the three eigenvalues are identical, then the rotations in eq.~\reef{eq:polar} act trivially. As a result, $M_3\propto \mathbb{I}$ which implies that the original state was actually pure, and hence this case is not really of interest here.}

Therefore we conclude that the complexity is optimized on the boundary of the allowed parameter space. However, there we found that one of the two ancillae decoupled and the optimal purification reduced to that found with a single ancilla in section \ref{warmup}. That is, with two ancillae, the minimum complexity in the diagonal basis is achieved with a purification where the physical degree of freedom is only entangled with one of the ancillae, and the second ancilla remains unentangled. We provide an additional numerical check of this result in appendix \ref{app:numerics}.

We might note that it appears that the first three cases in eq.~\reef{hospital} yield three distinct minima. However, we should note that the first two cases, \ie $\sin^2\theta =0$ and 1, differ only in the labelling of the eigenvalues and are in fact describing the same purifying states \reef{pure33}, where $x$ and $z$ are entangled while $y$ remains unentangled. The only difference in the third case, \ie $\sin\phi=0$ is that the purification entangles $x$ and $y$ while $z$ remains unentangled. Of course, both purifications yield the same optimal complexity. We might add that this degeneracy is a remnant of the $SO(2)$ symmetry implied by the discussion in subsection \ref{genre}.

While our analysis yields a clear result for a single physical oscillator, it would be difficult to extend this analysis to a mixed state with many degrees of freedom. Still, we are emboldened to interpret this result as an indication that adding extra ancillae will not improve the purification complexity for Gaussian mixed states in general. That is, throughout the following, we will assume that the optimal purification for a Gaussian mixed state for many oscillators is an essential purification, \ie the number of ancillae saturates eq.~\reef{rank} with $N_{\mA^c}=n_B$.

\subsubsection{Mode-by-Mode Purifications}
\label{MbyM}
\begin{figure}[tbph]
    \centering\includegraphics[width=4.5in]{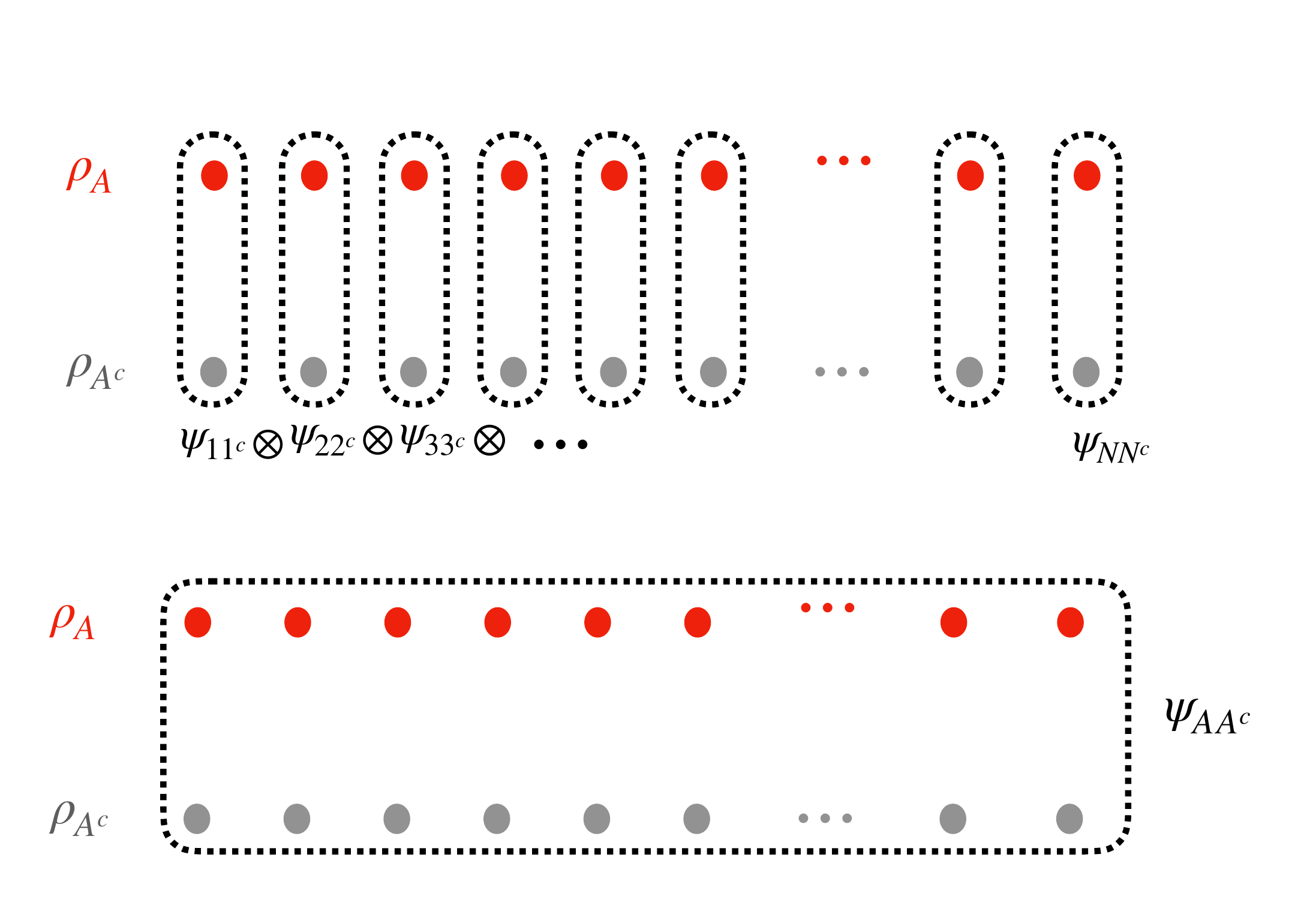}
    \caption{Illustration of the different ways to purify a multi-mode Gaussian state $\hat{\rho}_A$. We refer to the purifications of the form $\Psi_{11^c} \otimes \Psi_{22^c} \otimes \cdots \otimes \Psi_{NN^c}$ as mode-by-mode purifications.\label{more_modes}}
\end{figure}

In section \ref{sec:purificationX}, we identified the minimum number of ancillae required to purify a Gaussian mixed state \reef{density_fun_A}. Our approach involved finding a `diagonal' basis in which the density matrix $\rho_{\cal A}$ took a canonical form where each mode was separately either in a mixed or pure state. Each of the mixed state modes could then be purified by a single ancilla, using the construction presented in section \ref{sec:onemodepuri} for one-mode mixed states. We will refer to these purifications as {\bf mode-by-mode purifications}. Certainly, there are many ways to purify a mixed state on many degrees of freedom, as illustrated in figure \ref{more_modes}. The top panel indicates a simple mode-by-mode purification while the lower panel illustrates a general purification for a multi-mode Gaussian state. Implicitly, the general purification will have many more free parameters to optimize and so one would expect that this would allow for a smaller purification complexity for the corresponding mixed state. In this subsection, we will examine this question and identify the conditions for which a mode-by-mode purification provides the optimal purification for a Gaussian mixed state. To make our analysis both explicit and tractable, we focus on Gaussian mixed states for two degrees of freedom.

Hence, considering the two-mode system as an example, we can purify the two modes individually or together, as illustrated in the figure \ref{more_modes}. The corresponding pure states can be written as
\begin{equation}
\ket{\Psi}=  \ket{\Psi_{11^c}} \otimes \ket{\Psi_{22^c}} \qquad \text{or} \qquad  |\widetilde\Psi\rangle =\ket{\Psi_{12(12)^c}}  \,.
\end{equation}
That is, we have a mode-by-mode purification on the left and a general purification on the right.

To proceed with explicit calculations, let us begin with  Gaussian mixed states (for two modes) taking a simple product form,
\beqa
\label{eq:density-mixed}
\rho_{\mA}(\vq_{\mA}, \vqp_{\mA} )&=&\rho_{1}(x_1,x_1')\ \rho_{2}(x_2,x_2') \\
&=& \mathcal{N}_{\mA}\  \exp \[  -\frac{1}{2} (\vq_{\mA},\vqp_{\mA}) \left(
\begin{array}{cc}
\ A_{D} & - B_{D} \\
-B_{D}&\ A_{D} \\
\end{array}
\right) \left(
\begin{array}{c}
\vq_{\mA} \\
\vqp_{\mA} \\
\end{array}
\right)        \]\,,
\nonumber
\eeqa
where $\vq_{\mA}=(x_1,x_2)$,
\beq\label{gamble6}
A_{D}=\left(
\begin{array}{cc}
a_1 & 0 \\
0 & a_2 \\
\end{array}
\right)\qquad{\rm and}\qquad
B_{D}=\left(
\begin{array}{cc}
b_1 & 0 \\
0 & b_2 \\
\end{array}
\right)\,.
\end{equation}
Borrowing from the analysis in section \ref{sec:onemodepuri},
the mode-by-mode purification can be written as
\beq\label{gamble8}
\Psi_{\otimes \mA\mA^c} = \mathcal{N}_{\otimes \mA\mA^c}\ \exp \[  -\frac{1}{2} (\vq_{\mA},\vq_{\mA^c}) \left(
\begin{array}{cc}
\Gamma_D & K_D \\
K_D^T & \Omega_D \\
\end{array}
\right) \left(
\begin{array}{c}
\vq_{\mA} \\
\vq_{\mA^c} \\
\end{array}
\right)        \]
\eeq
where $\vq_{\mA^c}=(y_1,y_2)$,
\begin{equation}
\label{gamble}
\Gamma_D =\left(
\begin{array}{cc}
a_1+b_1 & 0\\
0 & a_2+b_2 \\
\end{array}
\right)\,,\quad
\Omega_D =\left(
\begin{array}{cc}
\frac{k_1^2}{2b_1}& 0\\
0 &  \frac{k_2^2}{2b_2}\\
\end{array}
\right)
\quad{\rm and}\quad
K_D =\left(
\begin{array}{cc}
k_1& 0\\
0 & k_2 \\
\end{array}
\right)\,.
\end{equation}
We then translate the above purification to the parameters introduced in section \ref{subsec:altdesc} using eq.~\reef{transform_paras}, \ie
\footnote{Note that we have set $\omega_1=\omega_2=\omega$ for both oscillators. Choosing different frequencies can be absorbed by redefining the squeezing parameters, $r_{1,2}$ and $s_{1,2}$.}
\begin{equation}
\begin{split}
a_1+b_1 &= \omega e^{2r_1} \cosh 2\alpha_1\,, \qquad  \frac{k_1^2}{2b_1} =\omega e^{2s_1} \cosh 2\alpha_1\,, \quad k_1= -\omega e^{r_1+s_1} \sinh 2\alpha_1 \,,\\
a_2+b_2 &= \omega e^{2r_2} \cosh 2\alpha_2\,, \qquad  \frac{k_2^2}{2b_2} =\omega e^{2s_2} \cosh 2\alpha_2\,, \quad k_2= -\omega e^{r_2+s_2} \sinh 2\alpha_2 \,.
\end{split}
\end{equation}
For this kind of purification, we only need to minimize two free parameters $s_1,s_2$ and the final complexity is given by the sum of the one-mode complexities of purification
\begin{equation}
\mC_{\otimes \mA\mA^c}^{\text{diag}}= \mC_{1}^{\text{diag}}\!\[\hat{\rho}_1(r_1,\alpha_1)\] +  \mC_{1}^{\text{diag}}\!\[\hat{\rho}_2(r_2,\alpha_2)\]\,,
\label{gamble4}
\end{equation}
where $\mC_{1}^{\text{diag}}$ is given in eq.~\eqref{complexity_one_mode}.

We can also consider the most general purification of eq.~\reef{eq:density-mixed}. The latter takes the form given in eq.~\reef{Gaussian_AB}, which we write here as
\beq\label{gamble2}
\Psi_{ \mA\mA^c} = \mathcal{N}_{ \mA\mA^c}\ \exp \[  -\frac{1}{2} (\vq_{\mA},\vq_{\mA^c}) \left(
\begin{array}{cc}
\Gamma & K \\
K^T & \Omega \\
\end{array}
\right) \left(
\begin{array}{c}
\vq_{\mA} \\
\vq_{\mA^c} \\
\end{array}
\right)        \]
\eeq
with
\begin{equation}\label{gamble2a}
\Gamma =\Gamma_D\,,
\quad
\Omega =\frac{1}{2} \left(
\begin{array}{cc}
\frac{f_2^2}{b_2}+\frac{k_1^2}{b_1} & \frac{f_2 k_2}{b_2}+\frac{f_1 k_1}{b_1} \\
\frac{f_1 k_1}{b_1}+\frac{f_2 k_2}{b_2} & \frac{f_1^2}{b_1}+\frac{k_2^2}{b_2} \\
\end{array}
\right)
\quad{\rm and}\quad
K =\left(
\begin{array}{cc}
k_1& f_1\\
f_2 & k_2 \\
\end{array}
\right)\,.
\end{equation}
Here we have used eq.~\eqref{constrain2} to constrain the pure state, but we do not demand that $\Omega$ or $K$ are diagonal as in eq.~\reef{gamble}. For this purification \reef{gamble2}, we have four free parameters $\(k_1,k_2,f_1,f_2\)$ and thus, the purification complexity is defined as
\begin{equation}
\mC_{\mA}^{\text{diag}}= \text{min}_{k_1,k_2,f_1,f_2}\( \mC_{\mA\mA^c}^{\text{diag}} \)\,.\label{gamble3}
\end{equation}
However, as discussed in subsection \ref{genre}, there is degeneracy amongst the possible optimal purifications. In particular, the purification complexity will be unchanged by the following $SO(2)$ transformation\footnote{We note that this rotation only acts on the ancillae and so leaves $\Gamma=\Gamma_D$ unchanged -- see eq.~\reef{newmat}.}
\begin{equation}
\hat{K}= \left(
\begin{array}{cc}
k_1& f_1\\
f_2 & k_2 \\
\end{array}
\right) \cdot \left(
\begin{array}{cc}
\cos \theta& -\sin \theta\\
\sin \theta &\ \cos \theta \\
\end{array}
\right)
=\left ( \begin{array}{cc}
\hat{k}_1& \hat{f}_1\\
\hat{f}_2 & \hat{k}_2 \\
\end{array}
\right)\,.
\end{equation}
Hence we can simplify the optimization by eliminating one of the parameters $(k_1,k_2,f_1,f_2)$, \eg we can choose $\tan \theta = f_1/k_1$ to set $\hat{f_1}=0$. That is, we know there will be an optimal purification in which $f_1=0$ and hence we can reduce eq.~\reef{gamble3} to
    \begin{equation}\label{two_modes}
    \mC_{\mA}^{\text{diag}}= \text{min}_{k_1,k_2,f_1=0,f_2} \( \mC_{\mA\mA^c}^{\text{diag}} \)\,.
    \end{equation}
However, performing the minimization numerically using Mathematica, we found that the optimal purification coincided with the mode-by-mode purification \reef{gamble} (\ie $f_2=0$). To be precise, we determined optimal purifications for mixed states described by $\alpha_i \in [0,5]$ and $r_i \in [-10,10]$,\footnote{Note that this corresponds to an exponentially large range for the parameters, $a_i$ and $b_i$, using eqs.~\reef{parameters} and \reef{hope}.} and were able to show that $\frac{\mC_{\otimes \mA\mA^c}^{\text{diag}}-\mC^{\text{diag}}_{\mA}}{\mC^{\text{diag}}_{\mA}} \lesssim 10^{-10}$. Hence the general purification complexity \reef{two_modes} reduces to the expression in eq.~\reef{gamble4}. This demonstrates that the mode-by-mode purification is indeed the optimal purification for two-mode Gaussian mixed states which factorize as in eqs.~\eqref{eq:density-mixed}-\eqref{gamble6}.

However, the previous conclusion does {\bf{not}} apply for the most general two-mode Gaussian mixed states, as we will now demonstrate. Let us replace the previous example \reef{eq:density-mixed}-\eqref{gamble6} with a general two-mode density matrix \reef{density_fun_A}
where
\begin{equation}\label{gamble7}
A=\left(
\begin{array}{cc}
a_1 & 0 \\
0 & a_2 \\
\end{array}
\right)
\qquad{\rm and}\qquad
B=\left(
\begin{array}{cc}
b_1 & g \\
g & b_2 \\
\end{array}
\right)\,.
\end{equation}
Comparing to the factorized case (\ie with eq.~\reef{gamble6}), one may also expect the off-diagonal components of $A$ to be nonvanishing in general. However, we can always perform an $SO(2)$ transformation to diagonalize $A$ leaving us with the expressions given above. In section \ref{sec:purificationX}, we showed that any Gaussian state can be decomposed into a product state using a general (\ie non-orthogonal) transformation as in eq.~\eqref{nonorthogonal_trans}.
One may then naively expect that the optimal purification of ${\rho}_{\mA}$ will be a simple mode-by-mode purification in this new `diagonal' basis, \ie the simplest solutions found in section \ref{sec:purificationX}. However, a more careful analysis is required since, as we stressed in eq.~\eqref{new_referenece}, this general transformation modifies the reference state so that it is no longer a product state in the new basis.

To test the hypothesis that the optimal purification takes the form of a mode-by-mode purification in the diagonal basis, we examined a variety of examples using similar numerical methods to those employed above. The purification takes the same form as in eq.~\reef{gamble2} where $\Gamma$, $\Omega$ and $K$ are constrained by eq.~\eqref{pure_constrians} using the $A$ and $B$ matrices given in eq.~\reef{gamble7}. Let us parameterize $K$ as in eq.~\reef{gamble2a} and then as in the previous example, we can use a rotation acting on the ancillary directions  $\vq_{\mA^c}=(y_1,y_2)$ to set $f_1=0$. Then as above, we found the optimal purification numerically for a variety of examples by minimizing over the three remaining free parameters $(k_1,k_2,f_2)$. Given the optimal purification, we can examine its form in the diagonal basis produced by the transformation in eq.~\reef{nonorthogonal_trans}. Since we have already diagonalized $A$ in eq.~\reef{gamble7}, this transformation reduces to
\beq\label{late1}
\vq_\mA= A^{-1/2}\cdot O_B
\cdot\tilde\vq_\mA\,.
\eeq

For a mode-by-mode purification, all three matrices, $\Gamma$, $\Omega$ and $K$, should be simultaneously diagonal in the new basis. The transformation \reef{late1} is chosen to make sure that $A$ and $B$ in the density matrix \reef{density_fun_A} are diagonal and hence with $\Gamma=A+B$ from eq.~\reef{pure_constrians}, the $\Gamma$ matrix is automatically diagonal in the new basis. Hence the question reduces to determining whether or not $\Omega$ and $K$ are both diagonal or rather simultaneously diagonalizable in the new basis. The latter refers to the fact that there are still the $SO(2)$ transformations \reef{newmat} which map amongst the optimal purifications. For example, let us begin with the original coordinates in eq.~\reef{Gaussian_AB},  and the optimal $K_{op}$ is found by the minimization among the three free parameters in the matrix
\begin{equation}
K =\left(
\begin{array}{cc}
k_1& 0\\
f_2 & k_2 \\
\end{array}
\right)\,.
\end{equation}
Now applying the (inverse of the) transformation in eq.~\reef{late1}, it becomes
\begin{equation}
\tilde{K}= O^T_B \cdot A^{-1/2} \cdot  K_{op} =
 \left ( \begin{array}{cc}
\tilde{k}_1& \tilde{f}_1\\
\tilde{f}_2 & \tilde{k}_2
\end{array}
\right)
\,. \label{late2}
\end{equation}
But then we would employ eq.~\reef{newmat} to see if we can find a rotation such that $\tilde K$ becomes diagonal, \ie
\beq \label{Khat}
\hat{K} =
 \left ( \begin{array}{cc}
\hat{k}_1& 0\\
0 & \hat{k}_2
\end{array}
\right)\overset{?}{=}\tilde{K}\, R_{\mA^c}
\eeq
with $R_{\mA^c} \in SO(2)$. This is a nontrivial constraint since $R_{\mA^c}$ rotates each of the rows of $\tilde K$ as a vector separately, and hence if $\tilde K$ is diagonalizable, then these row vectors must already be orthogonal in $\tilde K$. Therefore a necessary condition to have a mode-by-mode purification is that
\begin{equation}
\delta \tilde{K}= \tilde{k}_1\tilde{f}_2 + \tilde{k}_2\tilde{f}_1
= 0\,.\label{late3}
\end{equation}
In fact, $\delta \tilde{K}= 0$ is a necessary and sufficient condition. Since we have put $B$ in a diagonal form with the transformation \reef{late1}, the second constraint in eq.~\reef{constrain2} shows that if $K$ is also diagonal then $\Omega$ will also be diagonal in the same basis. Therefore we can see that when $\delta \tilde{K}= 0$, the optimal purification is indeed a mode-by-mode purification.
On the other hand, if $\delta\tilde{K} \neq 0$, the optimal purification must still have a more complicated form in the diagonal basis.

\begin{figure}[htbp]
    \centering\includegraphics[width=3.6in]{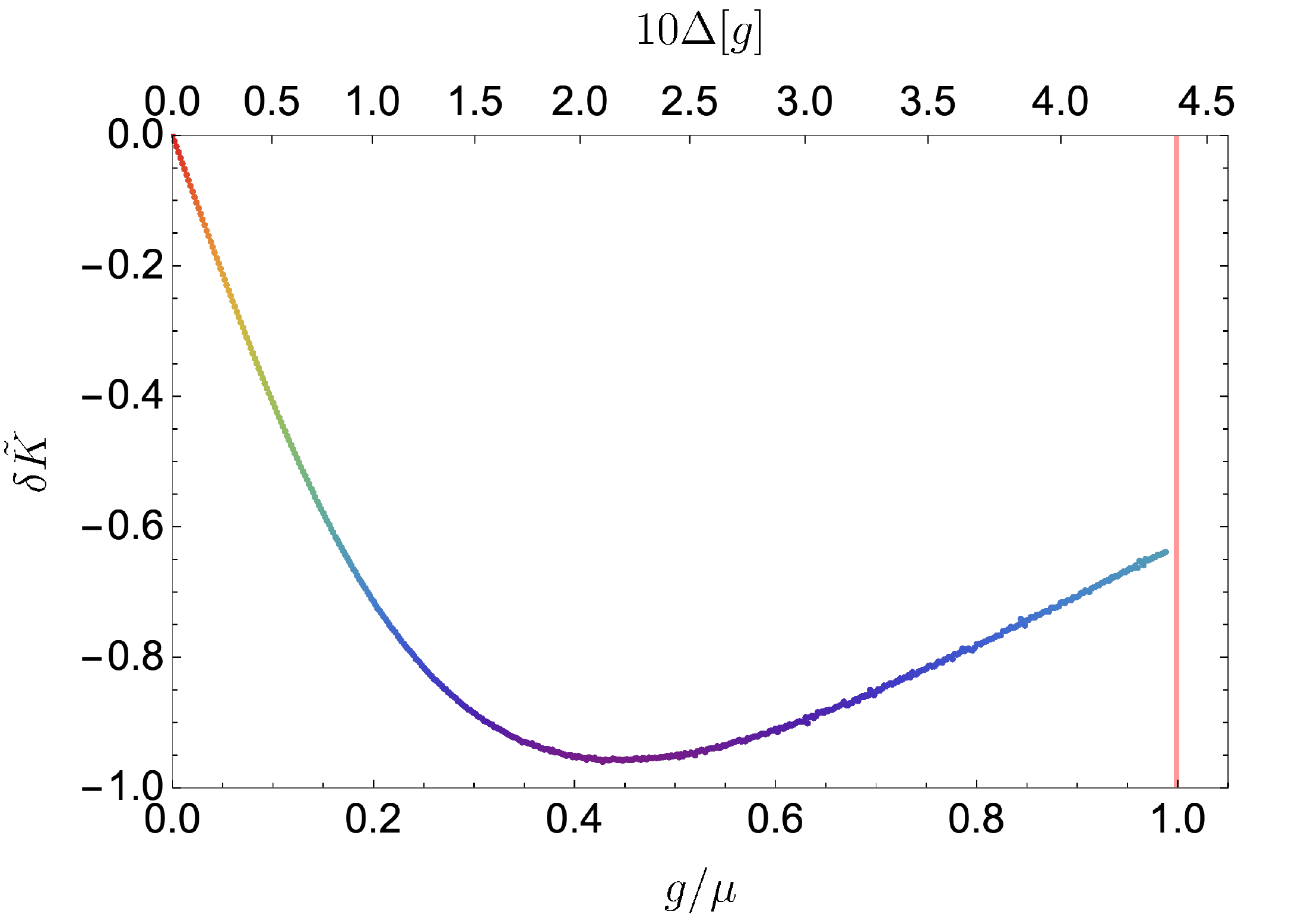}
    \caption{$\delta \tilde{K}$ (see eq.~\reef{late3}) as a function of $g$ (the off-diagonal component of $B$). In this example, ${a_1}/{\mu}=4, {a_2}/{\mu} =2, {b_1}/{\mu}= 2$ and ${b_2}/{\mu}= \frac{3}{2}$. The red vertical line indicates the upper bound ${g}/{\mu}=1$ for the parameter $g$, which is constrained by the positivity of the matrix $A-B$. Along the top axis, $\Delta$ is a measure which quantifies the deviation of $A$ and $B$ from being commuting --- see eq.~\reef{Delta1}.}
    \label{deltaK12}
\end{figure}

A typical plot for $\delta \tilde{K}$ is shown in figure \ref{deltaK12}. Recall that the optimal purification, \ie the optimal $\tilde K$, was found numerically following the scheme in eq.~\reef{two_modes}. Our numerical results support the hypothesis that the mode-by-mode purification is optimal when the $A,B$ matrices commute, or equivalently, when the density matrix can be factorized, as in eqs.~\eqref{eq:density-mixed}-\eqref{gamble6}.\footnote{For commuting $A,B$, we explored two  possibilities numerically: $a_1=a_2$ or $g=0$. The latter is the same as with the mode-by-mode purification. For the former, we considered the parameters in the ranges: $a_1=a_2 \in [2,6]\,,b_1,b_2 \in [1,3]\,, g\in [0,0.5]$. We found that $\delta \tilde{K}$ was fluctuating within the range $10^{-6}$ to $10^{-9}$.\label{hard}}
This result is not surprising because when $[A,B]=0$, it is always possible to find an orthogonal transformation that acts on the $\vq_\mA$ and brings the target state (explicitly) to the form of a factorized product of one-mode Gaussian states.  In this case, our numerical results support the previous conjecture about the optimal purification for such product states.

While we found in general that a mode-by-mode purification is not optimal, we would still like to show that such purifications (in the diagonal basis) produce a very good approximation to the optimal one in certain circumstances. In particular, for a variety of examples, as we detail below, we found the optimal purification numerically,
 but found that the associated complexity did not improve very much the complexity found by only optimizing over mode-by-mode purifications, \ie by restricting the purification to have the form in eqs.~\eqref{gamble8}-\eqref{gamble} in the $\tilde\vq_\mA$ basis of eq.~\eqref{late1} and minimizing the complexity of the two free parameters $\hat{k}_{1,2}$.\footnote{That is, we define a two-parameter family of purifications with
\begin{equation}
 K=\sqrt{A}\, O_B \, \hat{K} \qquad{\rm where}\quad  \hat{K} =
 \left ( \begin{array}{cc}
 \hat{k}_1& 0\\
 0 & \hat{k}_2
 \end{array}
 \right)\,,
\end{equation}
and then optimize the complexity in the diagonal basis $\mC^{\text{diag}}_1$, \ie \eqref{complexity_pure} with the forms of $K,\Gamma,\Omega$ matrices in the original basis  over the two free parameters $\hat{k}_i$. That is, the complexity is still defined  in the regular way but  our approximation is that $\tilde{\mC}$ is derived by limiting the optimization to only varying these two parameters. When the matrix $A$ is not diagonal, the mode-by-mode purification takes the form $K=O_A \cdot D_A^{-1/2}\cdot O_B \cdot \hat{K}$, where $O_A$ is the matrix that brings $A$ to the diagonal form $D_A$, see eq.~\eqref{nonorthogonal_trans}.}
In order to quantitatively measure the deviation of the $A$ and $B$ matrices from being commuting, we define
\begin{equation}\label{Delta1}
\Delta=\sqrt N\, \frac{\parallel\!\! [A,B]\!\!\parallel_\mt{F}}{\parallel\!\! A\!\!\parallel_\mt{F}\,\parallel\!\!B\!\!\parallel_\mt{F}}\,,
\end{equation}
where $\parallel\!\! A\!\!\parallel_\mt{F}$ denotes the Frobenius norm, \ie
\begin{equation}
\parallel\!\! A\!\!\parallel_\mt{F}\equiv  \sqrt{\operatorname{Tr}\left(A^\dagger\, A\right)} = \sqrt{\sum_{i=1}^m \sum_{j=1}^n |A_{ij}|^2}  \,,
\end{equation}
for an $m\times n$ matrix. We have chosen this definition \reef{Delta1} so that it does not change if we rescale the matrices $A$ and $B$ by an overall constant. Note that we have also included an overall factor of $\sqrt{N}$ in eq.~\reef{Delta1} where
$N$ is the number of oscillators in the original mixed state (\ie both $A$ and $B$ are $N\times N$ matrices). This ensures that if the matrices are chosen at random (\ie with all elements of order one), then $\Delta$ does not scale with $N$ (\ie it does not becomes arbitrarily small or large as the number of degrees of freedom becomes large, as in our QFT calculations).\footnote{When all elements are taken to be of order one, the Frobenius norm of a random matrix scales like $N$, but that of a commutator scales like $N^{3/2}$. This is because every element in the commutator is roughly  speaking the sum of $N$ random variables whose variance $\sigma^2\sim 1$. Hence, the variance of the sum is $\sigma^2\sim N$.}
These features eliminate any trivial effects from our measure of noncommutativity, assuring that we can still use it when dealing with a very large number of oscillators in the QFT calculations. For example, applying the definition \eqref{Delta1} to the two-mode case with the matrices $A$ and $B$ as defined in  eq.~\eqref{gamble7}, one finds
\begin{equation}\label{delta-bound}
\Delta = \frac{2 \,|\left(a_1-a_2\right) g\, |}{\sqrt{\left(a_1^2+a_2^2\right) \left(b_1^2+b_2^2+2 g^2\right)}} < \frac{ |a_1-a_2| }{\sqrt{\left(a_1^2+a_2^2\right)}} < 1\,,
\end{equation}
where the constrains are due to the positivity of the matrices $A$ and $B$.\footnote{In particular, the first inequality follows from $b_1^2+b_2^2+2 g^2=4 g^2+(b_1-b_2)^2+2(b_1 b_2-g^2)>4g^2$, where $b_1 b_2-g^2>0$ comes from the positivity of the matrix $B$.}

\begin{figure}[h]
    \includegraphics[width=0.5\textwidth]{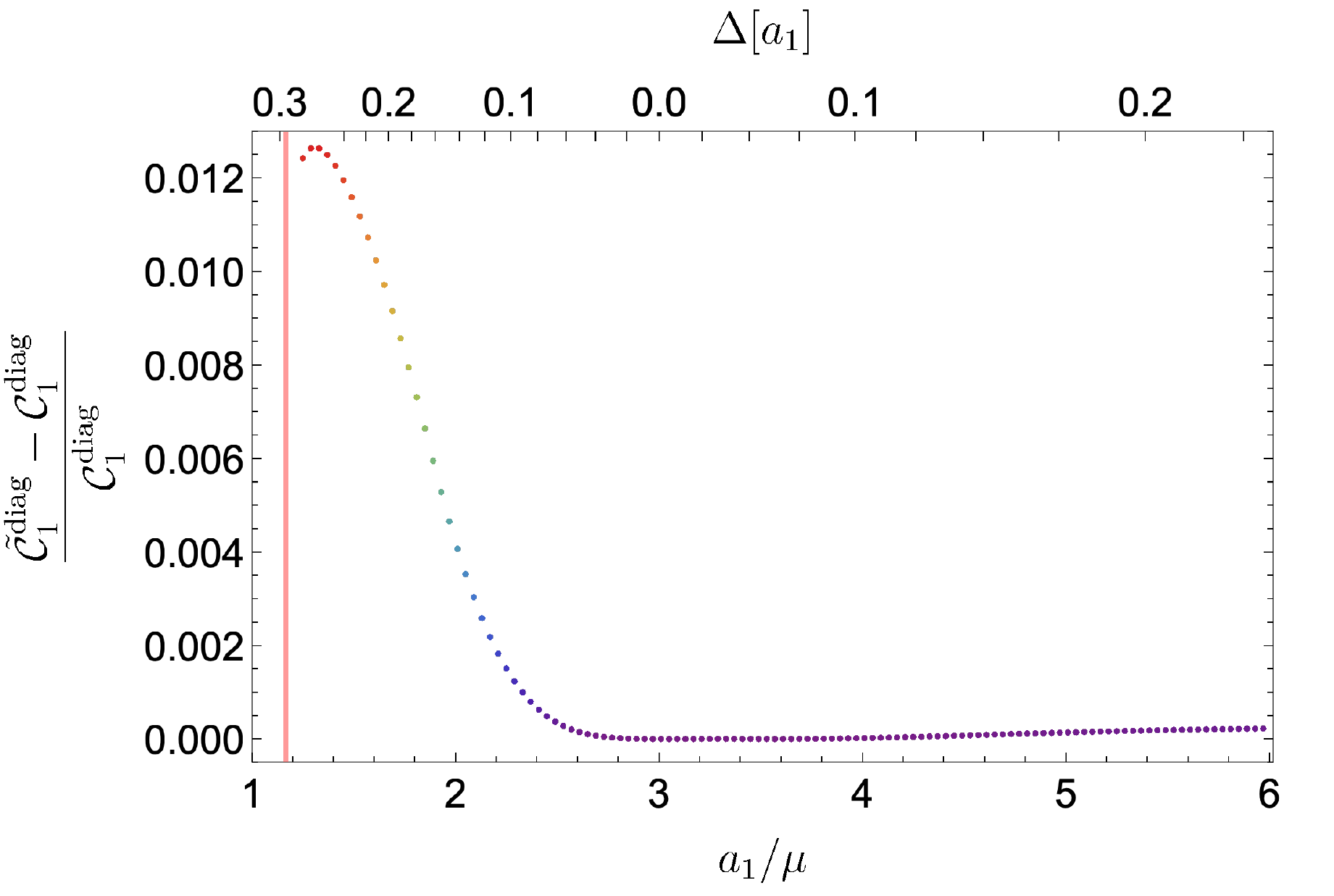} \hspace{0.01\textwidth}
    \includegraphics[width=0.5\textwidth]{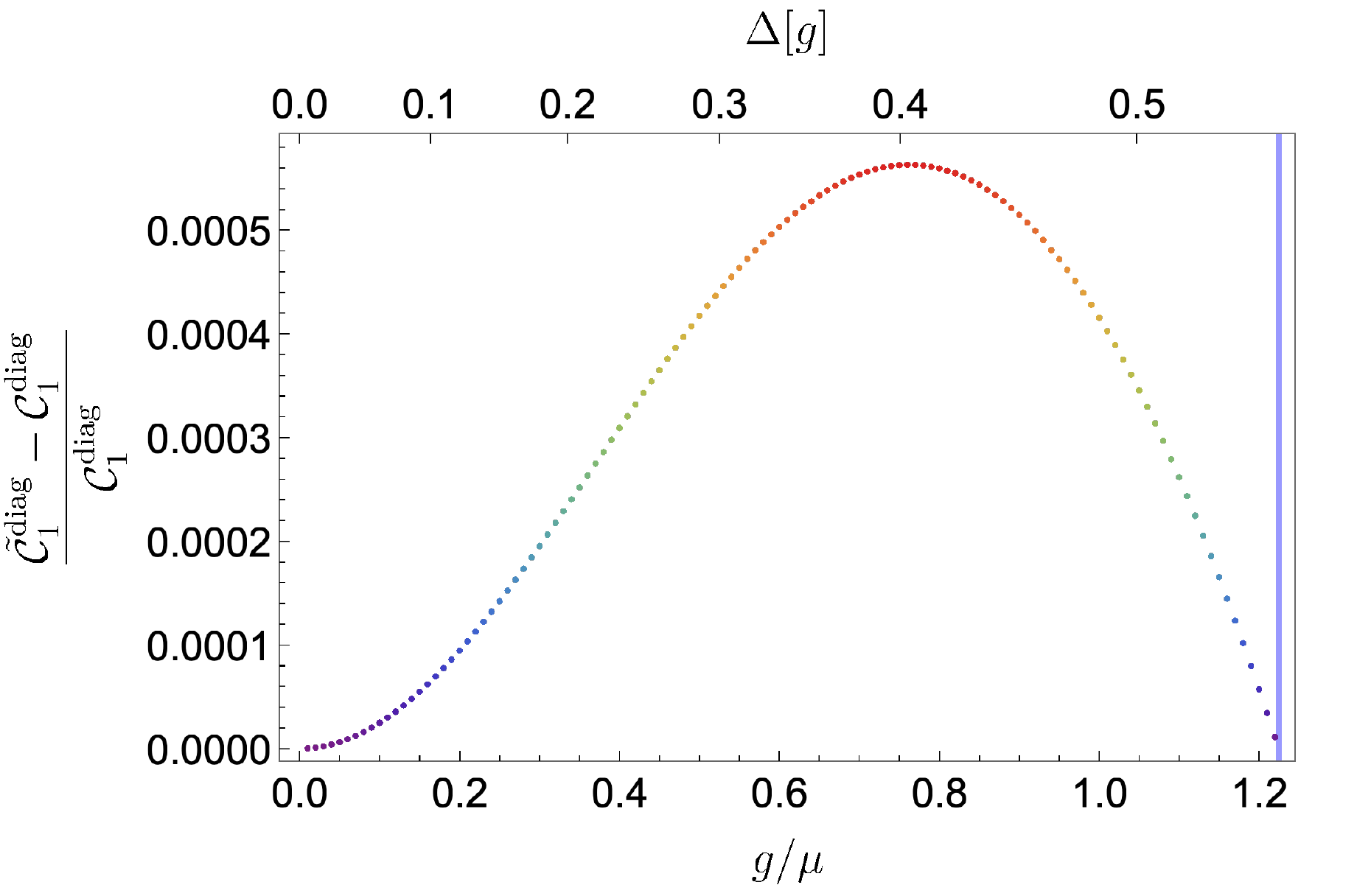}
    \caption{The relative difference between the mode-by-mode `complexity' $\tilde \mC_1^\mt{diag}$ and the optimal complexity $\mC_1^\mt{diag}$ for nearly commuting $A$ and $B$. Left panel: $a_2/\mu =3, b_1/\mu =1, b_2/\mu = 1.5, g_1/\mu =0.5$ and $a_1/\mu \in (1,6)$. The vertical red line indicates the lower bound $a_1/\mu= 7/6$, which is fixed by requiring that the matrix $A-B$ is positive. Right panel: $a_1 /\mu=5, a_2/\mu =2, b_1/\mu = 1, b_2/\mu= 1.5$ and $g/\mu \in (0,\sqrt{\frac{3}{2}}) $. The vertical blue line indicates the upper bound $g/\mu=\sqrt{\frac{3}{2}}$, which is  determined by requiring the positivity of the matrix $B$. At the upper bound $g/\mu=\sqrt{\frac{3}{2}}$, one of the eigenvalues of the matrix $B$ vanishes and we see that the relative complexity difference vanishes too. This is because in this case, the state $\hat\rho_{\mA}$ is mixed for only one of the two modes and its essential purifications will be by definition mode-by-mode purifications.} \label{deltaC_onebyone}
\end{figure}

\begin{figure}[h]
        \centering\includegraphics[width=4.0in]{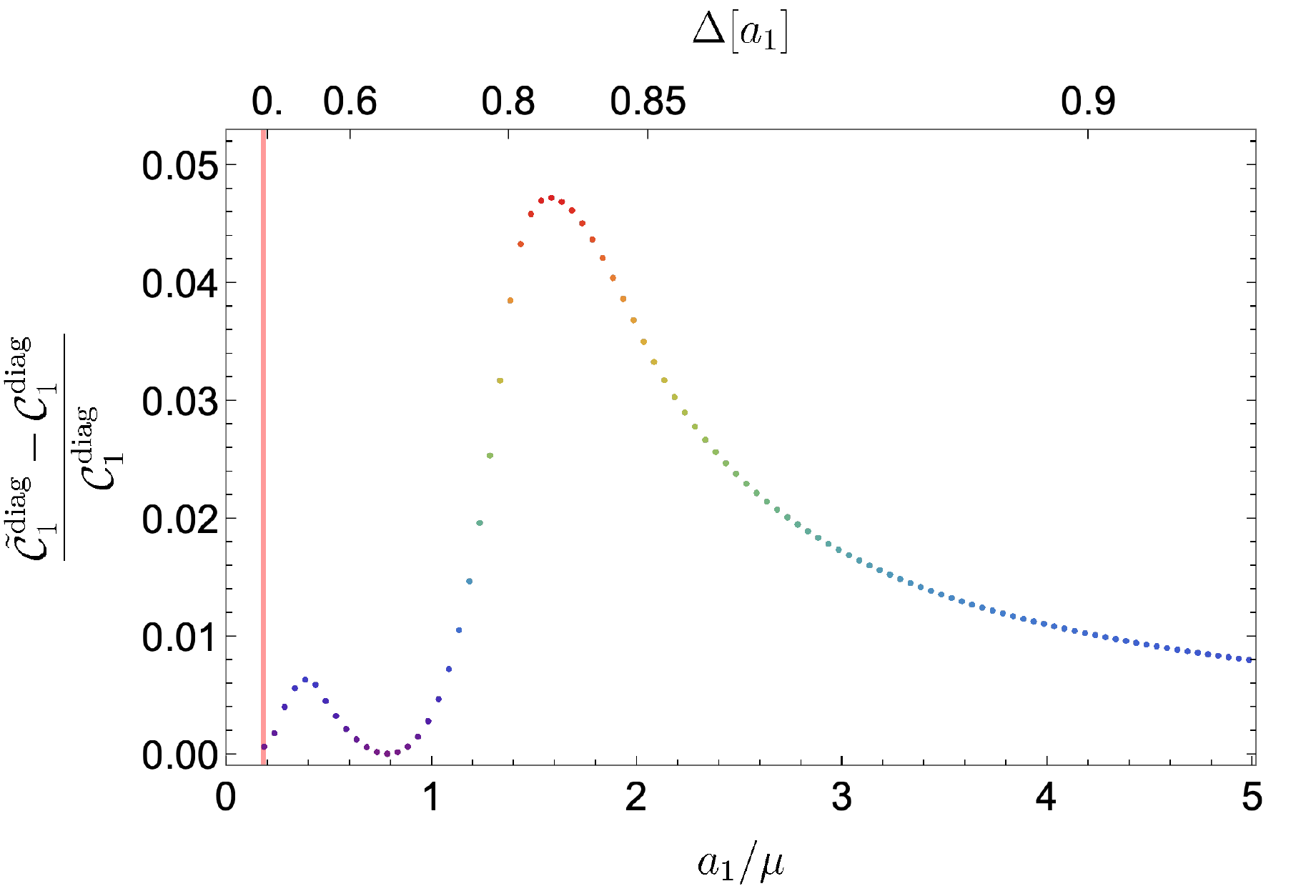}
    \caption{The relative difference between the mode-by-mode `complexity' $\tilde \mC_1^\mt{diag}$ and the optimal complexity $\mC_1^\mt{diag}$ for $A$ and $B$ with larger $\Delta$. Here we take $a_1/\mu \in (\frac{181}{1000},5)$ as the free parameter with fixed $a_2/\mu = \frac{1}{5}\,, b_1/\mu=b_2/\mu =\frac{1}{10}\,, g/\mu = \frac{9}{100}$. As $a_1/\mu \rightarrow \infty$, $\Delta$ approaches the upper bound,  \ie $1$ coming from the positivity of $A-B$ and $B$ matrices as shown in \eqref{delta-bound}.
The vertical red line indicates the lower bound $a_1/\mu =\frac{181}{1000}$ constrained by the positivity of $A-B$. The maximum relative difference with these parameter is $ 4.718\%$ at $a_1/\mu = 1.585$. Note that at $a_1/\mu=0.785$ the value of the relative difference of the complexity essentially vanishes (\ie $\frac{\tilde \mC^{\text{diag}}_1- \mC^{\text{diag}}_1}{\mC^{\text{diag}}_1}=6.583\times 10^{-8}$). This means that the mode-by-mode purification is optimal in this case, even though $A$ and $B$ do not commute and $\rho_{\mathcal{A}}$ is mixed in both modes. We might expect such ``coincidences'' from counting arguments similar to those that appear in the last paragraph of section \ref{genre} since the complexity is a scalar function of many parameters.} \label{deltaC_large}\end{figure}

Our numerical tests of the optimality of the mode-by-mode purifications compared to the complete minimization can be found in figures \ref{deltaC_onebyone} and \ref{deltaC_large}. Figure \ref{deltaC_onebyone}, demonstrates that the difference between the two complexities (mode-by-mode versus exact minimization) is very small when the matrices $A$ and $B$ are nearly commuting. Figure \ref{deltaC_large} explores a wider range of parameters, to include not nearly commuting matrices, \ie larger values of $\Delta$. We have scanned $a_1, a_2, \in [1,5]\,, b_1,b_2 \in [0.001, 3]$ and $g \in [0, 0.5]$ numerically, and found that with a large $\Delta$, the relative difference of complexity can rise up to about $5\%$, as shown in the figure \ref{deltaC_large}. We therefore see that at least in these cases, the mode-by-mode purifications provide a good approximation for the complexity.

To conclude this section, we are motivated by our numerical results for two-mode Gaussian states to make the second conjecture  that for the general $N_{\mA}$-mode Gaussian state $\rho_{\mA}$ whose density matrix elements satisfy
\begin{equation}
[A,B]=0\,,
\end{equation}
the optimal purification will be a mode-by-mode purification (in the diagonal basis). Further, when $[A,B]\ne0$ but these matrices are still close to commuting in the sense that $\Delta\ll1$, the mode-by-mode purification will still be a good approximation to the true optimal purification.

\subsection{Optimal Purification in the Physical Basis}\label{subsec3phys}

As we pointed out in section \ref{subsec:compure}, the $\mC_1$ complexity is basis dependent. In the previous subsection, we focused on the diagonal basis, and so here we would like to explore the sensitivity of our results to this choice. In particular, we will evaluate the purification complexity using, what we call, the physical basis. Recall that the diagonal modes are generally linear combinations of the physical degrees of freedom (in $\mA$ in eq.~\reef{Gaussian_AB}) and the auxiliary degrees of freedom (in  $\mA^c$) and further, these linear combinations are tuned in a way which depends on the state in question. Another natural basis would be one that separates the action of the fundamental gates~\eqref{gates} on the physical and ancillary degrees of freedom. More precisely, the generators might contain $\hat{x}_a$ ($\hat{p}_b$) which are linear combinations of positions (momenta) of physical oscillators or ancillae separately, but not both. Of course, we still require entangling gates which introduce entanglement between the two subsystems, \eg where $\hat{x}_a$ acts on $\mA$ and $\hat{p}_b$, on $\mA^c$. We denote this set of elementary gates, the physical basis.

In evaluating the $\mC_1$ complexity in the physical basis, we begin with the purification in eq.~\eqref{Gaussian_AB}. We then find the orthogonal transformation $O_{\mA\mA^c} = O_{\cal A} \otimes O_{\mA^c}$ (with $O_{\cal A} \in O(N_{\cal A})$,  $O_{\mA^c} \in O(N_{\mA^c})$) which diagonalizes the blocks $\Gamma$ and $\Omega$,
\beq
\label{diagonalize}
\Gamma' = O_{\cal A}\, \Gamma \,O_{\cal A}^T\,,\quad {\rm and} \quad \Omega' = O_{\mA^c}\, \Omega\,  O_{\mA^c}^T\,.
\eeq
The key difference from the diagonal basis is that this transformation leaves us with a nonvanishing off-diagonal matrix $K$, which captures the entanglement between the physical and ancillary subsystems. That is,
\beq
K' = O_{\cal A}\, K\, O_{\mA^c}^T \neq 0\,,
\eeq
and thus the purification takes the form
\begin{equation}\label{waffle}
\Psi_{\mA\mA^c} = \mathcal{N}_{\mA\mA^c}\ \exp\! \[  -\frac{1}{2} (\vec q_{\mA}{\!\!'}\ ,\vec q_{\mA^c}{\!\!\!\!'}\ ) \left(
\begin{array}{cc}
\Gamma' & K' \\
K'^T & \Omega' \\
\end{array}
\right) \left(
\begin{array}{c}
\vec q_{\mA}{\!\!'} \\
\vec q_{\mA^c}{\!\!\!\!'}\ \ \\
\end{array}
\right)        \]\,,
\end{equation}
which has diagonal blocks $\Gamma'$ and $\Omega'$ but nonvanishing off-diagonal blocks $K'$ and $K'^T$.
The physical basis complexity ${\cal C}^\mt{phys}_1$ is then given by eq. \eqref{rotate2pos}
\beq
\label{C1phys}
{\cal C}^\mt{phys}_1 = \sum_{a,b=1}^{N_{\cal A}+N_{\mA^c}} |H_{ab}|\,,
\eeq
where $H$ is the generator~\eqref{Hgen} producing the optimal trajectory in the physical basis.\footnote{It is important to keep in mind that the generator matrix $H$ is not diagonal in the physical basis. This matrix is diagonal only in the diagonal basis.} The generator matrix can be found by taking the matrix logarithm of the parameter matrix in eq.~\reef{waffle}, \ie
\beq\label{eq:Hphys}
H = \frac12\, \ln\!\left(\frac{M_\mt{T}}{\mu}\right)\qquad{\rm where} \quad  M_\mt{T} = \left(\begin{array}{cc}
    \Gamma' & K' \\
    K'^T & \Omega' \\
\end{array}\right)\,.
\eeq
We would like to stress that the original calculation of the pure state complexity was not optimized in this basis and so strictly speaking what we provide here is a bound on the physical basis complexity.

We now summarize how the results in section \ref{optimal} change for the physical basis.

\subsubsection{Degenerate purifications}

In section \ref{genre}, we discussed the $SO(N_{\mA^c})$ degeneracy of the purifications yielding equal complexities for any given mixed state. This degeneracy was characterized by orthogonal transformations $R_{\mA^c} \in SO(N_{\mA^c})$ of the ancillary degrees of freedom \eqref{transf99}. This degeneracy was due to the fact that a rotation of the degrees of freedom does not change the spectrum of the parameter matrix \eqref{newmat}, and the diagonal-basis complexity depends only on this spectrum and the reference scale $\mu$. Revisiting this question for the physical-basis complexity, we emphasize that this degeneracy is built into the definition of ${\cal C}^\mt{phys}_1$. Indeed, while the definition of physical-basis complexity~\eqref{C1phys} might not seem invariant under $SO(N_{\mA^c})$ transformations of the ancillary degrees of freedom at first sight, it is important to remember that the prescription to define the physical-basis complexity of any purification will give identical parameter matrix $M_\mt{T}$ after the canonical rotation required to diagonalize the blocks $\Gamma$ and $\Omega$. Consider any two purifications
\beq
 M_{\mt{T},1} = \left(\begin{array}{cc}
    \Gamma_1 & K_1 \\
    K_1^T & \Omega_1 \\
\end{array}\right) \,, \quad  M_{\mt{T},2} = \left(\begin{array}{cc}
\Gamma_2 & K_2 \\
K_2^T & \Omega_2 \\
\end{array}\right)\,,
\eeq
related by the transformation in eq.~\eqref{transf99}
\beq
\left(\begin{array}{cc}
    \Gamma_1 & K_1 \\
    K_1^T & \Omega_1 \\
\end{array}\right)
=
\left(\begin{array}{cc}
    \Gamma_2 & K_2 R_{\mA^c} \\
    R_{\mA^c}^TK_2^T & R_{\mA^c}^T \Omega_2 R_{\mA^c} \\
\end{array}\right)\,.
\eeq
Then the canonical transformations \eqref{diagonalize} diagonalizing the blocks $\Gamma_i$ and $\Omega_i$ will be related by
\beq
O_{1,\cal A} = O_{2,\cal A}\,, \quad O_{1,\mA^c} =  O_{2,\mA^c} R_{\mA^c}\,.
\eeq
The resulting physical-basis parameter matrix $M'_\mt{T}$ will be the same for both purifications, and consequently they will both have the same physical-basis complexity.

\subsubsection{Essential purifications}
In section \ref{threeA} using the diagonal basis, we showed that purifying a Gaussian mixed state \reef{dense1} for a single harmonic oscillator with two ancillae does not improve the purification complexity over the one found with a single ancilla. In the physical basis, we were not able to produce an analytical proof of the same result; however, our numerical results showed that again adding an extra ancillary degree of freedom did not improve the purification complexity for a wide range of single harmonic oscillator mixed states. In particular, we found that the purification complexity of the optimal purification with one ancilla and with two ancillae differed by $\frac{\Delta {\cal C}}{\cal C} \lesssim 10^{-13}$ for a wide range of mixed states.\footnote{The states considered numerically were of the form~\eqref{dense1} with $a\in [2\mu ,10\mu]$ and $b \in [\mu, a-\mu]$}
Moreover, when looking at the precise value of the parameters that minimize the complexity in the general case, we found that the purifications correspond to those which only entangle one ancilla to the physical oscillator, and the eigenvalue of the unentangled ancilla is simply the reference state scale $\mu$. These results seem to indicate that the conclusion of section \ref{threeA} applies to the physical-basis complexity as well. That is, we will assume that the optimal purification in the physical basis for a Gaussian mixed state with many degrees of freedom is again an essential purification.

\subsubsection{Mode-by-mode purifications}

Using more numerics, we examined the questions addressed in section \ref{MbyM} but here for the physical-basis complexity. In particular, we considered the conditions for the optimal purification of a mixed state of many degrees of freedom to be a mode-by-mode purification, and we found that the results are similar to those for the diagonal-basis complexity: the optimal purification of a Gaussian density matrix $\hat\rho$ for many modes is mode-by-mode when the density matrix is a product of single-mode density matrices (\ie $\hat\rho = \otimes \hat\rho_i$). More precisely, for a range
 of mixed states~\eqref{densityFun_A} of two harmonic oscillators characterized by commuting parameter matrices $A$ and $B$,\footnote{We considered states with parameter matrices of the form \eqref{gamble8}-\eqref{gamble} with $a_1 \in [2\mu , 4\mu]$, $a_2 \in [2\mu, 3\mu]$, $b_1 \in[\mu, a_1-\mu]$ and $b_2 \in[\mu, a_2-\mu]$.} we compared the complexity found by optimizing mode-by-mode purifications~\eqref{gamble8} with that found from more general purifications~\eqref{gamble2a}. Our numerical results showed essentially no difference, \ie $\frac{\Delta {\cal C}}{\cal C} \lesssim 10^{-12}$.

\begin{figure}[h]
    \includegraphics[width=0.5\textwidth]{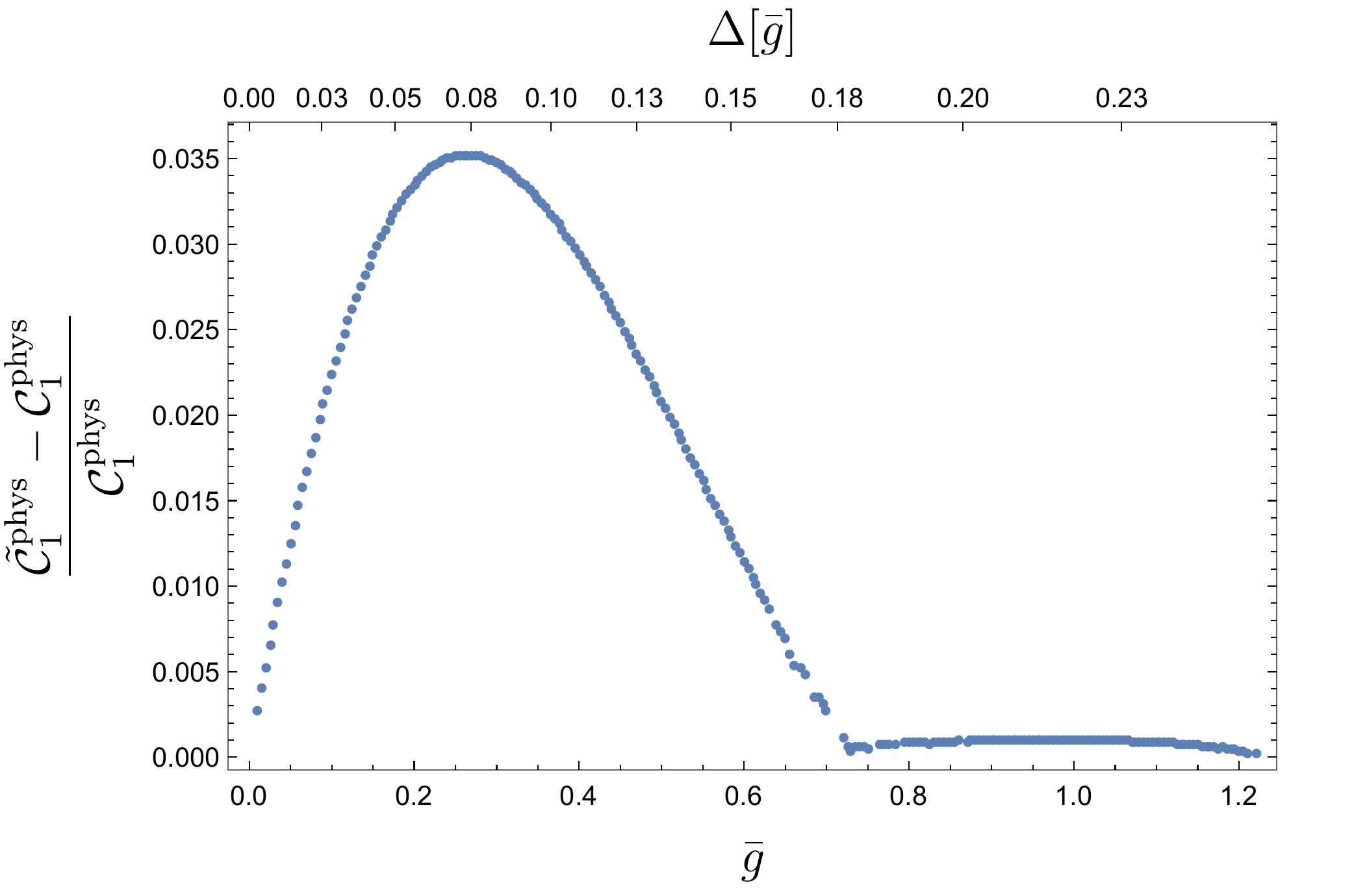}
    \includegraphics[width=0.5\textwidth]{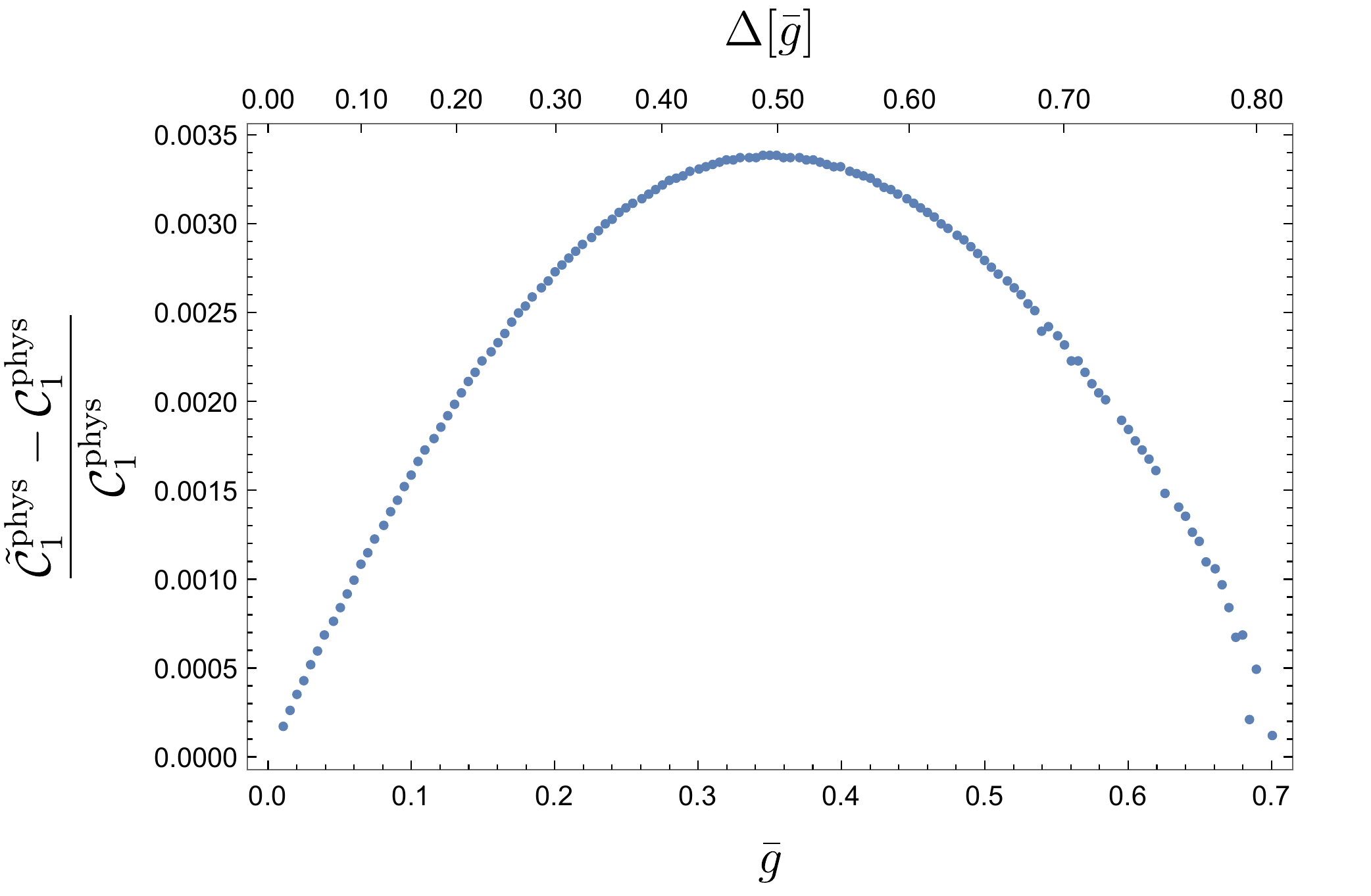}
    \caption{The relative difference between the mode-by-mode physical-basis `complexity' $\tilde \mC_1^\mt{phys}$ and the optimal physical-basis complexity $\mC_1^\mt{phys}$. In the left panel, we are focusing on nearly commuting $A$ and $B$ matrices with parameters: $\bar{a}_1=5\mu$, $\bar{b}_1=\mu$, $\bar{a}_2=3\mu$ $\bar{b}_2=1.5\mu$ and $\bar{g}\in [0,1.22\mu]$. In the right panel, we explore larger values of $\Delta$ with parameters: $\bar{a}_1=15\mu$, $\bar{b}_1=0.5\mu$, $\bar{a}_2=1.5\mu$ $\bar{b}_2=\mu$ and $\bar{g}\in [0,0.7\mu]$. The plots extend to the maximum allowed value for $\bar{g} $, which is determined by demanding $B>0$. } \label{deltaC_onebyone_position}
\end{figure}

For non-product density matrices (\ie where the parameter matrices $A$ and $B$ no longer commute), we compared the complexity found with general purifications to that found by only optimizing over mode-by-mode purifications.
We used the following convenient parametrization of the matrices $A$ and $B$ of the density matrix~\eqref{density_fun_A} in the physical basis\footnote{Note that this parametrization is not  the same as in section \ref{MbyM}, since $A$ is not diagonal here.}
\begin{equation}
A=\left(
\begin{array}{cc}
\bar{a}_1 & -\bar{g} \\
-\bar{g} & \bar{a}_2 \\
\end{array}
\right)
\qquad{\rm and}\qquad
B=\left(
\begin{array}{cc}
\bar{b}_1 & \bar{g} \\
\bar{g} & \bar{b}_2 \\
\end{array}
\right)\,.
\end{equation}
We again found that the difference is quite small -- see figures \ref{deltaC_onebyone_position}. The maximal difference obtained in these cases is about $3.5\%$. Note that in the cases examined there, we are fixing the parameters $\bar a_i$ and $\bar b_i$ while $\bar g$ varies. In this situation, there will be an upper bound on $\bar g$ which comes from requiring positivity of the parameter matrix $B>0$. As $\bar{g}$ reaches its maximum allowed value (\ie $\bar{g} \to 0.7\mu$ on the right), one of the eigenvalues of the  $B$ matrix approaches zero. Hence at this point, we are dealing with the purification of only one mode and as a result, the relative difference in complexity approaches zero. In the left panel of figure \ref{deltaC_onebyone_position}, the relative difference in complexity decreases earlier and gets close to zero across the entire range $\bar{g} \in \[0.725\mu,1.22\mu\]$. At present, we do not understand the reason for this usual behaviour,\footnote{We note that the left panel of figure \ref{deltaC_onebyone} seems to hint at similar behaviour.} but it may be related to the fact that one of the eigenvalues of $A-B$ also vanishes as $\bar{g} \to 1.22\mu$.

Let us add that the relative difference will not necessarily increase as $\Delta$ increases. In particular, it is also possible that the relative difference is very small, even for relatively large values of $\Delta$. As before, such ``coincidences'' could result from counting arguments similar to those that appear in the last paragraph of section \ref{genre} since the complexity is a scalar function of many parameters.

\section{Complexity of Thermal States in QFT} \label{apply01}
Now we wish to apply the techniques developed in the previous sections in order to evaluate the purification complexity for examples in quantum field theory (QFT). In particular, we start in this section with a thermal mixed state for a free scalar field theory.
As a simple exercise, we begin by considering the thermal state of a single harmonic oscillator. One question we ask here is while the thermofield-double (TFD) state for two harmonic oscillators provides a natural purification of the thermal state, is it ever the optimal purification for this state. Next, we briefly review the lattice regularization of a free scalar field theory, which reduces to a family of coupled harmonic oscillators. We then apply our results for the single oscillator case to examine the purification complexity for a thermal mixed state in the free scalar QFT, both in the diagonal basis and in the physical basis. In section \ref{sec:holo}, we follow up with a comparison of our results here with the analogous results from holographic complexity.

\subsection{Exercise: One-mode Thermal States} \label{exer4}

\begin{figure}[h]
\begin{center}
        \includegraphics[width=0.7\textwidth]{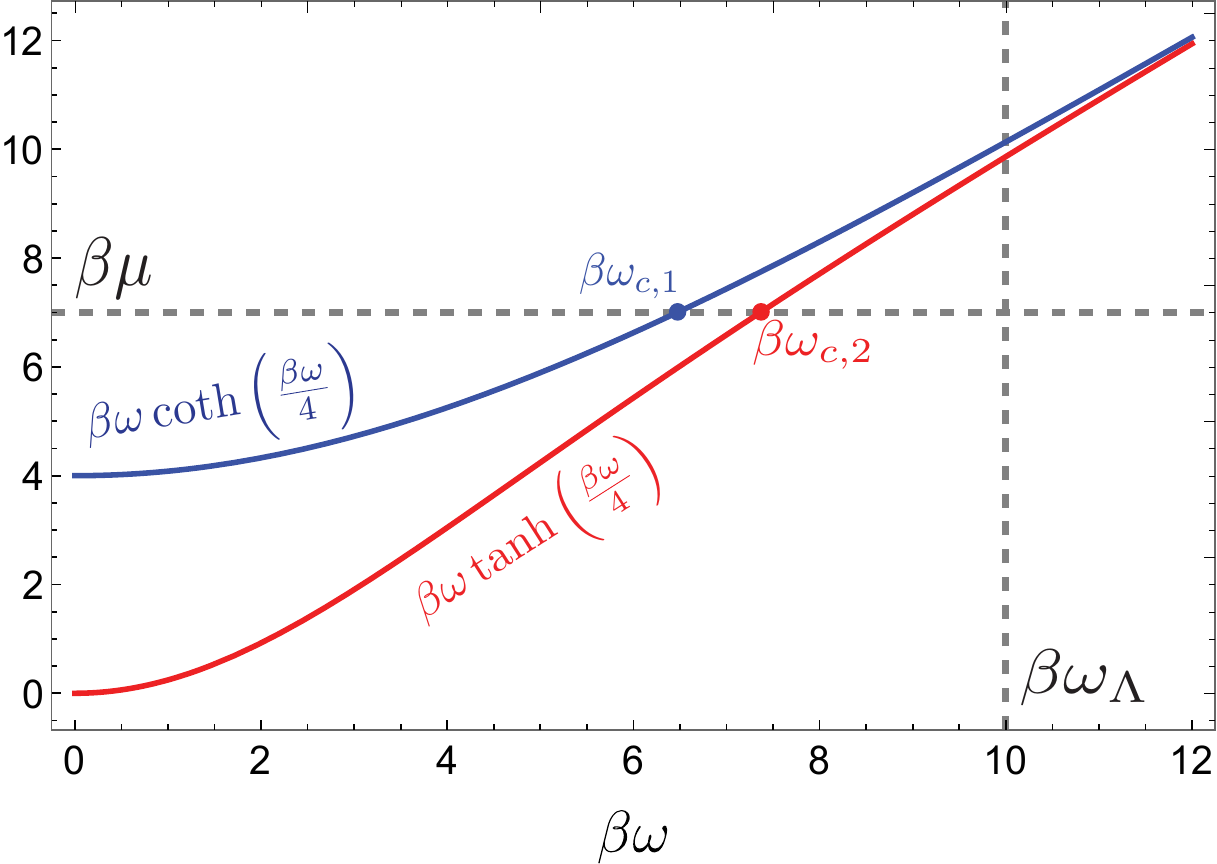}
\end{center}
          \caption{Different regimes of eq.~\eqref{complexity_thermal}: values of $\beta \mu$ above the blue curve, \ie $\beta \omega \coth(\beta\omega/4)$, correspond to the first regime in this equation; below the red curve \ie $\beta \omega \tanh(\beta\omega/4)$,  correspond to the third regime; while between the blue and red curves correspond to the second regime. We observe that when $\beta \mu\gg 1$, there is a very narrow range of frequencies $\beta \omega$ between the blue and red lines (since both curves converge towards $\beta\omega$) for which the intermediate regime applies. }
        \label{fig:Ffun}
\end{figure}

For simplicity, we start by analyzing the purification complexity of the thermal state for a single oscillator, \ie $\hat \upsilon_{{th}}(\beta,\omega)$ in eq.~\eqref{density_thermal}. For this exercise, we limit ourselves to considering the diagonal basis. In fact, this is a simple case of the one-mode mixed states \eqref{Gaussian_decom} studied in section \ref{sec:onemode}, where we set the squeezing parameter $r=0$. Hence the purification complexity is given by simply substituting $r=0$ into eq.~\eqref{complexity_one_mode},
\begin{equation}
\hspace{-16pt}\mC^{\mt{diag}}_{1,{th}}(\beta,\omega,\mu)=
\left\{
\begin{array}{lc}
 \frac 12 \ln \frac{\mu}{\omega} + \frac{1}{2}\ln \left(    \frac{{ \frac{\mu}{\omega}\,\coth(\beta \omega/2)-1}}{{\frac{\mu}{\omega}- \coth(\beta \omega/2)
}} \right)\, , &{\rm for}\ \   \coth(\frac{\beta \omega}{4}) \leq \frac{\mu}{\omega}\,,\\
\\
 \ln \coth\(\frac{\beta \omega}{4}\) \,  , &{\rm for}\ \   \tanh(\frac{\beta \omega}{4})\leq \frac{\mu}{\omega} \leq \coth(\frac{\beta \omega}{4})\,,\\
\\
\frac 12 \ln\frac{\omega}{\mu} +\frac{1}{2} \ln \left(    \frac{{\frac{\omega}{\mu}\,  \coth(\beta \omega/2)-1}}{{\frac{\omega}{\mu}-
\coth(\beta \omega/2)}} \right) , & {\rm for}\ \  \frac{\mu}{\omega} \leq \tanh(\frac{\beta \omega}{4}) \,.
\end{array}
\right.
\label{complexity_thermal}
\end{equation}
Here we have substituted $\bar r=\frac{1}{2}\ln \frac{\omega}{\mu}$ from setting $r=0$ in eq.~\eqref{rbar}, and we have used the definition of $\alpha$ given in eq.~\eqref{hope}. The interplay between the different regimes of eq.~\eqref{complexity_thermal} is explored in figure \ref{fig:Ffun}.

Of course, one well-known purification of the thermal state \eqref{density_thermal} is the TFD state, see eq.~\eqref{TFDpureonemode}. However, this is not necessarily the optimal purification which leads to a minimal complexity. Examining
eq.~\eqref{complexity_thermal}, it turns out that the optimal purification is in fact the TFD state for the intermediate regime, \ie
$\tanh({\beta \omega}/{4})\leq \frac{\mu}{\omega} \leq \coth({\beta \omega}/{4})$. This can be seen by observing that eqs.~\eqref{case2compl} and \eqref{rbar} yield $s=0$ when $r=0$ in this case and therefore the purification \eqref{Fock_psi12} reduces to the TFD state in eq.~\eqref{two_mode}. For example, this case will be of relevance when the reference frequency $\mu$ and the oscillator frequency $\omega$ are equal.

We may also consider two other interesting limits: First, for $\omega \coth
\frac{\beta \omega}{4} \ll \mu $, the first line in eq.~\eqref{complexity_thermal} applies and this limit yields
\begin{equation}
\mathcal{C}^{\mt{diag}}_{1,{th}}\simeq  \frac 12\, \ln\! \(\frac{\mu}{\omega}\, \coth
\frac{\beta \omega}{2}\)
\qquad {\rm with}\ \
{s} \simeq \frac 12\, \ln \!\(\frac{\mu}{\omega}\,\tanh
\frac{\beta \omega}{2}\)\,,
\end{equation}
see eq.~\eqref{case13r2} and \eqref{rbar}. Hence the optimal purification is far from being the TFD state, for which $s=0$. Next, in the opposite limit with $\mu  \ll \omega \tanh \frac{\beta \omega}{4}$,
 the third case in eq.~\eqref{complexity_thermal} applies. This limit then yields
\begin{equation}\label{thermal_complexity_case3}
\mathcal{C}^{\mt{diag}}_{1,{th}}
\approx  \frac 12\, \ln\! \(\frac{\omega}{\mu} \coth \frac{\beta \omega}{2}\) \qquad {\rm with}\ \
s\simeq \frac 12\, \ln \!\(\frac{\mu}{\omega}\,\coth
\frac{\beta \omega}{2}\)\,.
\end{equation}
Hence, the optimal purification is  again far from the TFD state.

While we have limited our attention to the diagonal basis here, the analogous results for the physical basis can be found by using $r=0$ in section \ref{fizz}.

\subsection{Discretization of the Free Scalar}\label{QFT}

In order to apply our results from the last several sections to a QFT, we follow \cite{qft1} and consider a free massive scalar theory with Hamiltonian
\begin{equation}\label{Ha_scalarQFT}
H=\frac{1}{2}\int d^{d-1}x\left[\pi(x)^2+(\vec\nabla\phi(x))^2+m^2\,\phi(x)^2\right]\,.
\end{equation}
We start by regulating the theory by placing it on a periodic `square' lattice with lattice spacing $\delta$ and where each side has a linear length $L$. Therefore the total number of sites is given by $N^{d-1}\equiv (L/\delta)^{d-1}$. The lattice Hamiltonian is then the Hamiltonian for $N^{d-1}$ coupled harmonic oscillators, which can be written as\footnote{The lattice sites are designated with $\vec{n}=n_i\,\hat{x}^i$, where $\hat{x}^i$ are unit normals along the spatial axes.}
\begin{equation}\label{ham88}
\begin{split}
H&=\sum_{\vec{n}}\left\{\frac{\bar p(\vec{n})^2}{2M}+\frac12 M \left[\bar{\omega}^2 \bar x(\vec{n})^2+\Omega^2\sum_i\( \bar x(\vec{n})-\bar x(\vec{n}-\hat{x}_i)\)^2\right]\right\}\,,
\end{split}
\end{equation}
where in the second line, we have defined $\bar x(\vec{n})=\delta^{d/2}\phi(\vec{n})$, $\bar p(\vec{n})=\delta^{(d-2)/2}\pi(\vec{n})$, $\bar{\omega}=m$ and $\Omega=1/\delta=M$, see, e.g., \cite{qft1}. Further, periodic boundary conditions are imposed with $\bar x(\vec{n}+ N\hat{x}_i)\equiv\bar x(\vec{n})$ for any $i$.
Next we rewrite the Hamiltonian in terms of the normal modes
\begin{equation}\label{Thefreq}
x_{\vec k}\equiv\frac{1}{N^{\frac{d-1}2}}\sum_{\vec n}\exp\!\left(\frac{2\pi i \vec k\cdot\vec n}{N}\right)\bar x(\vec n)\,, \quad \omega_{\vec k}^2= m^2 + 4\Omega^2 \sum_i \sin^2 \frac{\pi k_i}{N}\,,
\end{equation}
where $\vec k=(k_1,\cdots,k_{d-1})$ with $k_i=1,2,\cdots N$.   The Hamiltonian then becomes
\begin{equation}\label{normal_QFT}
H=\frac{1}{2M}\sum_{\vec k}\(|{p}_{\vec k}|^2+M^2{\omega}_{\vec k}^2\,|x_{\vec k}|^2\)\,,
\end{equation}
where we have used that $x^\dagger_{\vec k}= x_{-\vec k}$. This means that we can think of the system as a system of $N^{d-1}$ decoupled real harmonic oscillators with frequencies as indicated by eq.~\eqref{Thefreq} and with masses $1/\delta$.
Of course, the diagonalization process can also be performed directly for the continuum Hamiltonian and in the infinite volume limit,\footnote{Recall that there are two independent limits here. The continuum limit refers to taking the lattice spacing $\delta$ small compared to the other physical parameters in the problem, \eg $\delta m \to 0$ and $\delta/L\to0$. In that case, the sum over lattice points becomes an integral over positions on a square torus, given the boundary conditions under eq.~\reef{ham88}. The infinite volume takes the limit $L=N\delta\to\infty$ while holding $\delta$ fixed. Hence in this limit, $L$ is large compared to the other dimensionful parameters, \eg $mL\to\infty$ and $L/\delta\to\infty$.  Recall that the difference between adjacent values of the dimensionful momenta in eq.~\eqref{Thefreq} is $\Delta k=\frac{2\pi}{N\delta}= \frac{2\pi }{L}$, and hence the momentum sums are replaced with integrals in the infinite volume limit. The results of this section will  all involve both  the continuum and infinite volume limits, while those of section \ref{apply02} are given on the circle (\ie $d=2$) with finite $L$.\label{linecirclefoot}} in which case one obtains the eigenfrequencies $\omega_\vk=\sqrt{\vk^2+m^2}$ and the sum over the (dimensionless) $k_i$ is replaced by the (dimensionful) momentum integral $V_{d-1}\, \int\! \frac{d^{d-1}k}{(2\pi)^{d-1}}$. Here  $V_{d-1}=L^{d-1}$ was introduced as an IR regulator for the spatial volume of the system.

It is natural to interpret the reference state as the ground state of an ultralocal Hamiltonian of the form
\begin{equation}\label{ultra1}
H=\frac{1}{2}\int d^{d-1}x\left[\pi(x)^2+\mu^2\,\phi(x)^2\right]\,.
\end{equation}
That is, we have dropped the usual term with spatial derivatives here and so in the ground state, the field is not correlated at different spatial points.
On the lattice, this Hamiltonian \reef{ultra1} becomes
\begin{equation}\label{normal_QFT2}
H=\frac{1}{2M}\sum_{\vec k}\(|{p}_{\vec k}|^2+M^2\mu^2\,|x_{\vec k}|^2\)\,.
\end{equation}

Finally, recall that we have implicitly set the mass parameter $M$ to one in all our previous expressions, \eg in eqs.~\eqref{TargetGaussianPure} and \eqref{ref_state}. It is easy to restore the dependence on the mass by merely multiplying the frequencies by $M$. This does not influence the various expressions for the complexity since those were given in terms of ratios of frequencies.

\subsection{Purification Complexity in the Diagonal Basis}

As we noted above, the Hamiltonian \eqref{normal_QFT} consists of a sum of decoupled harmonic oscillators. As a consequence, the corresponding thermal density matrix for the QFT factorizes into a product of thermal density matrices, one for each mode. In other words, one can find the simple mixed state
\begin{equation}\label{density_QFT}
\hat\rho_{{th}}(\beta) =  \bigotimes_{\vec k}\ \hat{\upsilon}_{{th}}(\beta,\omega_{\vec k})\,,
\end{equation}
where $\hat \upsilon_{th}$ denotes the thermal density matrix of a single oscillator with frequency $\omega_{\vec k}$ and inverse temperature $\beta$, as defined in eq.~\eqref{density_thermal}. In proceeding with our evaluation of the purification complexity, we will focus here on the diagonal basis and save a discussion of the physical-basis complexity for section \ref{wonka}. Given a mixed state with a product structure as in eq.~\eqref{density_QFT}, we recall from section \ref{optimal} that we expect the optimal purification will be both an essential purification and a mode-by-mode purification.\footnote{To connect directly to the discussion in section \ref{sec:manyho}, we can write the thermal density matrix in the form given in eq.~\reef{density_fun_A} using the expressions in eq.~\reef{parameters} with $r=0$. In this form, we would find that $A$ and $B$ are commuting matrices with $A={\rm diag}(\omega_{\vec k}\, {\rm \coth} \beta\omega_{\vec k})$ and  $B= {\rm diag}(\omega_{\vec k}\, {\rm csch} \beta\omega_{\vec k})$.} Hence we expect that the final result for the purification complexity eq.~\eqref{density_QFT} is simply obtained by summing the complexities for the individual modes,
\begin{equation}\label{sum_complexity}
\mC_{1,th}^{\mt{diag,tot}}(\beta,\mu)= \sum_{\vec k} \mathcal{C}^{\mt{diag}}_{1,th}(\beta ,\omega_\vk,\mu) \,,
\end{equation}
where $\mathcal{C}^{\mt{diag}}_{1,th}(\beta ,\omega_\vk,\mu)$ is given in eq.~\eqref{complexity_thermal}.
Alternatively, in the continuum formulation, we have
\begin{equation}\label{sum_complexity02}
\mC_{1,th}^{\mt{diag,tot}}(\beta,\mu)=  V_{d-1}\ \int_{|\vec k|<\Lambda} \frac{d^{d-1}k}{(2\pi)^{d-1}} \ \mathcal{C}^{\mt{diag}}_{1,th}(\beta ,\omega_\vk,\mu) \,,
\end{equation}
where the momentum cutoff $\Lambda$ was introduced to regulate the system in the UV.\footnote{This regulator is different than the lattice regularization introduced above in that the momentum integration bound is a sphere, while the edge of the momentum integration of the lattice regularization is a cube given by the edges of the first Brillouin zone. The continuum limit corresponds to $\Lambda$ being much greater than any dimensionful parameter in the problem, \eg $\beta \Lambda  \to \infty$.}

To proceed, we define two critical frequencies with
\begin{equation}\label{dog3}
\omega_{c,1}: \quad \beta \mu=\beta \omega_{c,1} \coth \left(\frac{\beta \omega_{c,1}}{4}\right)\,, \qquad
\omega_{c,2}: \quad \beta \mu=\beta \omega_{c,2} \tanh \left(\frac{\beta \omega_{c,2}}{4}\right)\,.
\end{equation}
These correspond to the frequencies where there is a transition between the three different regimes in eq.~\reef{complexity_thermal} --- see the blue and red points indicated in figure \ref{fig:Ffun}. The critical frequencies are functions of $\beta$ and $\mu$, and of course, they can be converted to a corresponding momentum with $k_{c,1}^2=\omega_{c,1}^2-m^2$ and $k^2_{c,2}=\omega_{c,1}^2-m^2$.  Now we will evaluate eq.~\reef{sum_complexity02} for the three cases distinguished by the relation between the critical frequencies and the cutoff frequency $\omega_\Lambda \equiv \sqrt{\Lambda^2+m^2}$:
\begin{enumerate}
\item $\omega_\Lambda< \omega_{c,1}$:
\begin{equation} \label{dog35}
\hspace{-30pt}\mC_{1,th}^{\mt{diag,tot}}(\beta,\mu)=\frac{\Omega_{d-2}V_{d-1}}2\int_0^\Lambda \frac{k^{d-2}\,dk}{(2\pi)^{d-1}}\,\left[
\ln \frac{\mu}{\omega_\vk} + \ln \!\left(    \frac{{ \mu\,\coth(\beta \omega_\vk/2)-\omega_\vk}}{{\mu-\omega_\vk\, \coth(\beta \omega_\vk/2)
}} \right)\right]
\end{equation}
\item $\omega_{c,1}<\omega_\Lambda< \omega_{c,2}$:
\begin{equation}\label{dog4}
\begin{split}
\hspace{-40pt}\mC_{1,th}^{\mt{diag,tot}}(\beta,\mu)=&\frac{\Omega_{d-2}V_{d-1}}2\int_0^{k_{c,1}} \frac{k^{d-2}\,dk}{(2\pi)^{d-1}}\,\left[
\ln \frac{\mu}{\omega_\vk} + \ln \!\left(    \frac{{ \mu\,\coth(\beta \omega_\vk/2)-\omega_\vk}}{{\mu-\omega_\vk\, \coth(\beta \omega_\vk/2)
}} \right)\right]
 \\
 &\qquad+
 \Omega_{d-2}V_{d-1} \int_{k_{c,1}}^{\Lambda} \frac{ k^{d-2}dk}{(2\pi)^{d-1}} \,  \ln \coth\(\frac{\beta \omega_{\vk}}{4}\)
 \end{split}
\end{equation}
\item $\omega_{c,2}<\omega_\Lambda$:
\begin{equation}\label{dog5}
\begin{split}
\hspace{-40pt}\mC_{1,th}^{\mt{diag,tot}}(\beta,\mu)=&\frac{\Omega_{d-2}V_{d-1}}2\int_0^{k_{c,1}} \frac{k^{d-2}\,dk}{(2\pi)^{d-1}}\,\left[
\ln \frac{\mu}{\omega_\vk} + \ln \!\left(    \frac{{ \mu\,\coth(\beta \omega_\vk/2)-\omega_\vk}}{{\mu-\omega_\vk\, \coth(\beta \omega_\vk/2)
}} \right)\right]
\\
&\qquad+
\Omega_{d-2}V_{d-1} \int_{k_{c,1}}^{k_{c,2}} \frac{ k^{d-2}dk}{(2\pi)^{d-1}} \,  \ln \coth\(\frac{\beta \omega_{\vk}}{4}\)
 \\
 &+
 \frac{\Omega_{d-2} V_{d-1}}{2} \int_{k_{c,2}}^\Lambda \frac{k^{d-2}dk}{(2\pi)^{d-1}}\,\left[
 \ln \frac{\omega_{\vk}}{\mu}  + \ln\! \left(    \frac{ \omega_{\vk}\,\coth(\beta \omega_{\vk}/2)-\mu}{\omega_k-\mu\,\coth(\beta \omega_{\vk}/2)} \right)\right],
\end{split}
\end{equation}
\end{enumerate}
where $\Omega_{d-2}\equiv{2 \pi^{\frac{d-1}{2}}}/{\Gamma(\frac{d-1}{2})}$ is the volume of a unit ($d-2$)-sphere.

These results can be simplified in certain limits. In particular, here we will focus on the case of a massless scalar,
\ie $m=0$, in which case, the critical frequencies and momenta are equal to one another, \ie $k_{c,1}=\omega_{c,1}$ and $k_{c,2}=\omega_{c,2}$.
We also focus on the case where the reference frequency is much larger than the temperature, \ie $\beta \mu \gg 1$. Working in this regime, eq.~\reef{dog3} can be solved for the critical momenta in a perturbative expansion yielding
\begin{equation}
 k_{c,1}=\mu \left(1-2e^{-\frac{\beta \mu}{2}}+\cdots\right)\,  , \qquad
  k_{c,2}=\mu \left(1+2e^{-\frac{\beta \mu}{2}}+\cdots\right)\,.
\end{equation}
Hence we see that only the first case is relevant when $\mu \gtrsim \Lambda$ and that the third case becomes relevant as well when $\mu\lesssim\Lambda$. Further, since $k_{c,2}-k_{c,1}=4\mu e^{-\frac{\beta \mu}{2}} +\cdots$,
we see that the range of the integration in the second lines of eqs.~\reef{dog4} and \reef{dog5} is extremely small and the corresponding contributions are exponentially suppressed for $\beta \mu \gg 1$. Therefore, it is reasonable to ignore the contribution of these integrals to the complexity in the following.

Let us also comment on the behaviour of the various integrals near their limits of integration. First, near $k=0$, the integrands have at worst a logarithmic divergence in $d=2$, while this is suppressed by the factor of $k^{d-2}$ in higher dimensions, and so the integrals converge there. Logarithmic divergences also appear at $k_{c,1}$ and $k_{c,2}$, \ie $\ln(k_{c,1}-k)$ and $\ln(k-k_{c,2})$, and so the integrals are well behaved there. This leaves us with a UV divergence due to the terms proportional to $|\ln{\mu/\omega_{\vk}}|$. In fact, this contribution is identical to that for the vacuum state of the free scalar Hamiltonian \eqref{Ha_scalarQFT}, \eg see \cite{qft1}, and hence the UV divergence in the complexity is identical to that in the complexity of the vacuum state.

We note that the latter result is different from what happens for the TFD state for the same Hamiltonian  \eqref{Ha_scalarQFT}, where the UV divergence is precisely double that of the vacuum, \eg see \cite{Chapman:2018hou}. This doubling is natural if we think of the TFD state as an entangled state of two copies of the underlying QFT. In this case, the circuit constructing the state is introducing entanglement at short distance (\ie UV) scales in both copies of the QFT, which produces the UV divergences in the complexity. For the thermal mixed state, this short distance entanglement must be introduced for the physical degrees of freedom, but there is no need to do the same for the auxiliary degrees of freedom. Hence it is natural that the UV divergence in the purification complexity of the thermal state matches that in the complexity of the vacuum state.
We return to comment on this point and explicitly evaluate eqs.~\reef{dog35}-\reef{dog5} in section \ref{compare7}.

To close here, we note that the final result for the purification complexity (with $m=0$) can be shown to be proportional to $V_{d-1}\, T^{d-1}$, or equivalently to the thermal entropy, where the proportionality factor is a function of $\beta \Lambda$ and $\beta \mu$.
For later convenience, let us quote the result for the entropy of the thermal state for the massless theory,
\begin{equation}\label{green4}
\begin{split}
S\(\hat{\rho}_{\text{th}} \)\big|_{m=0}&=\frac{\Omega_{d-2}}{(2\pi)^{d-1}}\, \frac{\zeta(d)\Gamma(d+1)}{d-1}\,V_{d-1}\,T^{d-1}\,.
\end{split}
\end{equation}
We recall that ref.~\cite{Chapman:2018hou} showed that the complexity of formation for the TFD state is also proportional to the entropy when $m=0$.

\subsection{Purification Complexity in the Physical Basis}
\label{wonka}

Recall from sections \ref{sec:onemode} and \ref{fizz} that the complexity typically shows different properties in the diagonal and physical bases.  Hence we investigate the purification complexity for the thermal mixed state in the physical basis in this section. However, for the free scalar field theory where the density matrix takes the simple product form shown in eq.~\eqref{density_QFT},  we still expect that in the physical basis, the optimal purification will be an essential purification and also a mode-by-mode purification. So
the final result for the purification complexity  is again obtained by summing the complexities for the individual modes, \ie
\begin{equation}\label{sum_complexity98}
\mC_{1,th}^{\mt{phys,tot}}(\beta,\mu)= \sum_{\vec k} \mathcal{C}^{\mt{phys}}_{1,th}(\beta ,\omega_\vk,\mu) \,,
\end{equation}
where $\mathcal{C}^{\mt{phys}}_{1,th}(\beta ,\omega_\vk,\mu)$ is the purification complexity of the one-mode thermal density matrix, \ie of eq.~\eqref{Gaussian_decom} with $r=0$.   Alternatively, in the continuum formulation, we have
\begin{equation}\label{sum_complexity69}
\mC_{1,th}^{\mt{phys,tot}}(\beta,\mu)=  V_{d-1}\ \int_{|\vec k|<\Lambda} \frac{d^{d-1}k}{(2\pi)^{d-1}} \ \mathcal{C}^{\mt{phys}}_{1,th}(\beta ,\omega_\vk,\mu) \,,
\end{equation}
where the momentum cutoff $\Lambda$ regulates the UV portion of the integral.

Let us begin by examining $\mathcal{C}^{\mt{phys}}_{1,th}(\beta ,\omega,\mu)$, which is simply determined by setting $r=0$ or $\bar r=\frac12\ln(\omega/\mu)$ in the results of section \ref{fizz}.\footnote{We have dropped the subscript $\vk$ on the frequency here to reduce the clutter in our formulae for the time being. Further recall that the result for $\bar r$ follows from eq.~\reef{rbar}.} As shown in that section, we cannot find the full analytical results for the purification complexity in the physical basis. However, we can consider certain limits where the results are simplified. In particular, we now investigate the limit of small $\alpha $, which corresponds either to a low-temperature limit or a high-frequency limit, \ie $\beta \omega \gg 1$. In this limit, eq.~\reef{hope} yields $\alpha \simeq e^{-\beta\omega/2} \ll 1 $. Further, for small $\alpha$, the diagonal and physical bases are very close, \ie the orthogonal transformation in eq.~\reef{thetadef} is close to the identity. The latter follows from evaluating the expressions in eq.~\reef{rounder} with $\alpha\to 0$ and assuming $\sinh(\bar r-\bar s)=-\sinh s<0$, which yields\footnote{That is, we are assuming that the auxiliary squeezing parameter is positive, \ie $s>0$. Later, we see that this corresponds to $\mu>\omega$.
Footnote \ref{foot66} comments on the regime $s<0$, which corresponds to $\mu<\omega$.}
\beq\label{star1}
\theta\simeq \alpha/\sinh s +{\cal O}(\alpha^3)\,.
\eeq
Now since we want to expand our expressions for small $\alpha$, it is easiest to use $s$ as the optimization parameter in evaluating the purification complexity, in analogy to eq.~\reef{walk2}.\footnote{This contrasts with section \ref{fizz}, where we optimized with respect to $\theta$ as  in eq.~\reef{purification_C_pos}.}

In the physical basis, the single mode purification complexity is given by minimizing eq.~\reef{complexityPosition123}. Hence we must evaluate the expressions there in terms of $s$ and in a small $\alpha$ expansion using eqs.~\reef{eq:omegatotheta} and \reef{star1} as well as $r=0$.
We find\footnote{Note that  the first equation is exact because  $\bar{r}+ \bar{s}= 2\bar{r}+s$ with $r=0$.}
\begin{equation}\label{star2}
 \frac12 \ln \frac{\omega_+\omega_-}{\mu^2} = 2\bar{r} +s\,,
\qquad \frac12 \ln \frac{\omega_+}{\omega_-} = s + {\cal O}(\alpha^2)\,.
\end{equation} Now we see that eq.~\eqref{complexityPosition123} reduces to
\begin{equation}\label{lumber3}
\mC_{1,th}^{\mt{phys}} \left(\ket{\psi}_{12}\right) = |\bar{r}| +\left|\bar{r}+s \right|+\frac{2\, \alpha\,s}{\sinh s} +{\cal O}(\alpha^2)\,.
\end{equation}
At the leading order in $\alpha$, this is minimized when the second absolute value vanishes, which fixes $s=-\bar r=\frac12\ln(\mu/\omega)$ (which implies $\bar s=0$). Further, we note that consistency with our assumption that $s>0$ requires that we are in the regime $\mu>\omega$.\footnote{Let us add that if we assume $s<0$, we are lead to the following approximation
\begin{equation}
\begin{split}
\theta &= \frac{\pi}{2}- \frac{\alpha}{\sinh |s|} +{\cal O}(\alpha^3)\,,\qquad \text{with} \\
\frac12 \ln \frac{\omega_+\omega_-}{\mu^2} &= 2\bar{r} +s\,,
\qquad \frac12 \ln \frac{\omega_+}{\omega_-} = |s| + {\cal O}(\alpha^2)\,.
\end{split}
\end{equation}
The expression for the complexity in eq.~\reef{lumber3} remains unchanged, and it is again minimized by setting the second term to zero. Hence, we find  $s=-\bar{r} =\frac12\ln(\mu/\omega)$ as before,  but consistency with $s<0$ now requires that we are in the regime $\mu<\omega$. The final expression for the purification complexity \reef{small_alpha} also remains unchanged in this regime. \label{foot66}}
Hence in the region $\beta\omega\gg1$, we find that the purification complexity becomes\footnote{Note that the $\omega\to\mu$ limit of this expression agrees with the complexity of the thermofield double ${\cal C}_{1,th}^{\rm phys}\to 2\alpha$, as expected from the results of section~\ref{warmup}: namely, that the optimal purification for states with $\omega=\mu$ is the thermofield double.}
\begin{equation}\label{small_alpha}
\mC_{1,th}^{\mt{phys}} (\hat\upsilon_{th}) =\frac12\,\left| \ln\frac{\mu}{\omega} \right| +\frac{2 \alpha \,\ln\frac{\mu}{\omega}}{\sqrt{\mu/\omega}-\sqrt{\omega/\mu}} + {\cal O}(\alpha^2)\,.
\end{equation}
This result is very close to the complexity for the (pure) vacuum state of a single harmonic oscillator at frequency $\omega$, as expected.
Now let us turn to the purification complexity of the mixed thermal state for the free scalar field theory. As noted above, we expect that it takes the simple form given in eq.~\reef{sum_complexity98} or \reef{sum_complexity69} given the simple product structure of the thermal state \eqref{density_QFT}.
At this point, let us recall the definitions of our parameters for the thermal state
   \begin{equation}\label{termal_parameter}
   \alpha = \frac{1}{2}\, \ln \!\( \coth \frac{\beta\omega_\vk }{4}\) \,, \qquad   \bar{r}=\frac12\, \ln \frac{\omega_\vk}{\mu} \,.
   \end{equation}
As the combination $\beta\omega_\vk$ grows, the value of $\alpha$ rapidly decreases,
\eg $\frac{1}{2}\ln \( \coth \(10^{-2}\)\) \approx 2.3$, $\frac{1}{2}\ln \( \coth (10^2) \) \approx 10^{-87}$.  Now the momentum integral in eq.~\reef{sum_complexity69} is dominated by the phase space near the UV cutoff $|\vk|\sim \Lambda$ and hence with $\beta\Lambda\ll 1$, $\alpha$ will be very small over a majority of this integration.
Further, if the reference frequency $\mu$ is large enough, \eg near the cutoff $\Lambda$, we will have $-\bar r$ very large over the complementary part of the momentum integral. Hence, we can expect in a physically interesting setting that, over the entire integral, either $\alpha$ is small or $|\bar r|$ is large, and this is precisely the regime where the single-mode purification complexity in the physical basis is given by the simplified expression in eq.~\eqref{pos_ac}.
Hence we can simplify eq.~\reef{sum_complexity69} to the following
\begin{equation}\label{sum_complexity49}
\mC_{1,th}^{\mt{phys,tot}}(\beta,\mu)=  \Omega_{d-2}V_{d-1} \int_0^\Lambda \frac{ k^{d-2}dk}{(2\pi)^{d-1}} \, (\sin 2\theta_c + \cos 2\theta_c)\,\sinh ^{-1}\!\(\frac{\sinh 2 \alpha}{\sin 2\theta_c} \) \,,
\end{equation}
where both $\theta_c$ and $\alpha$ are implicitly functions of $k$ --- see eqs.~\reef{gumby7} and \reef{termal_parameter}.
However, it is still hard to explicitly do the remaining integral without any further assumptions. If we assume the small $\alpha$ limit is valid over most of the momentum integral, we can use eq.~\eqref{small_alpha} to simplify the purification complexity to
  \begin{equation}\label{laugh88}
  \mC_{1,{th}}^{\mt{phys,tot}}(\beta,\mu) \simeq \Omega_{d-2}V_{d-1} \int_0^\Lambda \frac{ k^{d-2}dk}{(2\pi)^{d-1}} \, \left[\frac12\,
 \left|  \ln\frac{\mu}{\omega_k} \right|+ \frac{ \ln\! \( \coth\! \frac{\beta\omega_{k} }{4}\) \,\ln\frac{\mu}{\omega_k}}{\sqrt{\mu/\omega_k}-\sqrt{\omega_k/\mu}} \right]\,,
  \end{equation}
where we use the notation $\omega_k=\sqrt{k^2+m^2}$ and where we have only dropped the higher order terms in the $\alpha$ expansion. Note that this approximation of the integrand is valid in the UV portion of the integration. In this case, the first term simply reproduces the vacuum complexity (\ie the zero temperature complexity) and hence the purification complexity has precisely the same UV divergences as the vacuum complexity (for one copy of the underlying QFT). Of course, this feature is identical to what we found for the diagonal basis. Further,
this approximation is valid more generally in the full range of integration in the situation where $\beta m\gg1$. In this case, the second term gives the leading finite temperature corrections to the vacuum complexity, which are suppressed by factors of $e^{-\beta m/2}$.

\subsection{Mutual Complexity of TFD States}
\label{compare7}

In this section, we compare the purification complexity of a thermal mixed state with the complexity of the corresponding TFD state, using a quantity known as the mutual complexity. We follow the nomenclature introduced by \cite{Ali:2018lfv} in considering the holographic complexity of subregions.

Consider a pure state $\ket{\Psi_{\mA\mB}}$ on a collection of degrees of freedom comprised of two subsystems, $\mA$ and $\mB$. There are two mixed states that are naturally constructed here, namely, the reduced density matrices,
\beq
\hat{\rho}_\mA={\rm Tr}_\mB(\ket{\Psi_{\mA\mB}}\bra{\Psi_{\mA\mB}})\,,\qquad
\hat{\rho}_\mB={\rm Tr}_\mA(\ket{\Psi_{\mA\mB}}\bra{\Psi_{\mA\mB}})\,.
\eeq
It is clear that each of the purification complexities for $\hat{\rho}_\mA$ and $\hat{\rho}_\mB$ is less than the complexity of the original pure state. That is, since $\ket{\Psi_{\mA\mB}}$ provides one particular purification of $\hat{\rho}_\mA$, it is unlikely to be the optimal purification and so we have the inequality
\begin{equation} \label{ramen1}
\mC\!\(\hat{\rho}_\mA\)  =\text{min} \ {\mC\!\( \ket{\Psi_{\mA\mA^c}}\)}  \le   \mC\!\( \ket{\Psi_{\mA\mB}}\)   \,,
\end{equation}
as well as the analogous inequality for $\hat{\rho}_\mB$. Implicitly, we chose the same cost function and basis to define the circuit complexity of the pure state $\ket{\Psi_{\mA\mB}}$.\footnote{Note the choice of basis is important in establishing the inequality for the $F_1$ cost function, which we are implicitly using here. For example, in eq.~\reef{ramen1}, we are {\it not} claiming that $\mC^{\mt{phys}}_1\(\hat{\rho}_\mA\)  \le   \mC^{\text{diag}}_1 \( \ket{\Psi_{\mA\mB}}\)$, even though $\mC^{\text{diag}}_1 \( \ket{\Psi_{\mA\mB}}\)$ may seem the natural definition for the complexity of the pure state. Of course, the basis choice does not play a role for covariant cost functions such as $F_2$. }

\begin{figure}[htbp]
        \centering\includegraphics[width=4.5in]{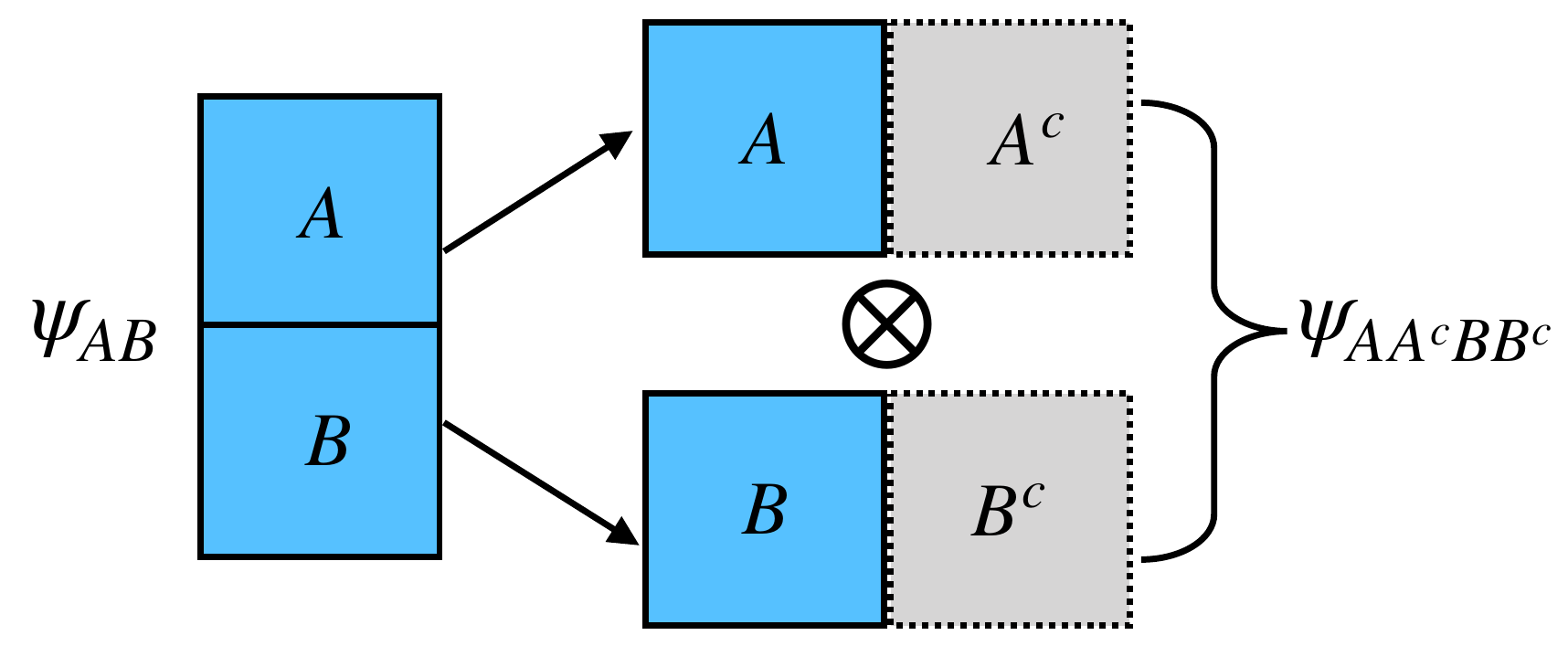}
        \caption{Illustration of the optimal purification of two mixed states in two complementary subsystems $\mA$ and $\mB$ of an original pure state $\ket{\Psi}_{\mA\mB}$. The state in the subsystem $\mA$ is purified by a state $\ket{\Psi}_{\mA\mA^c}$ and the one in the subsystem $\mB$ is purified by $\ket{\Psi}_{\mB\mB^c}$. Even though the direct product of the purifying systems $\ket{\Psi}_{\mA\mA^c} \otimes \ket{\Psi}_{\mB\mB^c}$ generally  has a larger number of degrees of freedom than the original state $\ket{\Psi}_{\mA\mB}$, the mutual complexity eq.~\eqref{ramen2A} can have either sign.}\label{purificationAB}
\end{figure}

As illustrated in figure \ref{purificationAB},  it is also obvious that in building the pure state, \eg $\ket{\Psi_{\mA\mA^c}}$, from the corresponding unentangled reference state, the circuit should only work hard enough to establish the correlations found in $\hat\rho_\mA$ amongst the physical degrees of freedom. However, it need not establish an analogous set of correlations (in particular, analogous UV correlations) amongst the ancillary degrees of freedom. Similarly, the correlations between $\mA$ and $\mA^c$  in $\ket{\Psi_{\mA\mA^c}}$ need not precisely mirror those between $\mA$ and $\mB$ in  $\ket{\Psi_{\mA\mB}}$.  As discussed in the introduction, the {\bf mutual complexity} is constructed to quantify the additional correlations in the original pure state with the following difference of complexities,
\begin{equation}\label{ramen2A}
\Delta\mC =  \mC\!\(\hat{\rho}_\mA\) + \mC\!\(\hat{\rho}_\mB\) - \mC\!\( \ket{\Psi_{\mA\mB}}\)   \,.
\end{equation}
This quantity was introduced in \cite{Ali:2018lfv}, where it was studied for subregions in the context of holographic complexity. The structure in eq.~\reef{ramen2A} was chosen to parallel that of the mutual information, which can be defined by a similar difference of entanglement entropies. However, whereas the mutual information is always positive (or zero),  we cannot prove that  $\Delta\mC$ is always positive or negative from the basic definitions of complexity and purification complexity. Hence the sign of the mutual complexity is nontrivial.

In the present case, the pure state of interest will be a TFD state, \ie $\ket{\Psi_{\mA\mB}}=\ket{\rm TFD}$, which can be regarded as an entangled state of two copies, \ie the left and right copies, of the underlying QFT. The corresponding mixed states will both be the thermal state \reef{density_QFT}, which is produced by tracing over either the left or right degrees of freedom, \ie $\hat\rho_\mA=\hat\rho_\mB=\hat\rho_{th}(\beta)$. That is, we will consider
\begin{equation}\label{ramen2}
\Delta\mC =  2\, \mC\!\(\hat\rho_{th}(\beta)\) - \mC\!\( \ket{\rm TFD}\)   \,.
\end{equation}
Again, while the TFD state provides one purification of the thermal mixed state, it will not generally be the optimal purification.\footnote{Let us point out that by examining figure \ref{fig:Ffun}, we find that there exist situations for which the TFD state is the optimal purification, but this requires $\beta \Lambda$ to be an order one number. However, we regard such a situation where the temperature is of the same order as the UV cutoff as unphysical.}

Another noteworthy feature of the mutual complexity \reef{ramen2} is that we expect it to be UV finite for the TFD state. This expectation arises from our previous observation that the UV divergences in the purification complexity of $\hat\rho_{th}(\beta)$ precisely matched those found in the vacuum state of one copy of the QFT, while the TFD state doubles the prefactors in those UV divergences. Hence we will see that these divergences cancel in our calculations below.

We refer to complexity models with the property that the mutual complexity is always positive as satisfying {\bf subadditivity} since in these cases the complexity of the combined state $\Psi_{\mA\mB}$ is less than the sum of the complexities of the two reduced density matrices, $\hat{\rho}_\mA$, and $\hat{\rho}_\mB$ \cite{BrianMixedComplexity} --- see also the discussion in \cite{Caceres:2018blh}. In the same way, we refer to complexity models as satisfying {\bf superadditivity} if $\Delta\mC$ is always negative. Further, in section \ref{sec:holo}, we will also see that the mutual complexity plays a role in distinguishing different holographic conjectures for the complexity of mixed states.

\subsubsection{Mutual complexity  in the diagonal basis}\label{subsubsubmcdb}
\begin{figure}[htbp]
 \center
 \includegraphics[width=0.7\textwidth]{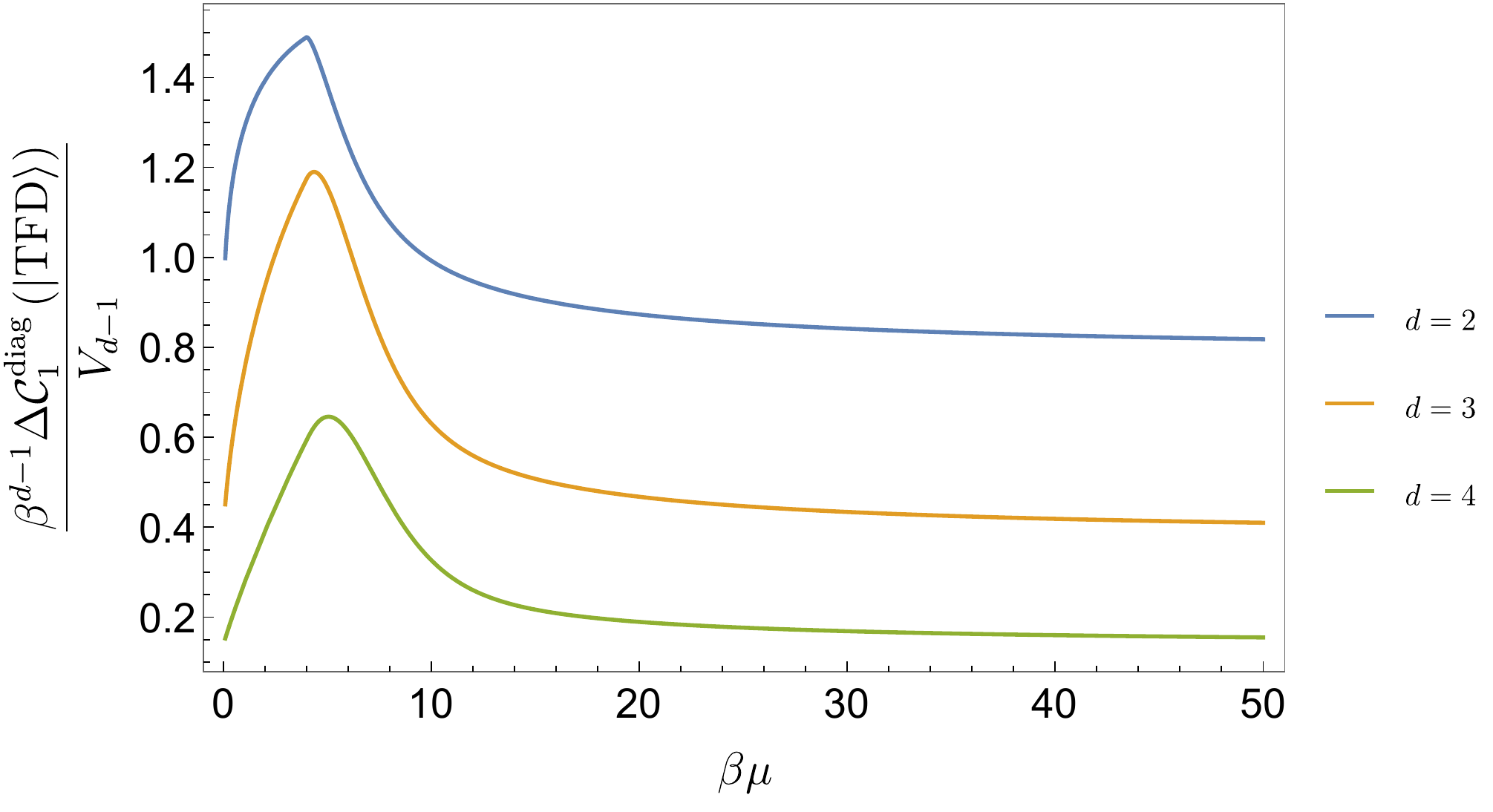}
 \caption{The integrated mutual complexity in the diagonal basis,~\ie $\Delta \mC_{1}^{\text{diag}} \(\ket{\text{TFD}}\) $ defined in eq.~\eqref{delta_diag} for a massless field theory in different dimensions.}\label{Delta_diag}
 \end{figure}

Let us begin with the TFD state entangling two modes. Eq.~\eqref{two_mode} shows that $\ket{\text{TFD}}_{12}$ is the two-mode squeezed state with $r=s=0$, and from eq.~\reef{eq:cases1}, we can see that its circuit complexity with the $F_1$ cost function in the diagonal basis reads \cite{Chapman:2018hou}
\begin{equation}\label{poll3}
\begin{split}
\mC_{1}^{\mt{diag}} \(\ket{\text{TFD}}_{12}\)&=
\left|  \frac 12 \ln \frac{\omega}{\mu}  + \alpha \right|
+\left|  \frac 12 \ln \frac{\omega}{\mu}  -  \alpha \right| \,,\\
&= \left\{
\begin{array}{llc}
\ln \frac{\mu}{\omega} & &{\rm for}\ \  \coth(\frac{\beta \omega}{4}) \leq \frac\mu\omega\,,\\
\\
\ln \coth\!\(\frac{\beta \omega}{4}\)  \quad &\ \ \ \ &{\rm for}\ \  \tanh(\frac{\beta \omega}{4})\leq \frac\mu\omega \leq \coth(\frac{\beta \omega}{4})\,,\\
\\
\ln \frac{\omega}{\mu} & &{\rm for}\ \ \frac\mu\omega \leq \tanh(\frac{\beta \omega}{4}) \,.
\end{array}
\right.
\end{split}
\end{equation}
Here we have expressed the three parameter regimes in the same way as they appears in eq.~\eqref{complexity_thermal} for the purification complexity of the thermal mixed state. Obviously, the results in the intermediate regime are the same in both cases because the optimal purification for the thermal state in this region coincides with the TFD state, as shown in section \ref{exer4}.
As noted in eq.~\reef{ramen2}, the two subsystems are described by the same mixed state, \ie $\hat{\rho}_{1,2}=\hat{\upsilon}_{{th}}$, and hence  the mutual complexity of this TFD state in the diagonal basis becomes
\begin{equation}\label{Delta_norm}
\Delta\mC_{1}^{\mt{diag}}\!\(\ket{\text{TFD}}_{12}\) = 2\,\mC_{1}^{\mt{diag}}\!\(\hat{\upsilon}_{{th}}\) - \mC_{1}^{\mt{diag}}\! \(\ket{\text{TFD}}_{12}\)\,.
\end{equation}
Combining eqs.~\eqref{complexity_thermal} and \reef{poll3}, we find
\begin{equation}\label{hunt1}
\Delta\mC_{1}^{\mt{diag}} (\ket{\text{TFD}}_{12})= \left\{
\begin{array}{llc}
\ln\! \left(    \frac{{ \mu\coth(\beta \omega/2)-\omega}}{{\mu-\omega \coth(\beta \omega/2)
}} \right) &&{\rm for}\ \   \omega\coth(\frac{\beta \omega}{4}) \leq \mu\,,\\
\\
\ln \coth\!\({\beta \omega}/{4}\)  &\ \ \ \ &
{\rm for}\ \ \omega\tanh(\frac{\beta \omega}{4})\leq \mu \leq \omega\coth(\frac{\beta \omega}{4})\,,\\
\\
\ln\! \left(    \frac{{\omega \coth(\beta \omega/2)-\mu}}{{\omega-
                \mu \coth(\beta \omega/2)}} \right) && {\rm for}\ \  \mu \leq \omega\tanh(\frac{\beta \omega}{4}) \,.
\end{array}
\right.
\end{equation}
It is straightforward to show that this result for $\Delta\mC_1^{\mt{diag}}\!\(\ket{\text{TFD}}_{12}\)$ is positive and decays exponentially with increasing frequency (yielding zero in the limit $\beta\omega\to\infty$). Using the nomenclature introduced above, we have found that in the diagonal basis, the $\mC_1$ complexity is  subadditive for these thermal states.
In order to be able to compare with the equivalent results in the physical basis which will appear in section \ref{sec:mcipb}, we plot $\Delta\mC_{1}^{\mt{diag}}\! \(\ket{\text{TFD}}_{12}\)$ in figure \ref{Delta_diag}.

Now let us evaluate the mutual complexity \reef{ramen2} of the TFD state in the free scalar theory.  Because of the product form of the TFD state and the corresponding thermal density matrices \eqref{density_QFT}, the mutual complexity simply requires summing eq.~\reef{hunt1} over all of the modes, \ie
\begin{equation}\label{rocket0}
\Delta \mC_{1}^{\text{diag}} \(\ket{\text{TFD}}; \beta,\mu\)  = \Omega_{d-2}V_{d-1} \int_{|\vec k|<\Lambda} \frac{k^{d-2} dk}{(2\pi)^{d-1}}\ \Delta\mC_{1}^{\mt{diag}} \(\ket{\text{TFD}}_{12}; \beta,\omega_{\vk},\mu\)\,.
\end{equation}
Using our previous results, it is easy to show that there are three possible expressions depending on the relation between the cutoff frequency $\omega_\Lambda$ and the critical frequencies, $\omega_{c,1}$ and $\omega_{c,2}$, defined in eq.~\reef{dog3}. We find
\begin{equation}\label{delta_diag}
\begin{split}
&\Delta \mC_{1}^{\text{diag}} \(\ket{\text{TFD}}\)\(\beta,\mu\) =\Omega_{d-2}V_{d-1} \int_0^\Lambda \frac{ k^{d-2}dk}{(2\pi)^{d-1}} \, \ln \coth\left(\frac{\beta \omega_{\vk}}{2}\right)  \\
&\ \ \ +\
\left\{
\begin{array}{lr}
\Omega_{d-2}V_{d-1} \int_0^\Lambda\! \frac{ k^{d-2}dk}{(2\pi)^{d-1}} \,I_1 & {\rm for}\ \  \omega_\Lambda \le  \omega_{c,1}\,,\\
\\
\Omega_{d-2}V_{d-1}  \(  \int_0^{k_{c,1}}\! \frac{ k^{d-2}dk}{(2\pi)^{d-1}} \,I_1 + \int_{k_{c,1}}^{\Lambda}\! \frac{ k^{d-2}dk}{(2\pi)^{d-1}} \,I_2\)& {\rm for}\ \   \omega_{c,1} \le \omega_\Lambda \le \omega_{c,2}\,,\\
\\
\Omega_{d-2}V_{d-1} \left(\int_0^{k_{c,1}}\!  \frac{ k^{d-2}dk}{(2\pi)^{d-1}} \, I_1
+\int_{k_{c,1}}^{k_{c,2}}\! \frac{ k^{d-2}dk}{(2\pi)^{d-1}} \, I_{2}
+\int_{k_{c,2}}^\Lambda\! \frac{ k^{d-2}dk}{(2\pi)^{d-1}} \, I_{3} \right) & {\rm for}\ \   \omega_{c,2} \le \omega_\Lambda\,.
\end{array}
\right.
\end{split}
\end{equation}
The first line is a ``universal contribution," which is common to all three cases, and the expression on the second line is determined by the relationship between the cut-off and the critical frequencies, with
\beqa
I_{1}&=&\ln \left( \frac{ \mu-\omega_{\vk}\tanh(\beta \omega_{\vk}/2)}{\mu-\omega_{\vk} \coth(\beta \omega_{\vk}/2)} \right)\,,\qquad
I_{2}=\ln \left(    \frac{\coth(\beta \omega_{\vk}/4)}{\coth(\beta \omega_{\vk}/2)} \right)\,,
\nonumber\\
&&\qquad\qquad I_{3}=\ln \left(    \frac{ \omega_{\vk}-\mu\tanh(\beta \omega_{\vk}/2)}{\omega_{\vk}-\mu\coth(\beta \omega_{\vk}/2)} \right)\,.
\eeqa
First, let us observe that as expected the mutual complexity $\Delta_{1}^{\text{diag}} \(\ket{\text{TFD}}\)$ is finite. In particular, the terms which could potentially produce UV divergences,  \ie $|\ln\frac{\mu}{\omega_{\vk}}|$, and which would appear in the complexity of the TFD state and the thermal state (as well as the vacuum state) separately, have been fully canceled in the mutual complexity.

In order to produce explicit results, let us focus on the massless field theory. For simplicity, we also assume that $\mu \gg\Lambda$ (as well as $\mu\beta\gg1$), which assures us that we are in the first regime, \ie $\omega_\Lambda< \omega_{c,1}$, in eq.~\reef{delta_diag}. Further, this assumption allows us to use $k/\mu$ as an expansion parameter in the second integral below. Now the universal contribution coming from the first line of eq.~\eqref{delta_diag} yields\footnote{Certain integrals relevant for the complexity can be evaluated analytically with $m=0$, \eg
           \begin{equation}\label{C_entropy}
           \begin{split}
           \int_0^{\infty} k^n\ \ln \coth({\beta k}/{2}) \,dk&= \frac{(2^{n+2}-1)\Gamma(n+2)\zeta(n+2)}{(n+1)(2\beta)^{n+1}}\,,\qquad {\rm for}\ n\geq 0\,,\\
                     \int_0^{\infty} \frac{k^n}{\sinh \beta k}  \,dk&= \frac{(2^{n+1}-1)\Gamma(n+1)\zeta(n+1)}{2^n\beta^{n+1}}\,,\qquad {\rm for}\ n\geq 1\,.
           \end{split}
           \end{equation}
}
\begin{equation}\label{gaga66}
\begin{split}
\Delta \mC_{1}^{\text{diag},\(0\)}(\ket{\text{TFD}}) \big|_{m=0} =& \, \Omega_{d-2}V_{d-1} \int_0^\Lambda \frac{ k^{d-2}dk}{(2\pi)^{d-1}} \, \ln\, \coth\!\left({\beta k}/{2}\right)
\\
 =& \, \frac{\Omega_{d-2}}{(4\pi)^{d-1}}\,(2^d-1)\zeta(d)\Gamma(d-1) \,V_{d-1} T^{d-1} \\
=& \, \frac{2^d-1}{2^{d-1}d}\,S\(\hat{\rho}_{\text{th}} \)\big|_{m=0}\,,
\end{split}
\end{equation}
where the expression for the thermal entropy was given in eq.~\reef{green4}.
Note that because the integral is UV finite, we have taken the upper limit of the integration to infinity. Turning to the second contribution, we find\footnote{The term we have neglected in the second line, \ie $\mathcal{O}\!\({k^{2}}/{\mu^2}\)$, is also proportional to $e^{-k\beta}$ when the momentum is large with respect to the temperature, which makes it convergent.}
\beqa
\Delta \mC_{1}^{\text{diag},\(1\)}(\ket{\text{TFD}}) \big|_{m=0}&=&\Omega_{d-2}V_{d-1} \int_0^\Lambda \frac{ k^{d-2}dk}{(2\pi)^{d-1}} \, \ln\! \left( \frac{ \mu-k\,\tanh(\beta k/2)}{\mu-k\, \coth(\beta k/2)} \right)
\nonumber\\
&\simeq& \Omega_{d-2}V_{d-1} \int_0^\Lambda \frac{ k^{d-2}dk}{(2\pi)^{d-1}} \,  \[ \frac{k}{\mu}\, \frac{2}{\sinh \beta k} +\mathcal{O}\!\({k^2}/{\mu^2}\) \]\label{gaga69}\\
\nonumber\\
&=&  \Delta \mC_{1}^{\text{diag},\(0\)}(\ket{\text{TFD}}) \big|_{m=0}\left[2(d-1)\,\frac{T}\mu +\mathcal{O}\!\({T^{2}}/{\mu^2}\)\right]\,.
\nonumber
\eeqa

Hence for the massless theory, the universal contribution \reef{gaga66} is proportional to the thermal entropy, while the second integral modifies this result with a series of corrections suppressed by powers of $T/\mu$. Note that both eq.~\reef{gaga66} and the leading correction in eq.~\reef{gaga69} are positive, and hence the mutual complexity of the thermofield double state exhibits subadditivity, for the massless scalar in the diagonal basis. Of course, this had to be the case since eq.~$\eqref{hunt1}$ is always positive.

For a small mass, we can also evaluate the integrals for the massive theory to find additional corrections suppressed by powers of $m/T$. The leading contribution comes from the universal correction, which can be rewritten as
\begin{equation}
\begin{split}
\Delta \mC_{1}^{\text{diag},(0)}(\ket{\text{TFD}})
&=  \Omega_{d-2} V_{d-1} \int \frac{k^{d-2} dk}{(2\pi)^{d-1}}\, \ln \frac{e^{\beta  \omega_{k}}+1}{e^{\beta  \omega_{k} }-1}\,,\\
&=  \frac{\Omega_{d-2}  }{(2\pi)^{d-1}}\,V_{d-1}T^{d-1}\int_{\beta m}^\infty dx ~ x \( x^{2}-\beta^2m^2\)^{\frac{d-3}{2}}\ln\,\coth\!\(x/{2}\)\,,
\end{split}
\end{equation}
where as usual, $\omega_{k}^2= k^2+m^2$, and in the second line, we defined $x\equiv\beta\omega_{k}$.
For $d=3$, the integral yields a relatively simple analytical answer
\beqa
\Delta \mC_{1}^{\text{diag},(0)}(\ket{\text{TFD}})\big|_{d=3} &=& \frac{V_2\, T^2}{2\pi} \bigg[-\beta^2m^2\(\frac13\,\beta m +i \frac\pi2\)
\nonumber\\
&&\qquad-\beta m \[\text{Li}_2\!\(e^{\beta m}\)+\text{Li}_2\!\(-e^{-\beta m}\)+\text{Li}_3\!\(e^{\beta m}\)-\text{Li}_3\!\(-e^{-\beta m}\)\]    \bigg]
\nonumber\\
&\simeq&  \frac{V_2\, T^2}{8\pi}  \[ {7\,\zeta(3)}+\frac{m^2}{T^2}\(2\,\ln
\!\(\frac{m}{2T}\)-1\) +\mathcal{O}\(m^3/T^3\)\]\,,
\label{gaga67}
\eeqa
where Li$_n$ denotes the polylogarithm function.
For $d>3$ (and $m/T\ll 1$ again), one finds
\beqa
\Delta \mC_{1}^{\text{diag},(0)}(\ket{\text{TFD}})\big|_{d} &\simeq&\frac{\Omega_{d-2}\,V_{d-1} T^{d-1}}{(2\pi)^{d-1}} \int_{m\beta}^\infty dx\[ x^{d-2} - \frac{d-3}{2}\beta^2m^2\,x^{d-4} \] \ln\,\coth\!\( x/{2}\)
\nonumber\\
&\simeq&   \frac{\Omega_{d-2}\,V_{d-1}T^{d-1}}{(4\pi)^{d-1}}\bigg[(2^d-1)\zeta(d)\Gamma(d-1)
\label{gaga68}\\
&&\qquad\qquad\quad -{(2^{d-1}-2)\zeta(d-2)\Gamma(d-2)}\,\frac{m^2}{T^2} +\mathcal{O}(m^3/T^3)\bigg]\,.
\nonumber
\eeqa
Of course, the leading contribution above (and in eq.~\reef{gaga67}) matches the universal result for $m=0$ in eq.~\reef{gaga66}. Note that the $m^2/T^2$ correction to the integrand in eq.~\reef{gaga68} vanishes for $d=3$. Hence in eq.~\reef{gaga67}, the correction at this order comes entirely from the modification to the lower limit of the range of integration. In contrast for $d>3$, the change in the lower limit of integration  yields a higher order correction of order $(\beta m)^{d-1}$, \ie this contribution is higher order than the $(\beta m)^2$ term retained in eq.~\eqref{gaga68}. We also note that for both $d=3$ and $d>3$, the leading correction is always negative. However, in this regime with $m/T\ll1$, the mutual complexity is still dominated by the leading term \reef{gaga66}, which is positive. Hence the complexity of the TFD state remains subadditive in this limit. Of course, this had to be the case given the positivity of eq.~\eqref{hunt1}.

\subsubsection{Mutual complexity in the physical basis}\label{sec:mcipb}

We now turn to evaluating the mutual complexity of the TFD state in the physical basis. For a single mode,
the TFD state \eqref{two_mode} is obtained from the general purification \eqref{Fock_psi12} by setting $r=s=0$. Using eqs.~\eqref{rbar} and \eqref{thetadef}, we can demonstrate that this corresponds to
\begin{equation}
\begin{split}
X_-= 1\,,\quad \theta=\frac{\pi}{4}, \quad \omega_{\pm}=\omega e^{\pm 2 \alpha},
\quad \frac 12 \ln  \frac{\omega_+}{\omega_-}= 2\alpha,  \quad  H =  \frac{1}{2}\left(
\begin{array}{cc}
\ln  \frac{\omega}{\mu} &  -2\alpha \\
-2\alpha & \ln \frac{\omega}{\mu}  \\
\end{array}
\right)\,.
\end{split}
\end{equation}
It is then straightforward to show that the complexity of the TFD state \eqref{two_mode} is given by
\begin{equation}\label{TFD_position}
\begin{split}
\mC_{1}^{\mt{phys}} \(\ket{\text{TFD}}_{12}\) &=  \left|  \ln \frac{\omega}{\mu} \right| + 2\alpha
= \left\{
\begin{array}{lr}
\ln \frac{\mu}{\omega} + \ln \coth \(\frac{\beta \omega}{4} \) ,  & \omega \le \mu\,,\\
\\
\ln \frac{\omega}{\mu}+ \ln \coth \(\frac{\beta \omega}{4} \) , & \omega \ge \mu \,.\\
\end{array}
\right.
\end{split}
\end{equation}
This result is consistent with the complexity derived in \cite{Chapman:2018hou} using the $F_1$ cost function --- see  eq.~(138) in \cite{Chapman:2018hou} with $C_1^{\mt{LR}}=|\ln \lambda|+2 |\alpha|$ and note that the physical basis was denoted as the LR basis there.

\begin{figure}[t]
        \includegraphics[width=0.5\textwidth]{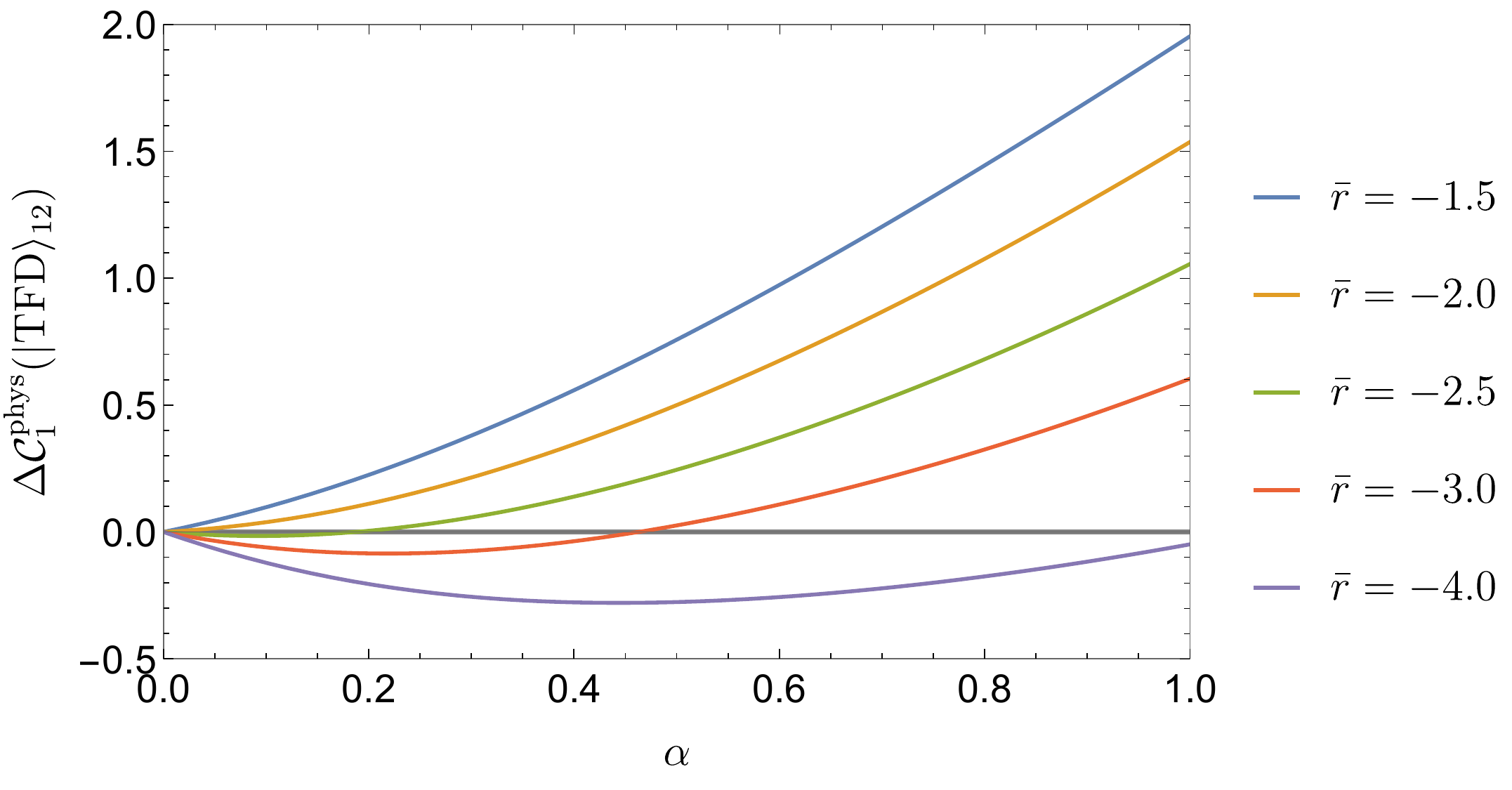} \hspace{0.01\textwidth}
        \includegraphics[width=0.55\textwidth]{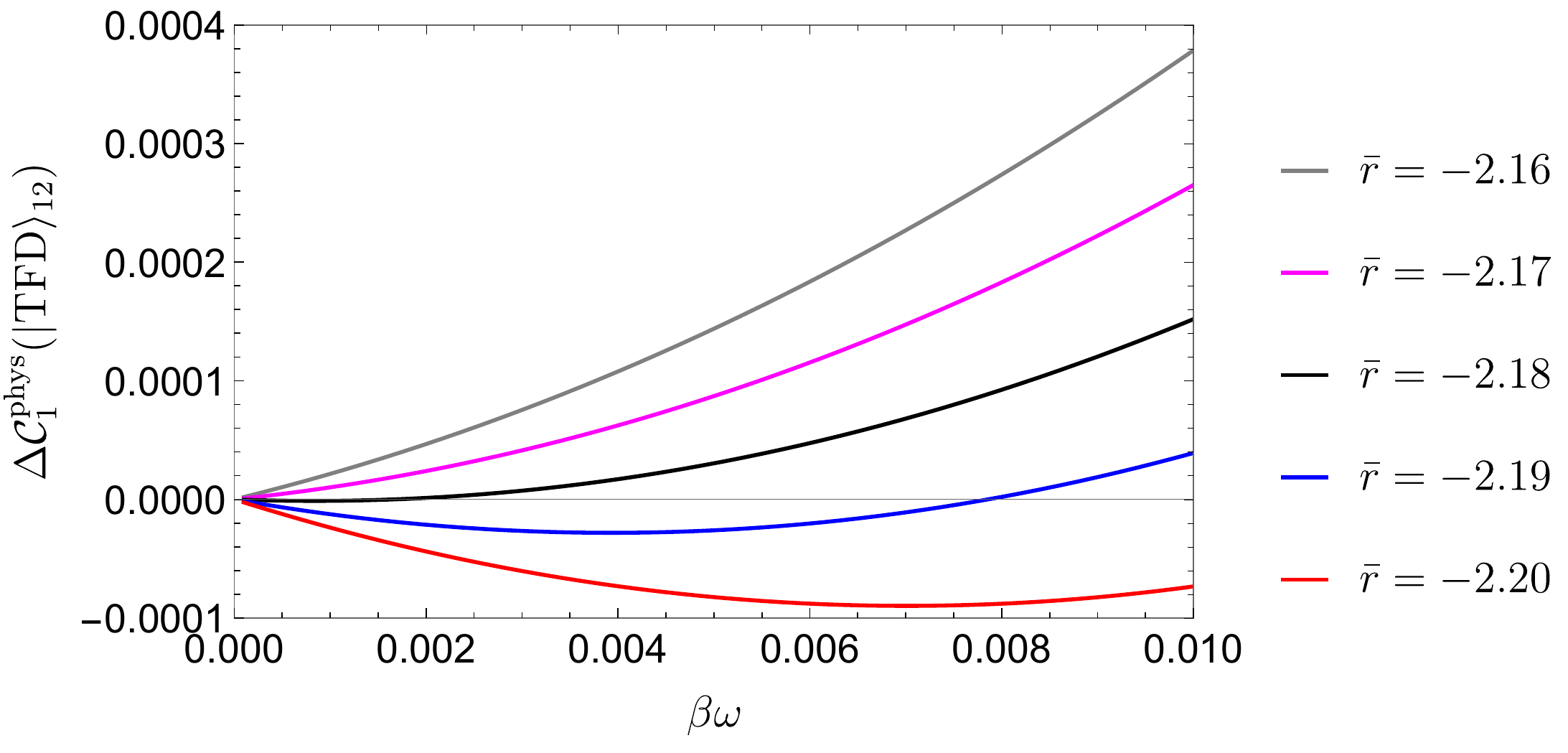}
        \caption{The mutual complexity $\Delta \mC_1^{\mt{phys}}\( \ket{\text{TFD}}_{12}\)$ as defined in eq.~\eqref{DeltaCpos} with fixed $\bar{r}=\frac{1}{2}\ln\frac{\omega}{\mu}<0$ as a function of $\alpha$. We find that the quantity $\Delta \mC_1^{\mt{phys}}$ can be either positive or negative. The right plot is the region with $\bar{r}$ near the transition point $\bar{r}=-2.177$.}\label{deltaCLR_thermal01}
\end{figure}

As before, the two reduced density matrices are $\hat{\rho}_{1,2}=\hat{\upsilon}_{{th}}$, and we wish to evaluate the mutual complexity of the TFD state but now in  the physical basis:
\begin{equation}\label{DeltaCpos}
\Delta \mC_1^{\mt{phys}} \( \ket{\text{TFD}}_{12}\) =  2\, \mC_1^{\mt{phys}} \( \hat{\upsilon}_{{th}}\)
- \mC_{1}^{\mt{phys}} \(\ket{\text{TFD}}_{12}\) \,.
\end{equation}
The purification complexity $\mC_{1}^{\mt{phys}}\!\(\hat{\upsilon}_{{th}}\)$ is defined using eq.~\eqref{purification_C_pos} and $\mC_{1}^{\mt{phys}} \(\ket{\text{TFD}}_{12}\)$ is given in eq.~\reef{TFD_position}. This expression is evaluated numerically in figure \ref{deltaCLR_thermal01}, and we note that in the physical basis, $\Delta \mC_1^{\mt{phys}} \(\ket{\text{TFD}}_{12}\)$ does not have a definite sign. That is, eq.~\reef{DeltaCpos} may be positive or negative depending on the parameters, which contrasts with the corresponding expression for the mutual complexity always being positive in the diagonal basis.

One can gain some analytical insight into the above result by focusing on the limit of small $\alpha$, \ie large $\beta\omega$. Combining eqs.~\reef{small_alpha} and  \eqref{TFD_position}, the single-mode mutual complexity \reef{DeltaCpos} becomes
\begin{equation} \label{board88}
\begin{split}
\Delta \mC_1^{\mt{phys}} (\ket{\text{TFD}}_{12})  &=2\,\mC_1^{\mt{phys}} (\hat{\upsilon}_{{th}}) - \mC_{1}^{\mt{phys}} \(\ket{\text{TFD}}_{12}\) \\
&= 2\alpha \(   \frac{2 \,\ln\frac{\mu}{\omega}}{\sqrt{\mu/\omega}-\sqrt{\omega/\mu}}-1  \) + \mathcal{O}(\alpha^2) \,.
\end{split}
\end{equation}
Comparing to figure \ref{deltaCLR_thermal01}, we see that this leading expression captures the linear behaviour in the vicinity of $\alpha=0$, and that the sign of the slope determines whether the corresponding mutual complexity will be negative over some range.
Further, eq.~\reef{board88} shows that the slope is determined by the ratio $\mu/\omega$ (or alternatively by $\bar r =\frac12\ln(\omega/\mu)$).
We also observe that this slope (\ie the function multiplying $2\alpha$) is invariant under $\frac{\mu}\omega\to\frac\omega\mu$. The transition between positive and negative values of the slope occurs at
\begin{equation}
2\,|\bar r_c|=\left| \ln \frac{\omega_c}{\mu} \right|\simeq 4.35464\cdots \,.
\end{equation}
That is, $\Delta \mC_1^{\mt{phys}} (\ket{\text{TFD}}_{12}) $ is entirely positive (for all values of $\alpha$) in the region  $0.01285\lesssim \omega/\mu\lesssim 77.84$, or alternatively $\left| \bar{r} \right| \lesssim 2.177$, and it has negative contributions (for small values of $\alpha$) outside of this range. Of course, these results precisely match those found numerically, as shown in figure \ref{deltaCLR_thermal01}.

\begin{figure}[htbp]
        \includegraphics[width=0.5\textwidth]{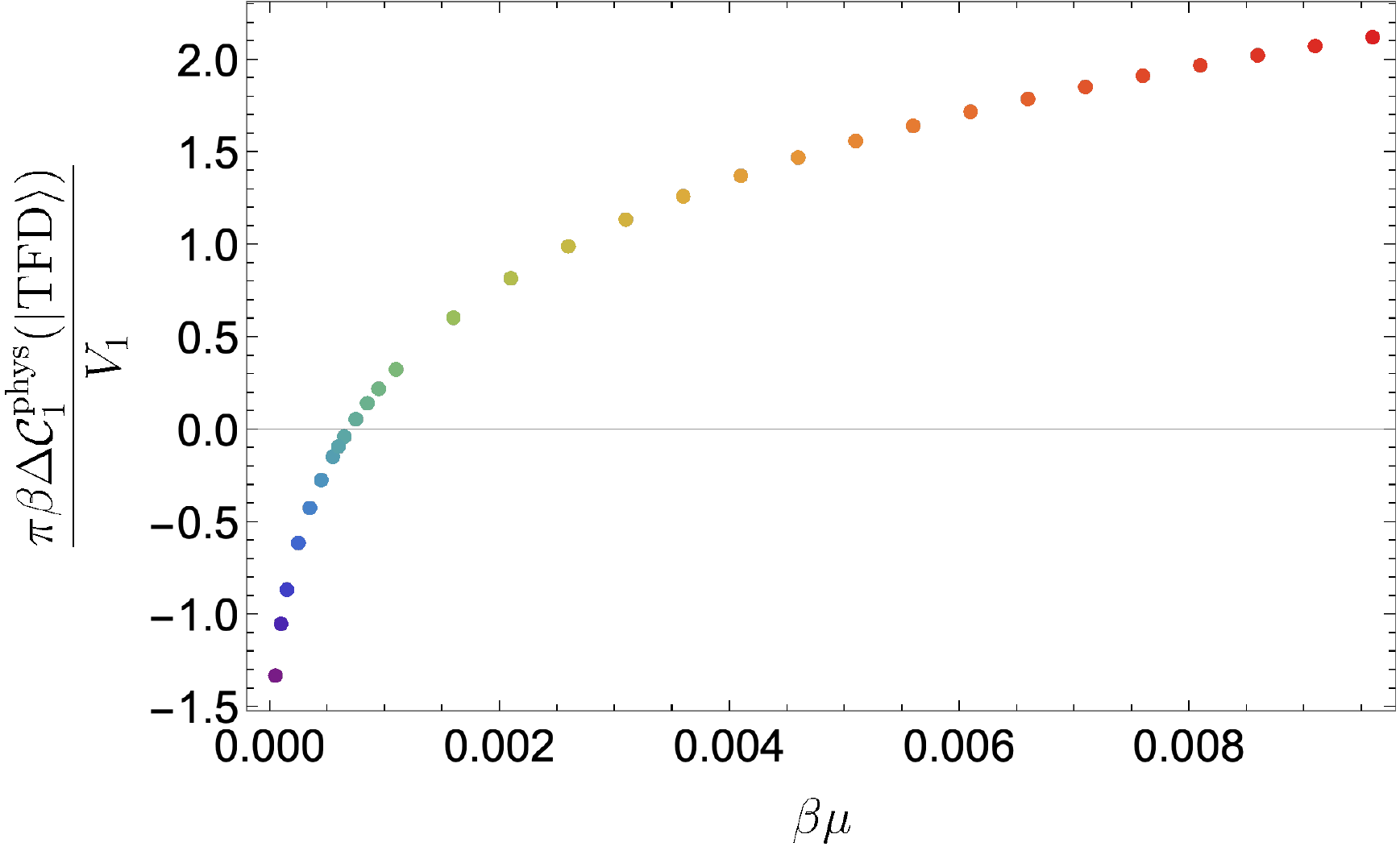} \hspace{0.01\textwidth}
        \includegraphics[width=0.48\textwidth]{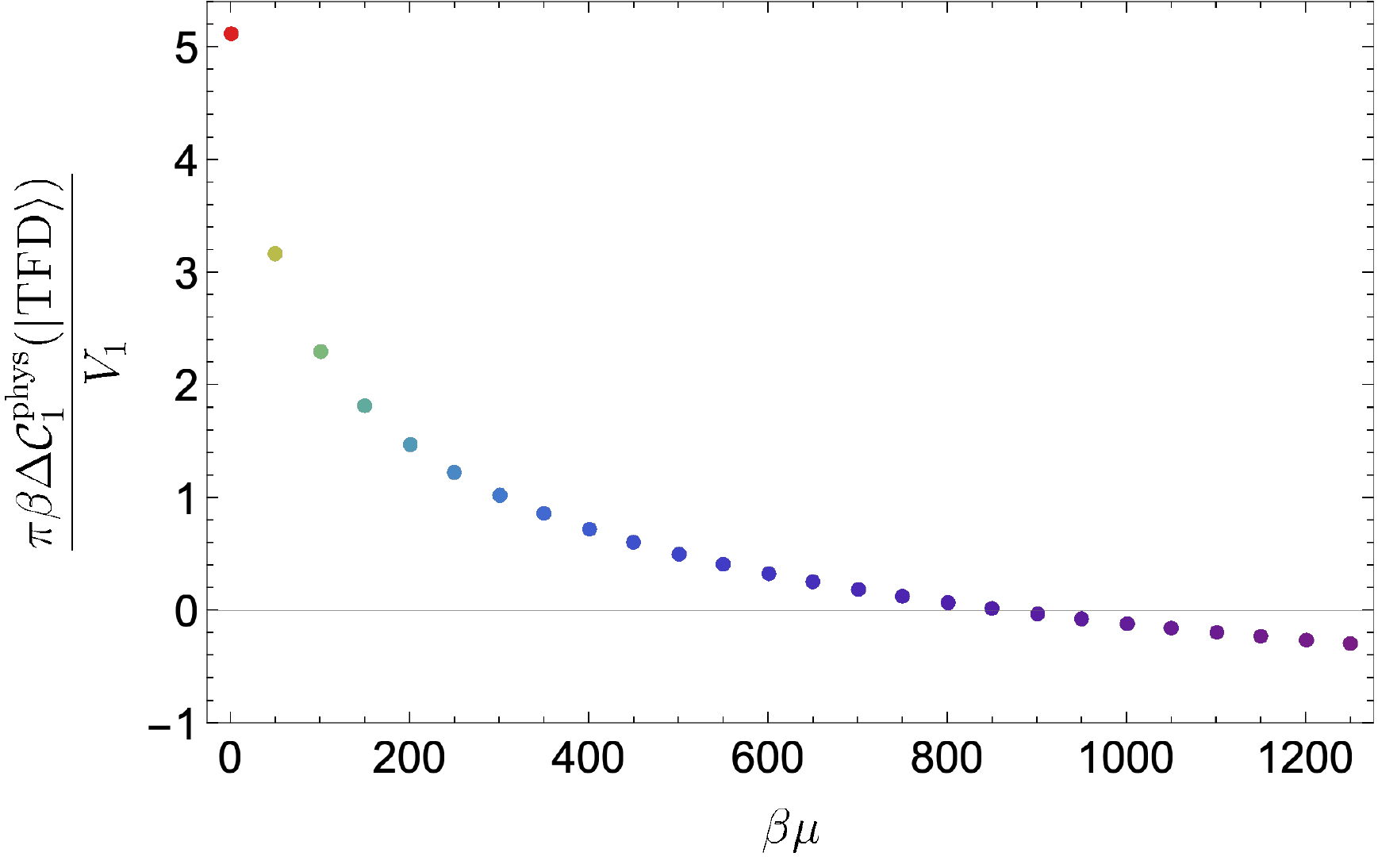}
        \caption{The integrated mutual complexity in the physical basis $\Delta \mC_1^{\mt{phys}}\( \ket{\text{TFD}}\)$ in eq.~\eqref{integral_phys} for a massless free scalar field theory in  $d=2$  as a function of $\beta \mu$. The two plots show different regimes of the parameter $\beta \mu$. The integrated mutual complexity is negative when  $\beta \mu$ is very small or very large. }\label{deltaC_phys}
\end{figure}

Now because of the factorization of the thermal state in free field theory, the corresponding mutual complexity is given by simply summing eq.~\reef{DeltaCpos} over each of the modes,
\begin{equation}\label{integral_phys}
\Delta \mC_1^{\mt{phys}} \( \ket{\text{TFD}}\)
= V_{d-1} \int \frac{d^{d-1}k}{(2\pi)^{d-1}} \left[ 2\,\mC_{1}^{\mt{phys}}\!\(\hat{\upsilon}_{{th}}\) - \mC_{1}^{\mt{phys}}\! \(\ket{\text{TFD}}_{12}\) \right]\,.
\end{equation}
It is possible to demonstrate that this expression for the mutual complexity in the physical basis is finite by considering the small $\alpha$ limit in eq.~\eqref{board88} which demonstrates that the mutual complexity is exponentially suppressed for large momentum, hence resulting in a convergent integral. Although evaluating this expression analytically is a challenge, it is straightforward to evaluate this mutual complexity numerically.  Figure \ref{deltaC_phys} shows the mutual complexity $\Delta \mC_1^{\mt{phys}}\( \ket{\text{TFD}}\)$ for a massless free scalar in $d=2$, as an example. Varying the reference frequency from IR scales to UV scales, we see that mutual complexity begins with negative values for $\beta \mu\ll1$, then rises to positive values at intermediate scales with $\beta \mu\sim 1$, and finally becomes negative again for $\beta \mu\gg1$. In other words, the mutual complexity $\Delta \mC_1^{\mt{phys}}\! \(\ket{\text{TFD}}\) $ can be {\bf negative} when the reference frequency is very large or extremely small. This again stands in contrast with the diagonal basis, where the corresponding mutual complexity was found to be positive for all values of the reference frequency.

Using a change of variables $\tilde k = \beta k$ in the integral in eq.~\eqref{integral_phys}, it is possible to extract an overall coefficient proportional to the entropy \eqref{green4} of the massless theory, \ie $V_{d-1}T^{d-1}\sim S_{\mt{th}}$. The remaining integral is a function of the dimensionless parameter $\beta\mu$. Finiteness of the result in the limit $\beta \mu\gg 1$ requires that this function will approach a constant.\footnote{Though it is not immediately obvious from the plot in the right panel of figure \ref{deltaC_phys}, we were able to confirm that in the limit of large $\beta \mu$, the result approaches a constant.} Hence, the resulting mutual complexity is proportional to the entropy in this limit.

\section{Complexity of Vacuum Subregions in QFT} \label{apply02}

In the previous section, we considered the purification complexity for thermal states of a free scalar QFT. In this section, we proceed with the QFT applications by considering mixed states on finite subregions of the vacuum state of a free scalar QFT. As in section \ref{QFT}, we regulate our field theory on a spatial lattice in order to obtain a finite result for the purification complexity. We evaluate the complexity and the mutual complexity both in the diagonal basis, and also in the physical basis, and comment on the sign of the mutual complexity in both cases. Our results are primarily evaluated numerically, and so we limit ourselves to considering the free scalar in two dimensions on a circular lattice. To illustrate the different bases relevant to this problem, in appendix \ref{app:4HO}, we study analytically examples of small lattices with two and four coupled oscillators and reduced density matrices associated with subregions consisting of half of the oscillators.

\subsection{Purification Complexity in the Diagonal Basis}
Here we study the diagonal basis complexity and mutual complexity of density matrices of different subregions of the vacuum state of a discretized free scalar theory in two dimensions. We focus on a circular lattice of oscillators.\footnote{See footnote \ref{linecirclefoot} on the distinction between a circular lattice and the line.} We state the problem in terms of matrices on this lattice, and then describe the algorithm we used in order to find the complexity numerically. We then present our results for the complexity and the mutual complexity. Further, in discussing our results, we focus on the case of a very small mass in order that the results might mimic those of a holographic CFT.

\subsubsection{Set-up}
We begin with the lattice of harmonic oscillators~\eqref{ham88} realizing a regularization of a free quantum field theory~\eqref{Ha_scalarQFT} on a one-dimensional circle of length $L$ with $N$ oscillators and lattice spacing $\delta = L/N$. The various oscillators are located at sites $\bar x_a$ where $a=1,\dots,N$ and we impose periodic boundary conditions $\bar x_{N+1}:=\bar x_1$. The Hamiltonian in normal mode coordinates $x_k$ defined in eq.~\eqref{Thefreq} is given by eq.~\eqref{normal_QFT2} and the complex coordinates are related according to $x^\dagger_k= x_{N-k}$.

The ground state wavefunction of this system of harmonic oscillators is straightforward to find in normal mode basis\footnote{Note that eq.~\eqref{eq:ground-normal} differs from~\eqref{TargetGaussianPure} in that we have the magnitude squared of $x_k$ instead of simply the squared of each $\tilde{x}_k$. This is because while we assumed $\tilde{x}_k$ is real, the transformation~\eqref{Thefreq} defining $x_k$ is complex. It is possible to use instead the real Fourier transformations involving trigonometric functions in which case we would find real normal modes $\tilde{x}_k$ and the ground state would be given by~\eqref{TargetGaussianPure}, but we opt instead to use the simpler transformation~\eqref{Thefreq} at the cost of having complex $x_k$.\label{footcomplexcoord}}
\begin{equation}
\label{eq:ground-normal}
\Psi_0(x_k) = \prod^{N}_{k=1} \left(\frac{\omega_k}{{\pi}}\right)^{1/4}\,\mathrm{exp}\!\left(-\frac{1}{2}\omega_k |x_k|^2\right)\,.
\end{equation}
This can be explicitly written in the physical basis using the transformation~\eqref{Thefreq}
\beq\label{eq:Mpos2}
\Psi_0(\bar{x}_a) = \left({\rm det}\left(\frac{M}{\pi}\right)\right)^{1/4} {\rm exp}\left[-\frac12 M_{ab} \bar{x}_a\bar{x}_b\right]\,,
\eeq
where
\beq
\label{eq:Mpos}
M_{ab} = \frac1N \sum_{k=1}^N \omega_k\, {\rm exp}\left[-\frac{2\pi i k}{N}(a-b) \right]\,.
\eeq

Next, we partition the system into two subregions ${\cal A}=\{\bar{x}_1,\,\bar{x}_2,\,\cdots,\,\bar{x}_J \}$ and ${\cal B}=\{\bar{x}_{J+1},\,\cdots,\,\bar{x}_N \}$ and decompose the matrix $M$ as in eq.~\eqref{Gaussian_AB}
\beq
M= \left(
\begin{array}{cc}
    \Gamma & K \\
    K^\dagger & \Omega \\
\end{array}
\right)
\eeq
where $\Gamma$ links the oscillators in the subregion ${\cal A}$ while $\Omega$ links the oscillators in subregion ${\cal B}$. The $K$ matrices link the two subregions and are responsible for the entanglement between ${\cal A}$ and ${\cal B}$. Tracing out the oscillators in ${\cal B}$ then gives us a density matrix of the form~\eqref{densityFun_A}-\eqref{density_fun_A}, where the matrices $A$ and $B$ are related to $M$ by~\eqref{pure_constrians}
\beq
\label{pure_constraints}
A=\Gamma-\frac12 K \Omega^{-1} K^\dagger\,,\quad B=\frac12 K \Omega^{-1} K^\dagger\,.
\eeq
If $K=0$ then $B=0$ and we have a pure state. This is to be expected since without $K$ there is no entanglement between the two regions and both wave-functions are pure: $\Psi_{\cal AB} = \Psi_{\cal A} \otimes \Psi_{\cal B}$.

In this section, our goal is to calculate the purification complexity of the density matrix~\eqref{densityFun_A} obtained by the procedure above. Although the numerical minimization for purification complexity is always possible in principle, the number of free parameters will increase rapidly with the size of the subsystem, which means that we will need much more time in order to perform the numerical minimization for a larger lattice. Instead, we have claimed in section~\ref{MbyM} that even for density matrices which are not simple products of single modes, mode-by-mode purifications can be used to provide a good approximation of the optimal purifications. Hence, here we have taken the strategy to focus on mode-by-mode purifications in the numerical minimization for the complexity of the mixed state in a given subregion $\hat{\rho}_\mA$.  We expect our results presented later will approximate the purification complexity ${\cal C}^{\rm diag}_1$ for subregions of the vacuum. We comment on the quality of this approximation in section \ref{commentsonApp}.

In order to find the purification complexity using a mode-by-mode approximation, we have followed the following algorithm.
We begin by computing the parameter matrix $M_{ab}$ in eq.~\eqref{eq:Mpos}. Next, given a partition of our system ${\cal A} \cup {\cal B}$, we compute $A$ and $B$ using~\eqref{pure_constraints}. We then diagonalize $A$ with an orthogonal transformation $O_A $ by $D_A=O_A A O_A^T$, and proceed to rescale the entries of $A$ by $D_A^{-1/2}$. We then diagonalize the $\tilde{B}= D_A^{1/2} O_A B O_A^T D_A^{1/2}$ matrix in this new non-orthogonal basis\footnote{The basis is non-orthogonal for non-commuting $A$ and $B$ because of the rescaling by $D_A$ between the two orthogonal transformations $O_A$ and $O_B$.} with an orthogonal transformation  $O_B$ by $D_B=O_B \tilde{B} O_B^T$. The density matrix in the non-orthogonal basis $\tilde{x} = O_B D_A^{1/2} O_A \bar{x}\equiv R\bar{x}$ now takes the following form
$$
\rho_{\cal A}(\tilde{x}_i,\tilde{x}'_i) = |{\rm det}R|^{-1} \sqrt{{\rm det}\left(\frac{A-B}{\pi}\right)} \prod_{i} {\rm exp}\left[-\frac12 (\tilde{x}_i^2+(\tilde{x}'_i)^2)+ b_i\tilde{x}_i \tilde{x}'_i\right]\,,
$$
where the number of non-zero eigenvalues $b_i$ indicates the number of ancillary oscillators which are necessary in order to purify the density matrix. We proceed to purify the mixed state $\hat{\rho}_\mA$ with a mode-by-mode purification in this  non-orthogonal basis, \ie
\beq\hat{\rho}_{\cal A} ={\rm Tr}_{\mA^c}\hat{\rho}_{\mA\mA^c}\,,\quad \hat{\rho}_{\mA\mA^c}=|\Psi_{\mA\mA^c} \rangle \langle \Psi_{\mA\mA^c}|\,,
\eeq
with
\beq
\label{eq:mbym}
\Psi_{\mA\mA^c}(\tilde{x}_i,y_{i}) = {\cal N} \prod_i {\rm exp}\left[-\frac12 (1+b_i)\tilde{x}_i^2- \frac{k_i^2}{4b_i}y_i^2-k_i \tilde{x}_iy_i \right]\,.
\eeq
We return to the orthogonal basis $\bar{x} = R^{-1}\tilde{x}$ with
    \beq
    \label{eq:AAc}
    \Psi_{\mA\mA^c}(\bar{x}_i,y_i) = {\cal N}' {\rm exp}\left[-\frac12 (\bar{x},y){\cdot}M_{\cal A} {\cdot}\left(
    \begin{array}{c}
    \bar{x} \\
    y \\
    \end{array}
    \right)  \right]\,,
    \eeq
    and find the eigenvalues $\lambda_i$ of $M_{\cal A}$.
Finally, we minimize the complexity ${\cal C}^{\rm diag}_1 = \frac{1}{2}\sum_i |{\rm ln}\frac{\lambda_i}{\mu}|$ over the free parameters $k_i$. For some of the subregions considered, this minimization has to be done over an ${\cal O}(10^2)$ number of parameters. Fortunately, in our problem at hand, dividing this minimization into a sequence of minimizations over ${\cal O}(1)$ parameters indeed reaches the global minima of the function to be optimized.\footnote{Indeed, even taking the minimization over one parameter at a time gives the global minima \emph{most} of the time. We found that minimizing over 2 or 3 parameters at a time gave accurate enough results without requiring too much more computational power.}

Obviously, we can follow the same process to derive the purification complexity for the complementary subregion $\hat{\rho}_\mB$. Following the analysis in section \ref{compare7}, we can define the mutual complexity for subregions in the diagonal basis as
\begin{equation}\label{mutual_diag_sub}
\Delta\mC_1^{\rm diag}\equiv {\cal C}^{\rm diag}_1(\rho_{\cal A}) + {\cal C}^{\rm diag}_1(\rho_{{\cal B}})- {\cal C}^{\rm diag}_1(|\Psi_0 \rangle ) \,.
\end{equation}

\subsubsection{Numerical results in the diagonal basis}
 Throughout the following discussion, we have set the mass to $mL=0.01$. Again, our aim is that by setting the mass to such a small value, our QFT results might resemble those found in holography where the boundary theory is conformal. A comparison of the results for the free scalar theory and for holography is considered in section \ref{sec:disc}.

\begin{figure}
    \center
    \includegraphics[width=0.7\textwidth]{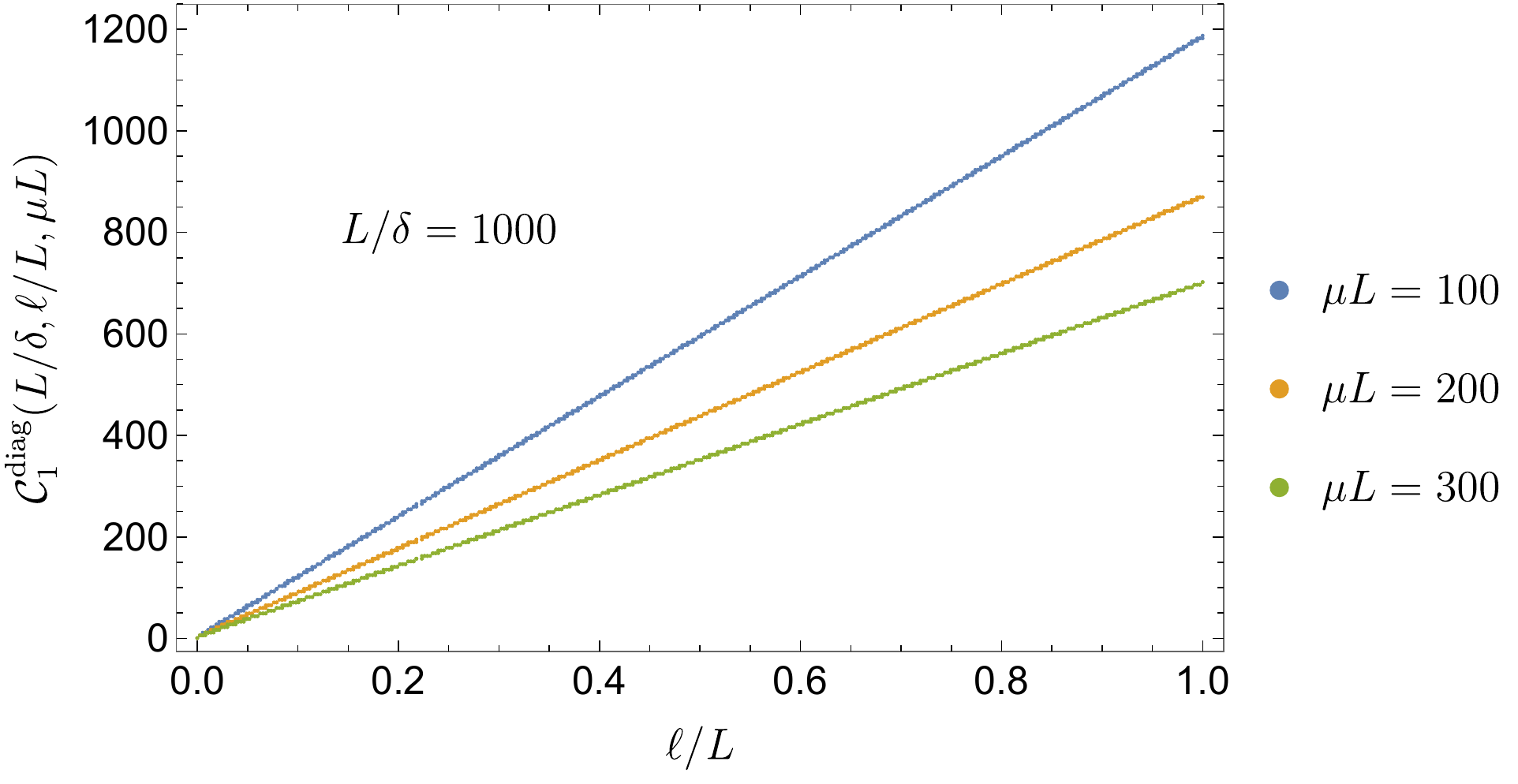}
    \caption{Purification complexity in the diagonal basis for subregions of the vacuum as a function of the subregion size. The cutoff was set to $N = L/\delta = 1000$ and the mass to $mL=0.01$. The purification complexity for the subregion with $\ell \to L$ agrees with the complexity of the ground state in diagonal basis.}\label{fig:size}
\end{figure}

\begin{figure}
    \includegraphics[width=0.44\textwidth]{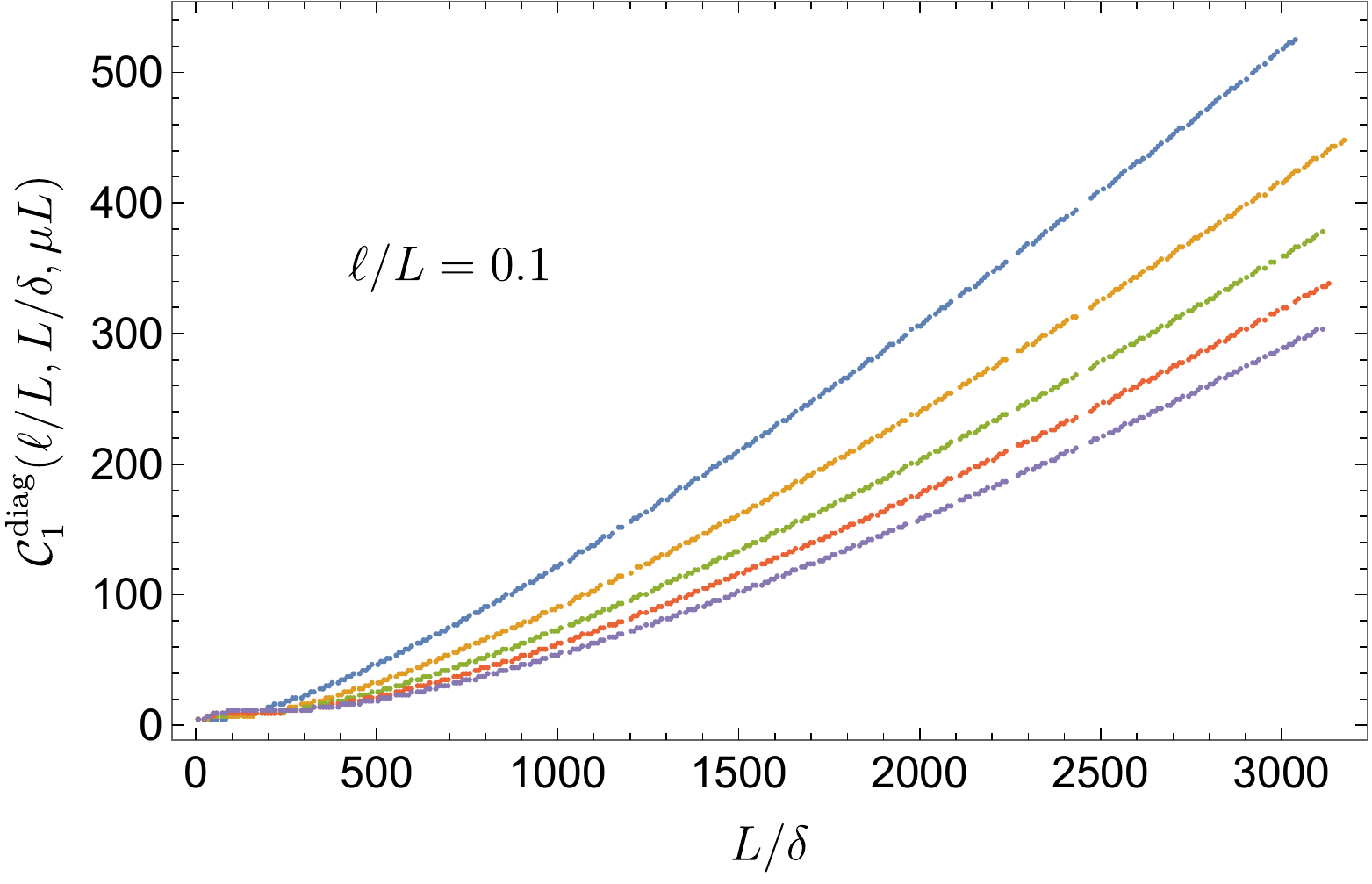}~
    \includegraphics[width=0.46\textwidth]{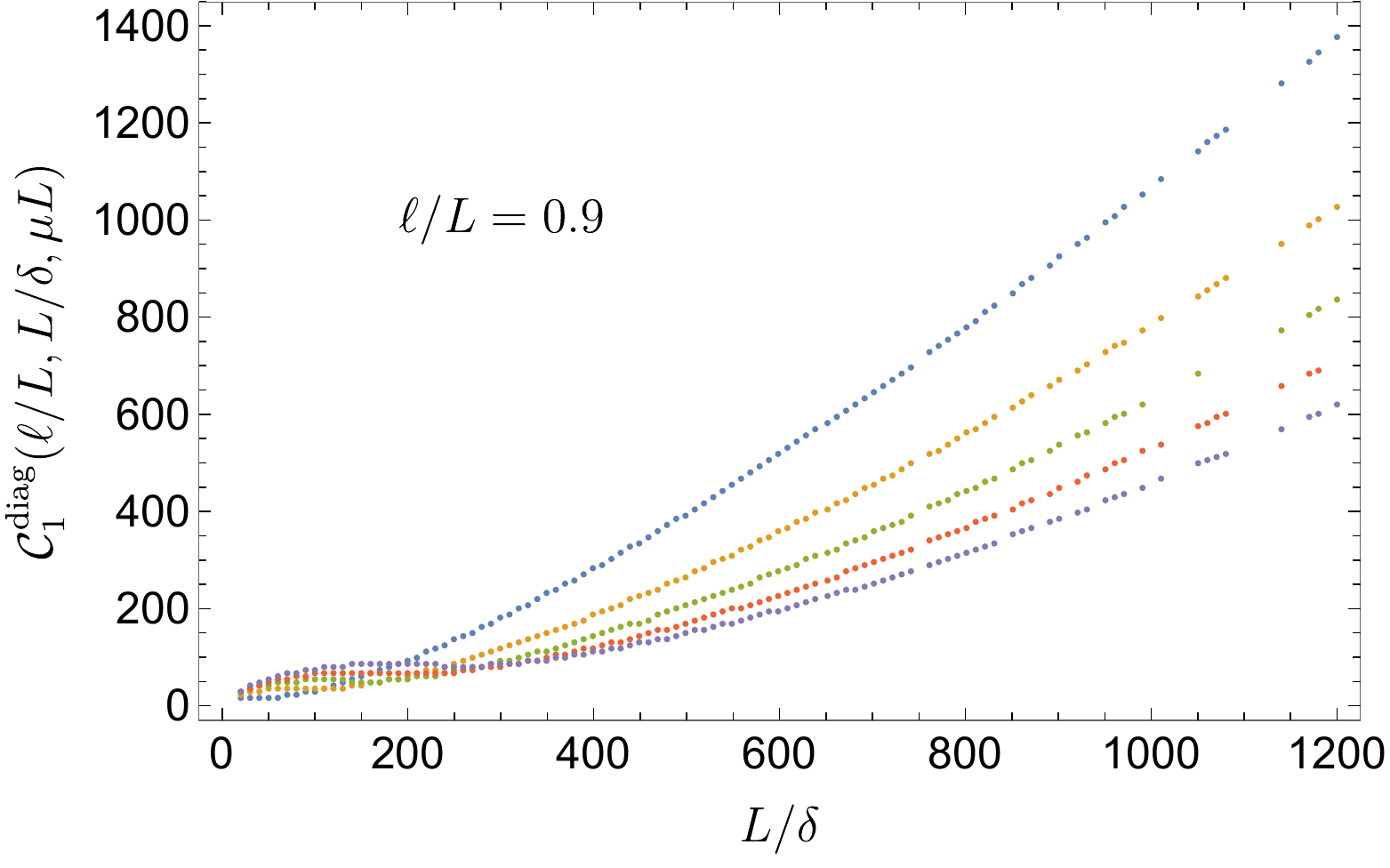}\\
    \includegraphics[width=0.45 \textwidth]{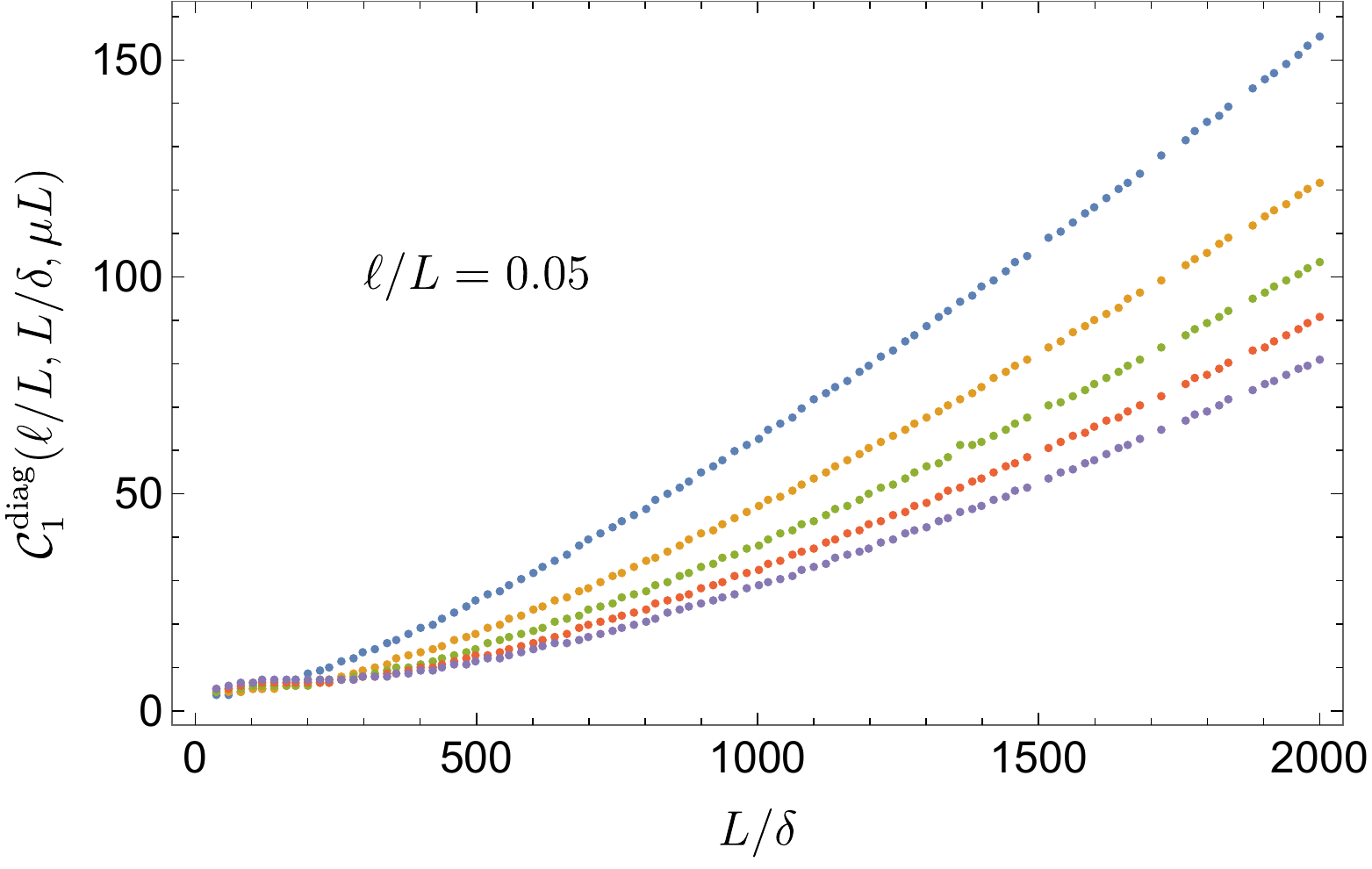}
    \includegraphics[width=0.55 \textwidth]{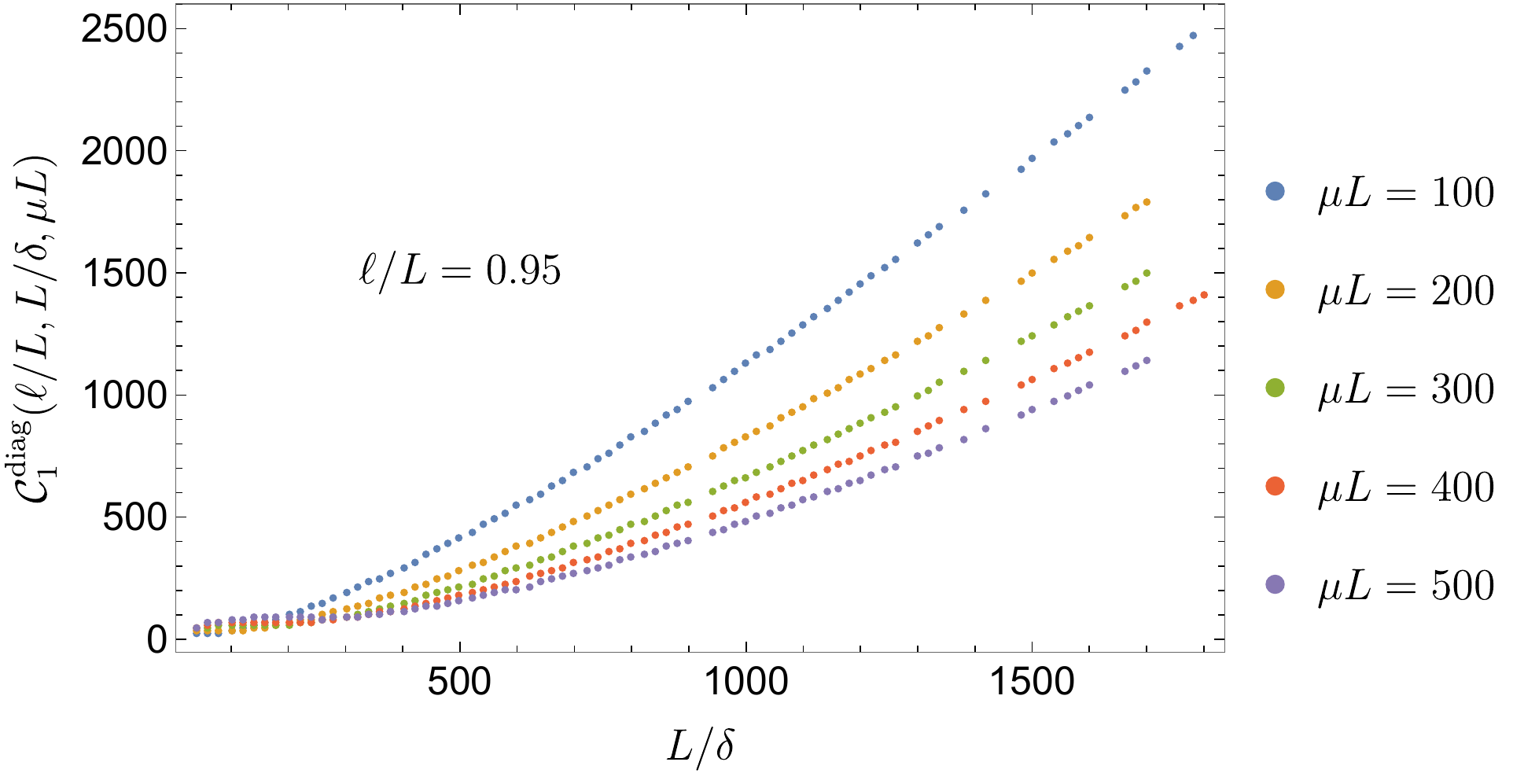}
\caption{Purification complexity in the diagonal basis for subregions of the vacuum as a function of the lattice cutoff. The mass was set to $m L =0.01$. The different plots correspond to different subregion sizes $\ell/L = 0.05,\,0.1,\, 0.9$ and $0.95$ as indicated and each plot contains five different reference frequencies of $\mu L = 100,\,200,\,300,\,400$ and $500$ respectively.}
\label{fig:subregion1}
\end{figure}

{\bf Dependence on the size of the subregion:} First, we find the subregion complexity as a function of the subregion size for a lattice of 1000  harmonic oscillators for different values of the reference frequency and plot the results in figure~\ref{fig:size}. For all cases, the complexity grows linearly with the subregion size up to the expected complexity of the vacuum. The slope of the plot decreases with increasing reference frequency.

{\bf Structure of divergences in purification complexity:} For subregions with fixed size, we plot the cutoff dependence of the purification complexity in figure~\ref{fig:subregion1}. The large $N$ (or equivalently, the small $\delta$) behavior of the subregion complexity with $\ell/L=1/20,\,1/10,\,9/10$ and $19/20$ is given by
\begin{equation} \label{fit}
\begin{split}
  \mC^{\rm diag}_1(\ell/L=0.05,\mu L,\delta/L)
 \approx & \, \frac{\ell}{2\, \delta} \ln \frac{1}{\mu \delta} + 0.232 \ln\frac{L}{\delta} + 0.307\, \mu \ell +2.08\,,\\
  \mC^{\rm diag}_1(\ell/L=0.10,\mu L,\delta/L)
 \approx & \, \frac{\ell}{2\, \delta} \ln \frac{1}{\mu \delta} + 0.241\ln\frac{L}{\delta}+ 0.312\, \mu \ell +2.11\,,\\
 \mC^{\rm diag}_1(\ell/L=0.90,\mu L,\delta/L)
 \approx & \, \frac{\ell}{2\, \delta} \ln \frac{1}{\mu \delta} + (0.542-0.304\mu\ell)\,\ln\frac{L}{\delta}+ 0.340    \, \mu \ell -0.308\,,\\
  \mC^{\rm diag}_1(\ell/L=0.95,\mu L,\delta/L)
 \approx & \,\frac{\ell}{2\, \delta} \ln \frac{1}{\mu \delta} +(0.383-0.147\,\mu \ell)\,\ln\frac{L}{\delta}+ 0.329\, \mu \ell +0.688\,.
 \end{split}
 \end{equation}
These suggest a divergence structure of the form\footnote{Note that the fits in eq.~\eqref{fit} were obtained using the data for large values of $L/\delta$, \ie $L/\delta>300$ in figure \ref{fig:subregion1}. Furthermore, we kept $\mu L$ fixed in these fits (and plots). Therefore, the fits correspond to a region where $\mu \delta$ is small. More generally, one could consider reference frequencies of the order of the cutoff, or even larger. The intuition from the pure state results (see footnote \ref{footvac}) leads to the conclusion that there should be an absolute value on the logarithmic factor, as we write in eq.~\eqref{Theamazingsubregionequation}.\label{foot67}}
 \beq\label{Theamazingsubregionequation}
 \mC^{\rm diag}_1(\ell/L,\mu L,\delta/L)
 \approx  \frac{\ell}{2\, \delta} \left|\ln \frac{1}{\mu \delta} \right|+f_1(\mu
 L,\ell/L)\ln\frac{L}{\delta}+f_2(\mu L,\ell/L)
 \eeq
where $f_1$ and $f_2$ are dimensionless functions, which are independent of the cutoff scale $\delta$. We note that the leading divergence matches the results found in~\cite{qft1,qft2} for the full system with $\ell\to L$.

In eq.~\eqref{Theamazingsubregionequation}, we have found the structure of divergences for our system with $mL=0.01$, which was chosen to emulate a massless field theory. In the case of a massive theory, \ie $mL\gtrsim 1$, we expect that the divergence structure is
again as in eq.~\eqref{Theamazingsubregionequation}, except that the coefficients $f_1$ and $f_2$ would now also depend on the additional mass parameter, \eg $f_1= f_1(\mu L, \ell/L, mL)$ and $f_2= f_2(\mu L, \ell/L, mL)$. On the other hand, we expect that the UV divergence in the first term is a universal volume term, as in the massless theory. This contribution represents the cost required to prepare the ground state entanglement at very short scales, while the other terms depend on the details of the QFT (\eg the mass).\footnote{In particular, we found that the complexity of the \emph{full} ground state is, using eqs.~\eqref{complexity_pure} and \eqref{Thefreq},
\beq\label{512}
{\cal C}^{\rm diag}_1(\hat{\rho}_0) = \frac{L}{2\delta} {\rm ln}\left(\mu\delta\right) + \frac12 {\rm ln}\left(\frac{1}{mL}\right) - \frac{m^2L^2}{48}+{\cal O}(m^4,m^2\delta^2)\,,
\eeq
where here we assumed $\mu \ge \sqrt{\frac{4}{\delta^2} + m^2}$ in order to obtain this simple analytic form. Alternatively, for $\mu<m$, the same result is obtained up to an overall minus sign. For the values we chose, $\frac12 {\rm ln}\left(\frac{1}{mL}\right)-\frac{m^2L^2}{48} \approx \frac12 {\rm ln}\left(\frac{1}{mL}\right)\approx 2.30$, although this zero mode contribution would diverge in the $m\to 0$ limit. For intermediate values of the reference frequency $m < \mu < \sqrt{\frac{4}{\delta^2}+m^2}$, numerical fitting show the same leading divergence and a subleading logarithmic divergence  ${\cal C}^{\rm diag}_1(\hat{\rho}_0) = \frac{L}{2\delta} | {\rm ln}\left(\mu\delta\right) | - \tilde f(\mu L)| \rm \ln(\mu \delta)|  +{\rm finite}$, with $\tilde f(\mu L) \approx 4.10 \times 10^{-7} (\mu L)^{1.85}>0$. We used the parameters $m L=0.01$, $\mu L = 20,\,40,\,60,\,80,\,100,\,200,\,300,\,400,\,500$ for data with $L/\delta = 1$ to $10^4$, and found fits for the large $L/\delta$ behaviour.  \label{footvac}}
The structure of UV divergences is similar for holographic complexity, as we examine in section \ref{sec:holo}. A detailed comparison of the QFT and holographic results is also discussed in section \ref{holod}.

\begin{figure}
    \center
    \includegraphics[width=0.6\textwidth]{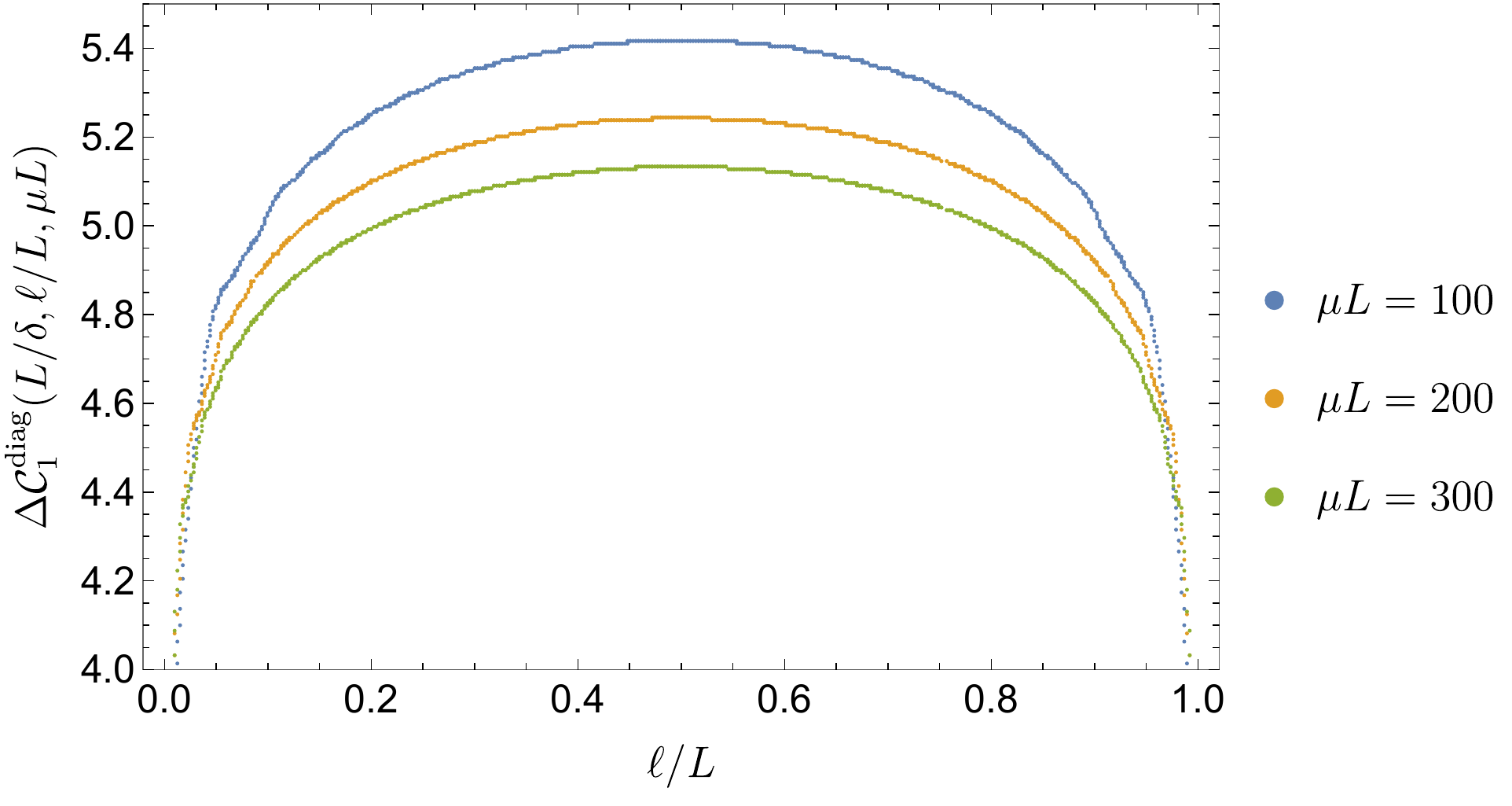}
\caption{Subregion size dependence of the mutual complexity in the diagonal basis $\Delta\mC_1^{\rm diag}$ for different reference frequencies  $\mu L = 100$, $200$ and $300$. The cutoff was set to $\delta/L = 1/N = 1/1000$ and the mass to $m L = 0.01$.}\label{fig:mutualcomp}
\end{figure}

\begin{figure}
\center
\includegraphics[width=0.43\textwidth]{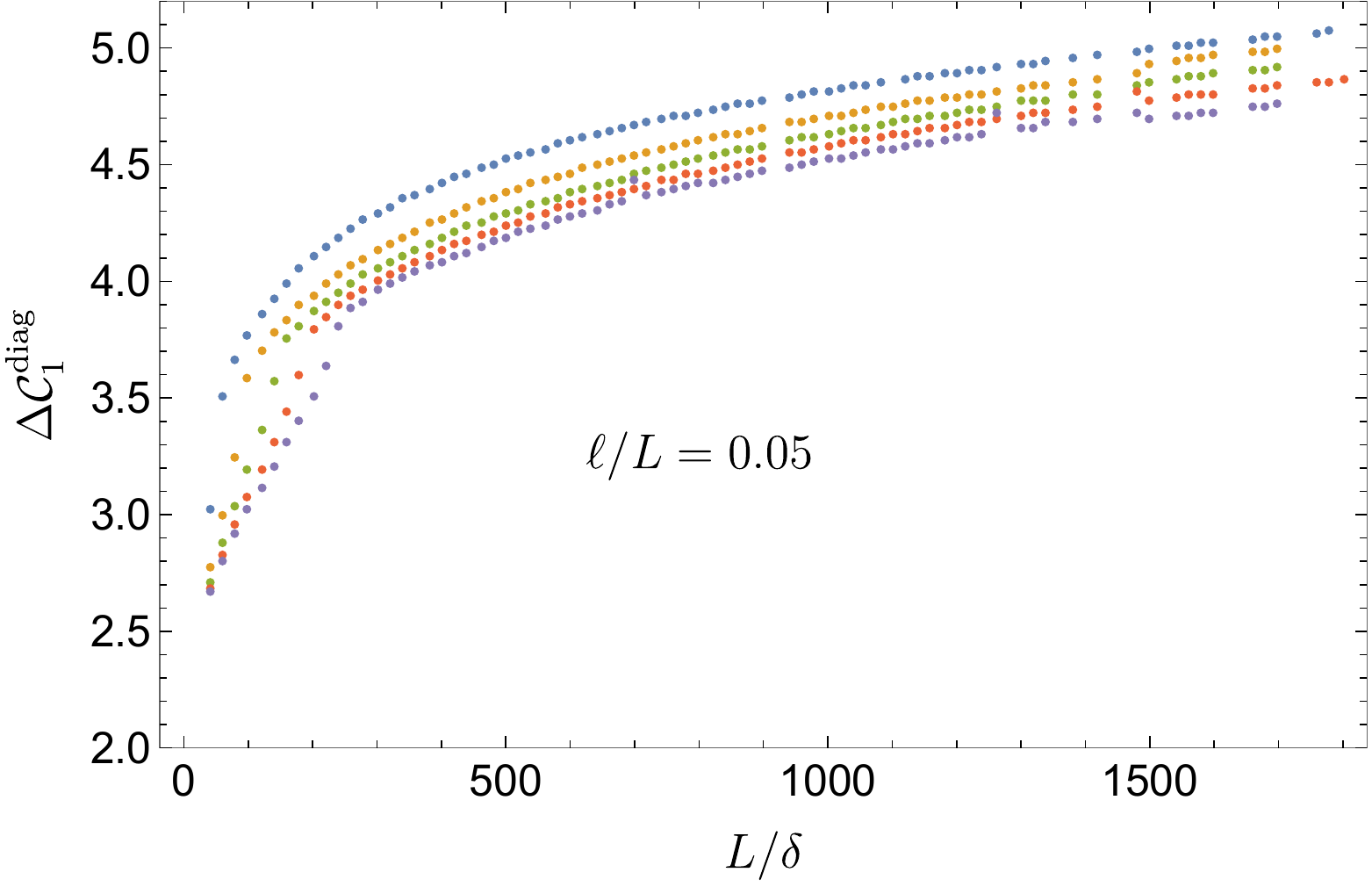}
\includegraphics[width=0.55\textwidth]{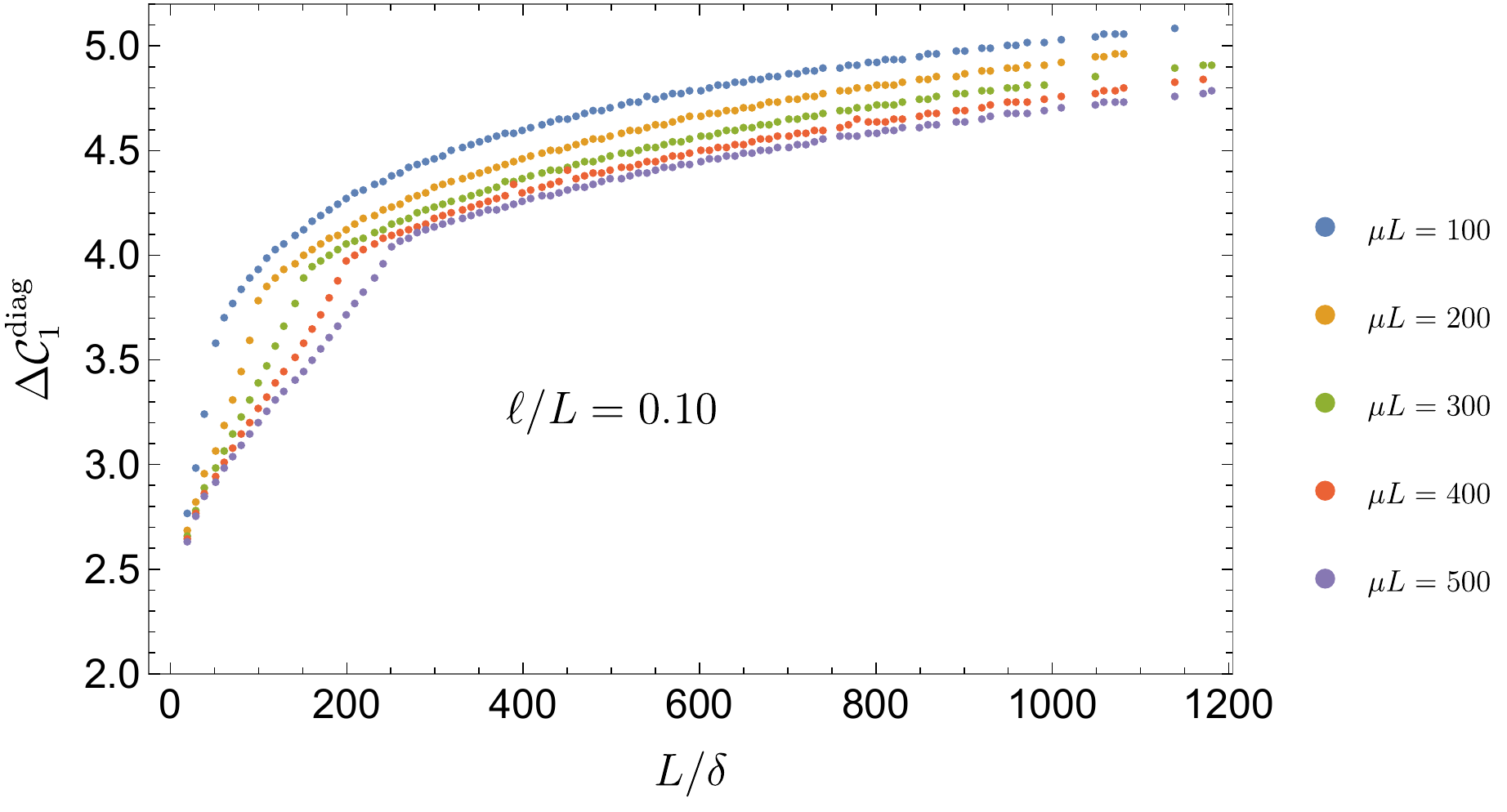}
\caption{Cutoff dependence of the mutual complexity in the diagonal basis $\Delta\mC_1^{\rm diag}$ for different reference frequencies  $\mu L = 100$, $200$, $300$, $400$ and $500$. The subregion sizes were fixed to $\ell/L = 0.1$ and $0.05$ and the mass to $m L = 0.01$.}\label{fig:mutualcomp2}
\end{figure}

{\bf Mutual complexity in the diagonal basis for subregions:} The numerical results for the mutual complexity \eqref{mutual_diag_sub} are shown in figures  \ref{fig:mutualcomp} and \ref{fig:mutualcomp2}. We observe that the mutual complexity in the diagonal basis is positive for all of the subregion sizes shown there. However, we do not have an analytic argument which proves that this should be the case in general. The mutual complexity rises dramatically for small subregion sizes in figure \eqref{mutual_diag_sub}, and then it continues to  increase as the subregion size grows until the subregion reaches half of the system. Further, $\Delta\mC$ is symmetric under $\ell\to L-\ell$. It has a positive logarithmic dependence on the cutoff which comes from the subleading divergence in the complexities. Looking at eq.~\reef{fit}, we observe that while $f_1(\mu \ell)+ f_1(\mu (L-\ell))$ becomes negative for large enough reference frequency, this contribution is offset by the negative coefficient of the logarithmic term in the vacuum complexity (see footnote~\ref{footvac}) to produce an overall  positive cutoff dependence in the mutual complexity, as can be seen in figure~\ref{fig:mutualcomp2}.

\subsection{Purification Complexity in the Physical Basis}

We introduced the physical basis purification complexity ${\cal C}^{\rm phys}_1$ in section \ref{fizz} and further investigated some of its properties in section \ref{subsec3phys}. In this subsection, we investigate the behaviour of ${\cal C}^{\rm phys}_1$ for subregions of the vacuum for a two-dimensional free scalar QFT on a circular lattice. The procedure to do this is very similar to the algorithm introduced in the previous section. In fact, the only difference comes after finding the purification matrix in the position basis in eq.~\eqref{eq:AAc}. From the purification matrix in the position basis
\begin{equation}\label{MAMAMAMAMAMA}
M_{\cal A} = \begin{pmatrix}
\Gamma^{\rm pos} & K^{\rm pos} \\ \left(K^{\rm pos}\right)^T & \Omega^{\rm pos}
\end{pmatrix}\,,
\end{equation}
we rotate the physical modes and the ancilla modes independently to diagonalize $\Gamma^{\rm pos}$ and $\Omega^{\rm pos}$ according to
\begin{equation}
M_{\cal A} \to M_{\cal A}^{\rm phys} = R_{\rm phys} M_{\cal A} R_{\rm phys}^T\,, \quad R_{\rm phys} = \begin{pmatrix}
R_{\cal A} & 0 \\ 0 & R_{{\cal A}^c}
\end{pmatrix}\,,
\end{equation}
where $R_{\cal A} \in SO(N_{\cal A},\mathbb{R})$ and $R_{{\cal A}^c} \in SO(N_{{\cal A}^c},\mathbb{R})$ such that $\Gamma^{\rm phys} = R_{\cal A} \Gamma^{\rm pos} R_{\cal A}^T$ and $\Omega^{\rm phys} = R_{{\cal A}^c} \Omega^{\rm pos} R_{{\cal A}^c}^T$ are diagonal.
Finally, the generator matrix $H^{\rm phys}$ can be found by taking the matrix logarithm of the parameter matrix in this basis as in~\eqref{eq:Hphys}
\beq
\label{eq:C1phys}
H^{\rm phys} = \frac12\, \ln\!\left(\frac{M_{\cal A}^{\rm phys}}{\mu}\right)\,.
\eeq
The physical basis complexity of these purifications is defined as in eq.~\eqref{C1phys}\footnote{Notice that eq.~\eqref{C1phys} does not have a minimization, while eq.~\eqref{eq:C1phys} includes a minimization over purifications. This is because~\eqref{C1phys} is the complexity in physical basis of~\emph{one particular purification}, while~\eqref{eq:C1phys} is the purification complexity of the density matrix $\rho_{\cal A}$, defined as the minimal complexity over all purifications  of $\rho_{\cal A}$.}
\begin{equation}
{\cal C}^\mt{phys}_1  \( \hat{\rho}_\mA\)= \text{min}\,\sum_{a,b=1}^{N_{\cal A}+N_{\mA^c}} |H^{\rm phys}_{ab}|\,,
\end{equation}
where we need to minimize the purification complexity  over the free parameters $k_i$ which were introduced in eq.~\eqref{eq:mbym}.

\subsubsection{Numerical results in the physical basis}

Again, we set $mL=0.01$ throughout the following. By setting the mass to such a small value, we expect that our QFT results might behave similar to those found for a holographic CFT.

\begin{figure}
\center
    \includegraphics[width=0.49\textwidth]{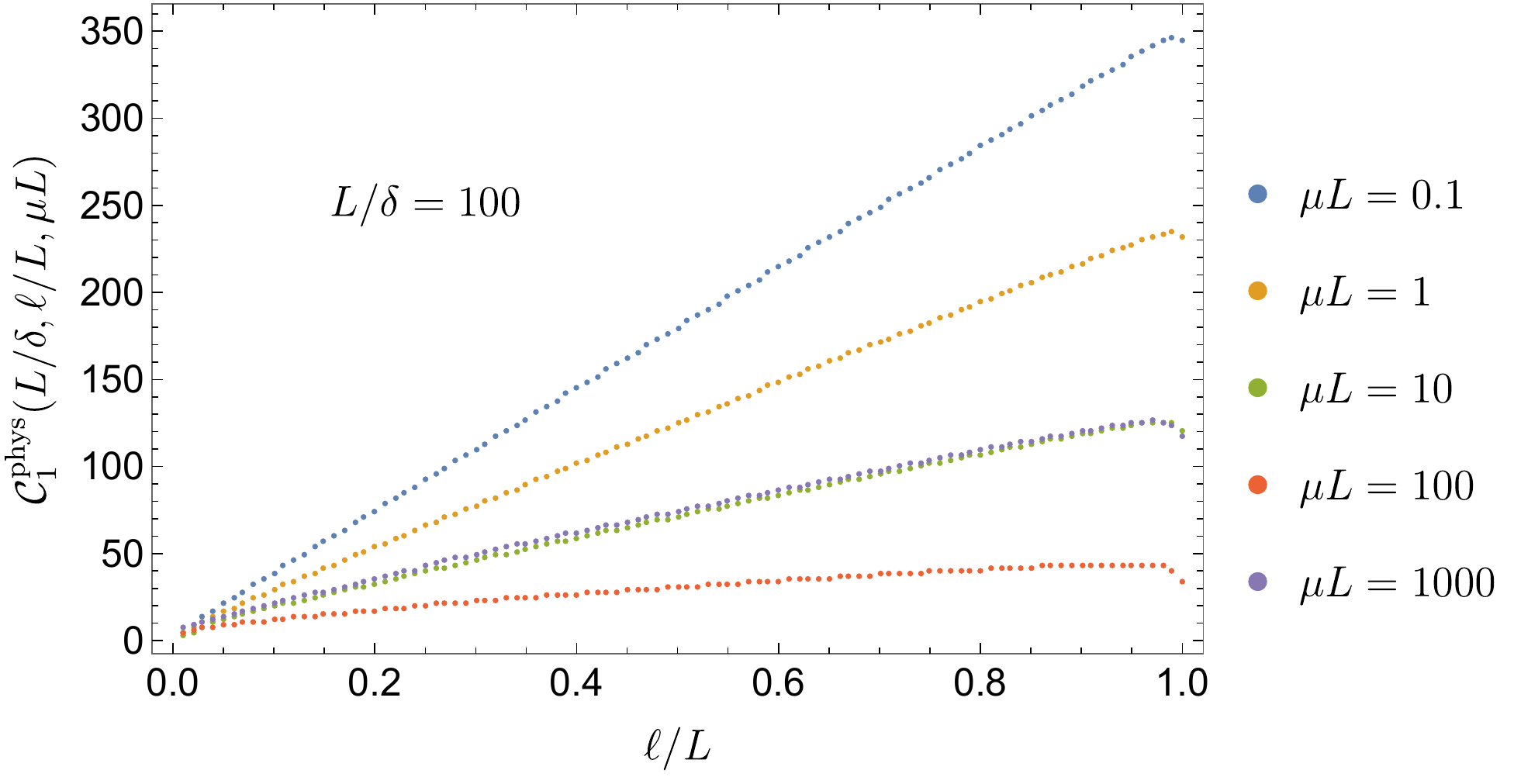}
    \includegraphics[width=0.49\textwidth]{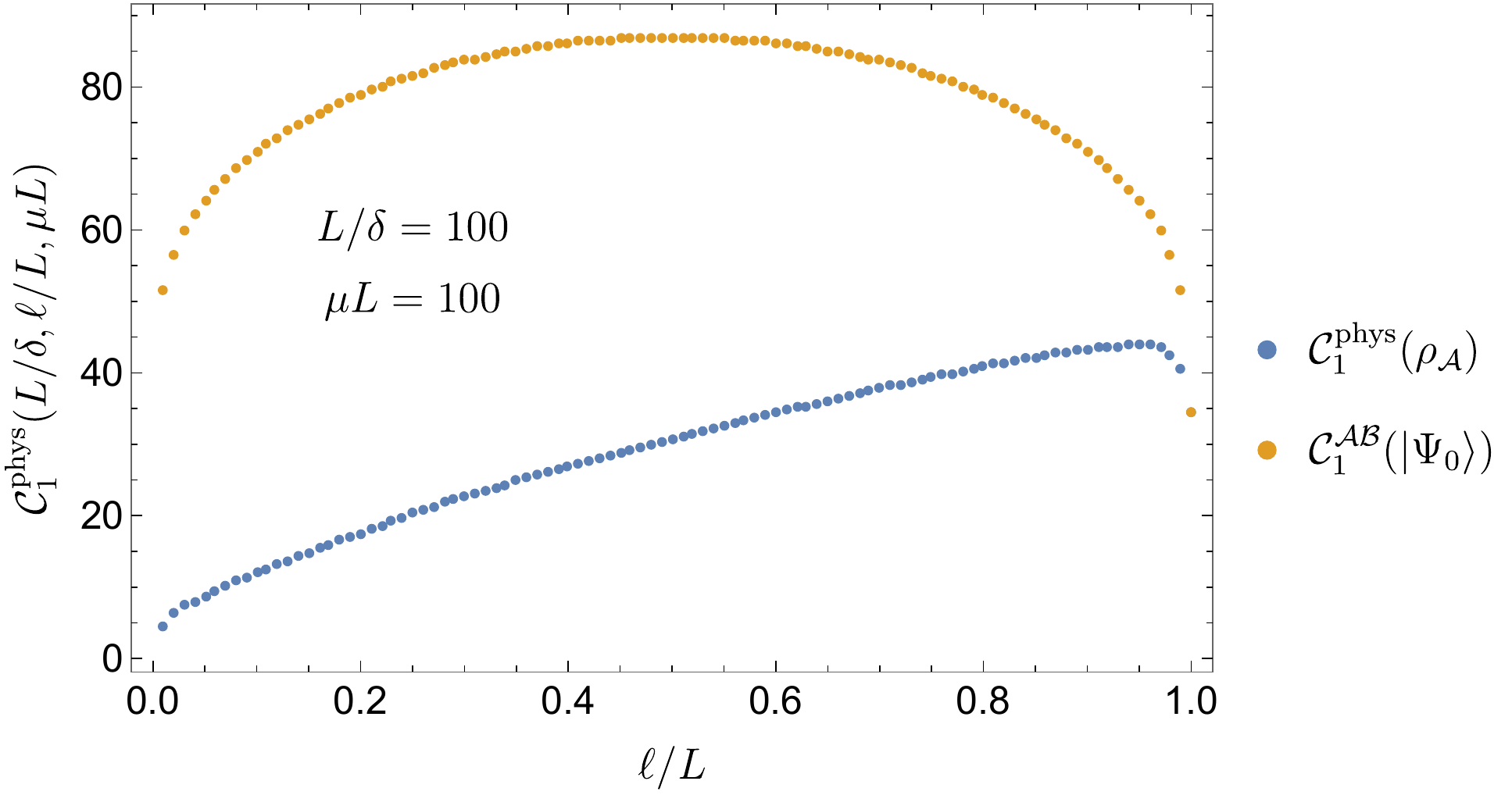}
    \caption{Left panel: subregion complexity as a function of the subregion size in physical basis for reference frequencies $\mu L = 0.1$, $1$, $10$, $100$, $1000$. Right panel: comparison of the subregion complexity to the complexity of the ground state in the physical basis for $\mu L = 100$. In both plots, the cutoff was set to to $L/\delta=N=100$ and the mass to $m L =0.01$.}
    \label{fig:subregion-phys}
\end{figure}

\textbf{Dependence on the size of the subregion:} We plot the purification complexity in the physical basis as a function of the subregion size for a lattice of 100 harmonic oscillators for different values of the reference frequency in figure~\ref{fig:subregion-phys}. Unlike the diagonal basis complexity, we find that for subregions approaching the full system, the physical basis purification complexity can increase beyond the complexity of the full system before decreasing rapidly to the full system complexity. At first sight, this might seem contradictory, since the ground state is one of the possible purifications over which the purification complexity is minimized. However, the complexity of the ground state in the physical basis partitioned by ${\cal A}$ and ${\cal A}^c$ can be greater than the complexity of the ground state itself. In fact, the purification complexity in the physical basis should be less than the complexity of the ground state~\emph{in that same basis}. In the right panel of figure~\ref{fig:subregion-phys}, we compare the purification complexity in the physical basis ${\cal C}^{\rm phys}_1(\rho_{\cal A})$ to the complexity of the ground state ${\cal C}^{\cal AB}_1(|\Psi_0 \rangle )$ in the basis which does not mix the degrees of freedom in the subsystem ${\cal A}$ with the modes in the complementary region $\mB$.  Indeed, we find that ${\cal C}^{\rm phys}_1(\rho_{\cal A}) \le {\cal C}^{\cal AB}_1(|\Psi_0 \rangle )$ for all subregions ${\cal A}$ and the inequality is only saturated when $\mA$ encompasses the entire system (\ie $\ell/L=1$). Note that  comparing $\mC^{\rm phys}_1(\rho_{\cal A})$ with the complexity of the ground state $\mC^{\rm diag}_1(|\Psi_0\rangle)$ in the diagonal basis,  we find that the above bound does not hold. In particular, the figure shows that for large subregions (\ie $\ell/L\gtrsim 0.6$), the subregion complexity exceeds that of the ground state in diagonal basis (but, of course, they coincide  at $\ell/L=1$). There is no contradiction in finding $\mC^{\rm phys}_1(\rho_{\cal A})>\mC^{\rm diag}_1(|\Psi_0\rangle)$ for some subregions since the two complexities are evaluated using different gate sets. As noted above, when the complexities are evaluated using the same basis, the subregion complexity is smaller than that of the vacuum.

\begin{figure}
    \includegraphics[width=0.44\textwidth]{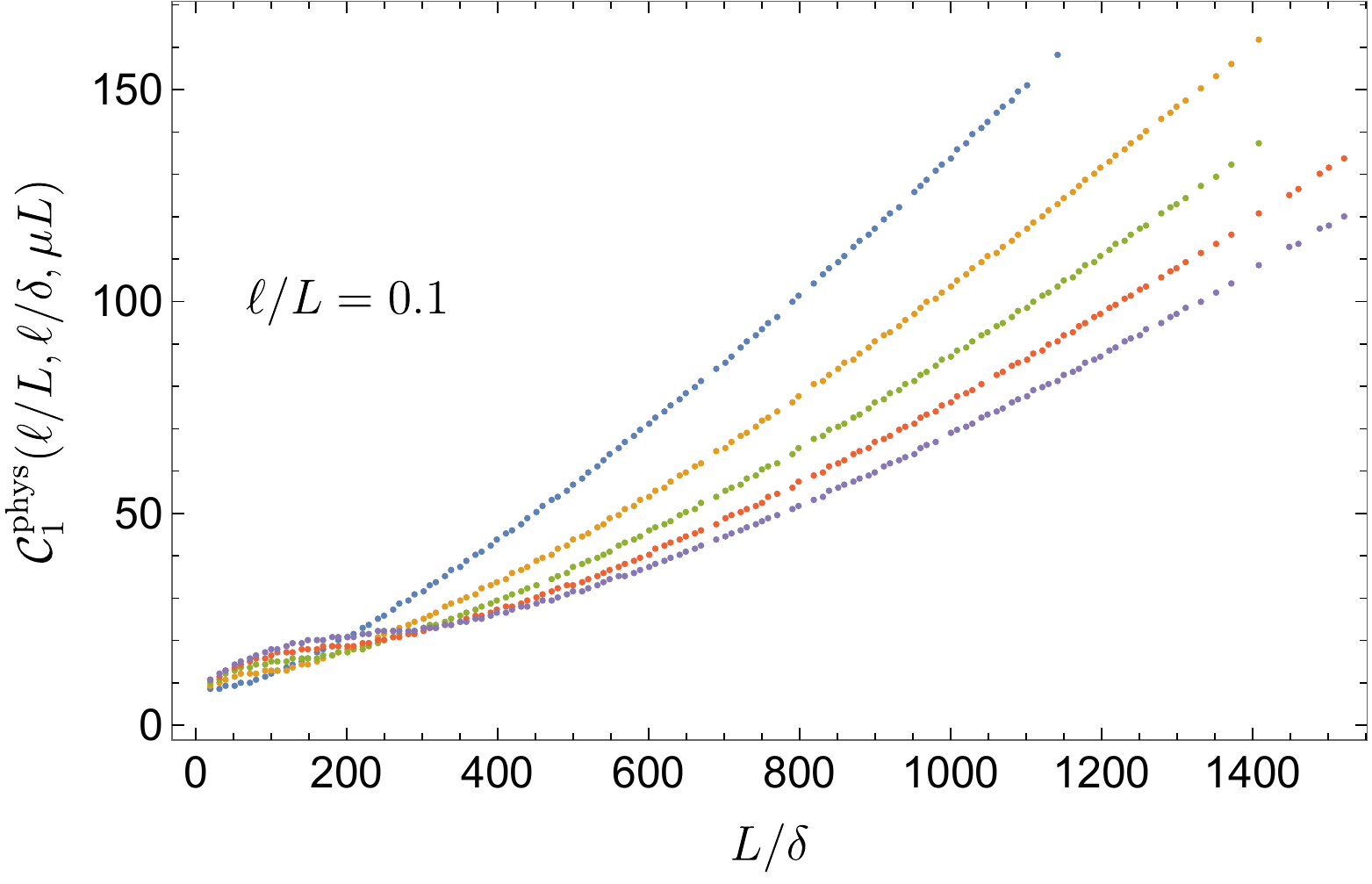}
    \includegraphics[width=0.45\textwidth]{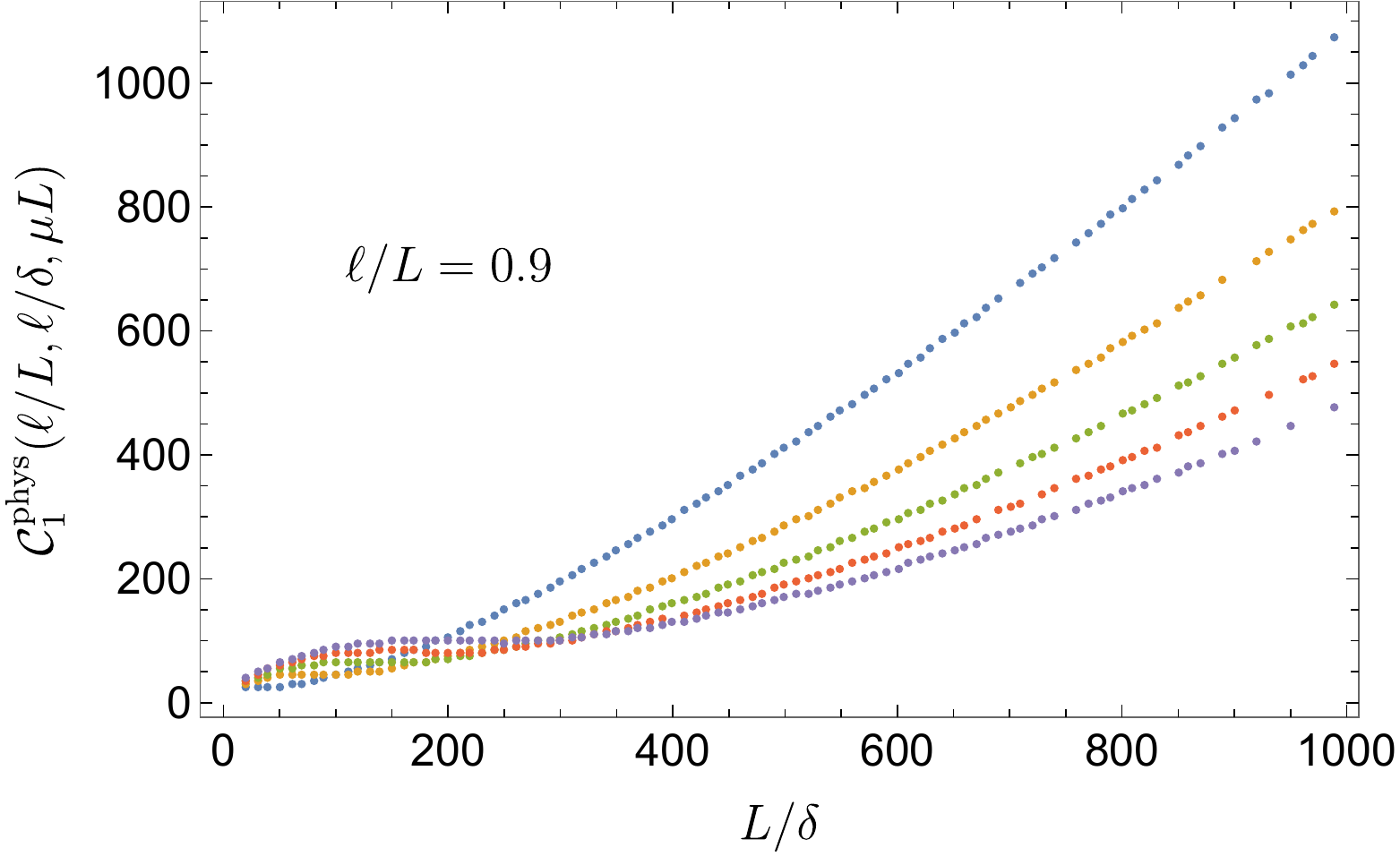}\\
    \includegraphics[width=0.44\textwidth]{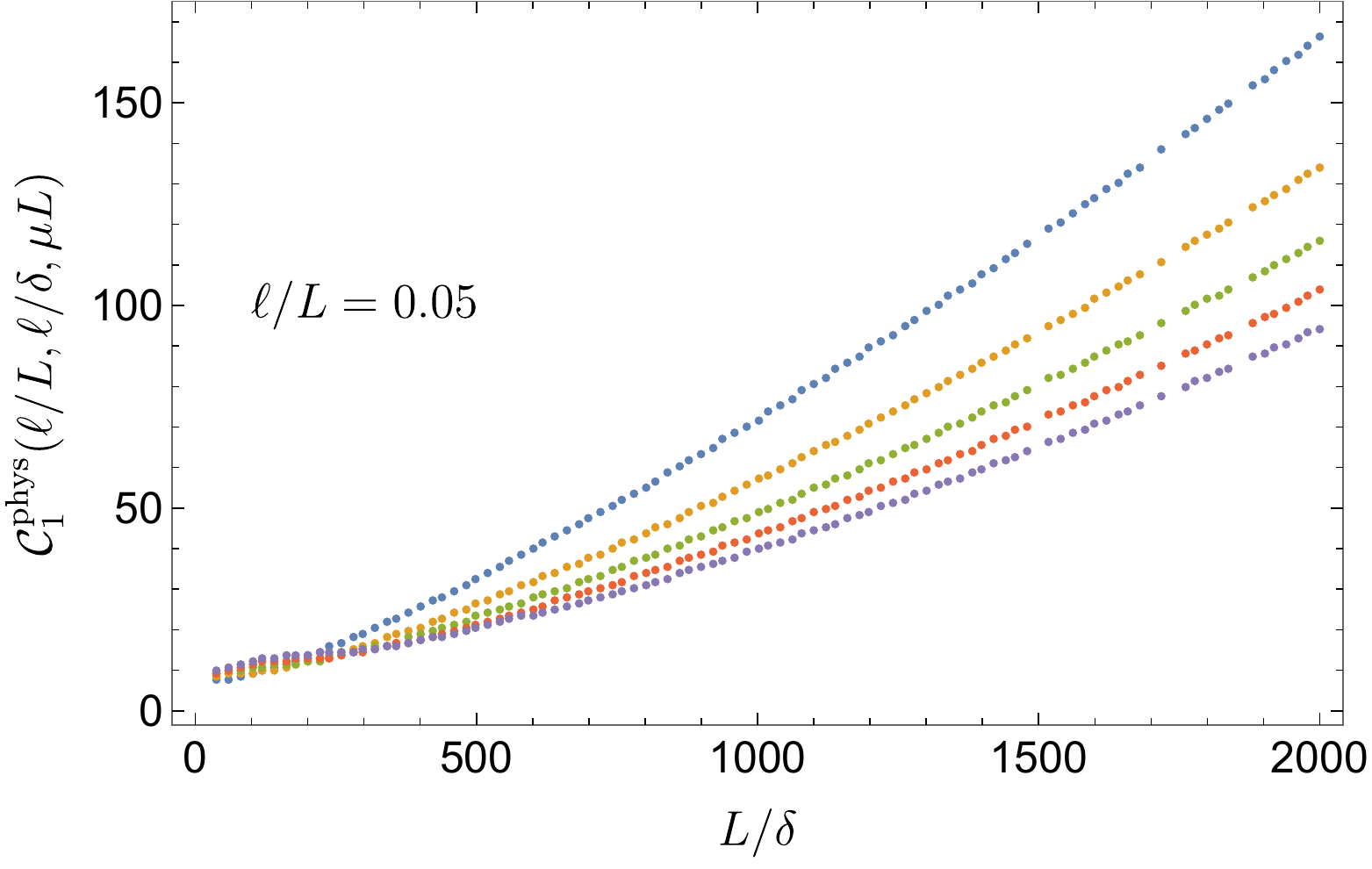}
    \includegraphics[width=0.55\textwidth]{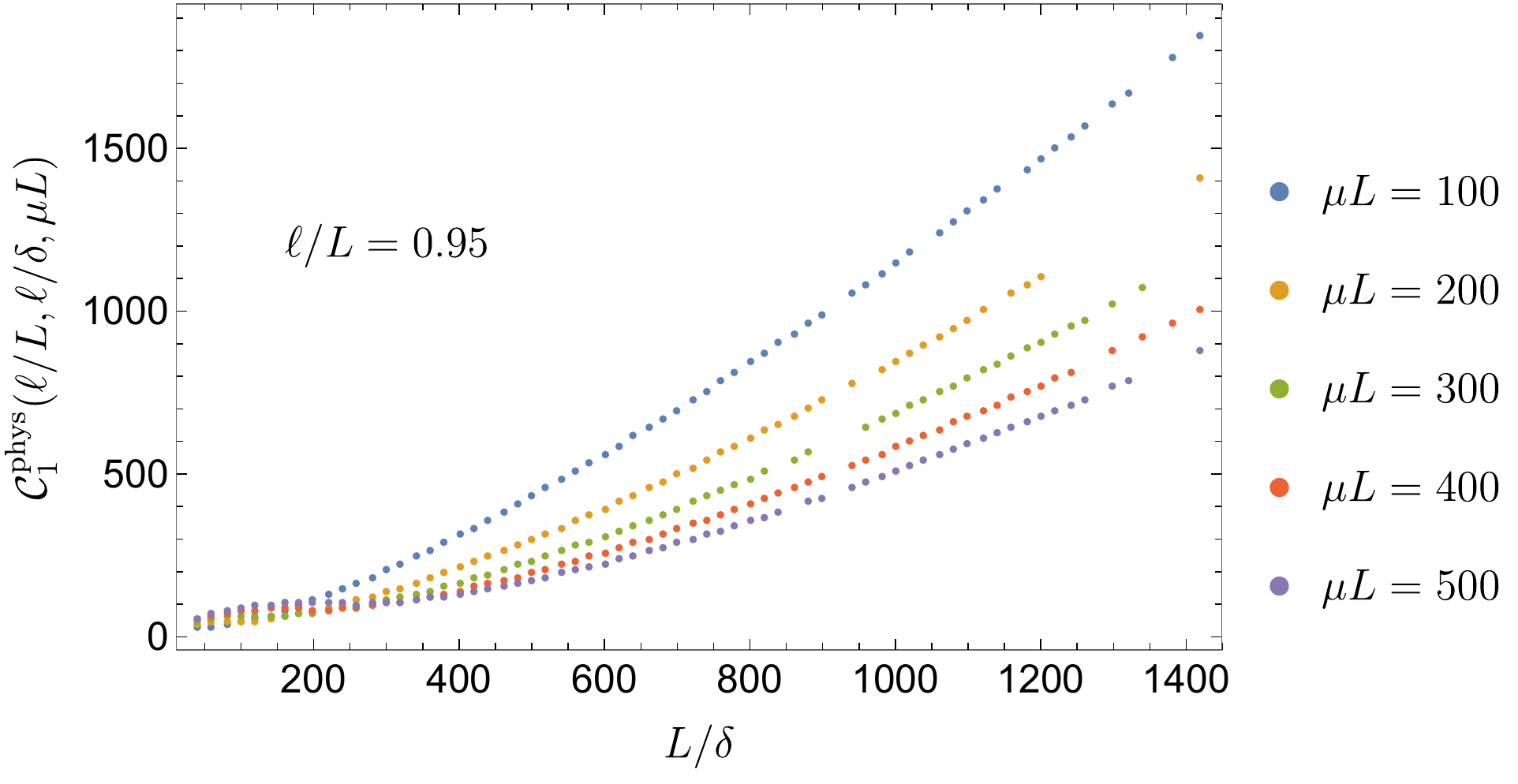}
    \caption{Subregion complexity in the physical basis as a function of the cutoff $N=L/\delta$ for $\ell/L = 0.05,\,0.1,\,0.9$ and $0.95$. The mass was set to $mL=0.01$.}
    \label{fig:subregion3-phys}
\end{figure}

\textbf{Structure of divergences in purification complexity:} For subregions with fixed size, we plot the cutoff dependence of the purification complexity in figure~\ref{fig:subregion3-phys}. The large $N$ (or equivalently, the small $\delta$) behavior of the subregion complexity with $\ell/L=1/10,\,9/10,\,1/20$ and $19/20$ is
\begin{equation}
\begin{aligned}
\label{fit2}
\mC^{\rm phys}_1(\ell/L=0.05,\mu L,\delta/L)
&\approx  \frac{\ell}{2\, \delta}\ln \frac{1}{\mu \delta}+3.31 \ln \frac{L}{\delta} + 0.149\, \mu \ell -6.54\,,\\
\mC^{\rm phys}_1(\ell/L=0.10,\mu L,\delta/L)
&\approx  \frac{\ell}{2\, \delta} \ln \frac{1}{\mu \delta} + 3.60 \ln \frac{L}{\delta}+ 0.253\, \mu \ell -5.79\,,\\
\mC^{\rm phys}_1(\ell/L=0.90,\mu L,\delta/L)
&\approx \frac{\ell}{2\, \delta}\ln \frac{1}{\mu \delta} +4.74 \ln \frac{L}{\delta} + 0.343\, \mu \ell -13.1\,,\\
\mC^{\rm phys}_1(\ell/L=0.95,\mu L,\delta/L)
&\approx \frac{\ell}{2\, \delta} \ln \frac{1}{\mu \delta} +5.04 \ln \frac{L}{\delta} + 0.333\, \mu \ell -14.5\,.
\end{aligned}
\end{equation}
These fits suggest a divergence structure for the subregion complexities in the physical basis of the form\footnote{As mentioned in footnote~\ref{foot67}, our fits were made for small $\delta/L$ with $\mu L $ fixed. In general, we expect the leading term to be the absolute value of the logarithmic term. Our resolution in the physical basis fits was not high enough to rule out a term of the form $f_0(\mu L,\ell/L)\, \frac{\ell}{\delta}$ where $f_0(\mu L,\ell/L) \lesssim {\cal O}(10^{-2})$.}
\beq
\label{divergences2}
\mC^{\rm phys}_1(\mu L,\delta/L)
\approx  \frac{\ell}{2\, \delta} \left|\ln \frac{1}{\mu \delta}\right| +f_1(\mu L,\ell/L) \ln \frac{L}{\delta}+f_2(\mu L,\ell/L)\,.
\eeq
Similarly to the discussion for the diagonal basis, we expect the structure of divergences in the physical basis to be the same as in eq.~\eqref{divergences2} for more general cases, except that the coefficients $f_1$ and $f_2$ will depend on the other parameters of the system. For example, for a massive scalar QFT, we expect $f_i=f_i(\mu L,\ell/L,mL)$.

\begin{figure}
    \center
    \includegraphics[width=0.45\textwidth]{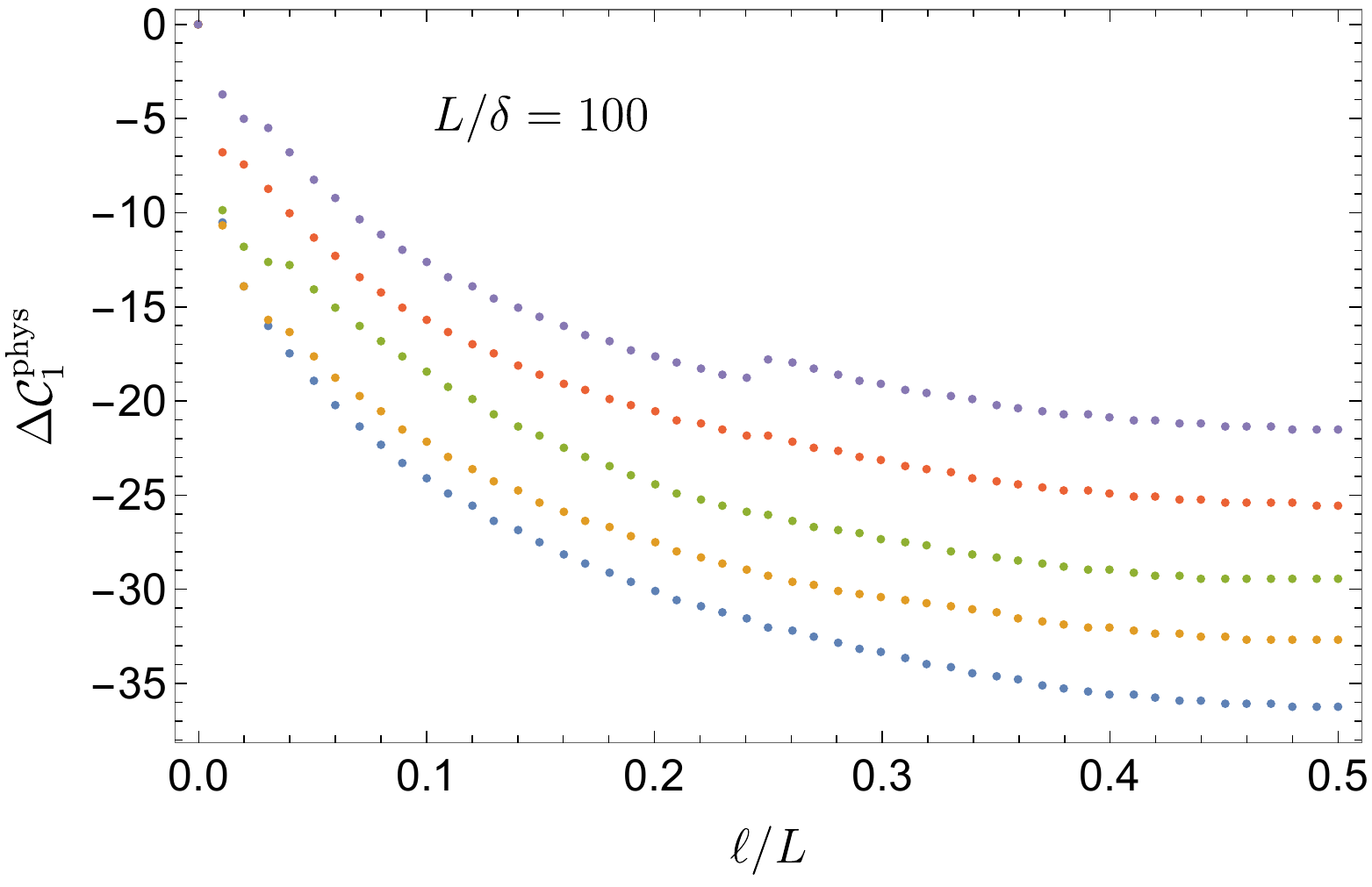}
    \includegraphics[width=0.54\textwidth]{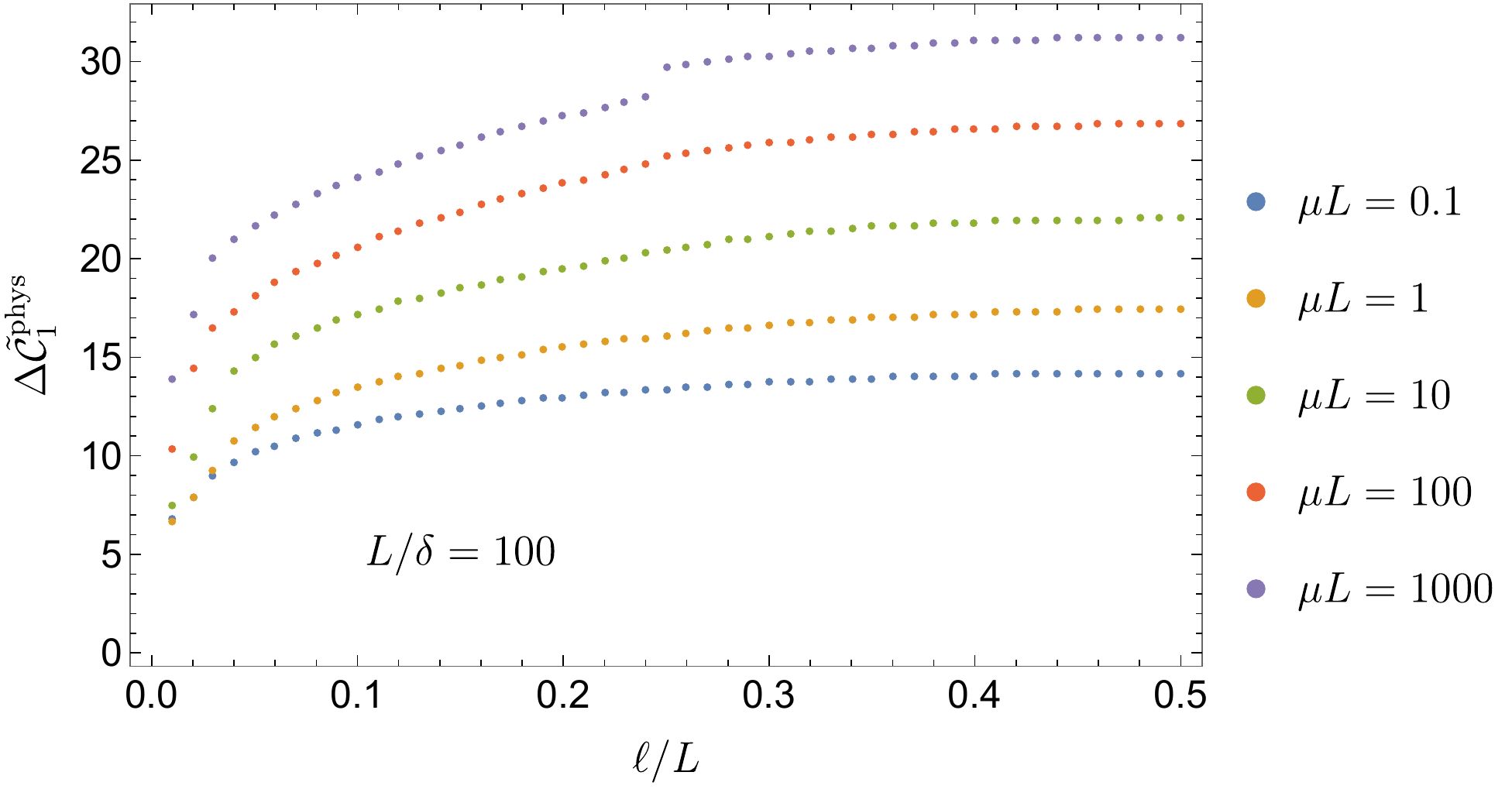}
    \caption{The two definitions of mutual complexity in the physical basis $\Delta\mC_1^{\rm phys}$ and  $\Delta\tilde{\mC}_1^{\rm phys}$ as a function of the subregion size for various reference frequencies. The cutoff was set to $L/\delta=N=100$ and the mass to $m L =0.01$.}
    \label{fig:subregion2-phys}
\end{figure}

\begin{figure}
    \center
    \includegraphics[width=0.44\textwidth]{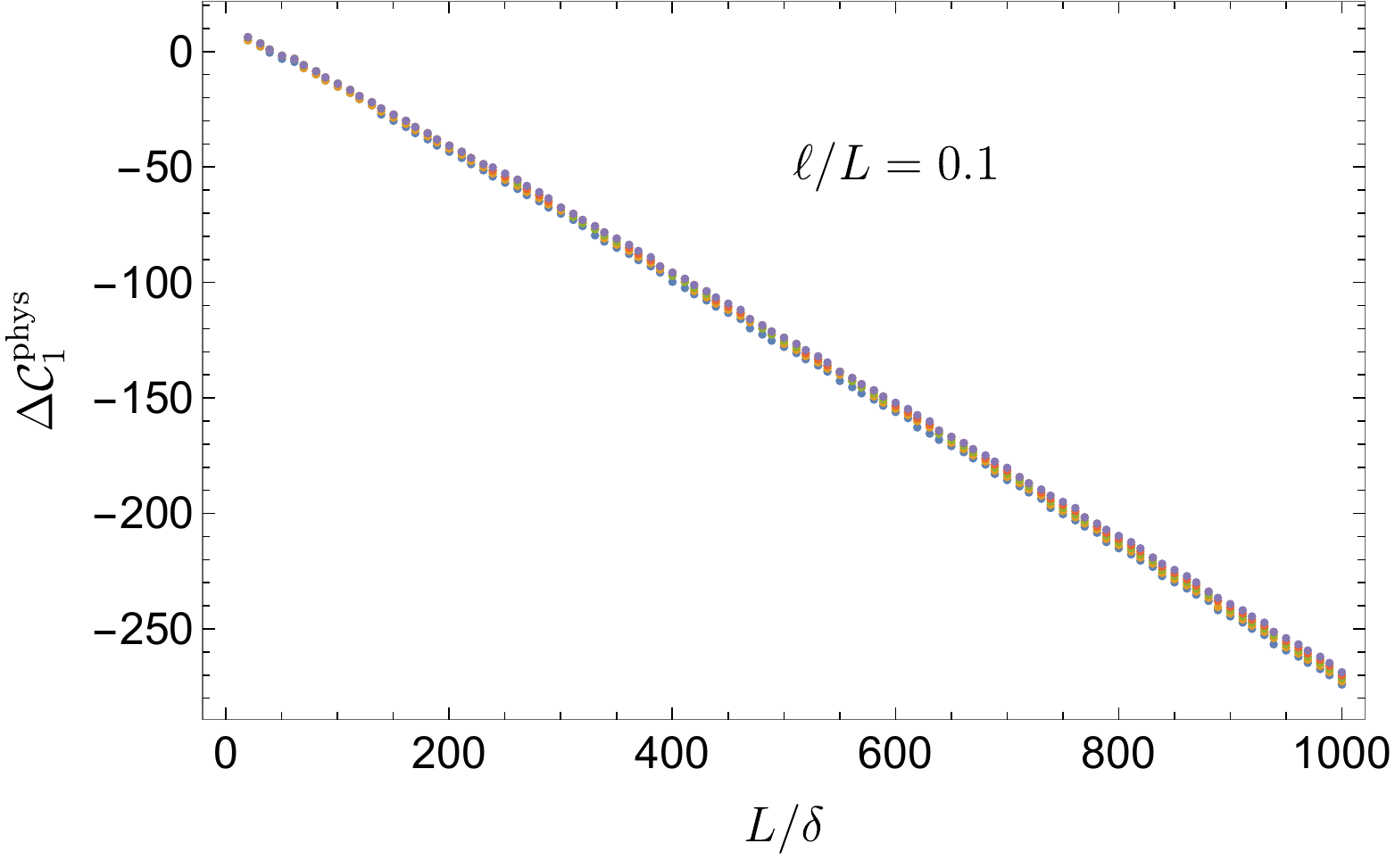}
    \includegraphics[width=0.55\textwidth]{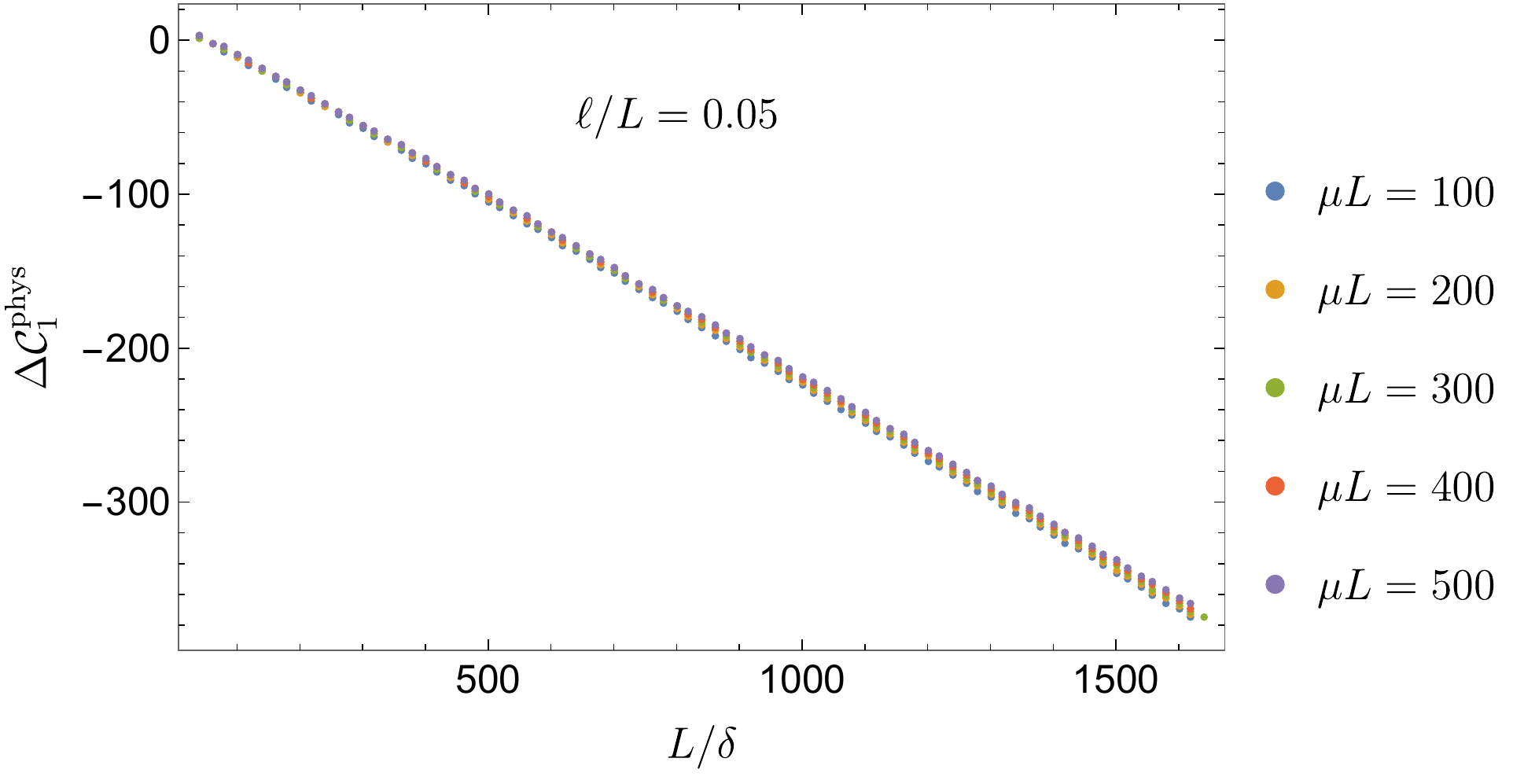}
    \includegraphics[width=0.44\textwidth]{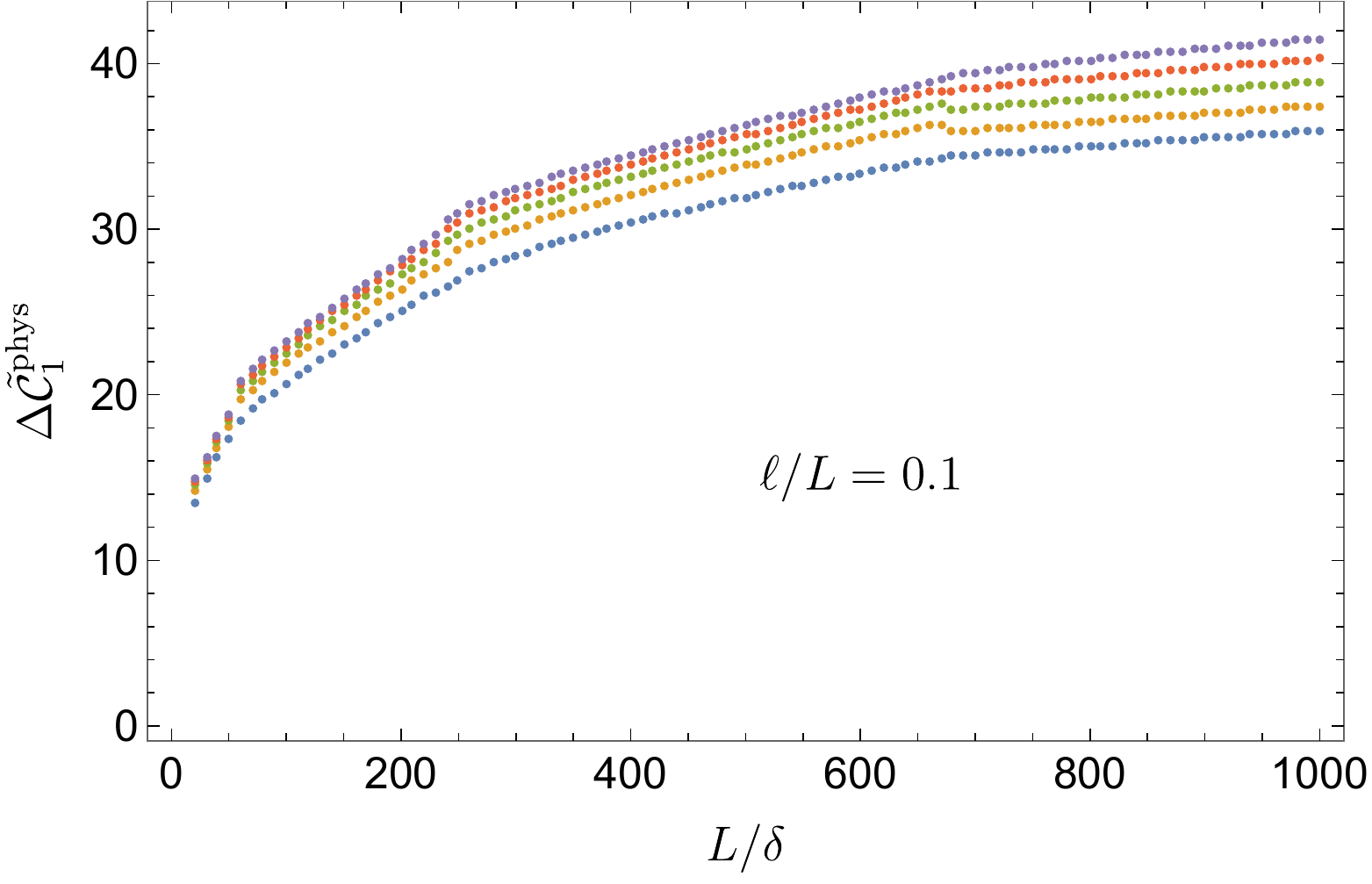}
    \includegraphics[width=0.55\textwidth]{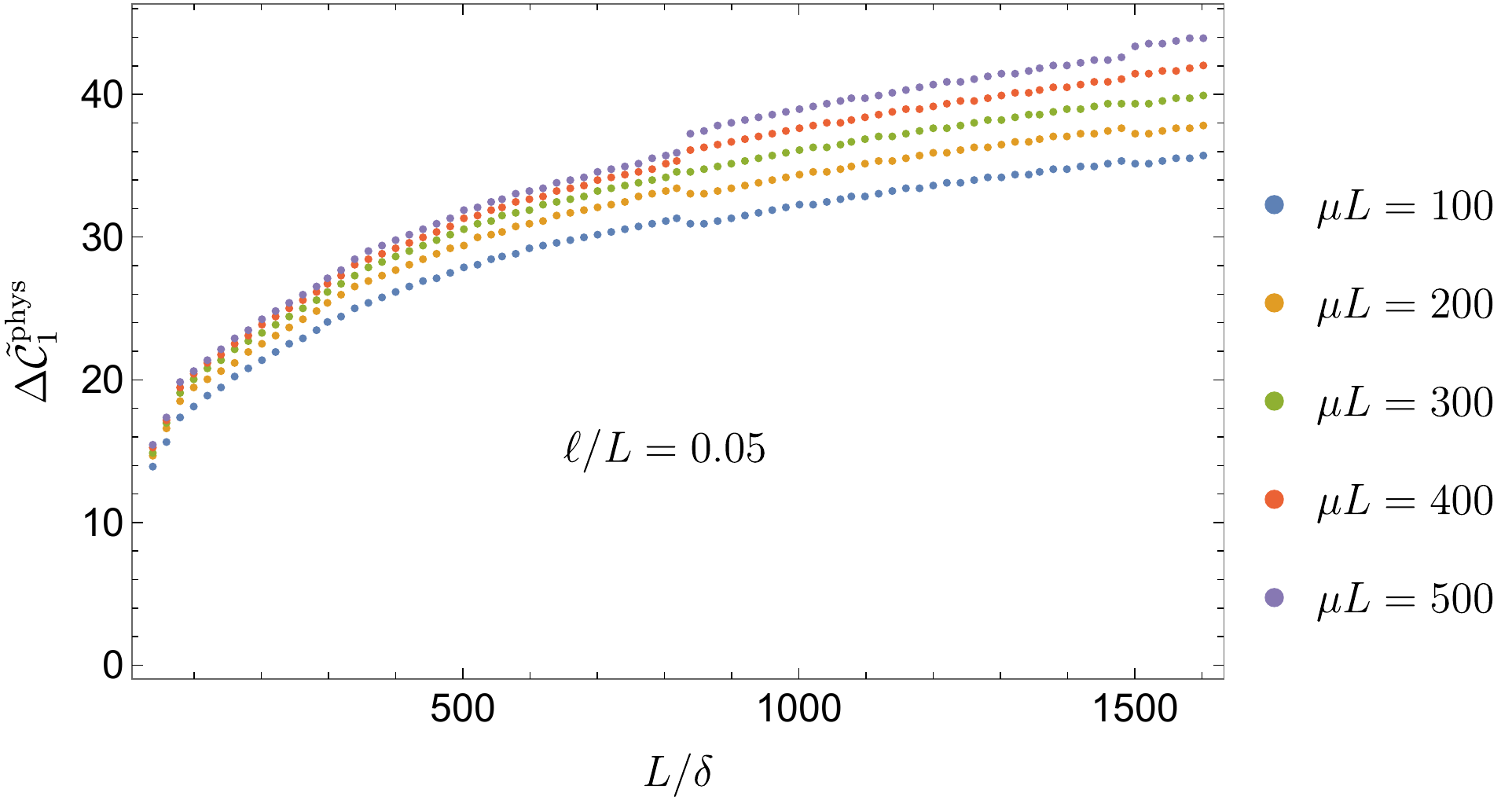}
    \caption{The two definitions of mutual complexity in the physical basis $\Delta\mC_1^{\rm phys}$ and  $\Delta\tilde{\mC}_1^{\rm phys}$ as a function of the cutoff for various reference frequencies $\mu L = 100$, $200$, $300$, $400$ and $500$. The mass was set to $m L =0.01$.}
    \label{fig:subregion4-phys}
\end{figure}

\textbf{Mutual complexity in physical basis:} We plot the mutual complexity in the physical basis
\begin{equation}
\label{DeltaCphys}
\Delta\mC_1^{\rm phys}\equiv {\cal C}^{\rm phys}_1(\rho_{\cal A}) + {\cal C}^{\rm phys}_1(\rho_{{\cal B}})- {\cal C}^{\rm phys}_1(|\Psi_0 \rangle ) \,,
\end{equation}
in figures~\ref{fig:subregion2-phys} and \ref{fig:subregion4-phys}, which we observe to be negative for all of the subregion sizes shown there. However, some explanation is required here. The mutual complexity~\eqref{DeltaCphys} will be different depending on whether the physical basis for the three states considered is fixed to be one which separates ${\cal A}$ and/or ${\cal B}$ from the rest of the degrees of freedom, or if the physical basis is considered for each state independently. More precisely, the physical basis for $\rho_{\cal A}$ (and $\rho_{\cal B}$) will be a basis in which the ${\cal A}$ and ${\cal A}^c$ (and the ${\cal B}$ and ${\cal B}^c$, respectively) degrees of freedom are kept separate. However, for the ground state, there is no natural partition of the system into ${\cal A} \cup {\cal B}$ independently of the density matrices $\rho_{\cal A}$ and $\rho_{\cal B}$. Therefore, if the physical basis in the evaluation of the complexity of the ground state were to be considered independently of the other two complexities, we would find that the physical basis for the ground state corresponds to all of the degrees of freedom in the system, and the physical basis would coincide with the diagonal basis. Therefore, to be more explicit, we define two mutual complexities in the physical basis

\beq
\begin{aligned}
    \label{twoDeltaCs}
\Delta\mC_1^{\rm phys}\equiv {\cal C}^{{\cal AA}^c}_1(\rho_{\cal A}) + {\cal C}^{{\cal BB}^c}_1(\rho_{{\cal B}})- {\cal C}^{\cal AB}_1(|\Psi_0 \rangle ) \,,\\
\Delta\tilde{\mC}_1^{\rm phys}\equiv {\cal C}^{{\cal AA}^c}_1(\rho_{\cal A}) + {\cal C}^{{\cal BB}^c}_1(\rho_{{\cal B}})- {\cal C}^{\rm diag}_1(|\Psi_0 \rangle ) \,,
\end{aligned}
\eeq
where
${\cal C}^{\cal AB}$ denotes the physical basis complexity of a state given a partition of the system into  ${\cal A} \cup {\cal B}$. It is natural to expect that $\Delta{\cal C}_1^{\rm phys} < \Delta \tilde{\mC}_1^{\rm phys}$, since the difference between the two definitions in eq.~\eqref{twoDeltaCs} is the subtraction of the vacuum complexity in two different bases. More precisely, the ${\cal C}^{\cal AB}_1(|\Psi_0 \rangle )$ evaluates the complexity of the ground state subject to the additional constraint that the ${\cal A}$ and ${\cal B}$ degrees of freedom remain separated. Being a minimization with additional constraints compared to ${\cal C}^{\rm diag}_1(|\Psi_0 \rangle )$, it follows that  ${\cal C}^{\cal AB}_1(|\Psi_0 \rangle ) > {\cal C}^{\rm diag}_1(|\Psi_0 \rangle )$ from which the above conclusion follows.

Just like the mutual complexity in the diagonal basis, we observe that both of the mutual complexities in the physical basis increase in magnitude as a function of the subregion size, reaching maximum at $\ell/L = 1/2$, and are symmetric about this point. The $\Delta \tilde{\mC}^{\rm phys}_1$ shows similar behaviour to the diagonal basis mutual complexity: it is positive and depends logarithmically on the cutoff. Again, this logarithmic dependence comes from the subleading logarithmic divergence of the complexities. The subleading divergence in the subregion complexities in the physical basis are positive, while the subleading divergence of the complexity of the ground state is negative for all cases studied here (see footnote~\ref{footvac}). On the other hand, the $\Delta \mC^{\rm phys}_1$ is negative and decreases linearly as a function of the cutoff. This contrasts with the logarithmic cutoff dependence of the mutual complexity in the diagonal basis in figure \ref{fig:mutualcomp2}. The negative linear dependence of $\Delta \mC^{\rm phys}_1$ on the cutoff is due to the vacuum complexity in the ${\cal AB}$ basis having a subleading positive linear divergence, which is not present for the diagonal basis.

\subsection{Comment on the Approximation}\label{commentsonApp}
Lastly, we comment on the accuracy of our approximation. As mentioned above, strictly speaking, the algorithms presented are only an upper bound for the complexity of subregions of the ground state of our QFT in the different bases. The reason for this is that we only minimized over possible \textbf{mode-by-mode purifications} (see section \ref{MbyM}), assuming that the complexity of the optimal mode-by-mode purification would be close to the optimal complexity obtained by exploring most general purifications. In section~\ref{MbyM}, we found that the optimal purification is indeed a mode-by-mode purification when the target density matrix is a product density matrix ($\hat{\rho} = \otimes_i \hat{\rho}_i$). For Gaussian density matrices of the form~\eqref{density_fun_A}, this is the case when $\[A,B\]=0$, and we introduced a measure of how close a matrix was to being a product state in~\eqref{Delta1}. For the subregions of the vacuum studied in this section, we find $\Delta = 0.5$ for $\ell > \ell_c$ with $\ell_c = {\cal O}(\delta)$. In contrast, for random matrices $A$ and $B$, we find $\Delta \approx \sqrt{2}$ for large~$N$.

\section{Holographic Complexity for Mixed States}\label{sec:holo}

In the previous two sections, we investigated the purification complexity for Gaussian mixed states in free scalar quantum field theory. In particular, we focused on two examples: the complexity of thermal states and the complexity of subregions in the vacuum state. The purpose of this section is to review and compare some general features of these results to those obtained using the proposals for holographic complexity.

In holography, there have been two different proposals for the gravitational dual of subregion complexity. These proposals are extensions of the complexity=volume (CV) \cite{Volume1,Volume2} and complexity=action proposals \cite{Action1,Action2}, and they were motivated by entanglement wedge reconstruction, \ie the understanding that the reduced density matrix of a boundary subregion encodes the dual entanglement wedge in the bulk \cite{EW1,EW2,EW3}.\footnote{The latter can be proven with the assumption that the bulk and boundary relative entropies are exactly equal \cite{Jafferis:2015del,Dong:2016eik}.} We denote the two proposals as the subregion complexity=volume (subregion-CV) \cite{Alishahiha:2015rta,Carmi:2016wjl} and the subregion complexity=action (subregion-CA) \cite{Carmi:2016wjl} conjectures. A third approach for holographic complexity was also proposed with the complexity = spacetime volume (CV2.0) conjecture \cite{Couch:2016exn}. Hence in the following, we also discuss the natural extension of this proposal for the case of subregions, which we designate the subregion-CV2.0 conjecture. Note that all three approaches for subregion complexity recover the corresponding original proposal for the holographic complexity of a pure state in the limit in which the subregion becomes the whole boundary.

Let us add that subregion complexity in holography has been widely explored. The studies include, to name only a few, general studies of the structure of divergences \cite{Alishahiha:2015rta,Carmi:2016wjl}, multiple subregions \cite{Abt:2017pmf}, subregions whose boundary includes kinks/corners \cite{Bakhshaei:2017qud,Caceres:2018luq}, subregions of systems with defects \cite{Chapman:2018bqj}, subregion complexity in eternal black hole backgrounds for subregions consisting of a single boundary \cite{BrianMixedComplexity} and the opposite limit of small subregions in eternal black hole geometry \cite{Bhattacharya:2019zkb}. We begin below with a brief review of the different approaches described above and their main properties. We will then review the results of evaluating these proposals for two examples, which are relevant for the comparison to the QFT results in the two previous sections: a subregion consisting of a single boundary of the TFD state (eternal black hole), where we are evaluating the complexity associated to the thermal state; and a boundary subregion of the CFT vacuum state (empty AdS).

\subsection{Review of the Holographic Proposals} \label{revvH}

The subregion-CV conjecture \cite{Alishahiha:2015rta, Carmi:2016wjl} suggests that the complexity associated to a boundary subregion $\mathcal{A}$ on a given time slice is given by the maximal spatial volume of a codimension-one surface, $\mathcal{R_\mathcal{A}}$ bounded by the boundary subregion and its Hubeny-Rangamani-Takayanagi (HRT) surface $\mathcal{E}_\mathcal{A}$ \cite{Ryu:2006bv,Ryu:2006ef,Hubeny:2007xt,Dong:2016hjy}:
\begin{equation}\label{eq:cv}
\mathcal{C}_V(\mathcal{A})=\max_{\del \mathcal{R_A}= \mathcal{A} \cup \mathcal{E}_\mathcal{A}} \left[\frac{\mathcal{V}(\mathcal{R_A})}{G_N\, \ell_{\text{bulk}}}\right]\,.
\end{equation}
The appearance of an arbitrary bulk  length scale,  $\ell_{\text{bulk}}$, is a somewhat undesirable feature. In the following, we assume that  $\ell_{\text{bulk}}=L$,  the AdS curvature radius. Note that while a more sophisticated prescription to define $\ell_{\text{bulk}}$ for black hole geometries was given in \cite{Couch:2018phr}, it still yields $\ell_{\text{bulk}}\sim L$ for the planar AdS black holes which we consider below, \ie see eq.~\reef{BH_metric}.

A second proposal is the subregion-CA conjecture \cite{Carmi:2016wjl}, which suggests that the subregion complexity is given by the on-shell gravitational action on a particular bulk region $\widetilde{W}_\mA$, which is defined as the intersection of the Wheeler-DeWitt (WDW) patch and the entanglement wedge of the boundary region $\mathcal{A}$ \cite{EW1,EW2,EW3}:
\begin{equation}\label{eq:ca}
\mathcal{C}_A(\mathcal{A})=\frac{I_{\text{grav}}(\widetilde{W}_{\!\mA})}{\pi  }\,.
\end{equation}
The gravitational action on regions with boundaries includes surface terms in addition to the usual bulk contribution. These surface terms include the Gibbons-Hawking-York (GHY) term \cite{Gibbons:1976ue,York:1972sj} for space-like or time-like boundaries, and an analogous contribution for null boundaries \cite{Lehner:2016vdi,Parattu:2015gga}. For null boundaries, one must also include the null counterterm introduced in  \cite{Lehner:2016vdi} to restore reparametrization invariance along the null generators. In addition to the codimension-one boundary surfaces, the boundary of  $\widetilde{W}$ also contains codimension-two joints at the intersection of the boundary surfaces. Their contributions have been addressed in \cite{Hayward1993,Brill:1994mb} for joints which do not involve null surfaces, and in \cite{Lehner:2016vdi} for joints which involve at least one null surface. The full prescription can be found in \cite{Lehner:2016vdi}, or in appendix A of \cite{Carmi:2016wjl}. Hence, in order to calculate $\mathcal{C}_A(\mathcal{A})$, we must evaluate
\begin{equation}
I_{\text{grav}}=I_{\text{bulk}}+ I_{\mt{GHY}}+I_{\text{null}}+ I_{\text{ct}}+ I_{\text{joints}}\,. \label{actor}
\end{equation}
Let us add that defining the counterterm $I_{\text{ct}}$ requires introducing a new arbitrary length scale $\ell_\text{ct}$ and the choice of this length scale influences various properties of the complexity. Comparing the structure of the UV singularities in holographic and QFT calculations of complexity leads to the suggestion that the choice of this length scale may be related to the choice of microscopic scales in defining the reference state and the gates in the complexity model of the boundary theory (\eg $\mu$ in our QFT construction) \cite{qft1,qft2,Vaidya2}.

The complexity = spacetime volume (CV2.0)\footnote{An update to the complexity = spacetime volume conjecture, denoted `CA2.0', was proposed in \cite{Fan:2018wnv}. However, for Einstein-Hilbert gravity with minimally coupled matter, this approach simply reduces to the CV2.0 proposal. As such, we will not consider it further here.} conjecture \cite{Couch:2016exn} simplifies the CA conjecture by proposing that the complexity can be determined by evaluating the spacetime volume of the WDW patch. The simplification still displays all of the properties expected of holographic complexity. Our subregion-CV2.0 conjecture is the natural generalization of this proposal to boundary subregions. That is, the complexity of a subregion $\mA$ is given by the spacetime volume of the region appearing in eq.~\reef{eq:ca}, \ie the intersection of the WDW patch and the entanglement wedge,\footnote{The units are naturally absorbed by the AdS curvature scale in the definition here following \cite{Couch:2018phr}. Their approach uses the relation $\mathcal{C}\sim {\cal P}\,\cal V_\text{WDW}$ where ${\cal P}= -\frac{\Lambda}{8\pi G_{N}}\sim 1/(G_N L^2)$ is the `bulk' pressure \cite{Kastor:2009wy}. Note that the application of these arguments is not straightforward for solutions with nontrivial scalar hair \cite{Fan:2018wnv}.}
\begin{equation}\label{eq:cv2}
\mathcal{C}_{V2.0}(\mathcal{A})=\frac{{\cal V}(\widetilde{W}_{\!\mA})}{G_N\, L^{2} }\,.
\end{equation}
As a pragmatic point, we note that in our calculations below, the integrand of the bulk action, \ie the Einstein-Hilbert term, is simply constant with $R-2\Lambda = -\frac{2d}{L^2}$. Hence, the complexity in eq.~\eqref{eq:cv2} and the bulk action evaluated for eq.~\reef{eq:ca} are simply related by
\begin{equation}\label{cv20main2}
\mathcal{C}_{V2.0}(\mA) = -\frac{8\pi}{d}\, I_{\text{bulk}}(\widetilde W_\mA)\,.
\end{equation}

{\bf Additivity properties:}
The various holographic proposals for subregion complexity differ in several important respects. $\mathcal{C}_V$ is  superadditive --- see section 2.1 of \cite{BrianMixedComplexity}. That is, let $\sigma$ be the Cauchy slice on which a pure state is defined, and divide this surface into a subregion $\mA$  and its complement $\mB$. Then the corresponding holographic complexities evaluated satisfy,
\begin{equation}\label{volineqholo}
\mathcal{C}_V(\mA) + \mathcal{C}_V (\mB) \le \mathcal{C}_V(\sigma=\mA\cup\mB)\,,
\end{equation}
\ie the mutual complexity \reef{ramen2A} is negative. Intuitively, superadditivity in $\mathcal{C}_V$ is the result of dealing with positive definite volumes and the fact that the requirement to pass through the HRT surface adds an additional constraint in maximizing the volume. Let us add that this inequality is saturated in simple examples where the boundary Cauchy slice defines a time-reversal symmetric state (for which the HRT surface for $\mA$ and $\mB$ lies within the corresponding extremal bulk surface).

Similarly, the subregion-CV2.0 conjecture yields superadditive results. This follows because the spacetime volume is always positive and further the intersection of the entanglement wedge and the WDW patch is a subregion within the WDW patch of $\sigma$. Hence it becomes evident that the mutual complexity \reef{ramen2A} will always be negative using this proposal. Let us emphasize that there are no obvious simple examples where the corresponding inequality would be saturated, \ie we {\it cannot} easily achieve $\Delta\mC_{V2.0}=0$, unless one of the subregions vanishes.

On the other hand, recall that the calculation of $\mathcal{C}_A$ in eq.~\reef{eq:ca} involves the length scale  $\ell_\text{ct}$ associated with the null boundary counterterm. Different values of this length scale result in  $\mathcal{C}_A$ being  subadditive or superadditive in different situations \cite{BrianMixedComplexity} --- see also \cite{Caceres:2018blh}. However, one should expect that the complexity, and hence the leading divergence, is positive, which partially fixes this ambiguity and further results in  $\mathcal{C}_A$ being superadditive.

{\bf Structure of divergences:}  All three proposals have a leading UV divergence proportional to the volume of the boundary subregion, \ie $V(\mA)/\delta^{d-1}$ but the subleading divergences are quite different. The subregion-CA conjecture yields subleading divergences with any power of $\delta$. In particular, in \cite{Carmi:2016wjl}, a class of subleading divergences associated with the boundary of the subregion were identified for the subregion-CA approach, \eg $V(\partial\mA)/\delta^{d-2}$.
Similarly, subleading divergences with any power of delta appear for subregion-CV2.0, as is easily inferred from the results of \cite{Carmi:2016wjl} and the relation in eq.~\reef{cv20main2}.
In contrast, it was shown that the subregion-CV approach yields power-law divergences involving only odd or even powers of the cutoff $\delta$ for an even- or odd-dimensional boundary theory, respectively.
Hence the $V(\partial\mA)/\delta^{d-2}$ term does not appear with the subregion-CV approach.

Before closing let us add that one could easily modify the three proposals in eqs.~\reef{eq:cv}, \reef{eq:ca} and \reef{eq:cv2} by including additional surface terms on the boundaries associated with the entanglement wedge. Because these bulk boundaries vanish when the subregion expands to fill the entire Cauchy slice on the holographic boundary, these surface contributions would disappear, and one would still recover the original proposal for holographic complexity of a pure state. For example, in the subregion-CV conjecture, one could add an extra term proportional to the volume of HRT surface $\mathcal E_A$ to produce the revised conjecture,
\begin{equation}\label{eq:cvX}
\mathcal{C}'_V(\mathcal{A})=\mathcal{C}_V(\mathcal{A}) +\eta\,\frac{\mathcal{V}(\mathcal{E_A})}{4 G_N}\,,
\end{equation}
where $\mathcal{C}_V(\mathcal{A})$ is the maximal volume expression in eq.~\reef{eq:cv} and $\eta$ is a (dimensionless) constant which remains to be determined. Our normalization of the second term makes clear that we are simply adding a term proportional to the entanglement entropy of the subregion $\mA$, \ie $\mathcal{C}'_V(\mathcal{A})=\mathcal{C}_V(\mathcal{A}) +\eta\,S_\text{EE}(\mA)$. With this revised proposal, the form of the UV divergences becomes closer to that found with the subregion-CA and the subregion-CV2.0 approaches, \ie new subleading divergences associated with the boundary of $\mA$ appear. Further, choosing a negative $\eta$ will ensure that the inequality in eq.~\reef{volineqholo} is never saturated with $\mathcal{C}'_V(\mathcal{A})$. On the other hand, if $\eta$ is chosen to be positive, this revised proposal \reef{eq:cvX} will typically be superadditive (because the mutual complexity will be dominated by the subleading divergence associated with the $S_\text{EE}(\mA)$ contribution). We reiterate that similar boundary terms could also be introduced to modify the subregion-CA and subregion-CV2.0 proposals, but the effect would be less important. We discuss this proposal \reef{eq:cvX} further in section \ref{holod}.

\subsection{Complexity of Thermal States}\label{holography_thermal}

Here, we apply these holographic prescriptions to evaluate the complexity of the thermal state, \ie where the subregion is taken to be one boundary of an (uncharged) eternal black hole, and to evaluate the mutual complexity of the corresponding thermofield double state. This system was already studied in \cite{BrianMixedComplexity} and we review their results here.\footnote{Note that our notation, \eg in eqs.~\reef{thermalCV} and \reef{eq:thermalCA}, is not identical to that in \cite{BrianMixedComplexity}, however, our results are in complete agreement with theirs. The only exception is that we have  accounted for a factor of 4 typo in the second term in eq.~(2.17) of \cite{BrianMixedComplexity}.} The holographic calculation is performed for  a two-sided AdS$_{d+1}$ black hole with the boundary dimension $d\geq2$ and with metric
 \begin{equation}\label{BH_metric}
 ds^2= \frac{L^2}{z^2} \( -f(z)\, dt^2 + \frac{dz^2}{f(z)} +  d\vec{x}^{\,2} \)\,,\qquad{\rm where}\ \   f(z)= 1-\(\frac{z}{z_0}\)^{d}\,.
 \end{equation}
Note that the boundary and horizon geometries are taken to be flat in this geometry. This eternal black hole in the bulk is dual to a thermofield double state in the boundary theory with temperature $T=\frac{d}{4\pi z_0}$.  As noted above, we choose the subregion to be a constant time slice on one of the boundaries and so the corresponding reduced state in the boundary theory is the thermal mixed state with the same temperature. With this choice, the HRT surface is simply the bifurcation surface on the horizon (which is reached with $z\to z_0$ holding $t$ fixed), and the entanglement wedge is simply the static patch outside of the horizon, \ie $z\ge z_0$.

{\bf Subregion-CV:} The result for subregion-CV \reef{eq:cv},
obtained in eq.~(2.16) of \cite{BrianMixedComplexity}, is
\begin{equation}\label{thermalCV}
\mathcal{C}_V(\mA) = \frac{L^{d-1}}{(d-1)G_N}\,\frac{L}{\ell_{\text{bulk}}}\,\frac{V(\mA)}{\delta^{d-1}} + {b}(d)\, \frac{L}{\ell_{\text{bulk}}}\, S \,,
\end{equation}
where $\ell_{\text{bulk}}$ is the extra length scale appearing in eq.~\eqref{eq:cv}, and  $V(\mathcal{A})$ is the spatial volume of the boundary theory. Further, ${b}(d)$ is a  positive dimension-dependent coefficient  given by
\begin{align}
{b}(d)=2\sqrt{\pi}\,\frac{d-2}{d-1}\,\frac{ \Gamma(\frac{d+1}{d})}{\Gamma(\frac{d+2}{2d})}\,.
\end{align}
Hence the finite term in eq.~\reef{thermalCV} is positive and proportional to $S=\frac{L^{d-1}}{4G_N\,z_0^{d-1}}\,V(\mB)$, the black hole entropy. Of course, $S$ can also be interpreted as the entropy of the thermal state in the boundary theory.

In the simplest situation where $t_\text{L}=t_{\text{R}}=0$,\footnote{Here, $t_\text{L}$ and $t_{\text{R}}$ denote the times on the left and right boundaries, respectively.} the mutual complexity \reef{ramen2A} vanishes, \ie
\begin{equation}\label{mutualCV}
\Delta \mC_{V} \equiv \mathcal{C}_V(\mL) + \mathcal{C}_V (\mathcal{R}) -\mathcal{C}_V (\mL\cup \mathcal{R})=0\,,
\end{equation}
because of the symmetry of the two-sided geometry.
Hence, in this case, the inequality \eqref{volineqholo} is exactly saturated. More generally, the same result arises if we choose $t_\text{L}+t_{\text{R}}=0$, which ensures that the full boundary state is still the TFD state without any additional time evolution. On the other hand, if we allow for some time evolution with $t_L,\,t_{R}>0$, then $\mathcal{C}_V(\mL)$ and $\mathcal{C}_V (\mathcal{R})$ remain invariant while $\mathcal{C}_V (\mL\cup \mathcal{R})$ increases. Therefore the mutual complexity becomes negative, and the complexity of the time-evolved TFD state is superadditive.

{\bf Subregion-CA:} The final result for subregion-CA \reef{eq:ca} is\footnote{Compare to  eq.~(2.14) of \cite{BrianMixedComplexity}.}
\begin{equation}\label{eq:thermalCA}
\mathcal{C}_A (\mA) = a(d)\,\frac{L^{d-1}}{16\pi^2 G_N} \,\frac{V(\mA)}{\delta^{d-1}}-\frac{a(d)+g_0}{4\pi^2}\, S
\end{equation}
where the constants, $a(d)$ and $g_0$, are given by
\beqa\label{eq:thermalCAhelp}
a(d)&=&4\, \ln\! \left[\frac{\ell_{\text{ct}}}{L} (d-1)\right]\,,
\nonumber\\
g_0&=& 2\left[\psi_0(1)-\psi_0\left(\frac{1}{d}\right) \right] \,,
\label{g0}
\eeqa
with $\psi_0(z)=\Gamma'(z)/\Gamma(z)$.
Note that $g_0$ is positive for $d>1$ (while, of course, it vanishes for $d=1$).
The constant $a(d)$ involves the scale $\ell_{\text{ct}}$ appearing in the boundary counterterm in the gravitational action \reef{actor} ---  see also eq.~\eqref{eq:ct}.
Note that we must choose that $\ell_{\text{ct}}>L/(d-1)$ to ensure that $a(d)$, and hence the complexity $\mC_A(\mB)$, is positive. Therefore, the finite contribution in eq.~\eqref{eq:thermalCA} is negative and proportional to the entropy of the thermal state.

Using the subregion-CA approach, the mutual complexity \eqref{sushi} for the TFD state with  $t_\text{L}=t_{\text{R}}=0$  becomes\footnote{Again, we may choose $t_\text{L}+t_{\text{R}}=0$ more generally. This result appears in eqs.~(2.7)-(2.8) of \cite{BrianMixedComplexity}.}
\begin{equation}\label{exs222}
\Delta \mC_{A} \equiv \mathcal{C}_A(\mL) + \mathcal{C}_A (\mathcal{R}) -\mathcal{C}_A (\mL\cup \mathcal{R}) =  - \frac{g_d}{2\pi^2}\,S
\end{equation}
where
\begin{equation}
g_d =  a(d) +g_0 +4\pi\,\frac{d-1}{d}\,.
\end{equation}
Since each of the terms contributing to $g_d$ is itself positive, the mutual complexity is negative and hence the complexity of the TFD state is superadditive. If we evolve the system forward in time  with $t_\text{L},\,t_{\text{R}}>0$, then $\mathcal{C}_A(\mL)$ and $\mathcal{C}_A (\mathcal{R})$ are again invariant while generally $\mathcal{C}_A (\mL\cup \mathcal{R})$ increases. A detailed analysis \cite{Carmi:2017jqz} shows that the complexity remains constant up to a critical time, at which point it briefly dips down slightly before beginning to grow linearly. We show in appendix \ref{app:notthesamelabel} that the mutual complexity will remain negative even in this short time period where $\mathcal{C}_A (\mL\cup \mathcal{R})$ decreases from its value at $t=0$ and therefore the complexity of the time-evolved TFD state is always superadditive as well.

{\bf Subregion-CV2.0:} It is easy to extract the results for the subregion-CV2.0 using eq.~\eqref{cv20main2}.  Some results for the bulk portion of the gravitational action appear in eqs.~(2.26), (B.10) and (B.16) of \cite{BrianMixedComplexity}. After accounting for the relevant proportionality factor, we obtain
\begin{equation}\label{CV20ref1}
C_{V2.0}(\mathcal A) = \frac{2\, V(\mathcal A) L^{d-1}}{d (d-1)G_N} \left(\frac{1}{\delta^{d-1}}-\frac{1}{z_0^{d-1}}\right),
\end{equation}
for the complexity of the thermal state, and
\begin{equation}\label{CV2.0entropy}
\begin{split}
\Delta C_{V2.0}& =-\frac{16 }{d}  \left(\frac1{d-1}+\frac{\pi}{d}\, \cot\frac{\pi}{d}\right)S\,,
\end{split}
\end{equation}
for the mutual complexity. This result for the mutual complexity is once again negative for $d\geq 2$, and this means that the complexity of the TFD state according to the subregion-CV2.0 proposal is again superadditive. We also note that, as with the other proposals, the mutual complexity is proportional to the entropy.

\subsection{Complexity of Vacuum Subregions } \label{holosub}

Below we summarize the results from all three approaches for a subregion of the CFT vacuum in two dimensions, \ie an interval in the boundary of AdS$_3$. These are the holographic results which are most relevant for the comparison with the QFT results in section \ref{apply02}. We also consider a disk-shaped subregion in the CFT vacuum in three dimensions, \ie on the boundary of AdS$_4$, to gain some intuition about the behaviour with an odd number of boundary dimensions. The general formulae for an arbitrary $d$ appear in appendix \ref{app:appsubCA4}.

{\bf Subregion-CV:}
With the subregion-CV approach for the case of AdS$_3$, both in global coordinates and in the Poincar\'e patch, we have
\begin{equation}\label{ads3subcv}
\text{AdS}_3,\text{G/P}\,:\quad\mathcal{C}_V(\mA) = \frac{2c}{3} \left( \frac{\ell}{\delta} -\pi \right)
\end{equation}
where $c=3L/(2\Gn)$ is the central charge of the two-dimensional boundary CFT \cite{Brown:1986nw}, $\ell$ is the size of the interval and $\delta$ is the UV cutoff. For global coordinates in AdS$_3$, this result comes from \cite{Abt:2017pmf}, and for the Poincar\'e patch, it was found in \cite{Alishahiha:2015rta}. The relevant formulae for the derivation in the Poincar\'e patch are summarized in appendix \ref{app:appsubCA4}, see eq.~\eqref{subCVamazing}. The constant term (\ie $-\pi$) is a topological term studied in \cite{Abt:2017pmf}.

For a ball-shaped subregion with radius $R$ on the boundary of AdS$_{d+1}$, the calculation of $\mathcal{C}_V$ is outlined in eqs.~(5) and (7) of \cite{Alishahiha:2015rta} --- see also eq.~(4.9) of \cite{Carmi:2016wjl} and our eq.~\eqref{subCVamazing}. For example, for the case of a disk on the boundary of $\text{AdS}_4$, one obtains
\begin{equation}\label{ads4subcv}
\text{AdS}_4,\text{P}\,:\quad\mathcal{C}_V(\mA)=\frac{\pi^4 c_T}{3}\left(\frac{  R^2}{2 \delta ^2}-\ln\! \left(\frac{R}{\delta} \right)-\frac{1}{2}\right)
\end{equation}
where $c_T=3 L^2/(\pi^3 \Gn)$ is the central charge appearing in the OPE of two stress tensors in the boundary theory, \eg see \cite{Buchel:2009sk}.

{\bf Subregion-CA:}
Next, we turn to the subregion-CA results.
For the case of a flat boundary (in the Poincar\'e patch), the divergence structure of the subregion complexity in vacuum AdS was studied in \cite{Carmi:2016wjl}.
However, these results did not include the boundary counterterms $I_{\text{ct}}$, which restores the reparametrization invariance on the null surfaces. We evaluate the contribution of $I_{\text{ct}}$ in our calculations in appendix \ref{app:appsubCA4}. We have also corrected a number of typos in the original calculation of \cite{Carmi:2016wjl}, and explicitly demonstrated the cancellation of the normalization constants of the null normals. Combining eqs.~\eqref{Ibulksubapp}, \eqref{Isjsubapp} and \eqref{CACAtotapp} for the case of AdS$_3$ yields
\begin{align}\label{3P}
 \text{AdS}_3,\text{P}\,:\quad   \mathcal{C}_A(\mA) &= \frac{c}{3 \pi^2}\Bigg( \frac{\ell}{2\delta}\ln \left(\frac{\ell_{\text{ct}}}{L} \right)
     -  \ln \left(\frac{2 \ell_{\text{ct}}}{L}\right) \ln \left( \frac{\ell}{\delta}\right) + \frac{\pi^2}{8}\Bigg) \, ,
\end{align}
where $\ell$ is again the size of the boundary interval. Further, we note that the UV divergences were regulated in the above calculation by anchoring the WDW patch at the UV cutoff surface. Repeating these calculations in global coordinates \cite{Chapman:2018bqj,future1}, we find\footnote{We note that this result can be obtained either by anchoring the WDW patch at the cutoff surface, or by anchoring it at the boundary of AdS$_3$ (as in \cite{Chapman:2018bqj}) but adding the usual counterterms of the kind often used in holographic renormalization (\eg see \cite{Emparan:1999pm}) on the cutoff surface. We return to the idea of adding holographic counterterms in regulating holographic complexity in \cite{future1}.}
\begin{equation}\label{toderive}
\text{AdS}_3,\text{G}\,:\quad\mathcal{C}_{A}(\mA) =
{c\over  3\pi^2}\left( \frac{\ell}{2 \delta}\, \ln\! \left( {\ell_{\text{ct}}  \over L} \right)  - \ln\!\left(\frac{2\ell_{\text{ct}}}{L}\right) \ln\!\left( \frac{C}{\delta}\right) \right)+f(\ell/C)\,,
\end{equation}
where $C$ is the circumference of a time slice on the boundary. Here, $f(\ell/C)$ is some finite contribution, whose precise form we did not determine analytically. However, we do know that in the limit $\ell/C\to 0$, eq.~\reef{toderive} should reduce to the previous expression in eq.~\reef{3P} and hence
\beq\label{lim0}
\frac{\ell}{C}\ll1\ :\quad
f(\ell/C)\simeq {c\over  3\pi^2}\(
 \ln\!\left(\frac{2\ell_{\text{ct}}}{L}\right) \ln\!\left( \frac{C}{\ell}\right) +
\frac{\pi^2}{8}\) + {\cal O}(\ell/C)\,.
\eeq
We return to examine this finite part in more detail in section \ref{holod}.

For a disk-shaped region (of radius $R$) on the boundary of AdS$_4$ using Poincar\'e coordinates, we obtain
\begin{align}\label{4P}
\hspace{-0.4cm}   \text{AdS}_4,\text{P}:~\mathcal{C}_A(\mA)&=
\frac{\pi^2 c_T}{12}\Bigg( \frac{ R^2}{\delta^2}\ln\! \left(\frac{2 \ell_{\text{ct}}}{L} \right)
     - \frac{2 R}{\delta}  \ln\! \left(\frac{4 \ell_{\text{ct}}}{L} \right)
     + 2\ln\frac{R}{\delta} + \ln\! \left(\frac{ \ell_{\text{ct}}}{2L} \right)\Bigg) .
\end{align}
This calculation can also be seen as the smooth limit of the result obtained in \cite{Caceres:2018luq} for subregions with kinks/corners, \ie compare with eq.~(5.8) of \cite{Caceres:2018luq}.

{\bf Subregion-CV2.0:} Again, it is straightforward to extract the results for the subregion-CV2.0 proposal using eq.~\eqref{cv20main2}. We have the results for the bulk portion of the gravitational action in eq.~\eqref{Ibulksubapp} for AdS$_3$ in  Poincar\'e coordinates (\ie $d=2$) and so after accounting for the relevant proportionality factor we obtain
\begin{equation}
\label{CV203P}
\text{AdS}_3,\text{P}\,:\quad \mC_{V2.0}(\mA) = \frac{4\,c}{3}\left(\frac{\ell}{2\delta}-\ln\frac{\ell}{\delta}-\frac{\pi^2}{8}\right)\,.
\end{equation}
Further the analogous result for AdS$_3$ in global coordinates \cite{Chapman:2018bqj,future1},\footnote{This result was obtained by anchoring the WDW patch at the cutoff surface, as we will describe in \cite{future1}. The result for another regularization scheme where the WDW patch is anchored at the boundary of AdS$_3$ can be read from eq.~(B.18) of \cite{Chapman:2018bqj}
\begin{equation}\label{anotherreg}
\text{AdS}_3,\text{G}\,:\quad\mC_{V2.0}(\mA) =\frac{4}{3}c \left(\frac{\ell}{\delta}-\ln \frac{\ell}{\delta}+\text{finite}\right),
\end{equation}
where we notice that the leading divergence has changed by a factor of 2, however, the universal logarithmic piece remains unchanged.}
\begin{equation}
\label{CV203G}
\text{AdS}_3,\text{G}\,:\quad\mC_{V2.0}(\mA) = \frac{4\,c}{3} \left(\frac{\ell}{2\delta}-\ln \frac{C}{\delta}\right)+\tilde f(\ell/C)\,,
\end{equation}
where $C$ is again the circumference of a time slice on the boundary and $\tilde f(\ell/C)$ is a finite contribution. We return to examine this contribution in more detail in section \ref{holod}. However, let us observe here that in the limit $\ell/C\to 0$, eq.~\reef{CV203G} must reduce to the previous expression in eq.~\reef{CV203P} and hence we expect to find
\begin{equation}\label{lim1}
\frac{\ell}{C}\ll1\ :\quad
\tilde f(\ell/C)\simeq \frac{4\,c}{3}
\(   \ln\!\left( \frac{C}{\ell}\right) -
\frac{\pi^2}{8}\) + {\cal O}(\ell/C)\,.
\end{equation}

We can also use eq.~\eqref{Ibulksubapp} to evaluate the complexity for a disk-shaped region on the boundary of AdS$_4$ in  Poincar\'e coordinates,
\begin{equation}
\label{CV204P}
\text{AdS}_4,\text{P}\,:\quad \mC_{V2.0}(\mA)  = \frac{\pi^4 c_T}{9}\left(\frac{R^2}{\delta^2}-\frac{2R}{\delta}-4 \ln \frac{R}{4 \delta}+1\right)\,.
\end{equation}

With all three proposals, the leading divergence is proportional to the volume of the boundary region $V(\mA)$, \ie $V(\mA)=\ell$ with $d=2$ while $V(\mA)=\pi R^2$ with $d=3$. However,  the subleading divergences are quite different for subregion-CV compared to subregion-CA and subregion-CV2.0. With either of the latter two, the subleading contribution is a negative term proportional to the area of the boundary of $\mA$, \eg  $V(\partial\mA)=2\pi R$ with $d=3$. In contrast, no comparable contribution appears in the subregion-CV results. Similar boundary contributions with a negative sign were found in \cite{Caceres:2018luq} for subregion-CA.  Such subleading divergences appear to be a generic feature of both the subregion-CA and subregion-CV2.0 approaches, and can be understood as a contribution to the complexity proportional to the entanglement entropy \cite{AgonCX} -- see also the discussion around eq.~\reef{eq:cvX}.

{\bf Mutual Complexity:} 
Now we can use the previous results together with the results for the complexity of the full boundary time slice to evaluate the mutual complexity. The first observation is that in our examples here, we are considering the vacuum state and subregions of the vacuum for the boundary CFT on a constant time slice. Hence for the CV and subregion-CV proposals, the maximal volume slices also all lie in the constant time slice in the bulk. Hence the two bulk volumes corresponding to a subregion and its complement precisely add up to equal the volume for the full vacuum state. That is, we are in a situation where we saturate the inequality in eq.~\reef{volineqholo} and the mutual complexity vanishes.\footnote{As for the previous discussion of the CV proposal for the TFD state with $t_L=0=t_R$.} Of course, if we choose to examine the vacuum state on a more general Cauchy slice in the boundary, we expect the mutual complexity to be negative, \ie the complexity would be superadditive. It would be interesting to understand the precise form of $\Delta\mC_V$ in these situations.

The results are more interesting for the CA and CV2.0 proposals. Here we will focus our discussion on the case of a flat boundary, \ie with Poincar\'e coordinates in the bulk, since they are easily generalized to higher dimensions. We illustrate the discussion with the example of AdS$_4$, where we begin by evaluating the complexity of the full vacuum state, using eqs.~\eqref{eq:Cafullappd} and \eqref{eq:Ibulkfs},
\beqa \label{D4P}
\text{AdS}_4,\text{P}\,:\qquad\mathcal{C}_A(\text{vac})&=&
\frac{\pi\, c_T}{12}\frac{V(\Sigma)}{\delta^2}\,\ln\! \frac{2 \ell_{\text{ct}}}{L}
     \,,
\nonumber\\
\text{AdS}_4,\text{P}\,:\quad \mC_{V2.0}(\text{vac})  &=& \frac{\pi^3 c_T}{9}\frac{V(\Sigma)}{\delta^2}\,,
\label{full4P}
\eeqa
where $V(\Sigma)$ is the spatial volume of the entire time slice in the boundary.\footnote{In fact, the time slice is two-dimensional and so $V(\Sigma)$ is an area in this specific example.} Next, we gave the results for a disk-shaped region in eqs.~\reef{4P} and \reef{CV204P} for the subregion-CA and subregion-CV2.0, respectively, which we re-express here as
\beqa
\text{AdS}_4,\text{P}\,:\qquad\mathcal{C}_A(\mA)&=&
\frac{\pi\, c_T}{12}\left( \frac{ {V}(\mA)}{\delta^2}\,\ln\! \left(\frac{2 \ell_{\text{ct}}}{L} \right)  - \frac{{V}(\partial\mA)}{\delta}  \ln \left(\frac{4 \ell_{\text{ct}}}{L} \right)
     + 2\pi\,\ln\frac{L}{\delta} + \text{finite}\, \right)\,,
\nonumber\\
\text{AdS}_4,\text{P}\,:\quad \mC_{V2.0}(\mA)  &=& \frac{\pi^3 c_T}{9}\left(\frac{{V}(\mA)}{\delta^2}-\frac{{V}(\partial\mA)}{\delta}-4\pi\, \ln \frac{L}{\delta}+\text{finite}\, \right)\,,
\label{A4P}
\eeqa
where $V(\mA)=\pi R^2$ is the area of the disk and $V(\partial\mA)=2\pi R$ is the circumference of the boundary of the disk.
This leaves us to evaluate the complexity of the exterior of the disk, which we denote $\mB$. While this calculation may seem more formidable because $\mB$ has an infinite extent in this flat boundary geometry, the geometric interpretation of the two leading singularities would be precisely as in eq.~\reef{A4P}. Further, we would have
$V(\mA)+V(\mB)=V(\Sigma)$ and $V(\partial\mA)=V(\partial\mB)$ and hence the mutual complexity becomes
\beqa
\text{AdS}_4,\text{P}\,:\qquad\Delta\mathcal{C}_A&=&- \frac{\pi\, c_T}{6}\,\ln \left(\frac{4 \ell_{\text{ct}}}{L} \right)  \, \frac{{V}(\partial\mA)}{\delta}
     + \cdots \,,
\nonumber\\
\text{AdS}_4,\text{P}\,:\quad \Delta\mC_{V2.0} &=&- \frac{2\pi^3 c_T}{9}\,\frac{{V}(\partial\mA)}{\delta}+\cdots\,.
\label{mut4P}
\eeqa
In fact, this result can be extended to any (smooth) bipartition of the two-dimensional time slice in the boundary theory, and $V(\partial\mA)$ will denote the length of the boundary between the subregion $\mA$ and its complement $\mB$. Given the sign of the results above, we see that the complexity of the vacuum is superadditive for both the subregion-CA and subregion-CV2.0 approaches. We might also note that the leading singularity in eq.~\reef{mut4P} has the same form as that in the entanglement entropy for the same bipartition. Hence, at least to leading order here, the mutual complexity is again proportional to the entanglement entropy between the two subregions.

Using the results of appendix \ref{app:appsubCA4} and of \cite{Carmi:2016wjl}, these calculations are easily extended to higher dimensions, where we find for $d>2$
\beqa \label{deltaC-subregion}
\text{AdS}_{d+1},\text{P}\,:\qquad\Delta\mathcal{C}_A&=&-\frac{L^{d-1}}{2\pi^2 (d-2) G_N}\, \ln\! \left(\frac{2(d-1)\ell_{\text{ct}}}{L}\right)\,\frac{{V}(\partial\mA)}{\delta^{d-2}}+\cdots\,,
\nonumber\\
\text{AdS}_{d+1},\text{P}\,:\quad \Delta\mC_{V2.0} &=&- \frac{4L^{d-1}}{ d(d-1)(d-2)G_N }\,\frac{{V}(\partial\mA)}{\delta^{d-2}}+\cdots\,.
\label{mutdP}
\eeqa
Of course, using our previous results for subregions on the boundary of AdS$_3$, these calculations are easily extended to $d=2$. In this case, we find that the mutual complexity becomes
\begin{equation}
\begin{split}
\text{AdS}_3,\text{P}\,:\qquad\Delta\mathcal{C}_A=&\,-\frac{2\,c}{3\pi^2 }\, \ln\! \left(\frac{2\ell_{\text{ct}}}{L}\right)\,\ln\frac{\ell}{\delta}+\cdots\,,\\
\text{AdS}_3,\text{P}\,:\quad \Delta\mC_{V2.0}=&\,- \frac{8\,c}{ 3}\,\ln\frac{\ell}{\delta}+\cdots\,.
\label{holomutualcomp}
\end{split}
\end{equation}
Hence these general results again show that the mutual complexity is negative and hence that the complexity of the vacuum state is superadditive. We may also note that to leading order, the mutual complexity is proportional to the entanglement entropy of the subregions.

\section{Discussion}\label{sec:disc}

In this paper, we focused on the mixed-state complexity of Gaussian mixed states. Our approach focused on a definition dubbed the {\bf purification complexity} in \cite{BrianMixedComplexity}. That is, we considered the minimal complexity amongst the possible pure states which purify the desired mixed state $\hat\rho_\mA$. Let us point out, however, that our perspective differs slightly from that of the authors of \cite{BrianMixedComplexity} in that the latter only consider essential purifications. The reason for this restriction was that they wanted the definition to collapse to the usual pure state complexity definition when the target state is pure. In section \ref{sec:manyho}, we found that essential purifications are actually optimal, at least for the Gaussian mixed states considered there, and as a consequence this assumption becomes redundant. It would be interesting to explore whether including extra auxiliary degrees of freedom which appear in a simple tensor product in the final pure state could actually reduce the complexity of mixed states (or pure states) in more complicated situations.\footnote{We are reminded here of coherent (pure) states \cite{coherent}, where, in simple examples, the reference and target states had a tensor product structure which was not respected by the intermediate states.}

We might add that the purification complexity discussed here and in \cite{BrianMixedComplexity} is closely aligned with the standard approach developed in quantum information theory, \eg \cite{watrous2009quantum,Aharonov:1998zf}. However, in this setting, the auxiliary degrees of freedom are regarded as another resource required for the preparation of the desired mixed states, and hence an additional cost is associated with adding more ancillae. This cost was not considered in our analysis nor in \cite{BrianMixedComplexity}. This would be another feature that would favour essential purifications as the optimal purifications. For example in section \ref{threeA}, where we found the same complexity using either one or two ancillae, the essential purification with one ancilla would clearly become the optimal one if we added an extra penalty for each ancilla that is introduced. Still, it would be interesting to investigate whether this simple result extends to, e.g., the case of interacting quantum field theories.

\subsection{Other Proposals for Mixed State Complexity}
Before proceeding with a further discussion of our results, we would first like to briefly review the other proposals for mixed-state complexity made in \cite{BrianMixedComplexity} and possible connections to our current work:

{\bf Spectrum and Basis Complexity:} One alternative \cite{BrianMixedComplexity} is to break the problem of preparing mixed states into two parts --- creating the spectrum and creating the basis of eigenvectors.  The spectrum complexity $\mC_S$ is defined as the minimal purification complexity of some mixed state $\hat\rho_{\text{spec}}$ which has the same spectrum as $\hat\rho_\mA$, where one also optimizes over the possible $\hat\rho_{\text{spec}}$. Since one possible $\hat\rho_{\text{spec}}$ with the required spectrum is simply $\hat\rho_\mA$, we conclude that $\mC_S \le \mC_P$, where $\mC_P$ denotes the purification complexity of $\hat\rho$. In our analysis, the spectrum is defined by the eigenvalues of the matrix $B$ in eq.~\eqref{matrix_B}.

The basis complexity can be defined in different ways: The first suggestion in \cite{BrianMixedComplexity} is simply the difference $\mC_P-\mC_S$. The second suggestion is to define ${\mC}_{B}$ as the complexity (\ie minimal number of unitary gates) required to go from the optimal $\hat{\rho}_{\text{spec}}$ to our target state $\hat{\rho}$. The latter preparation can be made with unitary gates because the two mixed states share the same spectrum. We can easily demonstrate $\mC_P\leq \mC_S+ \mC_B$ since on the left-hand side, the preparation is constrained to pass through the intermediate state $\hat\rho_{\text{spec}}$.

Our construction using the physical basis seems closely related to this approach. To modify the spectrum, one must use ``mixed" entangling gates acting between $\cal A$ and ${\cal A}^c$, and so these would appear in the circuit preparing (the purification of) $\hat\rho_{\text{spec}}$. The gates acting only on the $\cal A$ degrees of freedom are modifying the basis, and the circuit preparing $\hat\rho_\mA$ from $\hat\rho_{\text{spec}}$ is comprised solely of these gates. However, it seems that there is no natural role for the gates acting only on $\mA^c$. In this framework then, not using these gates may be the reason for the difference in the complexities, \ie $\mC_P\leq \mC_S+  {\mC}_{B}$. Let us also note that both the spectrum complexity and the entanglement entropy are both insensitive to the action of the gates acting only on ${\cal A}$ or only on ${\cal A}^c$. Only the ${\cal A}{\cal A}^c$ entangling gates change these quantities. For example, considering two mixed states of a single harmonic oscillator with the same entanglement entropies, this implies that the spectrum complexities must also be equal. It would be interesting to understand to what extent this property generalizes to states over many degrees of freedom, \eg the thermal state of a free scalar, studied in section \ref{apply01}. We will explore some of the issues above in the future work \cite{wip99}.

{\bf Open System Complexity:} Open system complexity studies the complexity of circuits which move through the space of density matrices using general CPTP maps, rather than only unitary transformations. This requires characterizing these general maps in terms of elementary operations and then assigning a cost to the latter. Of course, as discussed in the introduction, the dilation theorems \cite{dilaton} imply that the most general CPTP maps acting on a system of qubits can be realized as unitary evolution of the system coupled to ancillary qubits \cite{watrous2009quantum}, which seems to bring this approach back to the framework used for the purification complexity. However, one potential difference for the open system complexity is that some of the ancillae may be introduced and traced out, \ie they are re-initialized, at every step. This would contrast with having a single reservoir of ancillae on which we can repeatedly act before tracing them at the very end of the unitary evolution, as described for the purification complexity.

{\bf Ensemble Complexity:} The ensemble complexity is defined using a decomposition of the mixed state over an ensemble of pure states as follows
    \begin{equation}
    \mC_E=\min_{\{p_i, |\psi_i\rangle\} } \sum_i p_i \, \mC(|\psi_i\rangle )\,,\qquad{\rm where}\quad
    \hat \rho = \sum_i p_i |\psi_i\rangle \langle \psi_i|\,.
    \end{equation}
Of course, this notion reduces to the pure state complexity when the state $\hat \rho_\mA$ is pure. Even with a Gaussian mixed state $\hat \rho_\mA$, we would generally have to explore ensembles which are not constructed solely from Gaussian states. In the case of the thermal state, a decomposition is available in terms of coherent states and this allows to put a bound on the ensemble complexity of thermal states ---  see section 3.5 of \cite{BrianMixedComplexity} for further details.

\subsection{Mutual Complexity in QFT}

In section \ref{compare7}, we considered beginning with the pure state $\ket{\Psi_{\mA\mB}}$, and then constructed the two reduced density matrices, $\hat\rho_\mA$ and $\hat\rho_\mB$. Then in eq.~\reef{ramen2A}, the mutual complexity was defined as the combination \cite{Ali:2018lfv},
\begin{equation}\label{ramen2Ax}
\Delta\mC =  \mC\!\(\hat{\rho}_\mA\) + \mC\!\(\hat{\rho}_\mB\) - \mC\!\( \ket{\Psi_{\mA\mB}}\)   \,,
\end{equation}
which quantifies the additional correlations between the subsystems $\mA$ and $\mB$.

Our first application of this quantity was to compare the complexity of the TFD state with the purification complexity of the thermal mixed state produced by tracing out either the left or the right degrees of freedom, \eg see eq.~\reef{ramen2}. As a warm-up exercise, we evaluated the mutual complexity for a two-mode TFD state and as shown in eq.~\reef{hunt1}, we found $\Delta\mC_{1}^{\mt{diag}}\(\ket{\text{TFD}}_{12}\) > 0$. More generally, we might evaluate the mutual complexity for general two-mode pure Gaussian states $\ket{\Psi}_{12}$. That is, integrating out each of the degrees of freedom in term yields two distinct mixed states, $\hat{\rho}_1$ and $\hat{\rho}_2$, and so one might compare the purification complexity of these two mixed states with that of the parent pure state, with the analogous expression to that in eq.~\reef{ramen2Ax}. In fact, using the results for the purification complexity of one-mode Gaussian states in eq.~\eqref{complexity_one_mode}, it is straightforward to show that subadditivity always holds for any two-mode pure Gaussian state, \ie
\beq\label{astorm}
    \Delta\mC_{1}^{\mt{diag}}\(\ket{\Psi}_{12}\)= \mC^{\text{diag}}_{1}\(\hat{\rho}_1\)+\mC^{\text{diag}}_{1}\(\hat{\rho}_2\) - \mC_{1}^{\text{diag}} \(\ket{\Psi}_{12}\) \ge 0\,.
\eeq
However, this inequality does not extend to the purification complexity calculated in the physical basis, as in section \ref{fizz}.     It would be interesting to investigate whether the above inequality can be made more restrictive, \eg where the mutual complexity is greater than some finite bound proportional to the entanglement entropy.

Since in section \ref{apply01}, the TFD state has a simple product structure for the free scalar field theory, the mutual complexity becomes simply a sum over the same quantity evaluated for each of the individual modes --- see eqs.~\reef{rocket0} and \reef{integral_phys}. Hence the positivity appearing in eq.~\reef{hunt1} for the two-mode TFD states in the diagonal basis extends to the TFD state of the full scalar QFT. That is, $\Delta \mC_{1}^{\text{diag}} \(\ket{\text{TFD}}\)>0$ irrespective of the values of the temperature, reference frequency or the mass of the scalar.

This positivity is not replicated for the mutual complexity when it is evaluated using the physical basis, as shown in figure \ref{deltaC_phys}. There we showed that for a massless two-dimensional scalar, $\Delta \mC_{1}^{\text{phys}} \(\ket{\text{TFD}}\)$ becomes negative when the reference frequency $\mu$ is much smaller or much larger than the temperature.

In section \ref{compare7} we found that with $\mu\gg T$, the mutual complexity of the TFD state is proportional to entanglement entropy between the left and right copies of the field theory. However, in general, there would be an overall proportionality constant which contains a temperature dependence through the (dimensionless) ratio $T/\mu$, as well as $T/m$ for a massive scalar. This behaviour is easily seen analytically in the diagonal basis using eqs.~\reef{gaga66} and \reef{gaga69}, but similar results also apply in the physical basis, see comments at the end of section \ref{sec:mcipb}. In any event, the appearance of the entanglement entropy in the regime $\mu\gg T$ reinforces the intuition that the mutual complexity in eq.~\reef{ramen2Ax} quantifies the correlations between the subsystems to which the pure state is reduced.

Before turning to subregions, let us briefly comment again that $\Delta \mC$ is UV finite for the TFD state. For the free scalar, we found that the leading UV divergence in the purification complexity of the thermal mixed state is the same for either the diagonal or physical basis, as determined in eqs.~\eqref{dog35}-\eqref{dog5} or eq.~\eqref{laugh88}, respectively. The precise form of this leading divergence can be found as
\begin{equation}\label{holorocket}
\mC(\hat\rho_{th}(\beta)) \simeq
\begin{cases}
    \frac{\Omega_{d-2} V_{d-1}}{2\,(2\pi)^{d-1} (d-1)}\,\Lambda^{d-1} \( \ln \frac{\mu}{\Lambda}  + \frac{1}{d-1} \) \,, & \mu \ge \Lambda\,,\\
    \frac{\Omega_{d-2} V_{d-1}}{2\,(2\pi)^{d-1} (d-1)}\,\Lambda^{d-1} \( \ln \frac{\Lambda}{\mu} +  \frac{2}{d-1} \( \frac{\mu}{\Lambda}\)^{d-1}  - \frac{1}{d-1} \)  \,,&\mu \le \Lambda \,.
\end{cases}
\end{equation}
Exactly, the same divergences also appear in the complexity of the vacuum state of the scalar field theory, \eg see appendix B of \cite{qft2}. These divergences are also exactly one-half of those found for the TFD state, and hence the subtraction in eq.~\reef{ramen2} yields $\Delta\mC\(\ket{\text{TFD}}\)$ which is UV finite (in either basis).
More precisely, all of the potentially divergent contributions cancel in the integrand of eq.~\reef{delta_diag} for the diagonal basis and of eq.~\reef{integral_phys} for the physical basis, and so all of the UV divergences cancel in the corresponding mutual complexities.

Of course, this UV finiteness is directly related to the fact that optimal purification of the thermal state $\hat\rho_{th}(\beta)$ is not the TFD state. Much of the preparation of the TFD state involves introducing short-distance correlations in {\it both} copies of the field theory. Even though the optimal purification of $\hat\rho_{th}(\beta)$ involves introducing a number of auxiliary degrees of freedom that is equivalent to introducing a second copy of the QFT, there is no need to prepare the purification with UV correlations amongst the ancillae since after they are integrated out, these will not affect the physical correlations of the thermal mixed state.\footnote{Similar comments appear in \cite{BrianMixedComplexity} using the basis and spectrum language, \ie preparing the TFD state  requires many gates which adjust the basis of the purifying system but which do not affect the mixed thermal state of the original system.} This is why the UV divergences in $\mC(\hat\rho_{th}(\beta))$ carry exactly a factor of one-half compared to $\mC\(\ket{\text{TFD}}\)$.\footnote{Given the optimal purification of $\hat\rho_{th}(\beta)$, it may be interesting to investigate the properties of $\rho_{\mA^c}$, \ie the mixed state found after tracing out the physical degrees of freedom. For example, one should find that it is much less entangled at short distances.}

In section \ref{apply02}, we considered the purification complexity of subregions of the vacuum. In this case, both the vacuum state and the mixed states produced by reducing to a subregion can again be written in a product form. However, the basis of states appearing in these products is not the same, \ie for the vacuum, we use momentum eigenstates (which are eigenstates of the Hamiltonian), while for the subregions, we use eigenstates of the corresponding modular Hamiltonian. Hence we can no longer apply eq.~\reef{astorm} to determine the sign of the mutual complexity of the vacuum divided into two complementary subregions, $\mA$ and $\mB$. However, we found that $\Delta \mC_1^{\rm diag}$ is still positive in the diagonal basis, as illustrated in figure \ref{fig:mutualcomp}. In the physical basis, we gave two definitions of the mutual complexity in eq.~\eqref{twoDeltaCs}, which differ by the basis in which the ground state complexity is evaluated. Our analysis indicates that $\Delta\mC_1^{\rm phys}$ is generally negative, while $\Delta\tilde{\mC}_1^{\rm phys}$ is positive, as illustrated in figure \ref{fig:subregion2-phys}. The sign difference between these two definitions is due to the vacuum complexity being much larger in $\Delta\mC_1^{\rm phys}$ than in $\Delta\tilde{\mC}_1^{\rm phys}$. The cutoff dependence of $\Delta \mC^{\rm phys}_1$ is related to the subleading divergences of the subregion complexities and the ground state complexity, which are all logarithmic. On the other hand, the cutoff dependence of $\Delta\tilde{\mC}_1^{\rm phys}$ is dominated by the subleading divergence of $\mC^{\cal AB}_1(|\Psi_0\rangle)$, which is linear in the cutoff.

At this point, let us note that for subregions of the vacuum, it is again the case that the original state, \ie the vacuum state, does not provide the optimal purification. If the vacuum was the optimal purification, then the subregion complexity would simply match the complexity of the ground state. As a result, the leading divergence of all of the subregion complexities would be $\mC\sim V(\Sigma)/\delta^{d-1}$ (where $V(\Sigma)$ is the volume of the global time slice) and the corresponding mutual complexity would also exhibit a volume-law divergence. Instead as shown in eqs.~\reef{Theamazingsubregionequation} and \reef{divergences2}, the leading divergences are instead proportional to $V(\mA)$, the volume of the subregion, and as discussed above, the mutual complexity is then controlled by the subleading divergences appearing in the individual complexities. Again, this reflects the fact that in the optimal purification, there is no need to prepare UV correlations amongst the ancillae. 
Moreover, we might note that the ground state would not even be an essential purification (with the minimal number of ancilla) for subsystems whose size is less than half of that of the full system.

We turn to the comparison of the mutual complexity from our QFT and our holographic calculations in the next subsection. However, before closing here, let us note that there is no reason why in calculating the mutual complexity, the initial state must be a pure state. That is, a simple generalization of eq.~\reef{ramen2Ax} would be
\begin{equation}\label{ramen2Ay}
\Delta\mC =  \mC\!\(\hat{\rho}_\mA\) + \mC\!\(\hat{\rho}_\mB\) - \mC\!\( \hat{\rho}_{\mA\cup\mB}\)   \,,
\end{equation}
where the combined system begins in a mixed state $\hat{\rho}_{\mA\cup\mB}$. We still expect that in this situation the mutual complexity \reef{ramen2Ay} quantifies the additional correlations between the subsystems $\mA$ and $\mB$.  Using our results, a simple example would be to consider two neighbouring (but not overlapping) subregions, $\mA$ and $\mB$, in the vacuum state. These combine to form the larger subregion $\mA\cup\mB$ (but note that we assume $\mA\cap\mB=0$). Building on eq.~\reef{deltaC-subregion} in the holographic context, we would find that the leading contribution to the mutual complexity becomes
\beqa
\text{AdS}_{d+1},\text{P}\,:\qquad\Delta\mathcal{C}_A&=&-\frac{L^{d-1}}{2\pi^2 (d-2) G_N}\, \ln\! \left(\frac{2(d-1)\ell_{\text{ct}}}{L}\right)\,\frac{{V}(\partial\mA\cap\partial\mB)}{\delta^{d-2}}+\cdots\,,
\nonumber\\
\text{AdS}_{d+1},\text{P}\,:\quad \Delta\mC_{V2.0} &=&- \frac{4L^{d-1}}{ d(d-1)(d-2)G_N }\,\frac{{V}(\partial\mA\cap\partial\mB)}{\delta^{d-2}}+\cdots\,.
\label{mutdPmix}
\eeqa
In this case, we observe that this leading divergence is comparable to that in the mutual information between the subregions $\mA$ and $\mB$. Of course, this suggests that in general one should think of the mutual complexity as being related to mutual information, rather than the entanglement entropy even when $\hat{\rho}_{\mA\mB}$ is a pure state. It would be interesting to investigate this generalization \reef{ramen2Ay} further in the case of disjoint (\ie non-neighbouring) subregions $\mA$ and $\mB$, where the mutual information is finite, and exhibits an interesting phase transition for holographic CFTs \cite{Headrick:2010zt,Hartman:2013mia,Faulkner:2013yia}. A similar setup studying purifications of two complementary subregions appears also in the context of the entanglement of purification \cite{pure0,pure1,pure2,terhal2002entanglement}. It would be interesting to investigate the relation between these two notions.

Further, we observe that the mutual complexity \reef{ramen2Ay} for mixed states would generally be nonvanishing (but UV finite) using the subregion-CV approach \reef{eq:cv}, even if the subregions lie in a constant time slice on the boundary. Another interesting issue to investigate would be if inequalities similar to the Araki-Lieb inequality \cite{Araki:1970ba} can be used to bound the difference in complexity between two complementary subsystems when starting with a mixed state.
Finally, to close here, let us comment on the case of partially overlapping subregions. In this case, one is naturally lead to consider the following generalization of the mutual complexity
\begin{equation}\label{ramen2Aaa}
\Delta\mC =  \mC\!\(\hat{\rho}_\mA\) + \mC\!\(\hat{\rho}_\mB\) - \mC\!\( \hat{\rho}_{\mA\cup\mB}\) - \mC\!\( \hat{\rho}_{\mA\cap\mB}\)   \,.
\end{equation}
With this difference of complexities, the leading divergences in the individual complexities cancel, and the sign of the result is nontrivial. It would be interesting to investigate the properties of this generalization further.

\subsection{Holographic Complexity} \label{holod}

Much of the motivation of our paper was to compare the results for the purification complexity in the free scalar QFT to those for the mixed state complexity found in holography. Hence we now compare the QFT results of sections \ref{apply01} and \ref{apply02} for the purification complexity of thermal states and subregions in the vacuum state to the analogous results found with the subregion-CV \eqref{eq:cv}, subregion-CA \eqref{eq:ca}, and subregion-CV2.0 \eqref{eq:cv2} prescriptions found in section \ref{sec:holo}. Recall that motivated by previous comparisons, we focused our analysis of the complexity in the QFT on the $F_1$ cost function \reef{func2}. For example, the structure of the UV divergences for the $\mC_1$ complexity in QFT was found to be similar to that for holographic complexity \cite{qft1,qft2}. However, the basis dependence of this measure was found to play an important role in evaluating the complexity of TFD states \cite{Chapman:2018hou}, and so we also evaluated our QFT complexities in both the diagonal and physical bases here.
One more observation, before turning to the results, is that the authors of \cite{NonLocal} have argued that the relevant gates in holographic complexity should be non-local. Of course, the original analysis of the QFT complexity \cite{qft1}, which we adapt here in our analysis, also involves non-local gates. Hence this is a common point for the complexity in both frameworks.

The leading UV divergence in any of the holographic prescriptions for the complexity of the reduced state on a subregion has the same volume-law form as found for a pure state. That is, all three prescriptions yield an expression of the form $\mC\simeq k_d\, V(\mA)/\delta^{d-1}+\cdots$ where $V(\mA)$ is the volume of the boundary subregion $\mA$ on which the mixed state is defined, and $k_d$ is some constant depending on the dimension, the central charge $c_T$ and the prescription chosen.  In the vacuum (or any pure state), the leading divergence is precisely the same except that $V(\mA)$ is replaced by $V(\Sigma)$, the volume of the entire Cauchy surface in the boundary theory. This volume-law behaviour is the same as found for the free scalar. For example, the leading divergence in the QFT complexity of the thermal state is shown in eq.~\reef{holorocket}. As noted there, this divergence is precisely the same as found for the vacuum state \cite{qft2}. Similarly, for subregions in the vacuum state we found a leading divergence proportional to the volume of the subregion, see eqs.~\eqref{Theamazingsubregionequation} and \eqref{divergences2}.

When considering subregions of the vacuum, an interesting feature which distinguishes the subregion-CA and subregion-CV2.0 proposals from the subregion-CV prescription is that the former two generate subleading divergences that are associated with the geometry of the boundary of the subregion, \eg as shown in eq.~\reef{A4P}. In contrast, no such contributions appear with the subregion-CV proposal, \eg see eq.~\reef{ads4subcv}.\footnote{
While this equation does exhibit a subleading logarithmic divergence, there is no `area-law' divergence proportional to $R/\delta$.} Of course, as discussed in section \ref{revvH}, we could modify the subregion-CV prescription by adding a term proportional to the volume of the HRT surface, as in eq.~\reef{eq:cvX}. This modified prescription would yield boundary contributions similar to those found with the subregion-CA and subregion-CV2.0 proposals. As this modification of the subregion-CV prescription highlights, at least to leading order, the boundary contributions are proportional to the entanglement entropy of the reduced density matrix on the subregion.

\begin{figure}[t]
\begin{center}
        \includegraphics[width=0.49\textwidth]{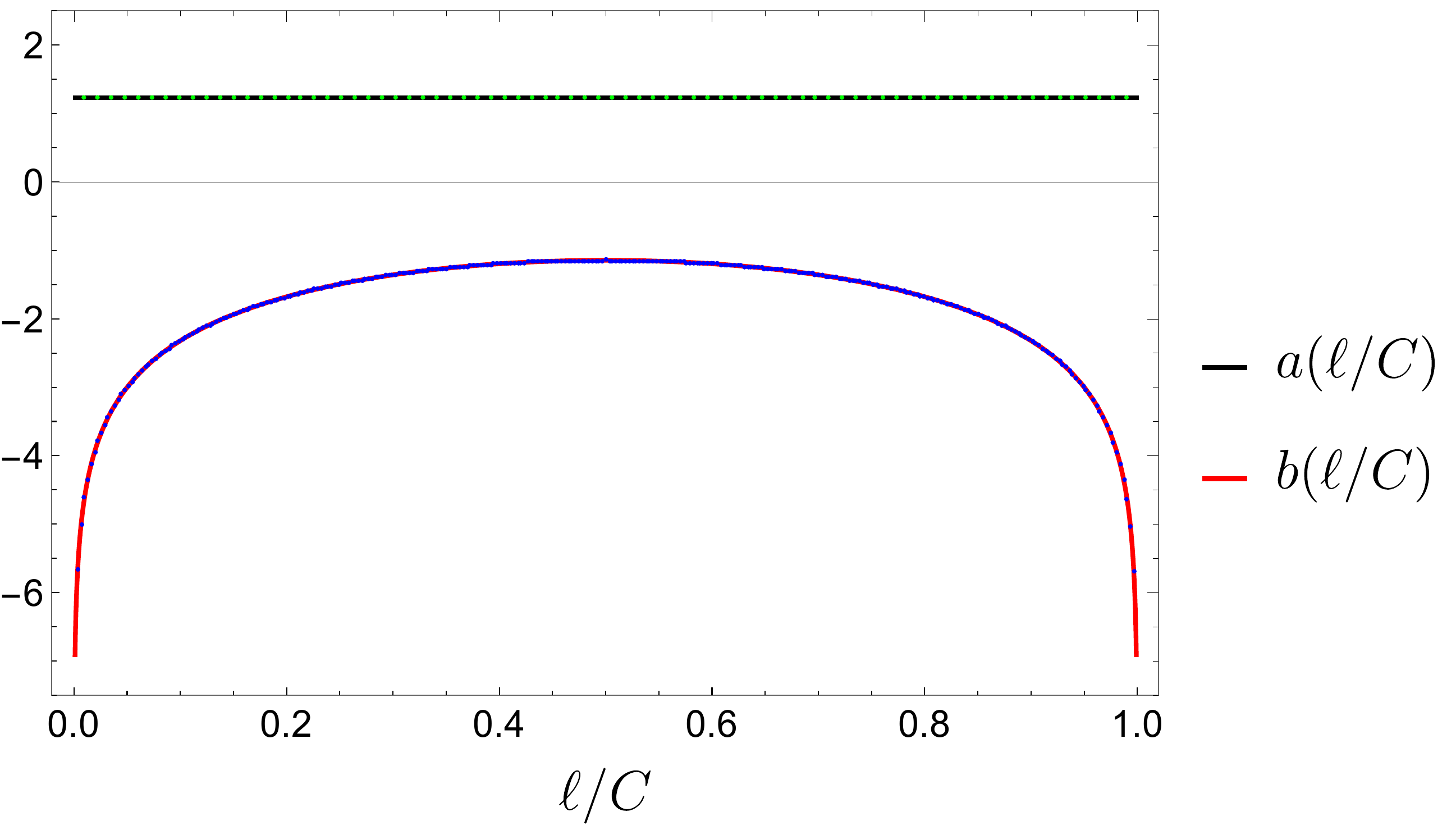}~~
        \includegraphics[width=0.43\textwidth]{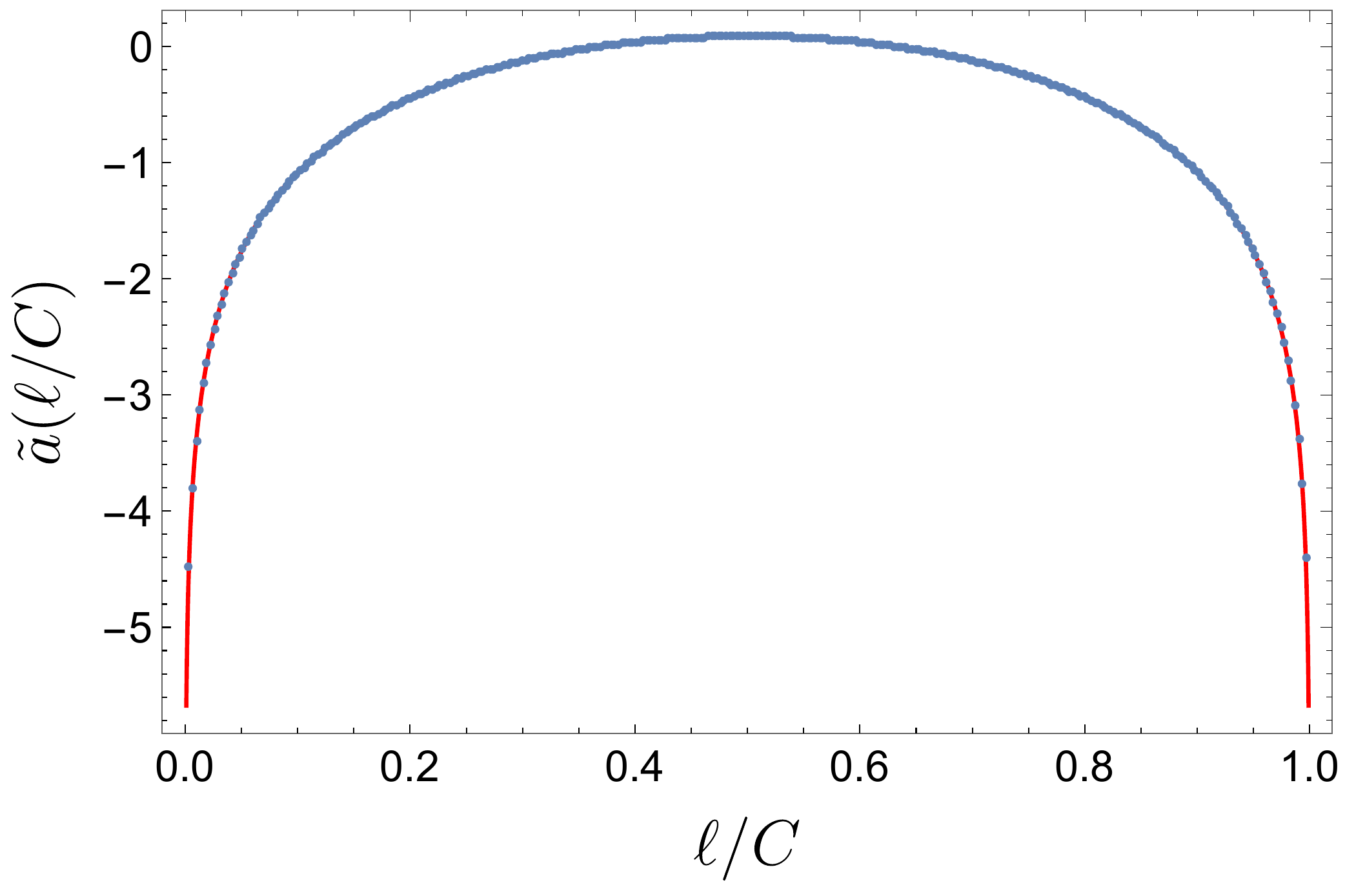}
\end{center}
        \caption{Finite part of the complexity using the subregion-CA (left) and subregion-CV2.0 (right) using global coordinates. For subregion-CA, we expressed $f(\ell/C)
\equiv {c\over  3\pi^2} \(- \ln\!\left(\frac{2\ell_{\text{ct}}}{L}\right) b(\ell/C)+a(\ell/C)\)$ and the left plot shows our numerical evaluation of $a(\ell/C)$ (green dots) and $b(\ell/C)$ (blue dots). These results are overlaid with the corresponding expressions suggested by the entanglement formula in eq.~\reef{ab}, \ie $a(\ell/C)=\pi^2/8$ and $b(\ell/C)=\ln\!\[ \frac{1}{\pi}\sin\(\frac{\pi\ell}{C}\)\]$ (black and red curves). For subregion-CV2.0, we expressed $\tilde f(\ell/C) \equiv -\frac{4\,c}{3}\, \tilde a(\ell/C)$ and the right plot shows our numerical evaluation of $\tilde a(\ell/C)$ (blue dots). These results are overlaid with the corresponding expression in eq.~\reef{at}, \ie $\tilde a(\ell/C)=\ln\!\[ \frac{1}{\pi}\sin\(\frac{\pi\ell}{C}\)\]+ \pi^2/8$ (red curve). In both cases, the numerical results fit almost perfectly with the predicted analytic expressions.}\label{numericsglobal}
\end{figure}

We would like to explore further the relation between the subleading divergences in the complexity and entanglement entropy by returning to our results of AdS$_3$ in section \ref{holosub}. Recall that using global coordinates in the bulk of AdS$_3$ corresponds to the two-dimensional boundary CFT living on a circle with a finite circumference $C$. Further our results for the subregion complexity for the subregion-CA and subregion-CV2.0 proposals were presented in eqs.~\eqref{toderive} and \eqref{CV203G} with a finite term, which we could not determine analytically. However, in the limit of small subregions, \ie $\ell/C\ll1$, we were able to predict the form of these finite functions $f(\ell/C)$ and $\tilde f(\ell/C)$ in eqs.~\eqref{lim0} and \eqref{lim1}, by comparing to the results coming from putting the boundary CFT on an infinite line. However, if we imagine that the boundary contributions to the subregion complexity are related to entanglement entropy, we should recall the formula for the entanglement entropy of an interval in CFT$_2$ on a finite circle: $S_{EE}=\frac{c}{3} \ln \(\frac{C}{\pi\delta}\sin\(\frac{\pi\ell}{C}\)\)$ \cite{Calabrese:2004eu,Calabrese:2005zw}. This formula suggests that $f(\ell/C)$ and $\tilde f(\ell/C)$ should be given by the following expressions,
\beqa
f(\ell/C)&=&{c\over  3\pi^2}
\(- \ln\!\left(\frac{2\ell_{\text{ct}}}{L}\right) \ln\!\left[ \frac{1}{\pi}\sin\(\frac{\pi\ell}{C}\)\right] +\frac{\pi^2}{8}\)\,,
\label{ab}\\
\label{at}
\tilde f(\ell/C) &=& -\frac{4\,c}{3}
\(   \ln\!\left[ \frac{1}{\pi}\sin\(\frac{\pi\ell}{C}\)\right] +
\frac{\pi^2}{8}\)\,.
\eeqa
Of course, the expressions above reduce to those in eqs.~\eqref{lim0} and \eqref{lim1} in the limit $\ell/C\to 0$. However, we note that eqs.~\reef{ab} and \reef{at} are symmetric about $\ell/C=1/2$, and so a similar logarithmic singularity appears in the limit $\ell/C\to 1$, \eg
$\tilde f(\ell/C)\simeq -\frac{4\,c}{3}
\,\ln [ \frac{C-\ell}{C}]$ in this limit. Figure \ref{numericsglobal} shows results for $f(\ell/C)$ and $\tilde f(\ell/C)$ obtained by numerical integration (see \cite{Chapman:2018bqj,future1} for further details) and compares these to the predictions in eqs.~\reef{ab} and \reef{at}. In both cases, the numerical results fit almost perfectly with the predicted analytic expressions. Hence it appears that the subleading logarithmic divergence in the complexities in eqs.~\eqref{toderive} and \eqref{CV203G}  takes precisely the same form as the corresponding entanglement entropy. This suggests a deep relation between the two quantities (at least for two-dimensional CFTs). It would be interesting to investigate this relation further, and to investigate if eqs.~\reef{ab} and \reef{at} can be derived analytically.

With the subregion-CA or subregion-CV2.0 proposals, the boundary divergences discussed above dominate the mutual complexity of the vacuum state, \eg see eqs.~\reef{mutdP} and \reef{holomutualcomp}. Hence, given a bipartition of the vacuum into subregions $\mA$ and $\mB$, the mutual complexity is UV divergent with the leading divergence taking the form $\Delta \mC\sim V(\partial\mA)/\delta^{d-2}$, where we have implicitly used that $\partial\mA=\partial\mB$. Of course, this divergence has precisely the same form as the celebrated area-law term \cite{Sorkin_1983,Bombelli_1986,Srednicki_1993} found in the entanglement entropy between $\mA$ and $\mB$. This again supports the claim that the mutual complexity characterizes the correlations between the two subsystems appearing in eq.~\reef{ramen2Ax}. Similar observations relating the mutual complexity and the entanglement entropy also appear in \cite{Caceres:2018luq}.

With a bipartition of the vacuum state on a fixed time slice, the mutual complexity precisely vanishes using the subregion-CV prescription. Of course, if we adopted the modified prescription for $\mathcal{C}'_V(\mathcal{A})$ in eq.~\reef{eq:cvX}, the resulting mutual complexity would, of course, be proportional to the entanglement entropy. Further, this construction emphasizes the observation below eq.~\reef{mutdPmix} that it is more appropriate to think of these mutual complexities as being proportional to the mutual information between the subregion and its complement. That is, applying eq.~\reef{eq:cvX} to evaluate eq.~\reef{ramen2Ay} clearly yields $\Delta\mathcal{C}'_V=\eta\,I(\mA,\mB)$ where
$I(\mA,\mB)=S_{\rm EE}(\mA) + S_{\rm EE}(\mB)-S_{\rm EE}(\mA\cup\mB)$ is precisely the mutual information of the two subregions.

The mutual complexity is, of course, an interesting quantity to compare between the holographic and QFT approaches. Our results for $\Delta\mC$ are summarized in table \ref{tab.comp} for all three holographic prescriptions calculated in section \ref{sec:holo}, as well as those for the free scalar QFT calculated in sections \ref{apply01} and \ref{apply02}.

One feature common to the holography and QFT is that the UV divergences in the complexity of the thermal state $\hat{\rho}_{th} (\beta)$ precisely match those found in the complexity of a single copy of the vacuum,\footnote{We return to this point below.} or alternatively, they are precisely one-half of those found for the TFD state. As a consequence, the mutual complexity of the TFD state is UV finite in both holography and the free QFT. Further, we demonstrated that the mutual complexity for the TFD state calculated for the free scalar in the diagonal basis is proportional to the thermal entropy in \eqref{gaga66}, where we have taken $m=0$ and also $\beta\mu\gg1$.  In the physical basis, we also expect that with the limit $\beta\mu\gg 1$ and $\beta m \ll 1$, the mutual complexity will be proportional to the entropy --- see comments at the end of section \ref{sec:mcipb}. Again, this matches the behaviour found in eqs.~\reef{exs222} and \reef{CV2.0entropy} for the subregion-CA and subregion-CV2.0 approaches.

Unfortunately, the holographic complexity is superadditive, while in the diagonal basis, the QFT complexity is subadditive, \ie $\Delta\mC({\rm TFD})<0$ for holography while $\Delta\mC({\rm TFD})>0$ for the free QFT using the diagonal basis.
However, the QFT mutual complexity in the physical basis was observed to be negative when the reference frequency $\beta \mu$ was either very small or very large, see the figure \ref{deltaC_phys}. Hence in these regimes, the physical basis results compare well with the holographic results, for the subregion-CA and subregion-CV2.0 proposals. Of course, for the $t_L=0=t_R$ time slice, the mutual complexity to the TFD state vanishes using the subregion-CV prescription. However, we could also apply the modified prescription in eq.~\reef{eq:cvX}, in which case we would find $\Delta\mC'_V({\rm TFD})=2\eta\, S$. In this case, the sign is determined entirely by the sign of the parameter $\eta$, and in particular, choosing $\eta>0$ would yield a subadditive result as found using the diagonal basis in the free QFT.

For subregions in the vacuum state of a two-dimensional free scalar field theory, using numerical fits, we inferred the general divergence structure of the purification complexity in the diagonal basis in eq.~\eqref{Theamazingsubregionequation} and in the physical basis in eq.~\eqref{divergences2}. The leading divergence is a volume term $\frac{\ell}{2\delta}|\ln \frac{1}{\mu\delta}\,|$, where the coefficient precisely matches that found in the vacuum.
In this respect, the QFT complexities show the same behaviour as found with the three holographic subregion complexity proposals, in eq.~\eqref{ads3subcv} for subregion-CV, eq.~\eqref{toderive} for subregion-CA and eq.~\eqref{CV203G} for subregion-CV2.0.\footnote{Note that our QFT results of section \ref{apply02} are valid for the circle and so should be compared to the holographic result in global coordinates, see footnote \ref{linecirclefoot}.}
The numerical fits for the QFT complexities (see eqs.~\eqref{Theamazingsubregionequation} and \eqref{divergences2}) did reveal a subleading logarithmic divergence proportional to $\ln(C/\delta)$,\footnote{Here we denote the total size of the system as $C(=L$ in section \ref{apply02}) to facilitate the comparison with the corresponding holographic results.} which was found in the holographic results for the subregion-CA and subregion-CV2.0 approaches (see eqs.~\eqref{toderive} and \eqref{CV203G}). However, our numerical results were not sensitive enough to resolve the precise form of the subleading contributions, \eg to find a form similar to that found for the corresponding holographic systems in eqs.~\reef{ab} and \reef{at}. It would be interesting to extend our QFT calculations to larger lattices, but also higher dimensional lattices where the subleading divergences become stronger.

\begin{figure}
    \center
    \includegraphics[width=0.6\textwidth]{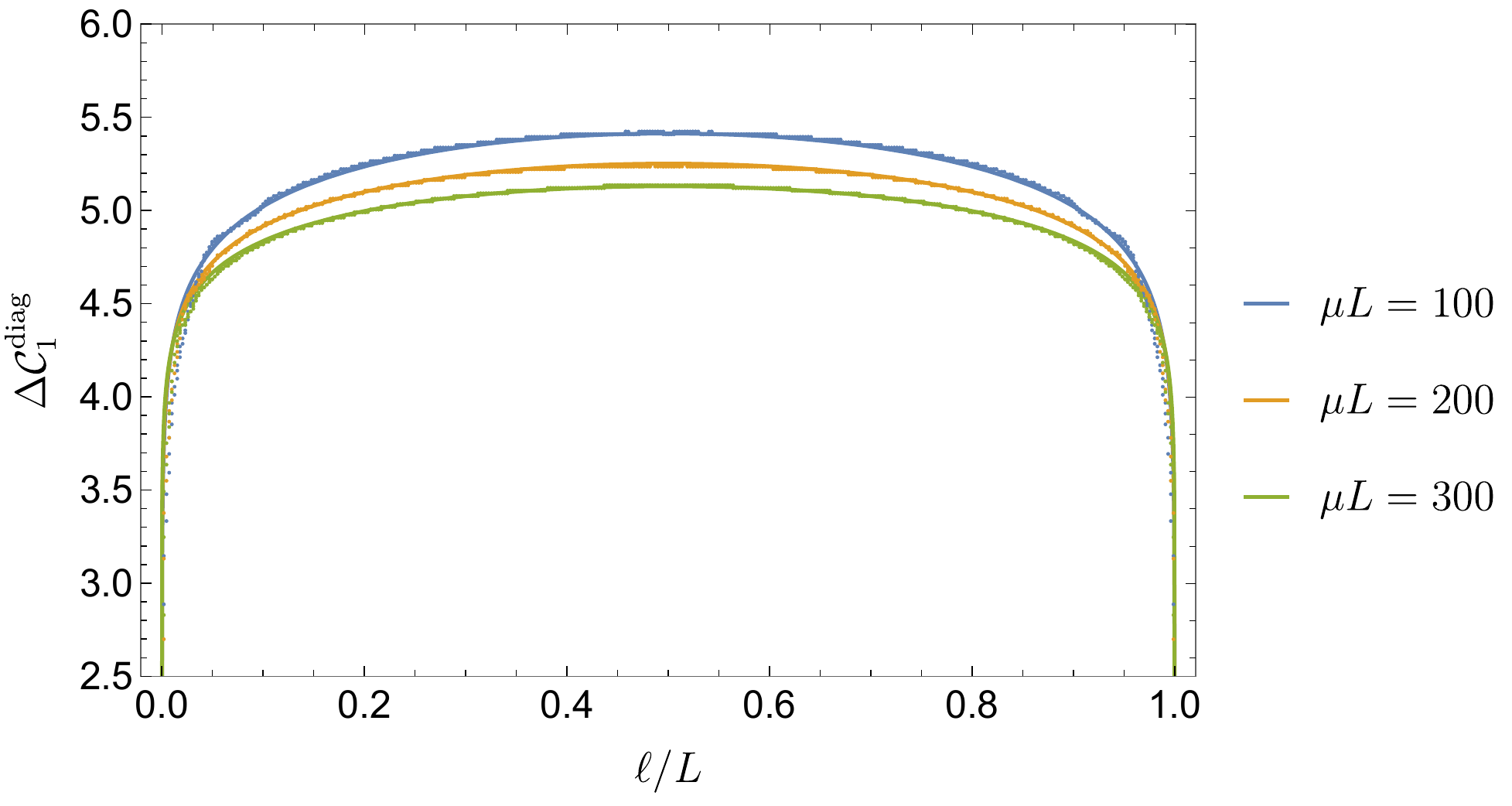}
\caption{Fits (solid curves) and data (points) of the size dependence of the mutual complexity in the diagonal basis $\Delta\mC_1^{\rm diag}$ for different reference frequencies  $\mu L = 100$, $200$ and $300$. The cutoff was set to $\delta/L = 1/N = 1/1000$. The solid lines correspond to the fit in eq.~\eqref{ccfit}.}\label{fig:mutualcompXYZ}
\end{figure}

Here, we might note that as discussed above, the subleading contributions in the subregion complexity are expected to dominate the corresponding mutual complexity. In this regard, the functional dependence of $\Delta\mC$ on $\ell/C$ compares well between the QFT and the holographic results on general grounds. That is, we may compare the free scalar QFT results
in figure \ref{fig:mutualcomp} for the diagonal basis and in figure \ref{fig:subregion2-phys} for both definitions in the physical basis with the form  appearing in figure \ref{numericsglobal} for the subleading contributions in the corresponding subregion-CA and subregion-CV2.0 results. In both cases, the mutual complexity rises dramatically for small $\ell/C$, has a broad maximum at $\ell/C=1/2$ and is symmetric under $\ell/C\to (C-\ell)/C$. A preliminary examination of the QFT results for the diagonal basis showed the following gave a good fit to our numerical results\footnote{Note that $L/\delta = 1000$ for all three curves.}
\beq
\label{ccfit}
\Delta \mC^{\rm diag}_1 \approx \frac{200}{500+\mu C}\left[ \ln\left(\frac{C}{\pi\delta}\sin\(\frac{\pi \ell}C\)\right) + 8.33+0.0214\,\mu C\right]\,.
\eeq
Figure \ref{fig:mutualcompXYZ} compares this function to our numerical results in figure \ref{fig:mutualcomp}. It would be interesting to investigate these fits in more detail and in particular, to produce the analogous fitting function for the physical basis results. The latter will require producing numerical results with much greater resolution than figure \ref{fig:subregion2-phys} which was produced with $N=C/\delta=100$.

Unfortunately, there was not a good match for the sign of these mutual complexities in comparing the holographic and free QFT results. In particular, for all three holographic approaches, the vacuum mutual complexity was generally superadditive, \ie $\Delta\mC<0$.\footnote{Of course, the modified subregion-CV approach \reef{eq:cvX} could yield either sign for the mutual complexity depending on the sign of the parameter $\eta$.} In contrast, using the diagonal basis in the free QFT produced a subadditive result for subregions of the vacuum. In the case of the physical basis, we actually proposed two definitions for the mutual complexity in eq.~\reef{twoDeltaCs}. With the first definition, where we introduce a partition of the vacuum degrees of freedom according to the arbitrary choice made for the subregions, $\Delta\mC_1^{\mt{phys}}<0$ which agrees with holography. However, the leading contribution in the QFT result appears to be linear, \ie proportional to $\ell/\delta$, whereas the leading term in the subregion-CA and subregion-CV2.0 results are proportional to $\ln(\ell/\delta)$. With the second definition, where we subtract the standard vacuum complexity, $\Delta\tilde\mC_1^{\mt{phys}}>0$ which disagrees with the holographic results. However, in this case, the leading contribution in the QFT result appears to be logarithmic, as shown in figure \ref{fig:subregion4-phys}.

If we compare the leading divergences noted above in the purification complexity and the holographic complexity from subregion-CA, we are lead to identify\footnote{Of course, the same identification comes from comparing leading divergences in the purification complexity of the thermal state, or even the complexity of vacuum state.}
\begin{equation}
\ln \left(\frac{\ell_{\text{ct}}}{L} \right) \sim    \big| \ln \( \mu\delta\) \big| =
\begin{cases}
\ln \left(\mu \delta \right) \,,&{\rm for}\ \mu\delta >1 \,,\\
\ln \left(\frac{1}{\mu \delta } \right) \,, &{\rm for}\ \mu\delta < 1\,.
\end{cases}
\end{equation}
We note that the definition of circuit complexity in the free scalar QFT introduces an new scale -- the reference frequency $\mu$, while the CA proposal for holographic complexity depends on the arbitrary length scale $\ell_{\rm ct}$, which is introduced by the null boundary counterterm \cite{Lehner:2016vdi}. The comparison of the divergences in these approaches motivates us to relate the ratio $\mu\delta$ in the QFT complexity to $\ell_{ct}/L$ in the CA proposal with $\ell_{\rm ct}/L \sim {\rm max}(\mu\delta,1/\mu\delta)$.\footnote{We are implicitly assuming that $\ell_{\rm ct}/L>1$ in order that the CA complexity is positive.} A similar identification was pointed out in \cite{qft1,qft2} and the discussion section of \cite{Vaidya2}.

We observe that this identification has interesting implications for the subregion-CA results since the coefficient $\ln(\ell_{ct}/L)$ also appears in terms beyond the leading contribution to the complexity. For example, an extra factor of $|\ln(\mu\delta)|$ would appear in the leading term in the mutual complexity in eq.~\reef{mutdP}. If $\mu$ and $\delta$ are independent scales, this would mean that this leading term no longer matches the area-law divergence appearing in the entanglement entropy. However, this interpretation can be restored if the reference frequency scales with the UV cutoff, \eg $\mu\delta = e^{-\sigma}$ so that the logarithmic factor simply introduces a new numerical factor, \ie  $|\ln(\mu\delta)|=|\sigma|$.

Our calculations also lend themselves to examining another interesting quantity, namely, the difference of the complexity of the thermal state and that of the vacuum state, \ie
\begin{equation}\label{ramenZ}
\begin{split}
\delta\mC =  \mC\!\(\hat\rho_{th}(\beta)\) - \mC\!\( \rm{vac}\)
=  \frac{\Delta\mC\(\ket{\text{TFD}}\)
+\widetilde{\Delta\mC}_{\text{formation}}}{2} \,.
\end{split}
\end{equation}
As we noted above, the UV divergences in $\mC\!\(\hat\rho_{th}(\beta)\)$ are precisely the same as in $\mC\!\( \rm{vac}\)$, and hence we are left with a UV finite quantity in $\delta\mC$. In the second expression in eq.~\reef{ramenZ}, we are expressing this quantity in terms of the mutual complexity of the TFD state (see eq.~\reef{ramen2}) and the ``complexity of formation'' of the TFD state \cite{Chapman:2016hwi,Chapman:2018hou}, \ie $\widetilde{\Delta\mC}_{\text{formation}} = \mC\(\ket{\text{TFD}}\)   - 2\mC\!\( \rm{vac}\)$.

 The quantity $\delta \mC$ will be positive in free scalar QFT using the diagonal basis. This can be seen by comparing  eq.~\eqref{complexity_thermal} with the corresponding the vacuum complexity for each mode,
\begin{equation}
\mC\!\( \rm{vac}\)=\frac{1}{2}\,\left|\ln \frac{\mu}{\omega}\,\right |\, .
\end{equation}
Hence the difference is positive for each mode, and summing over all modes, as in eq.~\reef{sum_complexity02}, we find a positive result. Further, we see that this integrand decays exponentially for large frequencies, \ie $\beta\omega\gg1$, and so the integral will be UV finite, as already noted above.

\begin{figure}[t]
\begin{center}
        \includegraphics[width=0.6\textwidth]{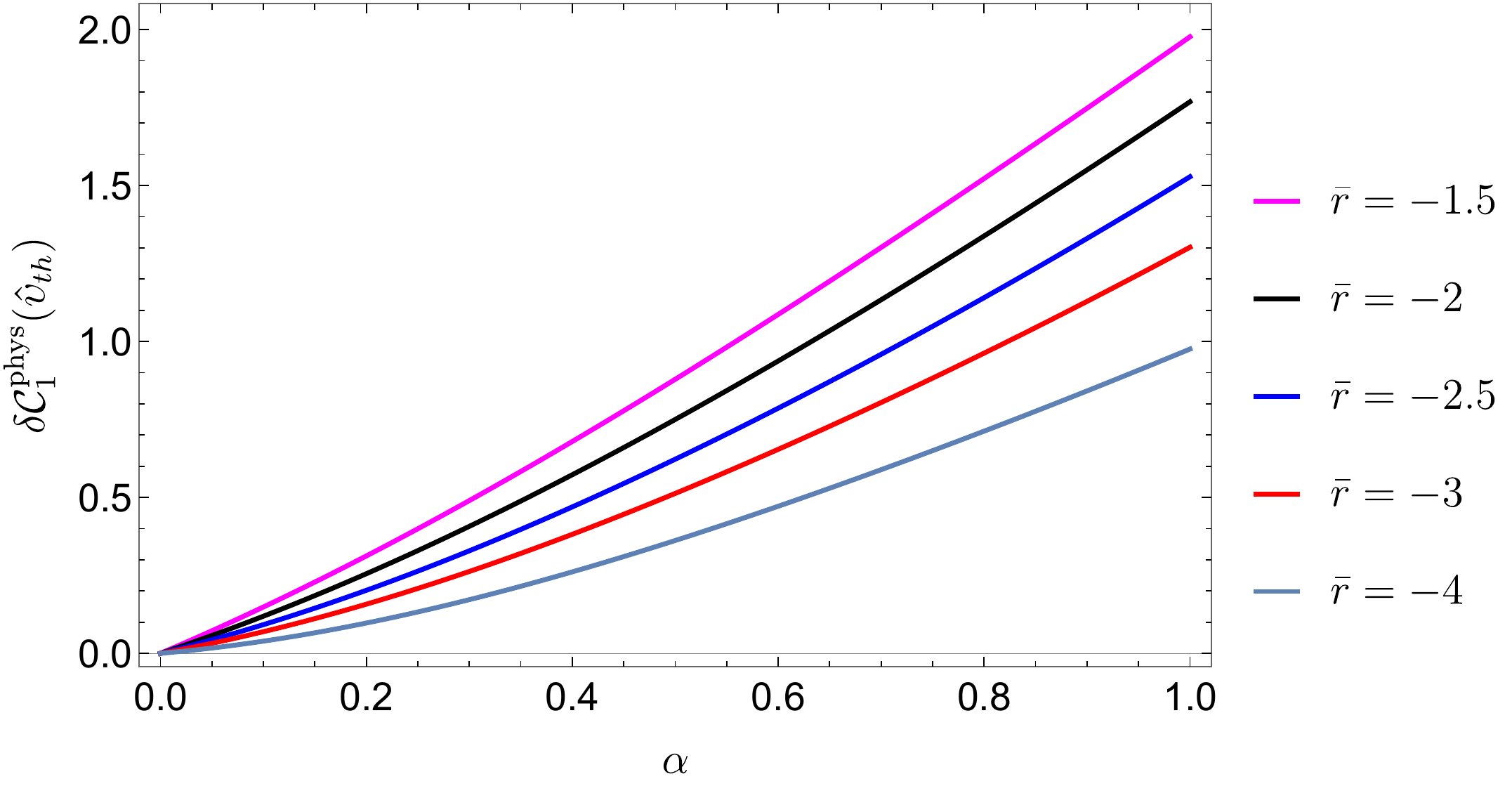}
\end{center}
        \caption{The quantity $\delta \mC_1^{\mt{phys}}\( \ket{\text{TFD}}_{12}\)$ as defined in eq.~\eqref{ramenZ} with fixed $\bar{r}=\frac{1}{2}\ln\frac{\omega}{\mu}<0$ as a function of $\alpha$. We find that the quantity $\delta \mC_1^{\mt{phys}}$ is always positive.}\label{deltaCLR_thermal01disc}
\end{figure}

In the physical basis, we can combine eq.~\eqref{board88} for the mutual complexity of the TFD state, together with the result that $\widetilde{\Delta\mC}_{\text{formation}}=2\alpha$ (see eq.~\eqref{TFD_position}) to show that in the limit $\beta\omega\gg 1$
\begin{equation}\label{deltaCslopdisc}
\delta\mC_1^{\mt{phys}}   = 2\alpha   \frac{ \,\ln\frac{\mu}{\omega}}{\sqrt{\mu/\omega}-\sqrt{\omega/\mu}}   + \mathcal{O}(\alpha^2) \,.
\end{equation}
The latter is again exponentially suppressed for large frequencies and so we expect the corresponding $\delta\mC$ to be UV finite when integrated over frequencies.
A plot of $\delta \mC_1^{\mt{phys}}$ for a single-mode is shown in figure \ref{deltaCLR_thermal01disc} and is also positive (this plot is simply obtained from the plot in figure \ref{deltaCLR_thermal01} by multiplying by a half and adding $\alpha$). Hence we also expect that $\delta\mC>0$ in the physical basis. Supporting evidence for this positivity can be found in observing that the slope in the plot for small $\alpha$ can be read from the coefficient in eq.~\eqref{deltaCslopdisc}, which again is always positive.

Let us add that in either the diagonal or physical basis, we find that $\delta C$ is proportional to the thermal entropy, at least for the limits of large $\mu \beta$ and small $m\beta$, as may be inferred from the discussion at the end of section \ref{sec:mcipb}.

Using any of the three holographic approaches, $\delta \mC$ is again a UV finite quantity because there are no boundary contributions in $\mC(\hat\rho_{th}(\beta))$, and the remaining UV divergences match those in $\mC({\rm vac})$. Recall that the mutual complexity of the TFD was vanishing for the CV proposal and so by eq.~\reef{ramenZ},  $\delta \mC$ is simply given by one-half of the complexity of formation. The latter was evaluated for planar geometries in eq.~(5.8) of \cite{Chapman:2016hwi}. Hence we obtain
\begin{equation}
\delta \mC_V = \frac{1}{2}\mC_V^{\text{formation}}= 2\sqrt{\pi}\,\frac{(d-2)\Gamma(1+\frac{1}{d})}{(d-1) \Gamma(\frac{1}{2}+\frac{1}{d})}\, S\,,
\end{equation}
which is positive for $d>2$ and vanishing for $d=2$. For the CA proposal,  eq.~\eqref{exs222} combines with the complexity of formation in eq.~(3.38) of \cite{Chapman:2016hwi} to yield
\begin{equation}
\delta \mC_A= -\frac{1}{4\pi^2}\left[a(d) +g_0+4\pi\(1-\frac{1}d-\frac{d-2}{2\,d}\, \cot\(\frac{\pi}{d}\)\) \right]\,S\,,
\end{equation}
where $a(d)$ and $g_0$ are defined in eq.~\eqref{eq:thermalCAhelp}. One may show that the sum of the three terms inside the square brackets above is always positive,\footnote{Note that for large $d$, the third term is actually large and negative (\ie $\sim -2d$), however, this behaviour is precisely canceled by the growth of $g_0\sim 2d$. Further recall that we argued that $\ell_{\rm ct}/L>1$ to ensure that the holographic complexity is positive and hence $a(d)$ is always positive.} and hence we find that $\delta\mC_A<0$.  Similarly, using the CV2.0 approach, we combine eq.~(3.35) of \cite{Chapman:2016hwi} with  eqs.~\eqref{cv20main2} and \eqref{CV2.0entropy} to find
\begin{equation}
\delta \mC_{V2.0}=-\frac{8 }{d(d-1)}\,S\,,
\end{equation}
which is once again negative. That is, according to both the CA and CV2.0 proposals, it is easier to prepare the mixed thermal state than the pure vacuum state.

Now comparing $\delta\mC$ for the free scalar QFT with that coming from holography, we observe that in all instances, this quantity is proportional to the entropy of the thermal state. As explained above, in the QFT result, this required considering the limits, $\mu \beta\gg1$ and $m\beta\ll1$. Of course, the latter is natural to compare the result to the boundary CFT in the holographic framework.
However, we must also note that while $\delta\mC$ is positive for the free scalar and for the CV approach, it is negative for the CA and CV2.0 approaches. Hence there is some tension between the results for the two last approaches and those for the free QFT.

Let us summarize our comparison of the purification complexity for the free scalar QFT with the various subregion proposals in holography: Our results do show that various general features are common to the two frameworks. However, a detailed comparison does not lead to any definite conclusions. Based on comparisons of the mutual complexity for the TFD and vacuum states, it seems that details of the QFT results using the diagonal basis are quite different from the corresponding holographic results. Recall that previous calculations of the complexity of formation for the free scalar \cite{Chapman:2018hou} already indicated that the diagonal basis did not produce results comparable to holography. The QFT results using the physical basis can be brought into closer alignment with the holographic results, at least in certain regimes, \eg $\beta\mu\gg1$ or $\beta\mu\ll1$ is required for the mutual complexity of the TFD state to be superadditive. These restrictions may be informing us about the microscopic model underlying holographic complexity. However, we are still left with apparent discrepancies for the mutual complexity of the vacuum state, as well as for the purification complexity of the thermal state \reef{ramenZ}, which may be warning us that these comparisons simply have limited applicability.

\subsection{Entanglement Entropy}
\label{sec_EE}

Much of the original motivation to study holographic complexity was trying to understand the structure behind the horizon which is not encoded in the entanglement entropy \cite{Volume3}. From the quantum information perspective, entanglement entropy is simply one of a broad array of diagnostics with which to characterize quantum entanglement. Hence while the Ryu-Takayanagi prescription \cite{Ryu:2006bv,Ryu:2006ef} for holographic entanglement entropy has provided many new insights on the connection between geometry and entanglement in quantum gravity, it is not surprising that the full picture will require drawing on additional observables, such as holographic complexity.

Here, we explicitly compare the information about a given reduced density matrix which is encoded in the complexity and the information which is encoded in the entanglement entropy. We begin with the density matrices for a single harmonic oscillator, as described in section \ref{warmup}.\footnote{This comparison was also examined in \cite{Camargo:2018eof}, with the conclusion that ``complexity is more.''} The entanglement entropy of such a density matrix can be read from the results of \cite{Srednicki_1993}
\begin{equation}\label{entropy_rho1}
S_1 = -\tr ( \hat{\rho}_1 \ln \hat{\rho}_1) = -\ln(1-u) -\frac{u}{1-u} \ln u\,,
\end{equation}
where using the parametrization in eq.~\eqref{Gaussian_decom}, we have $u=e^{-\beta \omega}$.\footnote{Using eqs.~\eqref{dense1} and \reef{parameters}, we may match the wavefunction parameters in \cite{Srednicki_1993} with our notation as $\cosh\beta\omega=\gamma/\beta$ and $\omega^2e^{2r}=\sqrt{\gamma^2-\beta^2}$.} Hence the entanglement entropy depends only on the combination $\beta\omega$, and is independent of the squeezing parameter $r$.
This is to be contrasted with our result for the complexity, \eg see eq.~\eqref{complexity_one_mode}, which depends on both of these parameters. That is, while the entanglement entropy is entirely fixed by the temperature, the purification complexity also contains information about the squeezing of the mixed state.

The natural extension of these observations to a general $N$-mode Gaussian state is as follows: First, note that general $N$-mode Gaussian states can be decomposed as\footnote{See appendix \ref{app:purification} for more details.}
\begin{equation}\label{laud}
\hat{\rho}_\mA = U_\mA \(\bigotimes_{i=1}^{N} \hat{\upsilon}_{\text{th}} (\beta_i, \omega_i) \) U_\mA^\dagger\,.
\end{equation}
However, acting on a density matrix $\hat{\rho}_\mA$ by a unitary operator as follows $U \hat{\rho}_AU^\dagger$ does not change its eigenvalues, and hence does not modify the entanglement entropy.
Hence the entanglement entropy of any such state \reef{laud} does not depend on the unitaries $U_\mA$ and, in fact, is a simple sum of the thermal entropies for each mode, \ie
\begin{equation}
S_\mA= \sum_{i=1}^N S_{\text{th}}(\beta_i \omega_i)
\quad{\rm where}\quad
S_{\text{th}}(\beta_i \omega_i)=\frac{\beta_i\omega_i}{e^{\beta_i\omega_i}-1}-\ln(1-e^{-\beta_i\omega_i})\,.
\end{equation}
That is, the entanglement entropy of the states $\hat{\rho}_\mA$ and $U_\mA^\dagger\,\hat{\rho}_\mA\, U_\mA=\bigotimes_{i=1}^{N} \hat{\upsilon}_{\text{th}} (\beta_i, \omega_i)$ are identical.
On the contrary, the purification complexity will generally depend on the unitary operator $U_\mA$,  as well as the choice of the reference state $\ket{\Psi_R}$. In the previous one-mode example, these two extra pieces of data are combined together and encoded in the single parameter $\bar{r} \equiv r+\frac{1}{2} \ln \frac{\omega}{\mu}$ appearing in eq.~\reef{complexity_one_mode}. For a more general mixed state, its entanglement entropy is only sensitive to its eigenvalues, while we expect that the purification complexity will capture some of the information about the unitary operator part $U_\mA$ in the density matrix which is absent in the entanglement entropy. In any event, this general example further emphasizes the conclusion that the purification complexity offers access to more information about mixed states than the entanglement entropy.

\subsection{Other Cost Functions}
In the main text, we defined the purification complexity by minimizing the complexity of pure states which purify a given mixed state using the $F_1$ cost function. Of course, the latter is only one choice amongst many possibilities, and so here we briefly explore the complexity of mixed states with other cost functions. We also give a short discussion of applying the Fubini-Study approach \cite{qft2} to this problem, but leave further details of this case to \cite{wip99}.

Before proceeding, we must note that the authors of \cite{Camargo:2018eof} have considered the purification complexity for one-mode Gaussian mixed states using the $F_2$ cost function in their appendix C, as we will do below. They considered the most general purification consisting of a six real (or three complex) parameter family of two-mode pure Gaussian states as purifications. The authors used numerical minimization to show that the purification complexity using the $F_2$ cost function is subadditive. Below, we go further analytically by restricting our attention to the three-dimensional space of real purifications, \ie eq.~\reef{Fock_psi12}.

{\bf Purification Complexity with $F_2$ and $\kappa=2$ cost functions:} Here, we will focus on the one-mode Gaussian state $\hat{\rho}_1$, \eg see eq.~\eqref{Gaussian_decom}, and consider the $F_2$ and $\kappa=2$ cost functions for which the complexity of the pure state \eqref{Fock_psi12} is defined as
\begin{equation}\label{F2_pure}
\begin{split}
\mathcal{C}_{\kappa=2}\(\ket{\psi}_{12} \)  &=\mathcal{C}_{2}\(\ket{\psi}_{12} \)^2  = \( \frac 12\ln \frac{\omega_+}{\mu}\)^2+ \( \frac 12\ln \frac{\omega_-}{\mu}\)^2  \\
&={\frac 12}  \left(\bar{r}+\bar{s}\right){}^2+ {\frac 12}\( \cosh^{-1}  \( \cosh 2\alpha \cosh (\bar{r}-\bar{s}) \) \)^2
\end{split}
\end{equation}
where the two normal frequencies $\omega_{\pm}$ are defined in eq.~\eqref{eigenvalues_2modes}. We can define the purification complexity of the mixed state $\hat{\rho}_1$ using the $\kappa=2$ and $F_2$ cost functions as the minimal value of eq.~\eqref{F2_pure} over all possible purifications
\begin{equation}
\mC_2 \( \hat{\rho}_1\) \equiv   \text{min}_{s}\,  \mC_{2} \left(\ket{\psi}_{12}\right) \,,  \quad \mC_{\kappa=2} \( \hat{\rho}_1\) =  \text{min}_{s}\,  \mC_{\kappa=2} \left(\ket{\psi}_{12}\right)= \mC_2 \( \hat{\rho}_1\)^2 \,.
\end{equation}
Here, the minimization is performed with respect to the free parameter $s$ (or equivalently $\bar{s}$ defined in eq.~\eqref{rbar}). In principle, we only need to find the extremal point by solving $\partial_{\bar{s}} \mathcal{C}_{2}= 0  = \partial_{\bar{s}}\mathcal{C}_{\kappa=2}$,
where the two cost functions share the same minimal point with respect to the free parameter $\bar{s}$.
For the special limit of the pure state we have $\alpha=0$ which leads to a minimum at $\bar{s}=0$ and $\mathcal{C}_{2}\big|_{\text{min}}=  \bar{r}$, as expected.
Unfortunately, the above minimizations cannot, in general, be performed analytically.
However, we are able to make more comments on the special case of the thermal mixed state $\hat{\upsilon}_{th}$ with $\bar r= \frac{1}{2}\ln\frac{\omega}{\mu}$, see eqs.~\eqref{Gaussian_decom} and \eqref{rbar}. Here, the purification complexity with the $F_2$ cost function reads
\begin{equation}
\begin{split}\label{kappa2222}
\mathcal{C}_{\kappa=2} \(\hat{\upsilon}_{\text{th}}\)
&=\text{min}_{s} {\frac 12}\(\left(s- \ln \frac{\mu}{\omega} \right)^2+ \( \cosh^{-1} \(\cosh 2\alpha \cosh s\) \)^2  \)\,.
\end{split}
\end{equation}
We are able to get an analytic solution for the purification complexity and for the mutual complexity in a number of special limits.
First, consider the limit of large frequency (or small temperature) $\beta \omega \gg 1$, where we have $\alpha \ll 1 $. Here it is easy to find that the minimal point at the $\mathcal{O}(\alpha^2)$ order approximately locates at
\begin{equation}
s\approx  \frac{1}{2} \ln \frac{\mu}{\omega} - \alpha^2 \frac{\mu^2 - 2\mu\omega \,{\rm ln}\frac{\mu}{\omega} -\omega^2}{(\mu-\omega)^2}\,,
\end{equation}
for which the complexity is
\beq
{\cal C}_{\kappa=2}\(\hat{\upsilon}_{\text{th}}\) = \frac14 \left({\rm ln}\frac{\mu}{\omega}\right)^2 + \alpha^2\, {\rm ln}\frac{\mu}{\omega}\, \left(\frac{\mu+\omega}{\mu-\omega}\right)+{\cal O}(\alpha^4)\,.
\eeq
On the other hand,  we can take the small frequency or large temperature limit $\beta \omega\ll 1$, or $\alpha \gg 1$, together with $\ln\frac{\mu}{\omega} -2\alpha \gg 1$ to find
\begin{equation}
2s\approx \ln \frac{\mu}{\omega} -\ln \cosh 2\alpha \approx \ln \frac{\mu}{\omega} + \ln 2 -2\alpha.
\end{equation}
The requirement $\ln \frac{\mu}{\omega} -2\alpha \gg 1$ ensures that $s$ is large and positive, allowing us to perform an expansion in $s$ and solve the resulting transcendental equation for the minimal point. This will not be satisfied, \eg for a small reference frequency.
The complexity of this purification is     then
\beq
{\cal C}_{\kappa=2} \(\hat{\upsilon}_{\text{th}}\)
 = \left(\alpha +\frac12 {\rm ln}\frac{\mu}{2\omega} \right)^2+ {\cal O}(1/\alpha)\,.
\eeq

Next, we turn to the mutual complexity. We will need the complexity of the TFD state which can be obtained by substituting $s=0$ in eq.~\eqref{kappa2222} and reads
\begin{equation}\label{Fk2_TFD}
\begin{split}
\mathcal{C}_{\kappa=2}\(\ket{\text{TFD}}_{12} \)  =
\frac{1}{2}  \( \ln \frac{\omega}{\mu} \)^2+ 2\alpha^2 \,.
\end{split}
\end{equation}
By numerical minimization, it is easy to show that the mutual complexity with the $\kappa=2$ cost function is subadditive, \ie
\begin{equation}
\Delta \mC_{\kappa=2} \( \ket{\text{TFD}_{12}}\) = 2\mC_{\kappa=2} \(\hat{\upsilon}_{\text{th}}\) - \mathcal{C}_{\kappa=2}\(\ket{\text{TFD}}_{12} \) \ge  0 \,.
\end{equation}

Except for the numerical proof, we can also show the subadditivity analytically in the various limits studied above.
For the case $\beta \omega \gg 1 $ we find
\begin{equation}
\begin{split}
\Delta \mC_{\kappa=2} \( \ket{\text{TFD}_{12}}\)
\approx
2\( \ln \frac{\mu}{\omega} \left(\frac{\mu+\omega}{\mu-\omega}\right) -1\) \alpha^2 + \mathcal{O}(\alpha^4) \ge 2\alpha^2 \ge 0\,.
\end{split}
\end{equation}
Similarly, in the opposite limit $\beta \omega \ll 1$ (but also $\ln\frac{\mu}{\omega} -2\alpha \gg 1$), we have
\begin{equation}
\begin{split}
\Delta \mC_{\kappa=2} \( \ket{\text{TFD}_{12}}\) \approx
2\alpha\, \ln\frac{\mu}{2\omega}- \ln 2\, \ln\frac{\mu}{\sqrt{2}\omega} \,,
\end{split}
\end{equation}
which is again positive in this limit. Note that the term proportional to $\ln^2(\mu/\omega)$ has been canceled in both these limits.

To close this section, we would like to mention the following inequality between purification complexities with the different cost functions
\begin{equation}
\mC_1^{\text{diag}}(\hat{\rho}_\mA)  \ge \mC_2(\hat{\rho}_\mA)= \sqrt{\mC_{\kappa=2}(\hat{\rho}_\mA)}\,.
\end{equation}
These relations are valid for any mixed state $\hat{\rho}_\mA$, and arise straightforwardly from the definition of complexity of pure states.
Although we do not have an analytic solution for $\mC_2(\hat{\rho}_\mA)$, this inequality can, of course, be tested numerically.
From the second equality, we can show that the subadditivity of purification complexity using the $F_2$ measure follows from the subadditivity for the $\kappa=2$ cost function because of their special relation and the fact that the inequality $A^2 +B^2 \ge C^2$ implies the inequality  $|A| + |B| \ge |C|$.

Finally, note that the analysis in this section is restricted to purifications of a single-mode Gaussian state by one additional auxiliary mode, and we have not addressed more general questions related to optimizing the purifications, when using the $F_2$ or $\kappa=2$ cost functions. For example, we have not clarified whether the optimal purifications should be essential purifications for these new cost functions. We will return to these issues in \cite{wip99}.

{\bf Fubini-Study approach:} Instead of approaching the complexity with Nielsen's geometric approach \cite{Nielsen:2006,nielsen2006quantum,nielsen2008}, the authors of \cite{qft2} developed a similar geometric approach based on the Fubini-Study metric, \ie quantum information metric for pure states as a distance measure on the space of pure states to derive complexity. It is straightforward to extend our discussion of purification complexity using this Fubini-Study method, as follows,
\begin{equation}\label{purificationFS}
\mC_{\rm{FS}} \( \hat \rho_\mathcal{A} \)  \equiv \text{min}_{\mA^c} \,  \mC_{\rm FS}\( \ket{\Psi_{\mA\mA^c}} \)\,.
\end{equation}
Here, the $\mC_{\rm FS}$ is the Fubini-Study complexity of the pure state $\ket{\Psi_{\mA\mA^c}} $, which purifies the target mixed state $\hat{\rho}_\mA$. Returning to our simple one-mode mixed state $\hat{\rho}_1$ in eq.~\reef{Gaussian_decom}, we can easily define
\begin{equation}
\mC_{\rm{FS}} \( \hat \rho_1 \)  \equiv \text{min}_{s} \,  \mC_{\rm FS}\( \ket{\psi_{12}} \)= \text{min}_{s} \,\mC_{2}\( \ket{\psi_{12}} \) \,,
\end{equation}
where we have used the fact that the Fubini-Study complexity of Gaussian states is the same as that evaluated using the $F_2$ cost function, \eg compare the results of \cite{qft1} and \cite{qft2}. Therefore, the above results for $\mC_{2}(\hat{\rho}_{1})$ also give the purification complexity for the Fubini-Study method.

As shown in this paper, the purification complexity for $N$-mode systems requires a careful treatment for the optimal purification. In order to avoid these complications, the future work \cite{wip99} extends the Fubini-Study method to mixed states by considering a quantum information metric, or quantum fidelity susceptibility, of mixed states in order to develop a measure of complexity for mixed states. Remarkably, it is found that the complexity of arbitrary Gaussian mixed states using this new approach is exactly equivalent to the purification complexity based on the Fubini-Study metric \eqref{purificationFS}. In other words, the quantum information metric provides a perfect measure for the purification complexity of mixed states, without implementing the purification, and hence without optimizing over the additional parameters associated with the auxiliary degrees of freedom.

\section*{Acknowledgments}
We would like to thank Cesar Ag\'on, Dorit Aharonov, Horacio Casini, Dongsheng Ge, Giuseppe Di Giulio, Giuseppe Policastro, Brian Swingle, Tadashi Takayanagi, Erik Tonni, Erik Verlinde, Jingxiang Wu and Ming-Lei Xiao for useful comments and discussions. Research at Perimeter Institute is supported in part by the Government of Canada through the Department of Innovation, Science and Economic Development Canada and by the Province of Ontario through the Ministry of Economic Development, Job Creation and Trade. JPH and SMR are supported in part by a Discovery Grant awarded to RCM by the Natural Sciences and Engineering Research Council of Canada. JPH is also supported by the Government of Ontario's Ministry of Advanced Education and Skills Development through a Queen Elizabeth II Graduate Scholarship in Science and Technology. RCM also received funding from the BMO Financial Group and from the Simons Foundation through the ``It from Qubit'' collaboration. Research by EC and JDC was supported by NSF grants PHY-1620610 and PHY-1820712. SC acknowledges funding from the European Research Council (ERC) under starting grant No.~715656 (GenGeoHol) awarded to Diego M. Hofman. EC, SC and RCM thank the Galileo Galilei Institute for Theoretical Physics for hospitality and the INFN for partial support where part of this work was carried out. JDC, JPH and SMR would like to thank IAS for their hospitality during PITP 2018 and for thought-provoking talks and conversation.

\appendix

\section{Alternative Parametrization for Gaussian States}\label{app:purification}
In section \ref{sec:purificationX}, we discussed general properties of the purification of mixed Gaussian states with $N_{\mA}$ modes. Here, we want to rephrase this discussion using (a generalization of) the notation introduced in section \ref{subsec:altdesc}, \ie using thermal density matrices and squeezing operators. Of course, all our conclusions will remain the same using this decomposition, however, let us point out that this description has certain numerical advantages since it scans the space of parameters much more efficiently than the naive decomposition in section \ref{sec:purificationX} due to the exponential relationship between the parameters, see \eg eq.~\eqref{parameters} and \eqref{transform_paras} for the case of a mixed state of a single oscillator.

Williamson's theorem implies that multi-mode Gaussian states with zero mean\footnote{A multi-mode Gaussian state is said to have zero mean if the expectation values of the positions $\hat x_i$ and momenta $\hat p_i$ of the various oscillators vanish in the state.} can be decomposed into a thermal part acted on by unitary operators \cite{RevModPhys.84.621,serafini2017quantum} (this discussion is often phrased in terms of covariance matrices). This is similar to the thermal decomposition of one-mode Gaussian states in eq.~\eqref{Gaussian_decom}.
Explicitly, starting with an $N_{\mA}$-mode Gaussian density matrix $\hat{\rho}_{\mA}$ we have
\begin{equation}\label{thermal_decomposition}
\hat{\rho}_{\mA} =    U_S \(\bigotimes_{i=1}^{N_{\mA}} \hat{\upsilon}_{th} (\beta_i,\omega_i) \) U_S^\dagger \,,
\end{equation}
where $U_S$ are unitary transformations which do not change the fact that $\hat{\rho}_{\mA}$ is zero-mean, and
\begin{equation}
\hat{\upsilon}_{th} (\beta_i,\omega_i)  \equiv      \(1- e^{-\beta_i\omega_i}\) \sum_{n=0}^\infty e^{-\beta_i \omega_i n} \ket{n(\omega_i)}\bra{n(\omega_i)}  
\end{equation}
are thermal density matrices for the different thermal modes with inverse temperatures $\beta_i$ and frequencies $\omega_i$ respectively.
The $i$-th mode will be pure if the associated temperature $T_i={1}/{\beta_i}$ vanishes.
The thermal decomposition of $\hat{\rho}_{\mA}$ is very suggestive of the fact that the minimal number of ancillary modes needed in order to purify the system is equal to the number of non-zero $T_i$-s. This corresponds to the rank of the matrix B in section \ref{sec:purificationX}.

The most general unitary $U_S$ which transforms among Gaussian states leaving $\langle \hat{x}_i\rangle = \langle \hat{p}_i\rangle =0$ invariant is spanned by exponentiating the $N_{\mA}(2N_{\mA}+1)$ generators of $SP(2N_{\mA},\mathbb{R})$:
$ i (\hat{x}_a \hat{p}_b +\hat{p}_b \hat{x}_a)$, $i \hat{x}_a \hat{x}_b$ and $i \hat{p}_a \hat{p}_b$. However, for the purposes of this paper where we restricted to the $N_{\mA}(N_{\mA}+1)$ parameter family of real density matrices $\hat{\rho}_{\mA}$ of the form~\eqref{density_fun_A} with $A$ and $B$ real and symmetric, the $GL(N_{\mA},\mathbb{R})$ subgroup generated by  $i (\hat{x}_a \hat{p}_b +\hat{p}_b \hat{x}_a)$ is enough for the purpose of constructing $U_S$ in the thermal decomposition~\eqref{thermal_decomposition}.

In section~\ref{MbyM} we made the distinction between Gaussian mixed states which take a product form in the different degrees of freedom \eqref{gamble6}
and more general Gaussian mixed states. We can make this distinction also in the language of the thermal decomposition in eq.~\eqref{thermal_decomposition}. The simple product state will be given by
\begin{equation}\label{rhoA_decomposition}
    \begin{split}
        \hat{\rho}_{\mA}
        &=   S(\vec{r}) \(\bigotimes_{i=1}^{N_{\mA}}     \hat{\upsilon}_{th} (\beta_i,\omega_i)\) S^\dagger(\vec{r})
        =\bigotimes_{i=1}^{N_{\mA}} \hat{\rho}_i\(\beta_i,r_i\),
    \end{split}
\end{equation}
where $S\(\vec{r}\) = \prod\limits_{i}^{N_{\mA}} S_i(r_i)$, with $\vec{r}\equiv (r_1,r_2,\cdots,r_{N_{\mA}})$, $S_i(r_i)$ are the one mode squeezing operators \eqref{onemode_squeezed} for the $i$-th oscillator and $\hat{\rho}_i $ is the one-mode Gaussian state defined in \eqref{Gaussian_decom} for the $i$-th oscillator. This corresponds to $U_S=S(\vec r)$ in eq.~\eqref{thermal_decomposition}.

However, the most general multi-mode Gaussian state cannot be written in this way as the tensor product of $N_{\mA}$ one-mode Gaussian states.
In general, we can always decompose $U_S$ in eq.~\eqref{thermal_decomposition} according to Euler/Bloch-Messiah decomposition \cite{Arvind:1995ab,PhysRevA.71.055801,RevModPhys.84.621}.
The decomposition implies that we can decompose the unitary $U_S$ as
\begin{equation}\label{eq:Euler}
    U_S= U_K \(\bigotimes_{i=1}^{N_{\mA}} S(r_i)   \) U_L  ,
\end{equation}
where $U_{K,L}$ are ``passive'' transformations.
The term ``active'' (``passive'') refers to the property that these transformation change (leave invariant) the photon number.
More concretely, they change (leave invariant) the following quantity
\begin{equation}\label{conserved}
 \sum_{i=1}^{N_{\cal{A}}} \biggr\langle \frac{1}{2\omega_i}\hat{p}_i^2 + \frac{\omega_i}{2}\hat{x}_i^2\biggr\rangle\,.
\end{equation}
The upshot of the Euler decomposition~\eqref{eq:Euler} is that when the first unitary $U_L$ and the multi-mode squeezing $S(\vec r)$
are non-trivial, the thermal decomposition~\eqref{thermal_decomposition} does not factorize into a product state~\eqref{rhoA_decomposition}.

Lastly, we comment on the restriction of the Euler decomposition~\eqref{eq:Euler} to the $GL(N,\mathbb{R})$ subgroup on which we focused in this paper. The condition that~\eqref{conserved} be conserved implies that the ``passive'' $GL(N,\mathbb{R})$ transformations are generated by the antisymmetric $i (\hat{x}_a \hat{p}_b - \hat{x}_b \hat{p}_a)$ generators, and correspond to the ``beam splitter transformations'' in the language of \cite{RevModPhys.84.621}. These are simply $SO(N,\mathbb{R})$ rotations which act simultaneously on $\hat{x}_a$ and $\hat{p}_a$. The matrices $U_K$ and $U_L$ in eq.~\eqref{eq:Euler} are therefore determined by $N_{\cal{A}}(N_{\cal{A}}-1)/2$ parameters each, and together with the $N_{\cal{A}}$ one mode squeezing  parameters $r_i$ in eq.~\eqref{eq:Euler} and $N_{\cal{A}}$ thermal parameters $\beta_i \omega_i$ in eq.~\eqref{thermal_decomposition} these constitute the $N_{\cal{A}}(N_{\cal{A}}+1)$ real parameters of the density matrix $\hat \rho_{\cal{A}}$ in eq.~\eqref{density_fun_A}. For completeness let us also specify that the ``active'' part of the $GL(N,\mathbb{R})$ subgroup is spanned by the two mode squeezing operators $i (\hat{x}_a \hat{p}_b + \hat{x}_b \hat{p}_a)$, similar to the one in eq.~\eqref{2squeezed_operator}.

\section{Numerics for Essential Purifications}
\label{app:numerics}
As discussed in section \ref{threeA}, one might naively expect that purifications with more than the minimal number of ancillae could lead to a lower value of the complexity for a given mixed state. However, we have proven that this is not the case for the $F_1$ cost function in the diagonal basis. In this appendix we parallel a part of that discussion using a thermal decomposition similar to the one in appendix \ref{app:purification}. Since this decomposition is more suitable for numerical analysis, we use this advantage to show numerical evidence confirming our previous conclusion by comparing the diagonal basis purification complexity of the one-mode mixed Gaussian state in eq.~\eqref{Gaussian_decom} using three-mode and two-mode purifications.

A general three-mode pure state (cf.~eq.~\eqref{pure33}) can be decomposed as
\begin{equation}\label{purified_state_3mode}
\begin{split}
\ket{\psi_{123}}
&= S(\alpha_{23})S(r_1)S(r_2)S\(r_3\)S(\alpha_{13})S(\alpha_{12})  \ket{0_1(\omega),0_2(\omega),0_3(\omega)},
\end{split}
\end{equation}
where $|0_i(\omega)\rangle$ is the ground state of the $i$-th oscillator with frequency $\omega$ and where we have suppressed the subscripts on the one and two-mode squeezing operators $S_i, S_{ij}$ (cf.~eq.~\eqref{onemode_squeezed} and \eqref{2squeezed_operator}) and used instead subscripts on their parameters $r_i, \alpha_{ij}$ in order to indicate which modes they act on. This is similar to the decomposition of the two-mode purification in eq.~\eqref{Fock_psi12}.
The corresponding wavefunction is Gaussian and takes the form
\begin{equation}
\begin{split}
\psi_{123}(x,y,z) \equiv  \langle{x,y,z}\ket{\psi_{123}}
=  \mathcal{N}_{123} \exp \(  -\frac{1}{2}\vec x^{\, T} \, A_{123} (r_i,\alpha_{ij})~\vec x  \)\,,
\end{split}
\end{equation}
where $\vec x^{\, T}=(x,y,z)$ and $A_{123}$ is a $3$ by $3$ real symmetric matrix.
The explicit form of $A_{123}$ can be found using the techniques of \cite{qft1} by representing the action of the squeezing operators on the matrix representation $A_{123}$ of the wavefunction in terms of matrix conjugation
\begin{equation}\label{MatrixRepQFT1}
A_{123} =  U_{123} \cdot A_0\cdot  U^T_{123}\,, \quad A_0\equiv\omega {\bf I}\,, \quad U_{123} = e^{\alpha_{23}g_{23}}e^{r_1 g_{11}}e^{r_2 g_{22}}e^{r_3 g_{33}}e^{\alpha_{13}g_{13}}e^{\alpha_{12}g_{12}},
\end{equation}
where the corresponding generators of the one-mode squeezing operators $S(r_1)$, $S(r_2)$, $S(r_3)$ and the two-mode squeezing operators $S(\alpha_{12})$, $S(\alpha_{13})$, $S(\alpha_{23})$ are given by
\begin{equation}\label{6generators}
\begin{split}
g_{11}&=\left(
\begin{array}{ccc}
1 & 0 & 0 \\
0 & 0 & 0 \\
0 & 0 & 0 \\
\end{array}
\right),~
g_{22}=\left(
\begin{array}{ccc}
0 & 0 & 0 \\
0 & 1 & 0 \\
0 & 0 & 0 \\
\end{array}
\right),~
g_{33}=\left(
\begin{array}{ccc}
0 & 0 & 0 \\
0 & 0 & 0 \\
0 & 0 & 1 \\
\end{array}
\right),~ \\
g_{12}&=\left(
\begin{array}{ccc}
0 & -1 & 0 \\
-1 & 0 & 0 \\
0 & 0 & 0 \\
\end{array}
\right),~
g_{13}=\left(
\begin{array}{ccc}
0 & 0 & -1 \\
0 & 0 & 0 \\
-1 & 0 & 0 \\
\end{array}
\right),~
g_{23}=\left(
\begin{array}{ccc}
0 & 0 & 0 \\
0 & 0 & -1 \\
0 & -1 & 0 \\
\end{array}
\right).
\end{split}
\end{equation}
The six parameter family of purifications \eqref{purified_state_3mode} is constrained to reproduce the desired mixed state $\hat{\rho}_1$ \eqref{Gaussian_decom} upon tracing out the two auxiliary oscillators which leaves us with a four parameters family over which we have to minimize the complexity. As explained in section \ref{genre} we can use $SO(2)$ rotations in the $y,z$ directions to eliminate a one parameter degeneracy among purifications with equal complexity.
The simplest choice is to eliminate $\alpha_{13}$, since then
\begin{equation}
\ket{\psi_{123}} \big|_{\alpha_{13}=0}= S(\alpha_{23}) S(r_1)S(r_2)S(r_3) \ket{\text{TFD}}_{12} \,,
\end{equation}
where the TFD state was defined in eq.~\eqref{two_mode} and is taken to have an inverse temperature $\beta_1$ where $\tanh \alpha_{12}= e^{-\beta_1 \omega/2}$.
Tracing out the second and third oscillators leaves us with
\begin{equation}\label{The trace}
\tr_{23} \(\ket{\psi_{123}} \bra{\psi_{123}}\) = S(r_1) \hat{\upsilon}_{\text{th}} (\beta_1 ,\omega )S^\dagger(r_1) = \hat{\rho}_1.
\end{equation}
In other words, the parameters $r_1,\alpha_{12}$ fully parametrize the mixed state $\hat{\rho}_1$ and the parameters $r_2,r_3, \alpha_{23}$ are totally free for the different purifications.
By choosing the simple reference state $\ket{\psi_R} = \ket{0_1(\mu),0_2(\mu),0_3(\mu)} $,
the purification complexity with the $F_1$ cost function in the diagonal basis is given by
\begin{equation}
\begin{split}\label{330}
\mathcal{C}_1^{\text{diag}}(\hat{\rho}_1;\psi_{123}) &= \text{min}_{r_2,r_3,\alpha_{23}} \, \frac{1}{2} \( \left|\ln \frac{\omega_1}{\mu}\right|+\left|\ln \frac{\omega_2}{\mu}\right| +  \left|\ln \frac{\omega_3}{\mu}\right| \)\,, \\
\end{split}
\end{equation}
where $\omega_i $ are the three eigenvalues of $A_{123}$ and the minimization is taken among the three free parameters $r_2$, $r_3$, $\alpha_{23}$.

We have performed this minimization numerically for $\bar{r}_1\equiv r_1  +  \frac{1}{2} \ln \frac{\omega}{\mu}=1$, see eq.~\eqref{rbar}, and $0\leq\alpha\leq 2$ and found
$|\mC_1^{\text{diag}}\left(\hat{\rho}_1;\psi _{123}\right)-\mC_1^{\text{diag}}\left(\hat{\rho}_1\right)|\leq 3 \cdot 10^{-8}$
and for $\alpha=1$ and $|\bar{r}_1|\leq 3$ and found
$|\mC_1^{\text{diag}}\left(\hat{\rho}_1;\psi _{123}\right)-\mC_1^{\text{diag}}\left(\hat{\rho}_1\right)|\leq 3 \cdot 10^{-7}$,
where $\mC_1^{\text{diag}}\left(\hat{\rho}_1\right)$ is the complexity obtained for a two mode purification in eq.~\eqref{complexity_one_mode}.
For the region with $\alpha > |\bar{r}_1|$, where the optimal purification is given by the TFD state the accuracy becomes extremely high and we have $|\mC_1^{\text{diag}}\left(\hat{\rho}_1\right)-\mC_1^{\text{diag}}\left(\hat{\rho}_1;\psi _{123}\right)| \le 10^{-15}$.
These small deviations from zero can be understood as resulting from the numerical accuracy of our minimization procedure.
The numerical simulations indicate that it is not necessary to introduce one more mode for the purification of one-mode Gaussian states in accord with our analytic proof in section \ref{threeA}.

\section{Complexity Basis Dependence}\label{app:4HO}
In this paper, we refer to two different bases for the definition of the ${\cal C}_1$ complexity: the \emph{diagonal basis} and the \emph{physical(-ancilla) basis}. In addition, for coupled harmonic oscillators representing a lattice quantum field theory, a natural basis to consider is the original \emph{position basis}, where each harmonic oscillator represents a position in the lattice. To help clarify the difference and relations between these bases, in this appendix we explicitly construct two examples of a discretized free scalar field theory on a lattice and write down the wavefunction matrix of the ground state in each of these three bases. We first look at the example of two coupled harmonic oscillators and find the complexity in the diagonal basis and in the physical basis. For the case of two coupled harmonic oscillators, there is no distinction between physical basis and position basis. In the next example of four coupled harmonic oscillators, we explicitly find the parameter matrices in the three bases. The position basis and the physical basis are different in this case, and we show that the ground state can be understood as the thermofield double of a two harmonic oscillator modular Hamiltonian, which we explicitly write. The physical basis modes are the eigenmodes of the modular Hamiltonians of each subregion.

Before going into the two specific examples, we explicitly rewrite some of the formulas in section~\ref{apply01} for the one dimensional case to describe the one-dimensional chain of $N$ harmonic oscillators. We begin with the lattice of harmonic oscillators~\eqref{ham88} realizing a regularization of a free quantum field theory~\eqref{Ha_scalarQFT} on a one-dimensional circle of length $L$ corresponding to the Hamiltonian\footnote{The following are the one dimensional versions of eqs.~\eqref{ham88}, \eqref{Thefreq} and \eqref{normal_QFT}.}
\begin{equation}
\label{eq:Hregulated}
\begin{split}
H
&=\frac{1}{2M}\sum^{N}_{a=1}[\bar p_a^2+M^2\bar\omega^2 \bar x_a^2+M^2\Omega^2( \bar x_a- \bar x_{a+1})^2]\,,
\end{split}
\end{equation}
where we have defined $\bar x_n\equiv\delta \phi(n)$, $\bar p_n\equiv\pi(n)$, $\bar\omega\equiv m$ and $\Omega=M\equiv 1/\delta$, see, \eg \cite{qft1}, and assumed periodic boundary conditions $\bar x_{N+1}:=\bar x_1$. The lattice spacing $\delta$ is related to the size of the system and the number of harmonic oscillators by $\delta = L/N$.
The Hamiltonian can be written in terms of normal modes as in eq.~\eqref{Thefreq}
\begin{equation}
\label{eq:Fourier}
x_k\equiv\frac{1}{\sqrt{N}}\sum^{N}_{a=1}\exp\left(\frac{2\pi ik}{N}a\right)\bar x_a, \quad \omega_k^2= \bar\omega^2 + 4\Omega^2 \sin^2 \frac{\pi k}{N}\,,
\end{equation}
where $k\in {1,\ldots N}$ (see, e.g., section 5.1 of \cite{Chapman:2018hou}). Using these degrees of freedom, the Hamiltonian reads~\eqref{normal_QFT}
\begin{equation}\label{gracht}
H=\frac{1}{2M}\sum^{N}_{k=1}\(|{p}_k|^2+M^2{\omega}_k^2|x_k|^2\),
\end{equation}
where we have used that $x^\dagger_k= x_{N-k}$.
The ground state wavefunction of this system of harmonic oscillators is straightforward to find in normal mode basis and is given by
eq.~\eqref{eq:ground-normal}. This can be explicitly written in the physical basis using the transformation~\eqref{Thefreq} and is given by eqs.~\eqref{eq:Mpos2}-\eqref{eq:Mpos}.

\subsection{Example 1: Two Coupled Harmonic Oscillators}
We will start by considering a simple toy model of two coupled harmonic oscillators with Hamiltonian\footnote{A similar toy model was considered in the context of complexity of pure states in reference \cite{qft1}.}
\begin{equation}\label{2_hamiltonian}
\begin{split}
H_{12} &= \frac{1}{2} \( p_1^2 +p_2^2 +\bar\omega^2(x_1^2+x_2^2) +\Omega^2(x_1-x_2)^2 \)  \\
&=   \frac{1}{2} \( p_+^2 +p_-^2 +\Omega_+^2 x_+^2+\Omega_-^2 x_-^2 \)  ,
\end{split}
\end{equation}
where the normal-mode coordinates are $x_\pm = \frac{1}{\sqrt{2}}(x_1 \pm x_2)$ and the normal-mode frequencies are $\Omega_+^2 =\bar\omega^2 < \Omega_-^2=\bar\omega^2 + 2\Omega^2$, and where we have set the mass of the oscillators $M$ to one, as in the bulk of the paper. This corresponds to the $N=2$ case of~\eqref{eq:Hregulated}. The corresponding ground state wave function is given by
\begin{equation}\label{pure_2HO}
\begin{split}
\psi_0(x_+,x_-) &=  \(\frac{\Omega_+\Omega_-}{\pi^2} \)^{1/4} \exp \( -\frac 12 \(\Omega_+ x_+^2 +\Omega_-x_-^2 \) \).
\end{split}
\end{equation}
Restoring the dependence of this wavefunction on the original coordinates $x_1$ and $x_2$ we obtain
\begin{equation}
\label{eq:groundreal}
\psi_{12}=\psi_0 = \left(\frac{\Omega_-\Omega_+}{\pi^2} \right)^{1/4} \exp \left(- \frac{\Omega_-+\Omega_+}{4} (x_1^2+x_2^2) +\frac{\Omega_--\Omega_+}{2} x_1 x_2\right)\,.
\end{equation}
We can then compare the wave function to the one in eq.~\eqref{wfunction1} (see also \eqref{transform_paras} and \eqref{rbar}) and find an easy translation to the notation of section~\ref{sec:onemode}
\begin{equation}\label{vacuum_parameters}
\bar{r}=\bar{s}= \frac14\ln \(\frac{\Omega_-\Omega_+}{\mu^2}\) \,, \quad  \alpha = \frac14 \ln \( \frac{\Omega_-}{\Omega_+}\)\,.
\end{equation}
For the reference state
\begin{equation}\label{reference_state}
\begin{split}
\psi_R(x_+,x_-) &=  \sqrt{\frac{\mu}{\pi}  }\exp \( -\frac 12 \(\mu x_1^2 +\mu x_2^2 \) \)\,,
\end{split}
\end{equation}
and using the entangling and scaling gates as the fundamental set of gates, the diagonal basis complexity of the ground state using Nielsen's geometric method with the $F_1$ cost function is (see  eq.~\eqref{complexity_pure} or \cite{qft1}),
\begin{equation}
\begin{split}\label{compureappC}
\mathcal{C}_{1}^{\text{diag}}\( \psi_0 \) = \frac 12 \left|\ln \frac{\Omega_+}{\mu}\right|+\frac 12\left|\ln \frac{\Omega_-}{\mu}\right| \,.
\end{split}
\end{equation}
In the following, we want to find the complexity of a subsystem given by the first harmonic oscillator after tracing out $x_2$.
Tracing out the second oscillator, we obtain a density matrix of the form~\eqref{dense1} for the first oscillator with parameters (see eq.~\eqref{ABA})
\beq
a=\frac{(\Omega_++ \Omega_- )^2+4\Omega_+\Omega_-}{4(\Omega_++\Omega_-)}\,, \quad b = \frac{(\Omega_+ - \Omega_-)^2}{4(\Omega_-+\Omega_-)}\,.
\eeq
As we already know from the discussion in section \ref{warmup}, the purification complexity for a subregion consisting of the first oscillator is not necessarily given by eq.~\eqref{compureappC} since the optimal purification is not necessarily the original pure state in eq.~\eqref{pure_2HO}. The purification complexity can be read by substituting the parameters $\alpha$ and $\bar{r}$ from  eq.~\eqref{vacuum_parameters} into eq.~\eqref{complexity_one_mode}.  The original ground state~\eqref{pure_2HO} will not be the optimal purification for the subregion except for $\Omega_->\mu>\Omega_+$. More precisely, we can compare the values of $\bar{r}= \frac14 \ln \frac{\Omega_-\Omega_+}{\mu^2}$ and $\alpha = \frac14 \ln \frac{\Omega_-}{\Omega_+}$ and we see that the three cases of section~\ref{sec:onemode} translate to
\beq
\begin{aligned}
	&{\rm case\ 1:}  \quad &0 \leqslant \alpha \leqslant - \bar{r} \quad & \to \quad &  \mu \geqslant \Omega_- \geqslant \Omega_+\,,\\
	&{\rm case\ 2:} \quad  & \alpha \geqslant |\bar{r}| \quad & \to \quad &  \Omega_- \geqslant \mu \geqslant  \Omega_+\,,\\
	&{\rm case\ 3:}  \quad &0 \leqslant \alpha \leqslant  \bar{r}\quad & \to \quad & \Omega_- \geqslant \Omega_+ \geqslant  \mu \,.
\end{aligned}
\eeq
Indeed, it is straightforward to confirm that when we are in case 2, the original ground states is the one with minimal complexity given by
\begin{equation}
{\cal C}_{1}^{\text{diag}}\( \rho_{\text{sub}} \) = 2\alpha=\frac12 \ln \frac{\Omega_-}{\Omega_+}= \mathcal{C}_{1}^{\text{diag}}\( \psi_0 \)\,,
\end{equation}
since $\bar{s} = \bar{r}=\bar{s}_{\text{min}}$ as was found in eq.~\eqref{case2compl}. In contrast, for $\mu $ outside of that region, for example in the case $\mu>\Omega_->\Omega_+$, the complexity of the optimal purification is
\begin{equation}
\begin{split}
{\cal C}^{\text{diag}}_{1}\( \rho_{\text{sub}}\)&= \frac 12 \ln \left(    \frac{{e^{-2 \bar{r}} \cosh 2\alpha-1}}{{1-e^{2 \bar{r}}\cosh 2\alpha}} \right) \\
&= \frac12 \ln \left[\frac{\mu^2}{\Omega_-\Omega_+} \frac{\Omega_-+\Omega_+-2\Omega_-\Omega_+/\mu}{2\mu-\Omega_--\Omega_+}\right]< \frac{1}{2} \ln \frac{\Omega_-\Omega_+}{\mu^2}\,,
\end{split}
\end{equation}
which means that the original pure state is not the optimal purification.

For completeness, we explicitly write the optimal purifications for each of the three cases. For case 2, as mentioned above, the optimal purification is the ground state of the two harmonic oscillator system itself. This corresponds to the state~\eqref{pure_2HO} or equivalently \eqref{eq:groundreal} in terms of the original $(x_1,x_2)$ coordinates.
For the other two cases, the optimal purification takes the following form, see eqs.~\eqref{house}, \eqref{rbar} and \eqref{case13r2},
\begin{equation}
\psi_{12} = \left(\frac{\Omega_-\Omega_+}{\pi^2}f \right)^{1/4} \exp \Bigg(-\frac{\Omega_-+\Omega_+}{4}\left(x_1^2 + f x_2^2\right) +\frac{\Omega_--\Omega_+}{2} \sqrt{f}\,x_1x_2 \Bigg)\,,
\end{equation}
where
\begin{equation}
f=\frac{\Omega_-+\Omega_+-2\,\mu}{2\,\Omega_- \Omega_+/\mu-\Omega_--\Omega_+}\,.
\end{equation}
In the case of two harmonic oscillators, there is no distinction between the physical-ancilla basis and the position basis. This is because the ``submatrices'' $\Gamma = \Omega = \frac{\Omega_-+\Omega_+}{2}$ in eq.~\eqref{Gaussian_AB} are simply a number and are therefore already diagonal. Thus, the ground state expressed in the position basis~\eqref{eq:groundreal} is also expressed in terms of the physical-ancilla modes.

Lastly, we mention the relation of these purifications to the TFD of the single harmonic oscillator. As observed in section~\ref{warmup}, in the comments around eq.~\eqref{scaling_rels}, the mixed state obtained after tracing out one of the oscillators corresponds to a thermal state with modified frequency $\omega'=\omega e^{2r} = \mu e^{2\bar r}=\sqrt{\Omega_-\Omega_+}$ at an inverse temperature of $\beta' \omega' = 2\, {\rm arcosh}\left(\frac{\Omega_-+\Omega_+}{\Omega_--\Omega_+}\right)$, see also eq.~\eqref{hope2}. Comparing the ground state parameter matrix~\eqref{eq:groundreal} with eqs.~\eqref{house} and~\eqref{hope}, the optimal purification can be seen to correspond to the TFD of two harmonic oscillator at this modified temperature and frequency when we are in case 2, that is, when $\Omega_- >\mu >\Omega_+$.

\subsection{Example 2: Four Coupled Harmonic Oscillators}

We restrict to the example of a lattice of four harmonic oscillators  with the goal of explicitly providing an example of the ground state in the normal mode basis, in position basis and in the physical(-ancilla) basis. We will express these in terms of the parameter matrix $M$ used throughout the main body of the paper. That is, we use $M_{\rm basis}$ to represent the state\footnote{We use the generalization of eq. \eqref{wavematrix} for a complex basis. This will be necessary since the Fourier transformation in eq.~\eqref{eq:Fourier} yields complex normal modes.}
\beq
\label{eq:Mbasis}
\Psi_0(x_{\rm basis}) = \left( \det \left( \frac{M_{\rm basis}}{\pi}\right) \right)^{1/4}
\exp \left[ -\frac12 x^\dagger_{\rm basis}
M_{\rm basis} x_{\rm basis}\right]\,.
\eeq
The state we are interested in is the ground state of the free QFT lattice Hamiltonian consisting of four coupled harmonic oscillators, \ie the $N=4$ case of~\eqref{eq:Hregulated}. This state was already written in normal mode basis in eq.~\eqref{eq:ground-normal}. For a lattice of four harmonic oscillators, the normal modes $x_k\equiv(x_1,x_2,x_3,x_4)^{T}$ are related to the original physical basis modes $\bar{x}_a \equiv(\bar x_1,\bar x_2,\bar x_3,\bar x_4)^{T}$ by eq.~\eqref{eq:Fourier}, namely
\beq
\label{eq:Rmat}
x = R \bar{x} \,,\quad {\rm where} \quad R= \frac12 \begin{pmatrix}
    i & -1 & -i & 1\\
    -1 & 1 & -1 & 1\\
    -i & -1 & i & 1\\
    1 & 1 & 1 & 1
\end{pmatrix} \,,
\eeq
or, explicitly
\begin{equation}
\begin{aligned}
x_1 &= \frac12 \left(i \bar{x}_1 - \bar{x}_2 - i \bar{x}_3 + \bar{x}_4\right)\,, \quad \quad \quad\, \, \, \,
x_2 = \frac12 \left(- \bar{x}_1 + \bar{x}_2 -  \bar{x}_3 + \bar{x}_4\right)\,,\\
x_3 &= \frac12 \left(-i \bar{x}_1 - \bar{x}_2 + i \bar{x}_3 + \bar{x}_4\right)\,, \quad\quad \quad
x_4 = \frac12 \left(\bar{x}_1 + \bar{x}_2 + \bar{x}_3 + \bar{x}_4\right)\,.
\end{aligned}
\end{equation}

Notice that, while the position basis degrees of freedom are real valued, this is not the case for the normal mode degrees of freedom where, in particular $x_1^* = x_3$ so that $x_{\rm normal}^\dagger = (x_3,x_2,x_1,x_4)$.\footnote{Recall that the Fourier transform obeys the identity  $x^\dagger_{ k}= x_{N- k}$, see comment below eq.~\eqref{gracht}.} The parameter matrix in normal mode basis can easily be read off   eqs.~\eqref{eq:ground-normal} and \eqref{eq:Fourier}
\beq
\label{eq:Mnormal}
M_{\rm normal} = \begin{pmatrix}
    \bar{\omega}_1 & 0 & 0 & 0\\
    0 & \bar{\omega}_2 & 0 & 0\\
    0 & 0 &\bar{\omega}_3 & 0\\
    0 & 0 & 0 & \bar{\omega}_4
\end{pmatrix}\,,
\eeq
where
\beq
\bar{\omega}_1=\bar{\omega}_3 = \sqrt{\bar\omega^2+2\sqrt{2}\Omega^2}\,, \quad \bar{\omega}_2 = \sqrt{\bar\omega^2+4\Omega^2}\,, \quad \bar{\omega}_4 = \bar\omega\,.
\eeq
The fact that the parameter matrix in normal mode basis is diagonal reflects the fact that there is no entanglement between normal mode degrees of freedom.\footnote{Note that substituting the parameter matrix \eqref{eq:Mnormal} into the bi-linear form in eq.~\eqref{eq:Mbasis} yields a wavefunction whose dependence on the $x_1$ and $x_3$ coordinates is of the form $\Psi_0 \propto \exp\left[ - \alpha (|x_1|^2 + |x_3|^2) \right] =   \exp \left[-2\alpha x_1 x_3\right]$, where $\alpha=\frac{1}{2} \bar\omega^2 + \sqrt{2} \Omega^2$. So although the form seems orthogonal in complex coordinates, it does not look orthogonal when reexpressing the conjugate coordinates in terms of the original ones. This is due to the fact that the normal mode basis given by eq.~\eqref{eq:Fourier} is not Hermitian. This awkward dependence on the product of seemingly different degrees of freedom can be removed by using a real Fourier transformation involving $\sin(\cdots)$ and $\cos(\cdots)$ instead of the complex exponentials in eq.~\eqref{eq:Fourier}. An equivalent way of getting rid of this dependence is to make a second transformation $x_k^{\rm real} =\frac12 \left(x_k+x_k^*\right)$ and $x_{N-k}^{\rm real} =\frac{1}{2i} \left(x_k-x_k^*\right)$ for those values of $k$ for which $x_k$ are not real.}

The physical basis parameter matrix can be found by applying the transformation~\eqref{eq:Rmat} to the normal mode basis parameter matrix~\eqref{eq:Mnormal}
\beq
M_{\rm pos} = R^\dagger M_{\rm normal} R
\eeq
or simply be read off eq.~\eqref{eq:Mpos}. Either way, for our four harmonic oscillator example it takes the form
\footnotesize
\beq\label{eq:Mpos4}
M_{\rm pos} =\frac14 \begin{pmatrix}
   \bar\omega +\bar{\omega}_2+ 2\bar{\omega}_1 & \bar\omega-\bar{\omega}_2 & \bar\omega+\bar{\omega}_2-2\bar{\omega}_1 &\bar\omega-\bar{\omega}_2 \\
    \bar\omega-\bar{\omega}_2 & \bar\omega+\bar{\omega}_2 + 2\bar{\omega}_1& \bar\omega-\bar{\omega}_2 & \bar\omega + \bar{\omega}_2-2\bar{\omega}_1\\
   \bar\omega + \bar{\omega}_2-2 \bar{\omega}_1 & \bar\omega-\bar{\omega}_2 &  \bar\omega+\bar{\omega}_2 + 2\bar{\omega}_1& \bar\omega-\bar{\omega}_2\\
 \bar\omega-\bar{\omega}_2 & \bar\omega+\bar{\omega}_2-2 \bar{\omega}_1 & \bar\omega-\bar{\omega}_2 &\bar\omega +\bar{\omega}_2 + 2\bar{\omega}_1
\end{pmatrix}\,.
\eeq
\normalsize
The form of the parameter matrix makes evident that the position basis degrees of freedom are entangled with each other. Furthermore, the entanglement decays for longer distances since $\bar\omega < \bar{\omega}_1 < \bar{\omega}_2$ implies $|\bar\omega-\bar{\omega}_2| > |\bar\omega+\bar{\omega}_2-2\bar{\omega}_1|$.\footnote{Recall that our lattice is periodic and so the sites $\bar x_1$ and $\bar x_4$ are nearest neighbors.} This is to be expected for entanglement being spread by nearest neighbor interactions coming from the discretized kinetic term (the last term in~\eqref{eq:Hregulated}).

The position and normal mode basis should be familiar to most readers; they are the lattice equivalents of the position and momentum bases in quantum field theory. The physical-ancilla basis is less familiar. In~\cite{Chapman:2018hou}, it appears under the name \emph{left/right} basis since it was used in the context of the TFD state, which is considered a natural purification of the thermal state where the left/right division corresponds to the physical degrees of freedom of the thermal system and the ancilla degrees of freedom introduced in order to purify it.\footnote{Of course, when talking about the TFD it is ambiguous which of the sides we should consider as the physical system and which side represents the ancillae since tracing out either side will reproduce the thermal density matrix.}

To define the physical-ancilla basis, we must partition the system into a physical subsystem and an ancilla subsystem. In other words, we consider the four harmonic oscillator ground state~\eqref{eq:Mbasis} as a purification of a mixed state of a subset of the oscillators. This is an important property of the physical-ancilla basis: it depends on a specific partition of the full system. In our example, we will choose to partition the system in two: the $\bar{x}_1$ and $\bar{x}_2$ oscillators as one subsystem and the $\bar{x}_3$ and $\bar{x}_4$ oscillators as the other subsystem. Which subsystem we call physical and which one ancilla depends on which degrees of freedom are traced out in order to construct the given two-mode mixed state.

With this partition in mind, we can decompose the physical basis parameter matrix~\eqref{eq:Mpos4}, as in eq.~\eqref{Gaussian_AB}, into\footnote{In section \ref{sec:purificationX} we introduced the decomposition~\eqref{Gaussian_AB}
    \beq
    M_{\rm pos} = \begin{pmatrix}
        \Gamma & K \\
        K^T & \Omega
    \end{pmatrix} \,,
    \eeq
    which has the unfortunate notation $\Omega$ for the lower right sub-matrix. In the following, we use instead the letter $\Sigma$ to denote this sub-matrix in order to avoid confusion with the oscillator coupling $\Omega$ in eq.~\eqref{eq:Hregulated}.}
\beq
M_{\rm pos} = \begin{pmatrix}
    \Gamma & K \\
    K^T & \Sigma
\end{pmatrix}
\eeq
where
\begin{equation}
\begin{split}
\Gamma = \,& \Sigma = \frac14 \begin{pmatrix}
    \bar\omega + \bar{\omega}_2 +2\bar{\omega}_1 & \bar\omega-\bar{\omega}_2 \\
   \bar\omega-\bar{\omega}_2  & \bar\omega + \bar{\omega}_2 +2\bar{\omega}_1\end{pmatrix} \,,
    \\
    \quad K =\,& \frac14 \begin{pmatrix}
    \bar\omega + \bar{\omega}_2 -2\bar{\omega}_1 & \bar\omega-\bar{\omega}_2 \\
    \bar\omega-\bar{\omega}_2  & \bar\omega + \bar{\omega}_2-2\bar{\omega}_1\end{pmatrix}\,.
\end{split}
\end{equation}

The physical-ancilla basis is defined as the basis which diagonalizes the sub-matrices $\Gamma$ and $\Sigma$ without mixing the two subsystems. More precisely, we look for transformations of the form
\beq
\label{eq:Rphys-anc}
R_{\rm phys-anc}= \begin{pmatrix}
    R_{\rm phys} & 0\\0 & R_{\rm anc}
\end{pmatrix}
\eeq
that diagonalize both $\Gamma$ and $\Sigma$. In our example, this transformation is given by
\beq
R_{\rm phys}= \frac{1}{\sqrt{2}} \begin{pmatrix}
    1 & 1 \\
    -1 & 1
\end{pmatrix}\,,\quad R_{\rm anc}= \frac{1}{\sqrt{2}} \begin{pmatrix}
1 & 1 \\
1 & -1
\end{pmatrix}\,,
\eeq
or, explicitly
\beq
\begin{aligned}
    x_1^{\rm phys} &= \frac{1}{\sqrt{2}}(\bar{x}_1+\bar{x}_2)\,, \quad \quad \quad x_2^{\rm phys} = \frac{1}{\sqrt{2}}(\bar{x}_2-\bar{x}_1)\,,\\
    x_3^{\rm phys} &= \frac{1}{\sqrt{2}}(\bar{x}_3+\bar{x}_4)\,, \quad \quad \quad x_4^{\rm phys} = \frac{1}{\sqrt{2}}(\bar{x}_3-\bar{x}_4)\,.
\end{aligned}
\eeq
The physical-ancilla basis parameter matrix can be found by applying the transformation~\eqref{eq:Rphys-anc} to the position basis parameter matrix~\eqref{eq:Mpos4}
\beq
\label{Mphys}
M_{\rm phys} = \frac12 \begin{pmatrix}
    \bar\omega + \bar{\omega}_1 & 0 &\bar\omega - \bar{\omega}_1 & 0 \\
    0 & \bar{\omega}_2 +\bar{\omega}_1 & 0 & \bar{\omega}_1 -\bar{\omega}_2\\
    \bar\omega - \bar{\omega}_1 & 0 & \bar\omega + \bar{\omega}_1 & 0 \\
    0 &\bar{\omega}_1 -\bar{\omega}_2 & 0 & \bar{\omega}_2 +\bar{\omega}_1
\end{pmatrix}\,.
\eeq
In this basis, there is no entanglement between the modes in each subsystem ($x_1^{\rm phys}$ is not entangled with $x_2^{\rm phys}$ and similarly for $x^{\rm phys}_3$ and $x_4^{\rm phys}$). However, the entanglement between the two subregions cannot be removed by transformations of the form~\eqref{eq:Rphys-anc}. Consequently, the modes between regions remain entangled. In our case, the state factorizes to a product state form where $x_1^{\rm phys}$ is entangled with $x_3^{\rm phys}$ and $x_2^{\rm phys}$ with $x_4^{\rm phys}$.
Bellow we will also see that the ground state is the TFD for a 2 harmonic oscillator modular Hamiltonian.

To see this, we compare the physical basis parameter matrix to the thermal parameters by using eqs.~\eqref{house} and \eqref{hope2} for each factor of the factorized state \eqref{Mphys}. First, focusing on the $x_1^{\rm \phys}$ and $x_3^{\phys}$ modes, we see that they are in a TFD state with inverse temperature $\beta_{13}$ and frequency $\omega_{13}$ given by
\beq
\beta_{13}\omega_{13} = 2\, {\rm arcosh}\left(\frac{\bar\omega+\bar{\omega}_1}{\bar{\omega}_1-\bar\omega}\right)\,, \quad \omega_{13} e^{2r_{13}} = \sqrt{\bar\omega\, \bar{\omega}_1}\,,
\eeq
and the $x_2^{\rm \phys}$ and $x_4^{\phys}$ modes are in a TFD state with inverse temperature $\beta_{24}$ and frequency $\omega_{24}$ given by
\beq
\beta_{24} \omega_{24} = 2\, {\rm arcosh}\left(\frac{\bar{\omega}_1+\bar{\omega}_2}{\bar{\omega}_2-\bar{\omega}_1}\right)\,, \quad \omega_{24} e^{2r_{24}} = \sqrt{\bar{\omega}_1\, \bar{\omega}_2}\,.
\eeq
For these to have the same inverse temperature $\beta_0$ we must fix\footnote{The temperature is a free parameter because the modular Hamiltonian can always be rescaled to change the value of $\beta_0$. However, the dimensionless products $\omega \beta_0$ will remain fixed.}
\beq
e^{-2r_{13}} = \frac{2}{\beta_0 \sqrt{\bar\omega\,\bar{\omega}_1}} \, {\rm arcosh}\left(\frac{\bar\omega+\bar{\omega}_1}{\bar{\omega}_1-\bar\omega}\right)\,, \quad
e^{-2r_{24}} = \frac{2}{\beta_0 \sqrt{\bar{\omega}_1\,\bar{\omega}_2}} \, {\rm arcosh}\left(\frac{\bar{\omega}_1+\bar{\omega}_2}{\bar{\omega}_2-\bar{\omega}_1}\right)\,,
\eeq
which leads to the following frequencies of the Rindler modes
\beq
\label{eq:freqs}
\omega_{13} = \frac{2}{\beta_0} \,{\rm arcosh}\left(\frac{\bar\omega+\bar{\omega}_1}{\bar{\omega}_1-\bar\omega}\right)\,, \quad
\omega_{24} = \frac{2}{\beta_0} \,{\rm arcosh}\left(\frac{\bar{\omega}_1+\bar{\omega}_2}{\bar{\omega}_2-\bar{\omega}_1}\right)\,.
\eeq
Lastly, we can take the small frequency limit (or equivalently, small cutoff limit) $\bar\omega \ll \Omega = 1/\delta$ to see that the $x_1^{\rm phys}$ and $x_3^{\rm phys}$ are the zero modes and that the $x_2^{\rm phys}$ and $x_4^{\rm phys}$ modes have frequencies  proportional to the temperature
\beq
\omega_{13} = \frac{2^{13/8}}{\beta_0} \sqrt{\frac{\bar\omega}{\Omega}}\,, \quad \omega_{24} = \frac{2}{\beta_0}\, {\rm arcosh}\left(\frac{2+2^{3/4}}{2-2^{3/4}}\right) \approx \frac{2\pi}{\beta_0}\,.
\eeq

Lastly, we can explicitly write the modular Hamiltonian of the $x_1^{\rm phys}$ and $x_2^{\rm phys}$ system from the expression of their frequencies~\eqref{eq:freqs}
\beq
\begin{aligned}
H_{\rm mod} = &\frac{1}{2M_0} \left(p_1^{\rm phys}\right)^2 + \frac{2\, M_0}{\beta_0^2}\, {\rm arcosh}^2\left(\frac{\bar\omega+\bar{\omega}_1}{\bar{\omega}_1-\bar\omega}\right) \left(x_1^{\rm phys}\right)^2\\
&+\frac{1}{2M_0} \left(p_2^{\rm phys}\right)^2 + \frac{2\, M_0}{\beta_0^2}\, {\rm arcosh}^2\left(\frac{\bar{\omega}_1+\bar{\omega}_2}{\bar{\omega}_2-\bar{\omega}_1}\right) \left(x_2^{\rm phys}\right)^2\,.
\end{aligned}
\eeq

\section{Holographic Subregion Complexity in the Poincar\'e Patch}\label{app:appsubCA4}
In this appendix, we summarize and extend the results in the literature regarding subregion complexity in holography. We start by summarizing the volume results from \cite{Alishahiha:2015rta,Carmi:2016wjl} for a ball-shaped subregion in general dimensions. After that, we discuss the subregion-CA complexity in the Poincar\'e patch, regulated in such a way that the WDW patch starts at the cutoff surface $z=\delta$ in Fefferman-Graham coordinates. This calculation was outlined in \cite{Carmi:2016wjl}. However, at the time the paper was written it was still not clear if the counterterm restoring reparametrization invariance is an essential ingredient of the complexity=action proposal. This later became clear, among other things, due to the fact that the counterterm is essential for obtaining the expected behavior in the presence of shocks, see \cite{Vaidya1,Vaidya2}. We briefly review the results of \cite{Carmi:2016wjl} and then extend them to include the counter term.

\subsection{Subregion-CV}
Here we summarize the results of \cite{Alishahiha:2015rta} (see eq.~(5)-(7)) as well as \cite{Carmi:2016wjl} (see eq.~(4.9)) for the subregion complexity using the CV conjecture for a ball shaped region on the boundary of AdS$_{d+1}$ in Poincar\'e coordinates. The bulk spacetime is described by the metric
\begin{equation}
ds^2 = \frac{L^2}{z^2}\left[dz^2 - dt^2 +d \rho^2 +\rho^2 d \Omega_{d-2}^2 \right].
\end{equation}
For a ball-shaped region on a constant time slice with $\rho\le R$, the complexity is given by performing the following integral
\begin{equation}\label{subCVamazing}
C_V = \frac{L^{d-1}\Omega_{d-2}}{(d-1) G_N} \int_\delta^R dz \frac{(R^2-z^2)^{\frac{d-1}{2}}}{z^d}
\end{equation}
where $\Omega_{d-2}=2 \pi^{\frac{d-1}{2}}/\Gamma\left(\frac{d-1}{2}\right)$ is the volume of the $S^{d-2}$ sphere and $R$ is the radius of the ball (or half the size of the interval for a two dimensional boundary).
The explicit results of this integration for $d=2$ (AdS$_3$) and $d=3$ (AdS$_4$) are presented in the main text in eqs.~\eqref{ads3subcv} and \eqref{ads4subcv}.

\subsection{Subregion-CA}
\begin{figure}[!ht]
	\centering
	\includegraphics[height=3in]{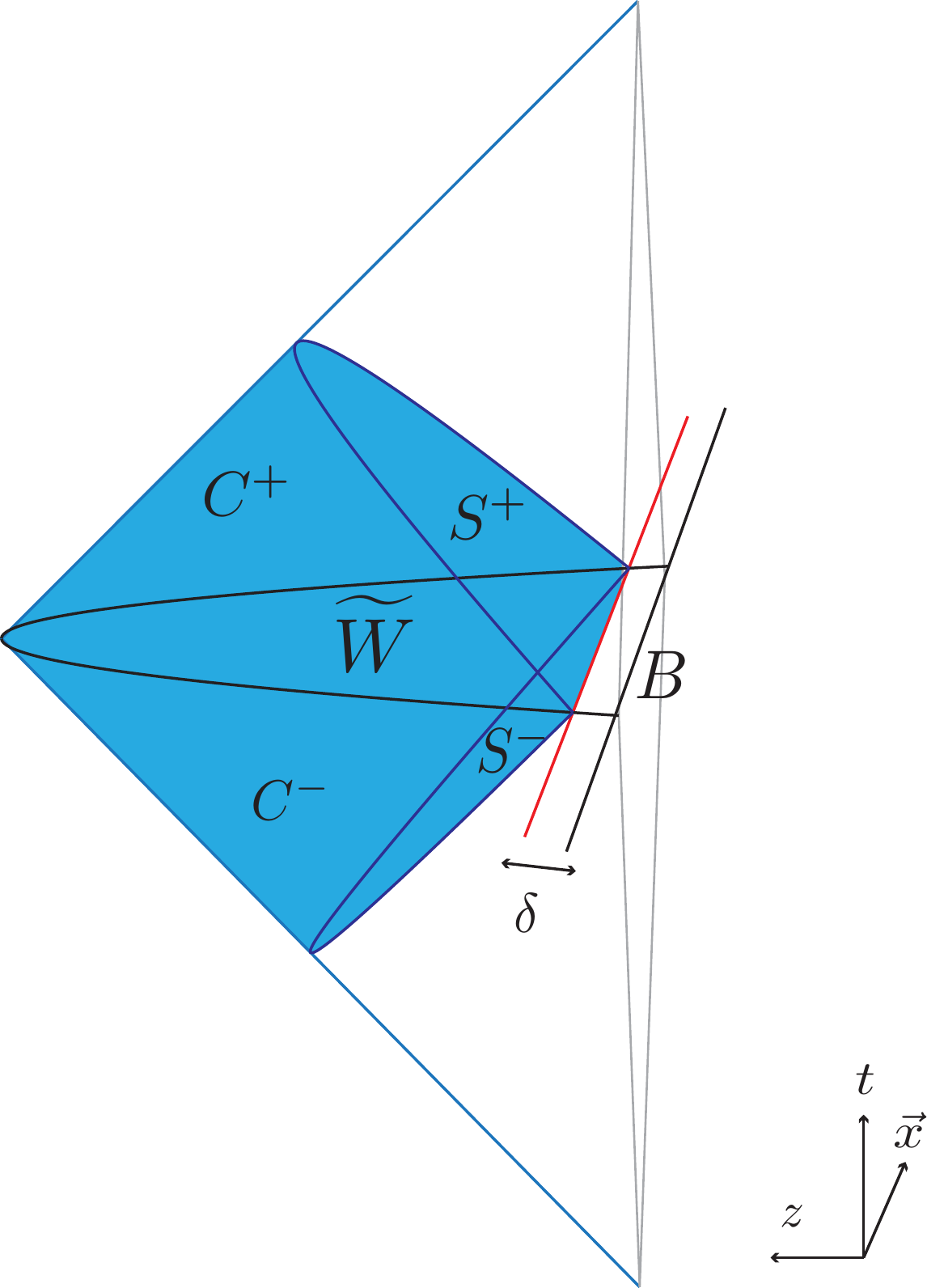}\hspace{2cm}
    \includegraphics[height=3in]{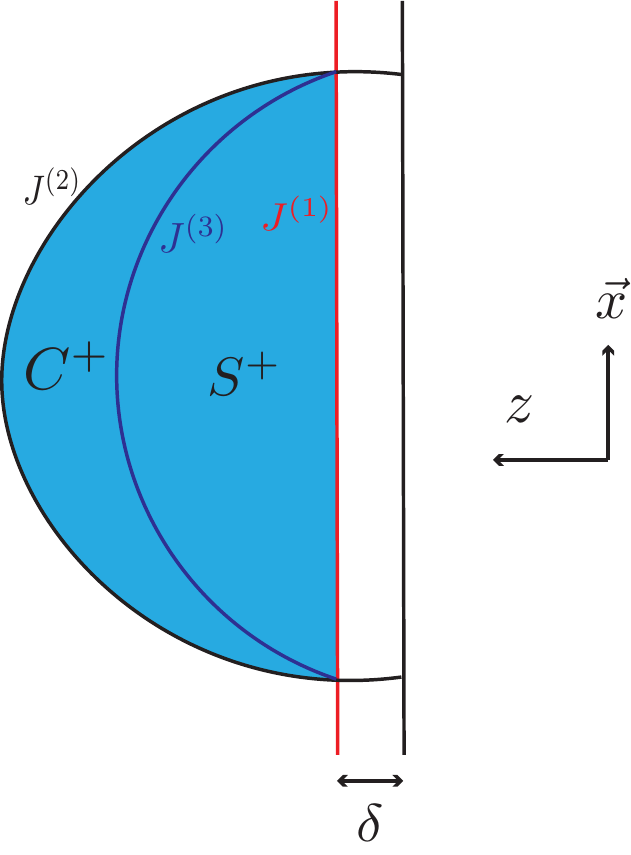}
	\caption{The intersection of the entanglement wedge and the WDW patch defines the  region $\mathcal{\widetilde{W}}$ that is   relevant for the evaluation of  $\mathcal{C}_A(B)$.}\label{fig:calV}
\end{figure}

The form of the intersection $\widetilde W$ between the WDW patch (starting at the cutoff surface) and the entanglement wedge is illustrated in figure \ref{fig:calV}, together with its projection on the $t=0$ time slice, where we label the various surfaces and joints required for the calculation. The region $\widetilde W$ is bounded by four surfaces. $S^{\pm}$ are the boundaries of the WDW patch and $C^{\pm}$ are the boundaries of the entanglement wedge. They are described by the following constraints
\begin{equation}\label{surfsurf}
S^{\pm}: \quad t= \pm (z-\delta), \qquad
C^{\pm}:\quad t=\pm (R-\sqrt{\rho^2+z^2}),
\end{equation}
where $R$ is the radius of the ball shaped subregion for which we evaluate the complexity. The affinely parameterized normals to the various surfaces are\footnote{Here we chose the direction such that the normal vectors are future oriented, in order to be consistent with the conventions of appendix C of \cite{Lehner:2016vdi}  which we use throughout the following.}
\begin{equation}\label{normalsnormals}
S^{\pm}: \quad k_{1,2}= \alpha (-dt \pm dz), \qquad
C^{\pm}: \quad k_{3,4}= \beta \left(-dt\mp \frac{\rho d\rho +z dz}{\sqrt{\rho^2+z^2}}\right).
\end{equation}

The subregion-CA conjecture consists of evaluating the gravitational action of the region $\widetilde W$. When the normals to the null surfaces are affinely parametrized the relevant contributions are: the bulk contribution $I_{\text{bulk}}$, the joints $J^{(1)}$, $J^{(2)}$ and (twice) $J^{(3)}$ (see figure \ref{fig:calV}), whose contributions we label $I^{(1)}$, $I^{(2)}$ and $I^{(3)}$, respectively, and finally the counterterm contribution required to render the result independent of the normalization constants $\alpha$ and $\beta$. This counterterm was first presented in appendix B of \cite{Lehner:2016vdi} and it reads
\begin{equation}\label{eq:ct}
I_{\text{ct}}=-\frac{1}{8 \pi \Gn}\int d \lambda \,  d^{d-1}x \sqrt{\gamma} \, \Theta \ln \left( \ell_{\text{ct}} |\Theta| \right),
\end{equation}
where the expansion parameter is $\Theta=\del_\lambda \ln \sqrt {\gamma}$, $\gamma$ is the metric on the light surface modulo light rays and $\ell_{\text{ct}}$ is an arbitrary constant representing the freedom in the definition of this counter term. The parameter $\lambda$ runs along the null generators of the light surface and has to be defined such that it matches our definition of the normal vectors $k^\mu = d x^\mu/ d\lambda$. Since the boundary of the entanglement wedge is a killing horizon with vanishing expansion \cite{Casini:2011kv,Faulkner:2013ica} we only have to include the counter term on the boundaries of the WDW patch $S^{\pm}$. Finally the complexity is given by
\begin{equation}
\mathcal{C}_A (B) = \frac{1}{\pi} \left(I_{\text{bulk}}+I^{(1)}+I^{(2)} + 2I^{(3)}+2 I_{\text{ct}} \right).
\end{equation}

Most of the contributions above were already evaluated in \cite{Lehner:2016vdi} and we quote the results here (fixing a few small typos).
For the bulk contribution we have
\begin{equation}\label{Ibulksubapp}
I_{\text{bulk}}=-\frac{d\, \Omega_{d-2} L^{d-1}}{4 \pi \Gn} \int_0^{\frac{R-\delta}{2}} dt \int_{t+\delta}^{R-t} \frac{dz}{z^{d+1}}\frac{((R-t)^2-z^2)^{\frac{d-1}{2}}}{d-1}.
\end{equation}
For the various joints we have\footnote{We have fixed the following factors: overall factor of $R$ in $I^{(2)}$ was missing, upper limit of integration in $I^{(3)}$ was changed to $\frac{R+\delta}{2}$.}
\begin{equation}
\begin{split}
I^{(1)} = & \, - \frac{ \Omega_{d-2} L^{d-1}}{4 \pi(d-1) \Gn} \frac{(R^2-\delta^2)^{\frac{d-1}{2}}}{\delta^{d-1}} \ln\left(\frac{\alpha \delta}{L}\right),
\\
I^{(2)} = & \, -\frac{L^{d-1} \Omega_{d-2} }{4 \pi \Gn} \int_\delta^R \frac{dz}{z^{d-1}} R (R^2-z^2)^{\frac{d-3}{2}}\ln\left(\frac{\beta z}{L}\right),
\\
I^{(3)} = & \,  \frac{L^{d-1} \Omega_{d-2} }{8 \pi \Gn} \int_\delta^{\frac{R+\delta}{2}} \frac{d\bar z}{\bar z^{d-1}} (R+\delta)^{\frac{d-1}{2}}
(R+\delta-2 \bar z)^{\frac{d-3}{2}} \ln\left(\frac{\alpha \beta \bar z^2 (R+\delta)}{2 L^2 (R+\delta-\bar z)}\right),
\end{split}
\end{equation}
where in $I^{(3)}$ we have relabeled the integration variable as $\bar z$ for reasons that will become clear in a moment. Recall that the boundaries of the WDW patch had vanishing expansion and so no counterterm was needed in order to cancel the dependence on the normalization constant $\beta$. To make this observation manifest let us use the following change of variables
\begin{equation}\label{zzbar}
\bar z = \frac{z (R+\delta)}{R+z}, \qquad z=\frac{R\, \bar z}{R+\delta-\bar z}\,,
\end{equation}
which relabels the various points on the joint $J^{(3)}$ by the corresponding value of $z$ on the joint $J^{(2)}$ along the same light ray originating from the point $z=\rho=0$, $t=R$. After this change of variables we are able to combine the contributions of the joints
$J^{(2)}$ and $J^{(3)}$ as follows
\begin{equation}
I^{(2)}+2 I^{(3)} = \frac{L^{d-1} \Omega_{d-2}}{4 \pi \Gn} \int_\delta^R \frac{dz}{z^{d-1}} R\, (R^2- z^2)^{\frac{d-3}{2}}  \ln\left(\frac{\alpha z (R+\delta)^2}{2(R+z) L R}\right),
\end{equation}
where we see explicitly that all the dependence on $\beta$ canceled out.

Next, we evaluate the contribution of the counterterm. For this purpose, we first identify the light-ray parameter $\lambda=-L^2/\alpha z$ which is consistent with the normal definition $k_1^{\mu} = dx^\mu/d\lambda$, see eq.~\eqref{normalsnormals}, along the surface $S^+$, see eq.~\eqref{surfsurf}. We then evaluate the expansion
\begin{equation}
\Theta = \del_\lambda \ln\sqrt{\gamma} = -\frac{\alpha (d-1) z}{L^2}.
\end{equation}
Finally the counter term contribution reads
\begin{equation}
I_{\text{ct}} = \frac{\Omega_{d-2}L^{d-1}}{8 \pi \Gn} \int_\delta^{\frac{R+\delta}{2}} \frac{d \bar z}{\bar z^d} (R+\delta)^{\frac{d-1}{2}}(R+\delta-2 \bar z)^{\frac{d-1}{2}} \ln\(\frac{\ell_{\text{ct}}\,\alpha (d-1)\bar z}{L^2}\).
\end{equation}
Once again, it will be useful to use the change of coordinates \eqref{zzbar}, which brings this contribution to the form
\begin{equation}\label{thecounterterm2}
I_{\text{ct}} = \frac{\Omega_{d-2}L^{d-1}}{8 \pi \Gn} \int_\delta^{R}  \frac{d z}{ z^d} R\, (R+z)^{\frac{d-3}{2}}(R- z)^{\frac{d-1}{2}} \ln\left(\frac{\ell_{\text{ct}}\,\alpha (d-1) z (R+\delta)}{L^2(R+z)}\right).
\end{equation}
Combining all the joints and the counter term and using integration by parts together with the identity $\int\frac{dz}{z^d}R^2 (R^2-z^2)^{\frac{d-3}{2}} = -\frac{1}{z^{d-1}}\frac{(R^2-z^2)^{\frac{d-1}{2}}}{d-1}$ finally yields
\begin{equation}
\begin{split}\label{Isjsubapp}
I_{\text{s,j,ct}} &\equiv 2 I_{\text{ct}}+I^{(1)}+I^{(2)}+ 2 I^{(3)}
=
\frac{\Omega_{d-2}L^{d-1}}{4 \pi \Gn} \int_\delta^R dz \frac{R\left(R^2 -z^2\right)^{\frac{d-3}{2}}}{z^{d-1}}
\times
\\
&~~~~~~~~~~~~~~~~~~~~~~~~~~~\times
\left[\frac{\left(R -z\right)
}{z  }
\left[\frac{1}{(d-1)}+\ln \left(\frac{\ell_{\text{ct}}(d-1)}{L}\right)\right] + \ln\left(\frac{R+\delta}{2 R}\right)\right],
\end{split}
\end{equation}
and we see that all the dependence on $\alpha$ has canceled.
The final result for the complexity is then given by combining eqs.~\eqref{Ibulksubapp} and \eqref{Isjsubapp}, \ie
\begin{equation}\label{CACAtotapp}
\mathcal{C}_A =\frac{1}{\pi}\left( I_{\text{bulk}}+I_{\text{s,j,ct}}\right).
\end{equation}
We evaluated this expression explicitly for the cases of $d=2$ and $d=3$ and the final results are given by eqs.~\eqref{3P} and \eqref{4P}.

\subsection{Full state CA in the Poincar\'e patch}
\begin{figure}[!ht]
	\centering
	\includegraphics[height=3in]{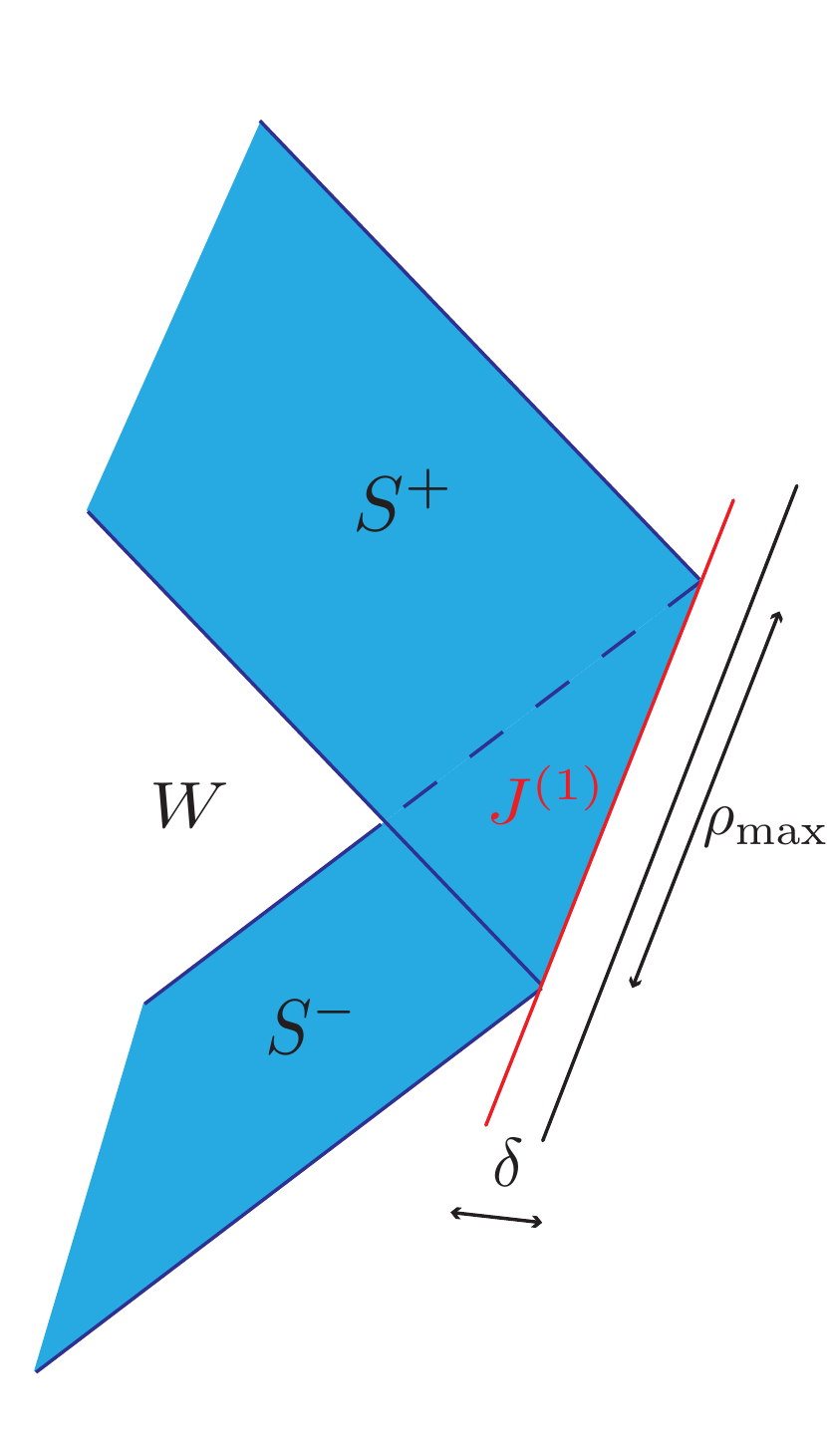}
	\caption{The WDW patch relevant for the evaluation of  $\mathcal{C}_A$ of the full region in the poincar\'e patch of AdS$_{d+1}$. We introduce an IR regulator $\rho_{\text{max}}$ in order to obtain a finite answer.}\label{fig:calAfull}
\end{figure}

For completeness, we also include the counter term contribution $I_{\text{ct}}$ in the full state CA calculation in the Poincar\'e patch of AdS$_{d+1}$. The other contributions to the full-state CA calculation appear in \cite{Carmi:2016wjl} and we review them below. The WDW patch starts at the cutoff surface $z=\delta$ and we use an IR regulator $\rho=\rho_{\text{max}}$ all the way through the bulk, see figure \ref{fig:calAfull}.

We start with the bulk contribution
\begin{equation}
\begin{split}\label{eq:Ibulkfs}
I_{\text{bulk}}&=-\frac{d\,  L^{d-1}\,\Omega_{d-2}}{4 \pi  G_N}\int_\delta ^\infty \frac{dz}{z^{d+1}}\int_0^{z-\delta} dt \int_0^{\rho_{\text{max}}} d\rho \,\rho^{d-2}\\
&=-\frac{ L^{d-1}\,\,\Omega_{d-2} }{4 \pi  G_N}\frac{\rho_{\text{max}}^{d-1}}{(d-1)^2} \frac{1}{\delta^{d-1}}\,.
\end{split}
\end{equation}
Next we consider the joint $J^{(1)}$, see figure \ref{fig:calAfull}, whose contribution reads
\begin{equation}
\begin{split}
I^{(1)} &= - \frac{ L^{d-1} \, \Omega_{d-2}}{4 \pi G_N (d-1)}\, \frac{\rho_{\text{max}}^{d-1}}{\delta^{d-1}}\, \ln\left(\frac{\alpha \delta}{L}\right)\label{eq:Ijfs}.
\end{split}
\end{equation}
Finally we include the counter term \eqref{eq:ct}\footnote{The surfaces $S^{\pm}$ only contribute the counter term since we chose their normals to be affinely parametrized.}
\begin{equation}
\begin{split}\label{eq:Ictfs}
I_{\text{ct}}&=\frac{L^{d-1}\Omega_{d-2}}{8\,\pi\, G_N} \rho_{\text{max}}^{d-1}\,\int_\delta^\infty \frac{dz}{z^d}\,  \,\ln\left(\frac{\ell_{\text{ct}} (d-1) \alpha}{L^2}\, z\right) \\
&=\frac{L^{d-1}\Omega_{d-2}}{8\,\pi\, G_N}\, \rho_{\text{max}}^{d-1}\frac{1}{(d-1)}\,\,\frac{1}{\delta^{d-1}}\left[\ln\left(\frac{\ell_{\text{ct}}(d-1)\,\alpha\,\delta}{L^2}\right)+ \frac{1}{d-1}\right]\,.
\end{split}
\end{equation}
Adding the bulk, joint and counterterm contributions in eqs.~\eqref{eq:Ibulkfs}-\eqref{eq:Ictfs}, we obtain
\begin{equation}
\begin{split}\label{eq:Cafullappd}
\mathcal{C}_A&= \frac{1}{\pi}\,(\, I_{\text{bulk}} + 2 I_{\text{ct}} + I^{(1)})
=\frac{L^{d-1}\Omega_{d-2}}{4\,\pi ^2 (d-1)\, G_N}\,\,\frac{\rho_{\text{max}}^{d-1}}{\delta^{d-1}}\,\ln\left(\frac{\ell_{\text{ct}}(d-1)}{L}\right).
\end{split}
\end{equation}

\section{Superadditivity of ${\cal C}_A(|\text{TFD}\rangle)$ at general times}\label{app:notthesamelabel}
Using the results of \cite{Carmi:2017jqz} we can demonstrate that the mutual complexity of the time evolved TFD state using the subregion-CA proposal is in general negative. As mentioned in the main text $\mC_A(\mL)$ and $\mC_A({\cal R})$ are invariant under time evolution and we therefore have
\begin{equation}\label{appFinqq}
\Delta \mC_{A}(t) \equiv \mathcal{C}_A(\mL) + \mathcal{C}_A (\mathcal{R}) -\mathcal{C}_A (\mL\cup \mathcal{R}) (t) =  \Delta \mC_{A}(t=0) - \delta \mC_A    \,,
\end{equation}
where $\Delta \mC_{A}(t=0)$ can be found in eq.~\eqref{exs222}
\begin{equation}\label{appFinqq2}
\Delta \mC_{A}(t=0) = -\frac{2S}{\pi^2} \ln \left(\frac{\ell_{ct}(d-1)}{L}\right)+\text{negative},
\end{equation}
and we have defined
\begin{equation}
\delta \mC_A = \mathcal{C}_A (\mL\cup \mathcal{R}) (t) - \mathcal{C}_A (\mL\cup \mathcal{R}) (0)\, .
\end{equation}
The most negative value obtained by $\delta \mC_A$ can be bounded using the results of \cite{Carmi:2017jqz} for the rate of change of the complexity of the TFD state. There the authors found that the rate of change of the complexity was vanishing for $t=t_{\mL}+t_{\mathcal{R}}<t_c$ where $t_c = 2(r^*_\infty-r^*(0))$ is the critical time  where the WDW patch leaves the past singularity,\footnote{For the definition of the critical time we have used the tortoise coordinate $r^*(r) = \int d r/f( r)$ as well as the blackening factor $f(r)= \frac{r^2}{L^2} +k -\frac{\omega^{d-2}}{r^{d-2}}$ and the mass parameter $\omega^{d-2} =r_{h}^{d-2} \( \frac{r^2_h}{L^2} +k  \)$, where $k=0,\pm 1$ correspond to the various possible  horizon geometries.} and after this time, the rate of change became negative for a brief amount of time and later on approached a positive constant proportional to the mass of the black hole. The explicit expression is give in eq.~(E.9) of  \cite{Carmi:2017jqz} and reads
\begin{equation}
\begin{split}
\hspace{-10pt}\frac{d C_{A}}{dt } \biggr|_{t>t_c} \hspace{-10pt}=\frac{\Omega_{k, d-1}(d-1) f\left(r_{m}\right)   }{16 \pi^2 G_{N}} \left[\frac{2 \omega^{d-2}}{ f\left(r_{m}\right)  }- r_{m}^{d-2} \left[\ln \frac{r_m^2}{L^2 |f(r_{m})|}-2 \ln \frac{(d-1) \ell_{\rm ct}}{L}\right]\right], \\
\end{split}
\end{equation}
where $r_m$ is the place where the null boundaries of the WDW patch meet behind the past horizon and is fixed according to the equation $ \frac{t-t_c}{2} + r^\ast(r_m) -r^\ast(0)=0$. This rate of change is negative for times $t \in (t_c, t_{c,2})$ corresponding to the region $r_m  \in (0, r_{c,2})$. Here, the second critical time, or the critical radius $r_{c,2}$, are found by solving the equation $\frac{d C_{\mA}}{dt} \big|_{t_{c,2}}=0$ and correspond to the time in which the rate of change in complexity becomes positive and the complexity starts increasing again. Of course, we have $r_{c,2} < r_{h}$.
In order to check that the time-evolved TFD state is  always superadditive, we need to consider the minimal value of the complexity for the TFD state which is decided by
\begin{equation}
\begin{split}\label{someappFeq}
\delta \mC_{A}^{\text{min}} &=  \int^{t_{c,2}}_{t_c}  \frac{d C_{\mA}}{dt } \, dt  = \frac{\Omega_{k, d-1}(d-1) }{8 \pi^2 G_{N}} \\
&\quad \times \int^{ r_{c,2}}_{0} \[ - \frac{2 \omega^{d-2}}{ f(r_m)}  +   r_m^{d-2} \( \ln \( \frac{r_m^2}{L^2|f(r_m)|} \)  - 2 \ln \( \frac{(d-1)\ell_{\rm ct}}{L} \)
	\) \] dr_m \\
& =  -\frac{\Omega_{k, d-1}  r_{c,2}^{d-1}}{4\pi^2 G_N} \ln \(\frac{(d-1)\ell_{\text{ct}}}{L} \)  +  \frac{\Omega_{k, d-1}(d-1) }{8 \pi^2 G_{N}} \times \text{positive}
\\
& >  -\frac{S}{\pi^2} \ln \(\frac{(d-1)\ell_{\text{ct}}}{L} \),
\end{split}
\end{equation}
where in the first equality we have used the relation $dt = -2 \frac{dr_m}{f(r_m)}$ to change the variable of integration to $r_m$ and where the last inequality follows from $ r_{c,2}< r_h$.
The extra piece in the third line of eq.~\eqref{someappFeq} is always positive. This can be demonstrated by using the explicit form of the blackening factor and the mass parameter as well as the relation $r_m \le r_{c,2}< r_h$.
Combining eqs.~\eqref{appFinqq}, \eqref{appFinqq2} and \eqref{someappFeq}, we arrive at the conclusion that the  mutual complexity of the time-evolved TFD state is negative as advertised, \ie
\begin{equation}
\Delta \mC_{A}(t) = \Delta \mC_{A}(t=0) - \delta \mC_A <  \Delta \mC_{A}(t=0) - \delta \mC_A^{\text{min}} <  0  \,.
\end{equation}

\bibliography{biographyMixed}
\bibliographystyle{utphys}

\end{document}